\documentclass[12pt]{report}
\usepackage[utf8]{inputenc}
\usepackage{hyperref}
\usepackage{graphicx}
\graphicspath{ {images/} }
\usepackage{caption}
\usepackage{subcaption}
\usepackage{multirow}
\usepackage{longtable}
\usepackage{float}
\usepackage[left=1in, right=1in, top=1in, bottom=1in]{geometry}
\usepackage{fancyhdr}
\usepackage{setspace}
\usepackage{amsmath}
\usepackage{amssymb}
\usepackage{amsfonts}
\usepackage{bm}
\usepackage{cite}

\usepackage{tikz}
\usetikzlibrary{positioning, fit, arrows, arrows.meta, shapes.geometric, matrix}
\tikzstyle{cnn} = [rectangle, minimum width=4.3cm, minimum height=0.55cm, text centered, draw=black]
\tikzstyle{cnnr} = [rectangle, minimum width=1.9cm, minimum height=0.5cm, text centered, draw=black, rotate=90]
\tikzstyle{arrow} = [thick, -{Stealth[scale=.75]}] 
\tikzstyle{revarrow} = [thick, {Stealth[scale=.75]}-]
\tikzstyle{circ} = [circle, minimum width=.0em, minimum height=0.55cm, text centered, draw=black]
\definecolor{azul}{RGB}{187, 233, 255}

\title{Cosmology at the Field Level with Probabilistic Machine Learning}
\author{Rouhiainen}
\date{Day December 2023}

\begin{document}
\pagenumbering{roman} 

\begin{titlepage}
    \begin{center}
        \vspace*{1cm}
        
        Cosmology at the Field Level with Probabilistic Machine Learning
        
        \ 

        \ 

        \ 
        
        By\\

        \ 
        
        Adam Rouhiainen

        \ 

        \ 

        \ 
        
        A dissertation submitted in partial fulfillment of\\

        \ 
        
        the requirements for the degree of\\

        \ 

        \ 

        \ 
        
        Doctor of Philosophy\\

        \ 
        
        (Physics)\\
    
        \ 

        \ 

        \ 
        
        at the\\

        \ 
        
        UNIVERSITY OF WISCONSIN-MADISON\\

        \ 
        
        2023\\

        \ 

        \ 

        \ 

        \ 

        \ 

        \ 

        \begin{flushleft}
            Date of final oral examination: 12/19/2023\\
            
            \ 
            
            The dissertation is approved by the following members of the Final Oral Committee: \\
            \setlength{\parindent}{10ex}
            Moritz M{\"u}nchmeyer, Assistant Professor, Physics\\
            Gary Shiu, Professor, Physics\\
            Keith Bechtol, Associate Professor, Physics\\
            Kangwook Lee, Assistant Professor, Electrical and Computer Engineering
        \end{flushleft}
        
    \end{center}
    
\end{titlepage}



\chapter*{Statement of Contribution}

Chapter~\ref{ch:nf_random}, ``Normalizing Flows for Random Fields in Cosmology,"~\cite{rouhiainen2021} is a conference proceeding at the Machine Learning and the Physical Sciences Workshop at NeurIPS 2021, in collaboration with Utkarsh Giri and Moritz M{\"u}nchmeyer.

Chapter~\ref{ch:denoising}, ``Denoising Non-Gaussian Fields with Normalizing Flows,"~\cite{rouhiainen2022} is a conference proceeding at the Machine Learning and the Physical Sciences Workshop at NeurIPS 2022, in collaboration with Moritz M{\"u}nchmeyer.

Chapter~\ref{ch:sr_diffusion}, ``Super-resolution of Large Cosmological Fields with a 3D Conditional Diffusion Model,"~\cite{Rouhiainen:2023ewv} is a conference proceeding at the Machine Learning and the Physical Sciences Workshop at NeurIPS 2023, in collaboration with Michael Gira, Moritz M{\"u}nchmeyer, Kangwook Lee, and Gary Shiu, with code written in collaboration with Michael Gira.

Chapter~\ref{ch:introduction}, Sec.~\ref{sec:persistence_images} is related to the work ``Learning from Topology: Cosmological Parameter Estimation from the Large-scale Structure,"~\cite{yip2023learning} a conference proceeding at the Synergy of Scientific and Machine Learning Modeling Workshop at ICML 2023, in collaboration with Jacky Yip and Gary Shiu.

\chapter*{Acknowledgements}

I thank my advisor, Moritz M{\"u}nchmeyer, for giving me the opportunity to study cosmology with him for the past three years. I am happy to be your first student, and I hope there are many more to come.

I thank Yurii Kvasiuk for being a valuable member of our research group with Moritz, for our friendly conversations, and for being a pleasure to share an office with. I thank Utkarsh Gira for taking the time to answer all of my various questions over the past few years, and for his patience while helping me install decades old software.

I thank Jacky Yip for engaging in detailed discussions on machine learning and topology, and for being the friendliest face in physics department. I thank Gary Shiu for his encouraging support, for introducing me to Moritz, and for being on the committee.

I thank Michael Gira for the tedious work he put into our super-resolution project, and Kangwook Lee for sharing his extensive knowledge of diffusion models and for being on the committee.

I thank Keith Bechtol for leading many interesting discussions in the cosmology journal club, and for being on the committee.

I thank Jeff Schmidt and Lisa Everett for teaching me more physics than I could ever hope to retain, and for genuinely caring about their students' success.

Finally, I thank Aryel for all her support and encouragement.







\chapter*{Abstract}

The large-scale structure in cosmology is highly non-Gaussian at late times and small length scales, making it difficult to describe analytically. Parameter inference, data reconstruction, and data generation tasks in cosmology are greatly aided by various machine learning models. In order to retain as much information as possible while solving these problems, this work operates at the field level, rather than at the level of summary statistics.

The probability density function (PDF) of the large-scale structure is learned with normalizing flows, a class of probabilistic generative models, in Chapters~\ref{ch:nf_random},~\ref{ch:nf_parameters}, and~\ref{ch:denoising}. Normalizing flows learn the transformation from a simple base distribution to a more complicated distribution, much like the matter content evolved to its present day complexities from a Gaussian field at early times. Considering two normalizing flow architectures, real NVP and Glow, we are able to model the late time matter content projected onto 2 dimensions, and we measure its accuracy at the level of the power spectrum and bispectrum. The PDF of matter fields conditional on parameters $\Omega_\text{M}$ and $\sigma_8$ is learned by modifying the Glow architecture to receive a parameter embedding in each of the flow transformation steps. With the conditional PDF, field level parameter inference of cosmological parameters is highly accurate, outperforming inference with the power spectrum alone. Considering the data reconstruction problem of removing uncorrelated noise from our fields, we take the flow's PDF of the matter content as a non-Gaussian prior distribution; we are thus able to denoise fields with greater accuracy than Wiener filtering.

While the normalizing flows have accurately modelled 2-dimensional projections of the matter content, we find that denoising diffusion models are well-suited for volumetric data in Chapter~\ref{ch:sr_diffusion}. A super-resolution emulator is developed for cosmological simulation volumes, generating high-resolution baryonic simulation volumes conditional on low-resolution dark matter simulations. The super-resolution emulator is trained to perform outpainting, and can thus upgrade very large cosmological volumes from low-resolution to high-resolution using an iterative outpainting procedure. As an application of our super-resolution emulator, we train the model on the Illustris TNG300 simulation having side length $205\ \text{Mpc}/h$. Then, we generate a super-resolution $410\ \text{Mpc}/h$ length box, with 8 times the volume of the entire training data volume. Our super-resolution emulator is accurate as measured by summary statistics including the power spectrum, bispectrum, and void size function.


\singlespacing
\tableofcontents
\listoffigures

\chapter*{Notation}
\begin{longtable}{cl}
    $\bm{r}$                  & Spacial coordinates\\
    $\bm{k}$                  & Fourier space coordinates\\
    $k$                       & Magnitude of $\bm{k}$\\
    $\bm{x}$                  & A vector of data\\
    $\bm{y}$                  & A vector of data conditional on other data $\bm{x}$\\
    $\bm{x}_{<i}$             & Elements of $\bm{x}$ with index less than $i$\\
    $\bm{u}$                  & Random Gaussian field\\
    $\bm{\theta}$             & Neural network model parameters\\
    $\mathbb{I}_{d\times d}$  & $d\times d$ identity matrix\\
    $\bm{0}_{d\times d}$      & ${d\times d}$ zero matrix\\
    $\mathcal{F}[\bm{x}]$     & Fourier transform\\
    $T(\bm{x})$               & Coordinate transformation\\
    $J$                       & Jacobian matrix\\
    $\mathbb{E}$              & Expectation value\\
    $\mathcal{N}(\bm{\mu},\bm{\sigma}^2)$ & Multivariate Gaussian distribution\\
    $p(...)$                  & Probability density function\\
    $\delta_\text{D}(\bm{k})$ & Dirac delta function\\
    $H(t)$                    & Hubble parameter\\
    $H_0$                     & Hubble constant\\ 
    $h$ & Reduced Hubble constant, $H_0/(100\ \text{km}\ \text{s}^{-1}\text{Mpc}^{-1})$\\
    $a(t)$                    & Cosmological scale factor\\
    $G$                       & Gravitational constant\\
    $c$                       & Speed of light\\
    $\Omega_\text{M}$         & Critical matter density\\
    $\Omega_\Lambda$          & Critical vacuum density\\
    $\Omega_\text{R}$         & Critical radiation density\\
    $\sigma_8$ & Root mean square matter fluctuations in an $8\ \text{Mpc}/h$ radius\\
    $z$                       & Redshift\\
    $\text{Mpc}$              & Megaparsec, approximately $3.1\times10^{10}\ \text{m}$\\
\end{longtable}

\doublespacing

\chapter{Introduction}
\label{ch:introduction}
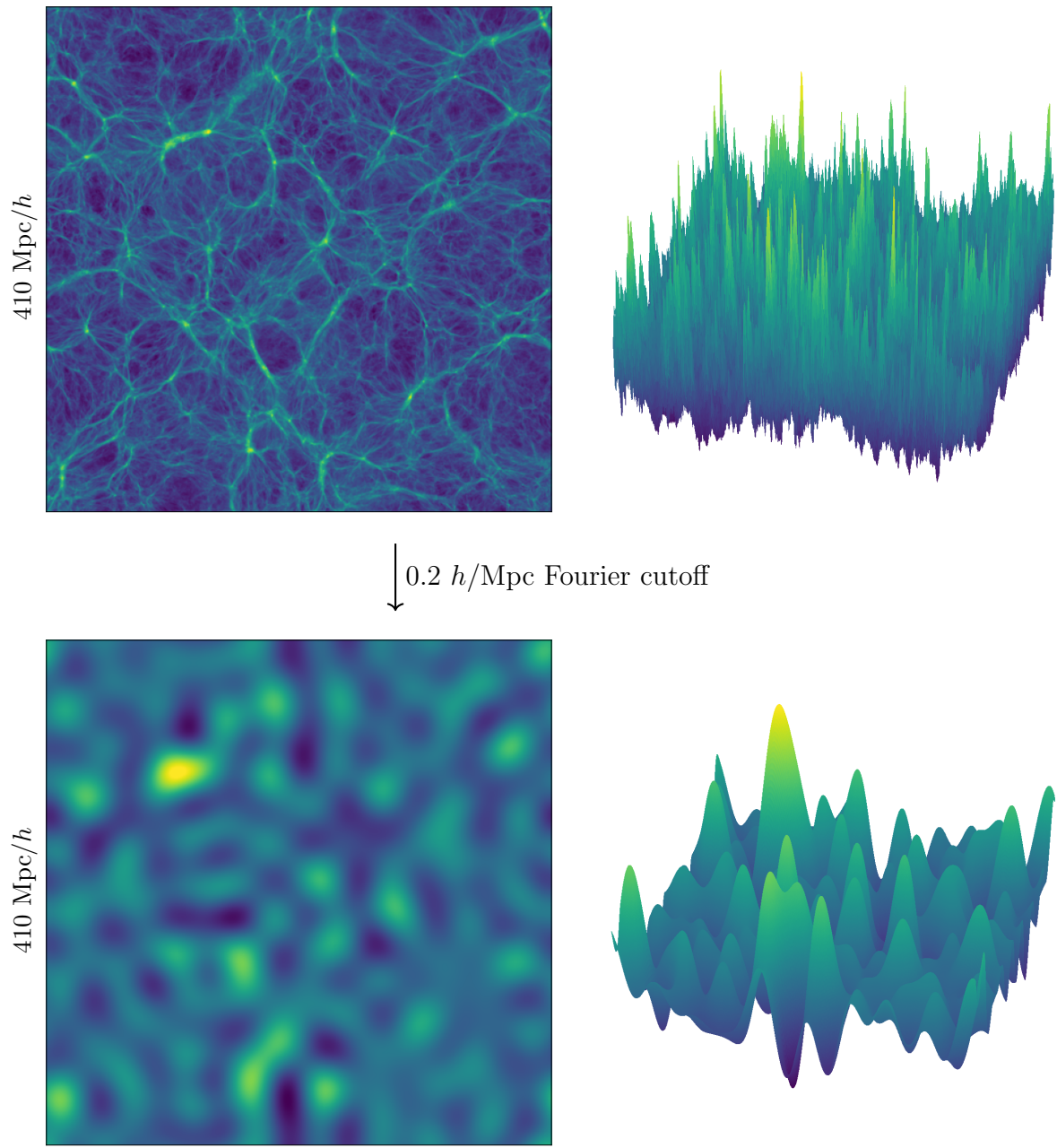
\begin{figure}
    \centering
    {\Large\hspace{0.6cm}Dark matter density, present day}

    \ 
    
    \rotatebox{90}{\small\hspace{3cm}$410\ \text{Mpc}/h$}\includegraphics[width=0.47\textwidth]{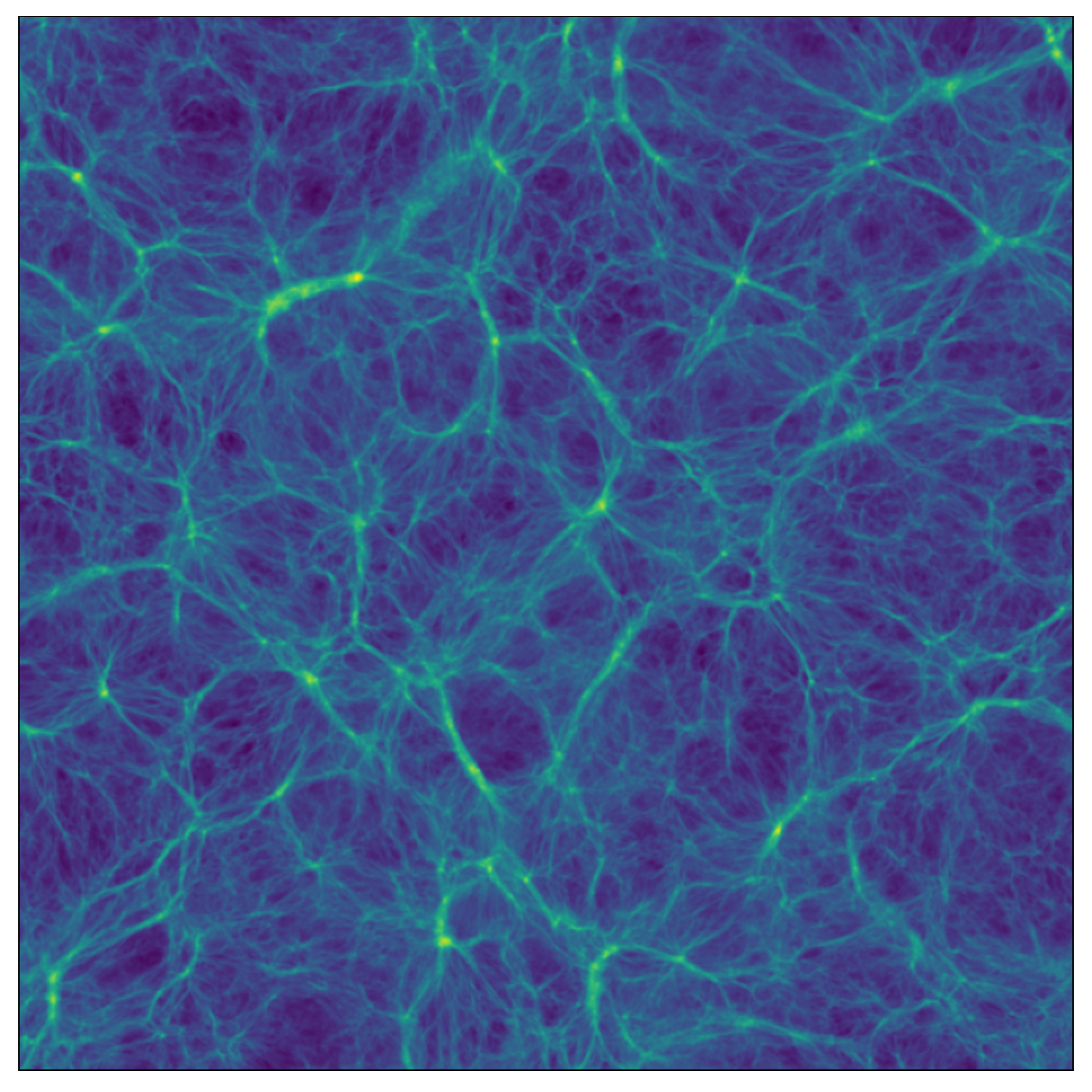}
    \includegraphics[width=0.47\textwidth, trim={5cm 5cm 5cm 5cm}, clip]{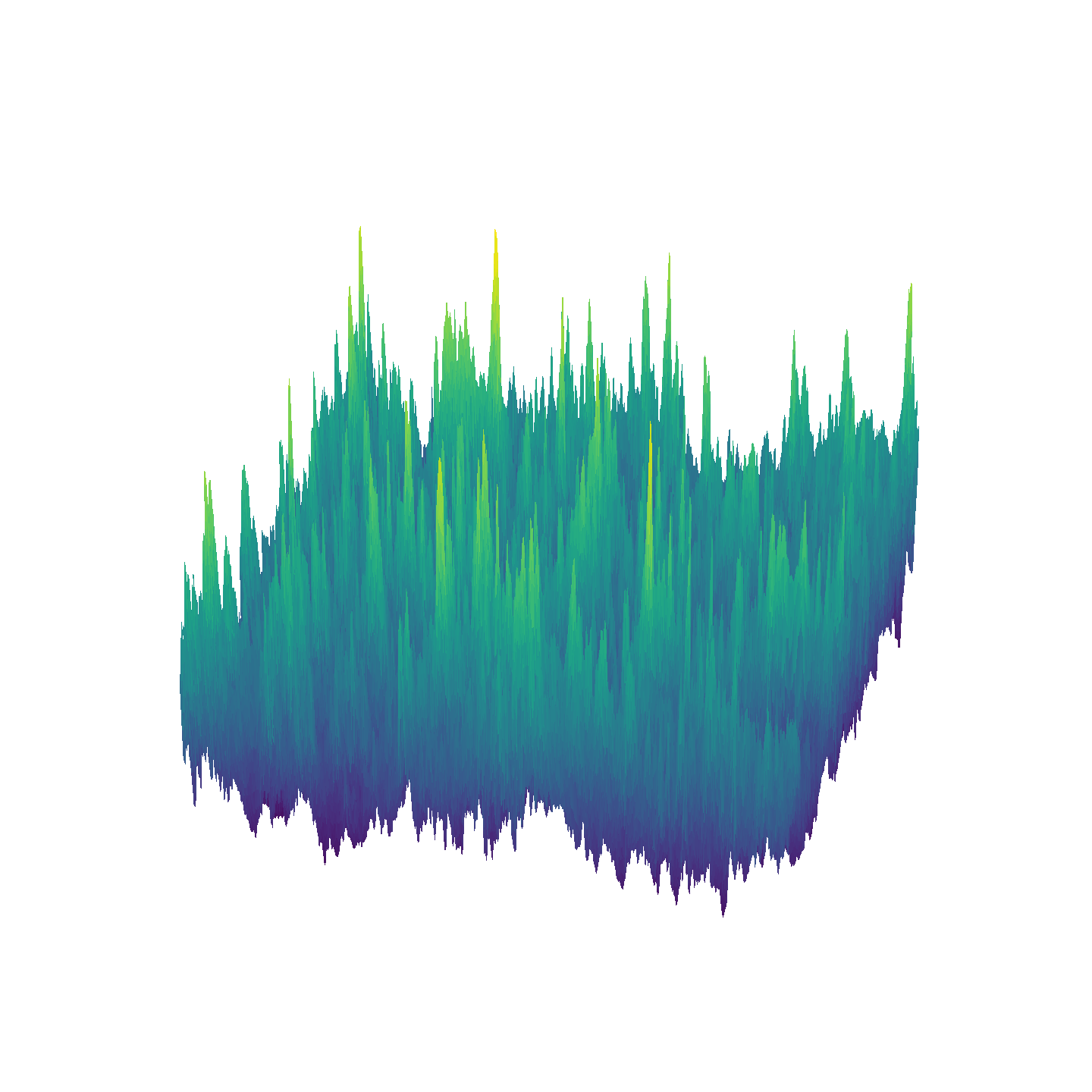}

    \begin{tikzpicture}
        \node  (A) {};
        \node [below=of A] (B) {};
        
        \draw [->, thick] (A) -- (B) node[midway, right] {$0.2\ h/\text{Mpc}$ Fourier cutoff};
    \end{tikzpicture}

    \rotatebox{90}{\small\hspace{3cm}$410\ \text{Mpc}/h$}\includegraphics[width=0.47\textwidth]{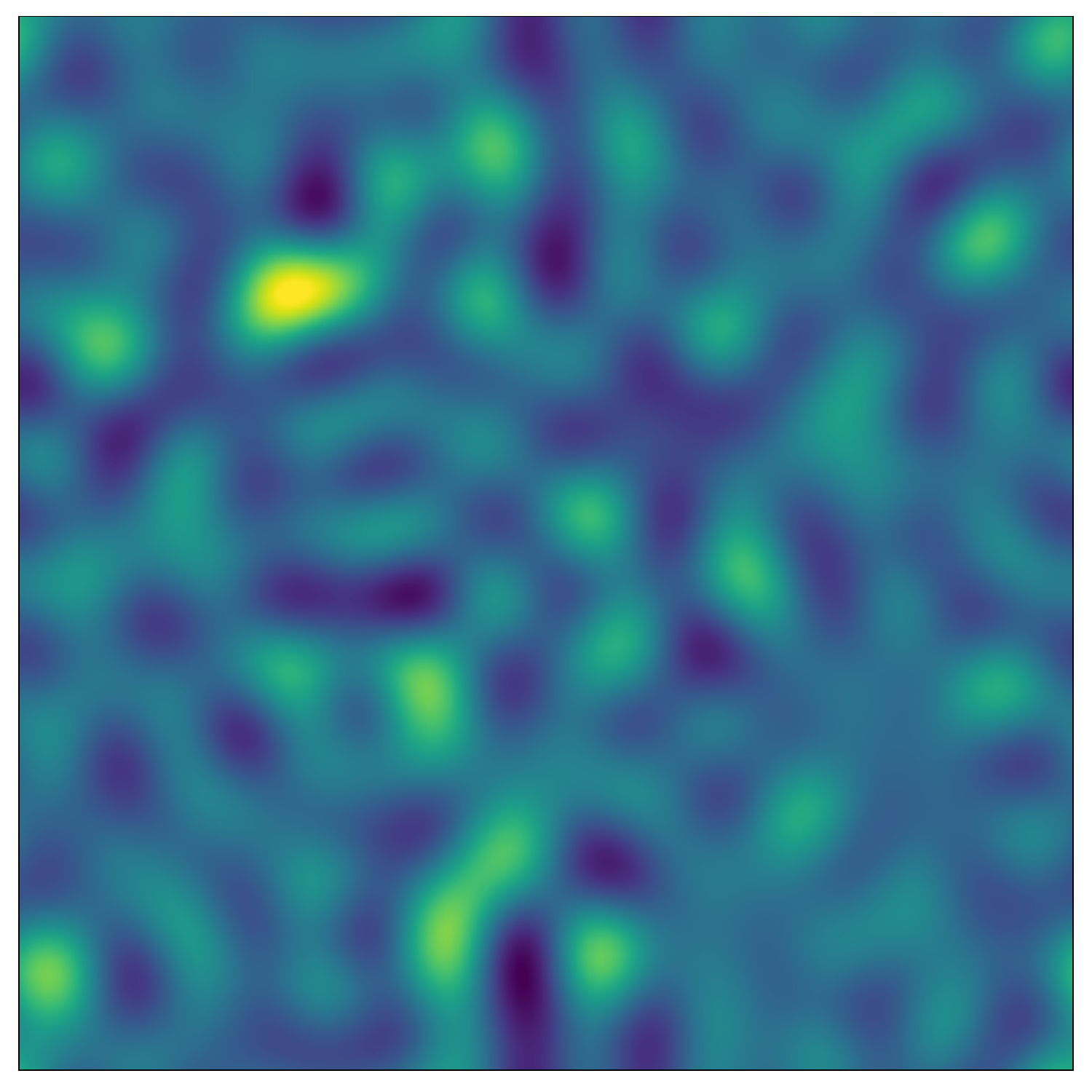}
    \includegraphics[width=0.47\textwidth, trim={5cm 5cm 5cm 5cm}, clip]{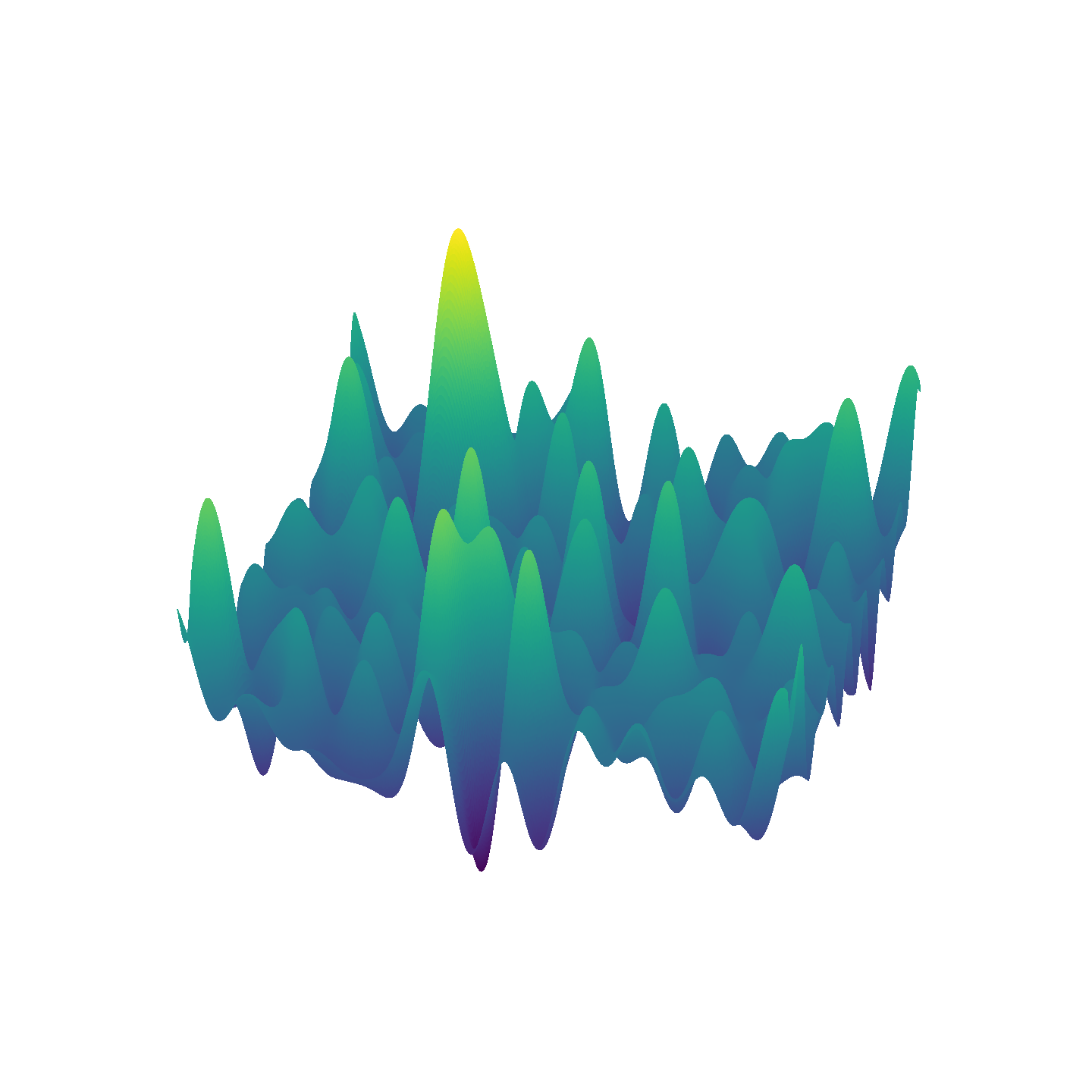}
    
    \caption[The large-scale structure as simulated with TNG300-Dark]{The large-scale structure near the present day. The TNG300-Dark simulation, run in a box of length $205\ \text{Mpc}/h\approx300\ \text{Mpc}$, is shown here at redshift $z=0.01$. (Left) The particles are visualized on a mesh with the cloud-in-cell mass assignment scheme, projected into 2 dimensions with $8.5\ \text{Mpc}/h$ depth. (Right) An isometric plot of the matter density fluctuations. (Bottom) A Fourier filter with cutoff $0.2\ h/\text{Mpc}$ (corresponding to a length of $31\ \text{Mpc}/h$) applied to the dark matter field reveals the Gaussian nature of the larger length scales.}
    \label{fig:lss_TNG300}
\end{figure}

\pagenumbering{arabic}
Cosmology is the study of all that is, or ever was, or ever will be. This area of physics considers length scales approaching the size of the universe down to interactions of fundamental particles, and time scales from the inflationary epoch to the dark energy dominated present and future. This work is concerned with the large-scale structure at late times. Evolving from a Gaussian field after recombination, at late times and small length scales, we find highly non-Gaussian structure, with matter forming into a complex web of halos, filaments, and voids, as shown in Fig.~\ref{fig:lss_TNG300}. Of great importance to us will be the distinction between the linear regime, where matter can be described analytically as a Gaussian field, and the nonlinear regime, where we resort to machine learning models to aid us in various data generation and inference tasks. 

Until recently, a field level analysis of cosmological data has been exceedingly difficult. A number of summary statistics have been standardized in cosmology to reduce the massive amount of data to computationally simpler forms. Cosmological parameter inference, for example, can then be done on the summary statistics. Only in the last several years have advances in machine learning led us to a much more information intensive approach.

The tools used in this work to understand the large-scale structure are probabilistic machine learning models, namely normalizing flows and denoising diffusion models. We will see that with normalizing flows, we can learn the analytically intractable probability density function (PDF) of the large-scale structure matter density, and may also make the PDF conditional on cosmological parameters. Moving towards a data generation task, the highly expressive denoising diffusion models are capable of stochastically generating large, high-resolution simulation volumes given a quickly simulated low-resolution volume.

This Chapter discusses the growth of the large-scale structure, from the linear regime in Section~\ref{sec:growth_of_structure} to nonlinear corrections in Section~\ref{sec:nonlinear_lss}. It is important to understand where the analytic theory breaks down to appreciate the need for simulation-based inference methods, which in turn will be discussed in Section~\ref{sec:simulation-based_inference}. After understanding the difficulties that come with likelihood-free inference with cosmological simulations, we begin to solve all of our problems with machine learning methods, discussed in Section~\ref{sec:deterministic_ml}. We make a distinction between ``deterministic`` and ``probabilistic`` machine learning models, the latter of which are broadly introduced in Section~\ref{sec:probabilistic_ml}, and then used throughout the following Chapters.

The remainder of this work will apply probabilistic machine learning models to three classes of problems in cosmology, data generation, parameter inference, and data reconstruction. In Chapter~\ref{ch:nf_random}, normalizing flows are used to learn the field-level PDF of large-scale structure data. In Chapter~\ref{ch:nf_parameters}, a normalizing flow is conditioned on cosmological parameters, from which a conditional PDF is used for parameter inference and conditional data generation. In Chapter~\ref{ch:denoising}, a learned prior distribution of non-Gaussian fields is used to reconstruct images of the large-scale structure. In Chapter~\ref{ch:sr_diffusion}, a denoising diffusion model is used to generate high-resolution cosmological fields conditional on low-resolution dark matter simulations. Many of the machine learning methods used throughout this work are described in Appendix~\ref{ch:ml}.

\section{Growth of structure}
\label{sec:growth_of_structure}

Much of this work is concerned with the large-scale structure at late times, where the matter density has become highly non-Gaussian on small length scales. This Section reviews some of the cosmological parameters that will be worked with in subsequent Chapters, and describes the growth of structure with linear theory. Nonlinear perturbation theory will be briefly discussed in Section~\ref{sec:nonlinear_lss}.

Our cosmology is a psuedo-Riemannian manifold~\cite{Bishop1968} with topology $\mathbb{R}\times\mathbb{R}^3$, homogeneous and isotropic at large length scales. From homogeneity and isotropy, and taking space to be flat, follows the Friedmann–Robertson–Walker metric~\cite{Robertson:1935zz,Robertson:1935jpx,Robertson:1936zz,Walker:1937}
\begin{equation}
\label{eq:metric}
    g_{\mu\nu}\text{d}x^\mu\text{d}x^\nu=-c^2\ \text{d}t^2+a^2(t)\ \text{d}\bm{r}^2.
\end{equation}
Gravitational dynamics is governed by the Einstein field equations~\cite{Einstein:1916vd},
\begin{equation}
    R_{\mu\nu}-\frac{1}{2}Rg_{\mu\nu}+\Lambda g_{\mu\nu}=\frac{8\pi G}{c^4}T_{\mu\nu}.
\end{equation}
The curvature of spacetime described by the Ricci tensor $R_{\mu\nu}$ and Ricci scalar $R$ is caused by dark matter, baryons, radiation, and neutrinos through the energy momentum tensor $T_{\mu\nu}$, and by dark energy through the cosmological constant $\Lambda$. Inputting the metric Eq.~\ref{eq:metric} into the field equations gives the Friedmann equations~\cite{Friedman:1922kd}. The $00$ index is the energy density component of the field equations, $T_{00}=\rho(t)$, which can be written as
\begin{equation}
    H(t)\equiv\frac{\text{d}a/\text{d}t}{a(t)}=H_0\sqrt{\frac{8\pi G\rho(t)}{3}+\frac{\Lambda c^2}{3}}.
\end{equation}
The trace of the field equations gives
\begin{equation}
    \frac{\text{d}H}{\text{d}t}+H^2(t)=\frac{\text{d}^2a/\text{d}t^2}{a(t)}=-\frac{4\pi G}{3}\left(\rho(t)+\frac{3p(t)}{c^2}\right)+\frac{\Lambda c^2}{3}.
\end{equation}
The density
\begin{equation}
    \rho(t)=\frac{\rho_{\text{M}}(0)}{a^3(t)}+\frac{\rho_{\text{R}}(0)}{a^4(t)}
\end{equation}
is made up of dark matter and baryonic matter density in $\rho_\text{M}(t)$, as well as radiation and neutrinos in $\rho_\text{R}(t)$. We only consider ``cold" dark matter, being collisionless particles that interact with matter only through gravitation~\cite{Peebles:1982ff}. Owing to the extra factor of $1/a$ in the radiation term due to the Doppler effect with the expansion of space, radiation is highly subdominant to matter at later times. So to does the matter density begin to lose dominance at late times, leading us to dark energy domination around the present day.

The densities $\rho$ and $H$ can be combined into a dimensional parameter, the critical matter density $\Omega_\text{M}$,
\begin{equation}
    \Omega_\text{M}\equiv\frac{8\pi G\rho_\text{M}}{3H^2}.
\end{equation}
Cosmic microwave background (CMB) measurements currently give $\Omega_\text{M}=0.315\pm0.007$~\cite{Planck:2018parameters}, in agreement with late time measurements from Type Ia supernova of $\Omega_\text{M}=0.334\pm0.018$~\cite{Brout:2022vxf}, and galaxy clustering and lensing giving $\Omega_\text{M}=0.339^{+0.032}_{-0.031}$~\cite{DES:2021wwk}.

The critical dark energy parameter $\Omega_\Lambda$ is
\begin{equation}
    \Omega_\Lambda\equiv\frac{\Lambda}{3c^2}.
\end{equation}
This work considers only a flat cosmology, and therefore
\begin{equation}
    \Omega_\text{R}+\Omega_\text{M}+\Omega_\Lambda=1.
\end{equation}
This assumption is highly accurate, as early time measurements give a curvature of $\Omega_K=0.001\pm0.002$~\cite{Planck:2018parameters}.

With the introduction of $\Omega_\text{R}$, $\Omega_\text{M}$, and $\Omega_\Lambda$, the energy Friedmann equation can be written as
\begin{equation}
\label{eq:Friedmann_a}
    \frac{\text{d}a/\text{d}t}{a(t)}=H_0\sqrt{\frac{\Omega_\text{R}}{a^4(t)}+\frac{\Omega_\text{M}}{a^3(t)}+\Omega_\Lambda},
\end{equation}
an equation which we will reference when discussing N-body simulations. We have taken a dark energy equation of state $w=-1$, an assumption made throughout this work. For the times where we can make the approximation of neglecting $\Omega_\text{R}/a^4$, this can be solved analytically for $a(t)$ for a few special cases. A matter dominated cosmology with $\Omega_\text{M}/a^3\gg\Omega_\Lambda$ gives a scale factor of $a(t)=\left(1+\frac{3}{2}H_0(t-t_\text{p})\right)^{2/3}$, with $t_\text{p}$ being the present time. For a dark energy dominated cosmology, $\Omega_\text{M}/a^3\ll\Omega_\Lambda$ gives $a(t)=e^{H_0(t-t_\text{p})}$. These two solutions hint at what we expect the dynamics of expansion to look like. At the time of matter dominance, well after recombination and before dark energy becomes significant, the scale factor increases as $a(t)\propto t^{2/3}$. At later times, as dark energy becomes significant relative to the matter density, the expansion accelerates, approaching $a(t)\propto e^{H_0t}$. This acceleration affects the growth of matter, as will be described by the growth factor, has been directly observed~\cite{SupernovaSearchTeam:1998fmf,SupernovaCosmologyProject:1998vns}.

Much of this work will be operating with matter overdensities
\begin{equation}
    \delta(\bm{r},t)\equiv\frac{\rho(\bm{r},t)-\langle\rho(\bm{r},t)\rangle}{\langle\rho(\bm{r},t)\rangle}.
\end{equation}
Here $\langle...\rangle$ is an ensemble average, and so $\langle\rho(\bm{r},t)\rangle$ is the mean value of $\rho(\bm{r},t)$.
To describe the matter content at times well after recombination, the Poisson equation of the gravitational potential can be written in two ways. In Fourier space, there is the familiar
\begin{align}
    k^2\Phi&=4\pi G a^2\rho\delta(\bm{k},a)\\
    &=\frac{3}{2}\Omega_\text{M}H_0^2\delta(\bm{k},a).
\end{align}
The second version of the Poisson equation comes from the continuity and Euler equations for $\delta$, which may be combined as~\cite{Weinberg2008}
\begin{equation}
    k^2\Phi=\frac{\text{d}}{\text{d}t}\left(a(t)\frac{\text{d}}{\text{d}t}\delta(t)\right).
\end{equation}
The previous two equations give a differential equation for $\delta$,
\begin{equation}
\label{eq:delta_t_difeq}
    \frac{\text{d}}{\text{d}t}\left(a^2(t)\frac{\text{d}}{\text{d}t}\delta(t)\right)=4\pi Ga^2(t)\rho\delta(t).
\end{equation}
It is standard to factorize the time dependence (located in $a$) from the $k$ dependence, introducing the growth factor $D_+(t)$ and transfer function $T(k)$ as
\begin{equation}
\label{eq:delta_factorized}
    \delta(\bm{k},a)=\frac{2k^2}{5\Omega_\text{M}H_0^2}\mathcal{R}(\bm{k})T(k)D(a).
\end{equation}
Primordial fluctuations are located in $\mathcal{R}$, with a primordial power spectrum that is most simply parameterized as
\begin{equation}
    P_\mathcal{R}(k)=\frac{2\pi^2\mathcal{A}_s}{k^3}\left(\frac{k}{k_p}\right)^{n_s-1}.
\end{equation}
The tilt of the primordial power spectrum is given by the scalar spectral index $n_s$, defined by
\begin{equation}
\label{eq:n_s_def}
    n_s\equiv\frac{\text{d}\ln{\left(P_\text{Lin}(k)\right)}}{\text{d}\ln{(k)}}.
\end{equation}
CMB measurements currently give $n_s=0.965\pm0.004$~\cite{Planck:2018parameters}.

\begin{figure}
    \centering
    \hspace{0.65cm}Growth factor for various cosmologies
    
    \includegraphics[width=0.495\textwidth]{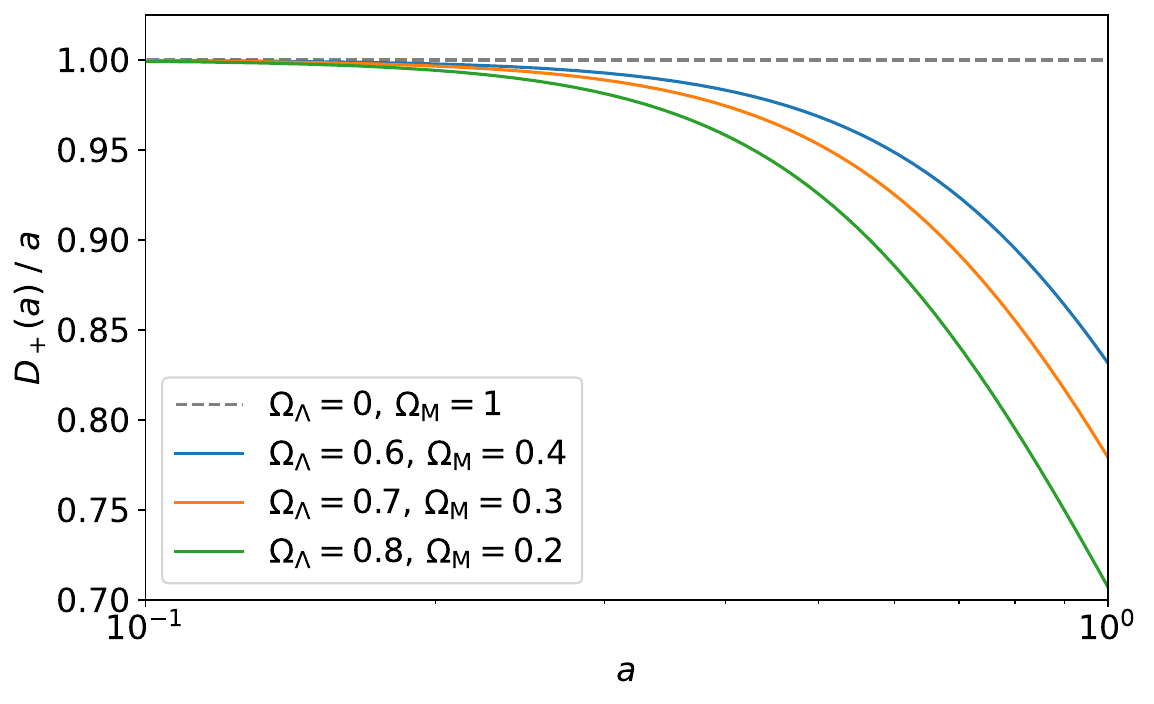}
    \caption[Growth factor]{The growth factor for several values of $\Omega_\Lambda$ and $\Omega_\text{M}$. Dark energy suppresses growth, with increased suppression at later times.}
    \label{fig:growth_function}
\end{figure}

The growth factor $D(a)$ is the solution to \ref{eq:delta_t_difeq} after dropping the $\bm{k}$ dependent factors, giving
\begin{equation}
    \frac{\text{d}}{\text{d}t}\left(a^2(t)\frac{\text{d}}{\text{d}t}D(t)\right)=4\pi Ga^2(t)\rho D(t)
\end{equation}
The solution needs to satisfy $D(t)\propto t^{2/3}$ at the time of matter dominance. In terms of $a(t)$, the solution can be written as~\cite{Weinberg2008}
\begin{equation}
    D_+(a)=\frac{5}{6}\left(\frac{\Omega_\Lambda}{\Omega_\text{M}}\right)^{-5/6}\sqrt{\frac{1}{a^3}+\frac{\Omega_\Lambda}{\Omega_\text{M}}}\int_0^{\frac{\Omega_\Lambda}{\Omega_\text{M}}a^3}\frac{\text{d}u}{u^{1/6}(1+u)^{3/2}}
\end{equation}
This integral is an incomplete beta function; writing in terms of the $\, _2F_1$ hypergeometric function, the growth factor evaluates to
\begin{equation}
    D_+(a)=a\, _2F_1\left(\frac{1}{3},1;\frac{11}{6};-\frac{\Omega_\Lambda}{\Omega_\text{M}}a^3\right).
\end{equation}
This result goes to $D_+(a)\rightarrow a\propto t^{2/3}$ in the small $\frac{\Omega_\Lambda}{\Omega_\text{M}}a$ limit, and decays to zero for large $\frac{\Omega_\Lambda}{\Omega_\text{M}}a$. A plot of $D_+(a)/a$ for several values of $\Omega_\Lambda$ and $\Omega_\text{M}$ is shown in Fig.~\ref{fig:growth_function}. A cosmological constant suppresses the growth of structure at all times, with increasing suppression at later times.

At the time of radiation-matter equality when $a_\text{eq}=\frac{\Omega_\text{R}}{\Omega_\text{M}}$, the mode $k_\text{eq}$ entered the horizon at
\begin{align}
\label{eq:k_eq}
    k_\text{eq}&=a_\text{eq}H(a_\text{eq})\\
    &=a_\text{eq}H_0\sqrt{\frac{\Omega_\text{R}}{a_\text{eq}^4}+\frac{\Omega_\text{M}}{a_\text{eq}^3}}\\
    &=\sqrt{2}\frac{\Omega_\text{M}}{\sqrt{\Omega_\text{R}}}H_0\\
    &\approx0.073\Omega_\text{M}h^2\ \text{Mpc}^{-1}
\end{align}
where $\Omega_\text{R}h^2=4.15\times10^{-5}$ and $H_0=\frac{1}{2998}\ h/\text{Mpc}$ are used for the final line. Using $\Omega_\text{M}=0.315$ and the late time measurements of $h=0.73$ gives
\begin{equation}
k_\text{eq}\approx0.017\ h/\text{Mpc},
\end{equation}
corresponding to a length of $373\ \text{Mpc}/h$. It will be helpful to keep this length scale in mind while considering the various simulation volumes used in subsequent Chapters.

\begin{figure}
    \centering
    \hspace{0.65cm}Transfer function for various cosmologies
    
    \includegraphics[width=0.495\textwidth]{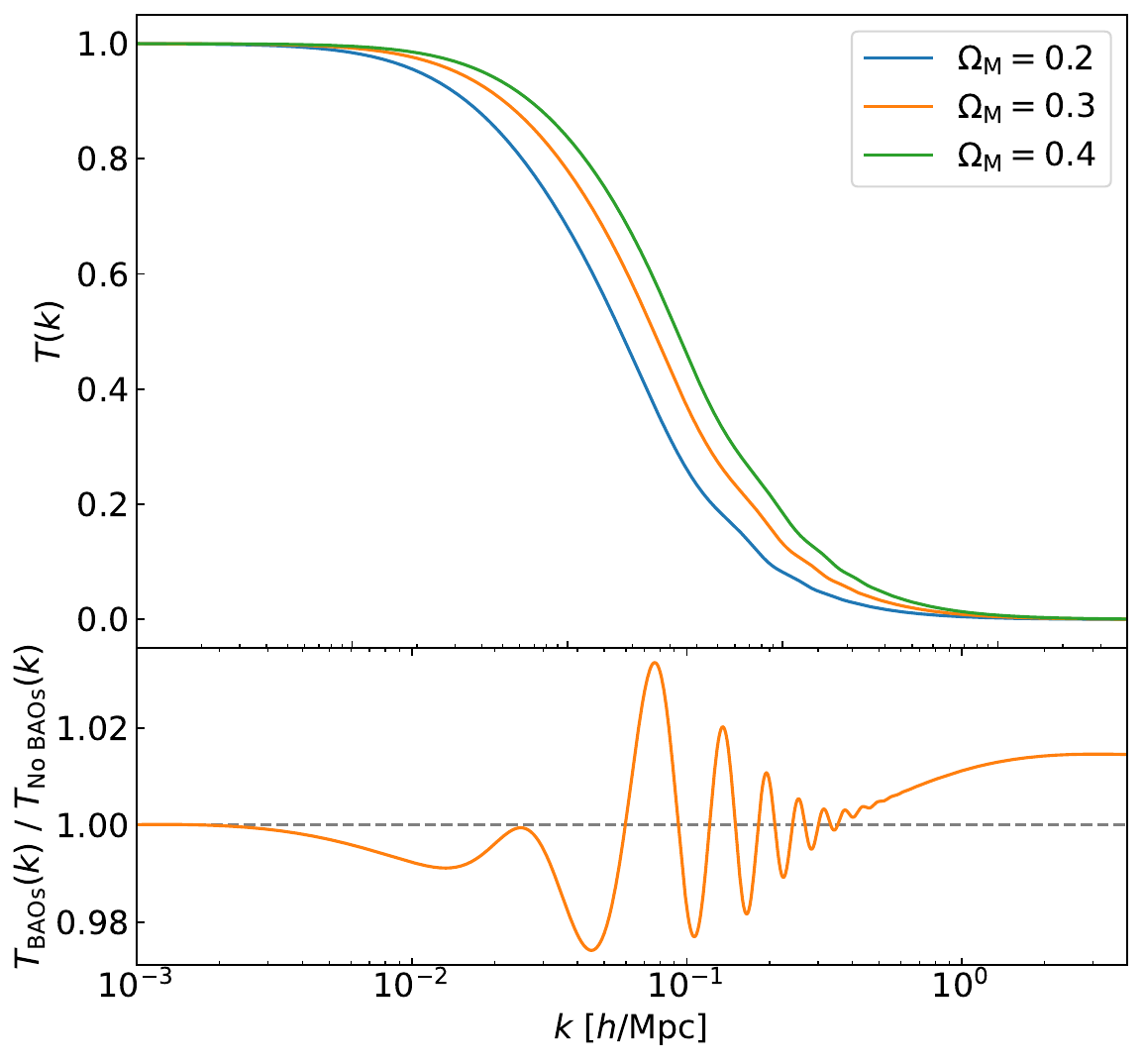}
    \caption[Transfer function]{(Top) The Eisenstein \& Hu transfer function~\cite{Eisenstein_1998} for various values of the matter density, computed with \texttt{nbodykit}\cite{Hand:2017pqn}. (Bottom) The $\Omega_\text{M}=0.3$ transfer function compared to a version of the Eisenstein \& Hu transfer function without BAOs.}
    \label{fig:transfer_function}
\end{figure}

Now looking at the $k$ factorization in Eq.~\ref{eq:delta_factorized}, the transfer function $T(k)$ is more difficult to derive. In the small $k$ limit, the transfer function approaches unity by construction. In the presence of baryons, fits to the transfer function are often written as a function of $q\propto k/\left(\Omega_\text{M}h^2-\Omega_\text{B}h^2\right)$ for a small baryons density ($\Omega_\text{B}\ll\Omega_\text{M}$). One such simple form in the presence of baryons, accurate at the $10\%$ level, is the BBKS transfer function~\cite{BBKS_1983,BBKS_1984},
\begin{equation}
    T_\text{BBKS}(k)=\frac{\ln{(1+2.34q)}}{2.34q}\left(1+3.89q+(16.1q)^2+(5.46q)^3+(6.71q)^4\right)^{-1/4},
\end{equation}
A transfer function derived by Eisenstein \& Hu~\cite{Eisenstein_1998} containing several dozen terms is accurate to percent level, considering Baryon acoustic oscillations (BAOs)~\cite{Eisenstein_2005,Cole_2005}, diffusion damping~\cite{Silk_1968}, and Compton drag. The Eisenstein \& Hu transfer function is plotted in Fig.~\ref{fig:transfer_function} for several values of $\Omega_\text{M}$, along with a comparison between BAOs and a version of the Eisenstein \& Hu transfer function without BAOs. Another fit to the transfer function at percent level accuracy was found with a genetic algorithm taking the simple form~\cite{Orjuela-Quintana:2022nnq}
\begin{equation}
    T_\text{GA}(k)=\left(1+\sum_{i=1}^4c_iq^{p_i}\right)^{-1/4},
\end{equation}
for constants $c_i$ and $p_i$. Considering the small $k$ and large $k$ limits, any of the aforementioned transfer functions have the form~\cite{Dodelson2003}
\begin{equation}
\label{eq:transfer_analytic}
    T(k)=
    \begin{cases}
    1&\text{for}\ k\ll k_\text{eq}\\
    12.\frac{k^2_\text{eq}}{k^2}\ln{\left(0.12\frac{k}{k_\text{eq}}\right)}&\text{for}\ k\gtrsim1\ h/\text{Mpc}.
    \end{cases}
\end{equation}

Perhaps the most studied summary statistic in cosmology is the power spectrum. The power spectrum $P_\text{Lin}(k)$ is a Fourier space two-point correlation function defined by
\begin{equation}
    (2\pi)^3P(k)\delta_\text{D}^3(\bm{k}+\bm{k}')=\langle\delta(\bm{k})\delta(\bm{k}')\rangle,
\end{equation}
where $\delta_\text{D}$ is the Dirac delta function. From homogeneity and isotropy, the power spectrum is a function of only the magnitude $k$; this can be demonstrated with
\begin{align}
    \langle\delta(\bm{k}_1)\delta(\bm{k}_2)\rangle&=\int\text{d}^3r_1\int\text{d}^3r_2\langle\delta(\bm{r}_1)\delta(\bm{r}_2)\rangle e^{-i\bm{k}_1\cdot\bm{r}_1}e^{-i\bm{k}_2\cdot\bm{r}_2}\\
    &=\int\text{d}^3r_1e^{-i(\bm{k}_1-\bm{k}_2)\cdot\bm{r}_1}\int\text{d}^3(r_2-r_1)\langle\delta(\bm{r}_1)\delta(\bm{r}_2)\rangle e^{-i\bm{k}_2\cdot(\bm{r_2}-\bm{r}_1)}\\
    &=(2\pi)^3\delta_\text{D}^3(\bm{k}_1+\bm{k}_2)\int\text{d}^3(r_2-r_1)\xi(|\bm{r}_2-\bm{r}_1|)e^{-i\bm{k}_2\cdot(\bm{r_2}-\bm{r}_1)}.
\end{align}
In the final line, the position space correlation function $\langle\delta(\bm{r}_1)\delta(\bm{r}_2)\rangle=\xi(|\bm{r}_2-\bm{r}_1|)$ is a function of the magnitude $|\bm{r}_2-\bm{r}_1|$, as required by homogeneity and isotropy. The matter power spectrum in the linear regime with the $a$ and $k$ factorization is
\begin{equation}
    P_\text{Lin}(k,a)=\frac{8\pi^2\mathcal{A}_sk^{n_s}}{25\Omega_\text{M}^2H_0^4k_p^{n_s-1}}T^2(k)D_+^2(a)
\end{equation}
At small $k$, as the transfer function approaches $1$, the power spectrum is proportional to $k^{n_s}$. At large $k$, using the transfer function in Eq.~\ref{eq:transfer_analytic}, the power spectrum falls as approximately $k^{n_s-4}\ln^2{(k)}$. The linear power spectrum with the estimates considered thus far is
\begin{multline}
\label{eq:ps_lin}
    P_\text{Lin}(k,a)=\frac{8\pi^2\mathcal{A}_s}{25\Omega_\text{M}^2H_0^4k_p^{n_s-1}}a^2\, _2F_1^2\left(\frac{1}{3},1;\frac{11}{6};-\frac{\Omega_\Lambda}{\Omega_\text{M}}a^3\right)\\
    \times
    \begin{cases}
        k^{n_s}&\text{for}\ k\ll k_\text{eq},\\
        144.k_\text{eq}^4k^{n_s-4}\ln^2{\left(0.12\frac{k}{k_\text{eq}}\right)}&\text{for}\ k\gtrsim1\ h/\text{Mpc},
    \end{cases}
\end{multline}
with $k_\text{eq}$ given by Eq.~\ref{eq:k_eq}. This linear power spectrum is plotted in Fig.~\ref{fig:ps_pt_class}, along with nonlinear corrections to be discussed in the next Section. The information in a Gaussian field $\delta$ is fully described by the power spectrum, as proven with Wick's theorem in Appendix~\ref{sec:wick}. Specifically, every $n$-point correlation function $\langle\delta(\bm{k}_1)...\delta(\bm{k}_n)\rangle$ is either zero for $n$ odd, or a sum of products of $\langle\delta(\bm{k}_i)\delta(\bm{k}_j)\rangle$ for $n$ even.

We now introduce another cosmological parameter derived from the linear matter power spectrum, which will be considered several times throughout this work. The overall amplitude of the matter power spectrum is determined by the size of fluctuations in the matter density. The RMS density in a sphere of radius $R$ is defined by the parameter $\sigma_R$ as
\begin{equation}
\label{eq:sigma_8_def}
    \sigma_R^2(a)=\frac{1}{2\pi^2}\int_0^\infty\text{d}k\ k^2P_\text{Lin}(k,a)\ W^2(kR),
\end{equation}
where $\text{d}^3k\rightarrow4\pi\ \text{d}k\ k^2$ has been taken due to the power spectrum being a function of the magnitude $k$. Here $W(x)$ is a top-hat filter, which in Fourier space is
\begin{equation}
    W(x)=\frac{3\left(\sin{\left(x\right)}-x\cos{\left(x\right)}\right)}{x^3}.
\end{equation}
It is standard to calculate $\sigma_R$ with a filter radius of $R=8\ \text{Mpc}/h$ (corresponding to a Fourier mode of $0.79\ h/\text{Mpc}$), defining $\sigma_8\equiv\sigma_{8\ \text{Mpc}/h}$. This length is about the typical radius of filaments~\cite{Bonjean_2020,Kuchner_2020}, and the upper end of galaxy cluster radii~\cite{Hansen_2005}. Considering how the time dependence factorizes as in Eq.~\ref{eq:ps_lin}, we can define a time-independent $\sigma_8$ with $\sigma_8(a)=\sigma_8D(a)$. Early time measurements give $\sigma_8=0.811\pm0.006$~\cite{Planck:2018parameters}. Late time measurements from galaxy clustering give $\sigma_8=0.722_{-0.036}^{+0.032}$~\cite{Philcox_2022}, and CMB lensing tomography give $0.707\pm0.035$~\cite{Chen_2022}, giving a roughly $3\sigma$ tension between early and late time measurements of $\sigma_8$.

\section{Nonlinear large-scale structure}
\label{sec:nonlinear_lss}

\begin{figure}
    \centering
    \hspace{0.8cm}Variance in matter density at present day
    
    \includegraphics[width=0.495\textwidth]{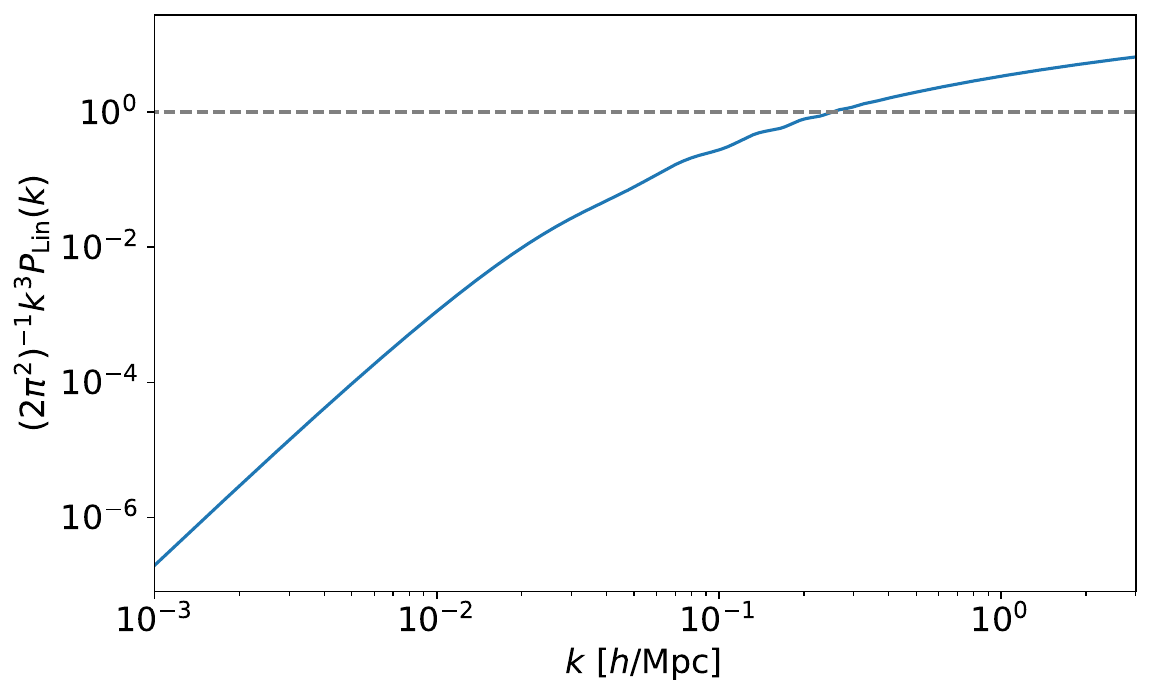}
    \caption[The linear matter power spectrum]{A plot of the dimensionless linear matter power spectrum $\Delta^2(k)$ at $z=0$. The variance in the matter density becomes significant around $k_\text{NL}\sim0.2h/\text{Mpc}$, where the field departs from Gaussianity.}
    \label{fig:kcubed_Pklin}
\end{figure}

Our present day universe is highly non-Gaussian at small length scales, as evident by the existence of filaments, voids, and galaxies. Nonlinearities occur above a mode $k_\text{NL}$ when the variance of the matter field becomes significant,
\begin{equation}
\label{eq:kcubed_Pklin}
    \Delta^2(k_\text{NL})=\frac{1}{2\pi^2}k_\text{NL}^3P_\text{Lin}(k_\text{NL})\sim1.
\end{equation}
This dimensionless version of the power spectrum $\Delta^2(k)$ is shown in Fig.~\ref{fig:kcubed_Pklin} at $z=0$, demonstrating that the nonlinear scale is around $k_\text{NL}\sim0.2h/\text{Mpc}$.
Approaching this mode, we can no longer accurately take the field as Gaussian. Consider the bispectrum $B(k)$, a three-point correlation function defined by
\begin{equation}
    (2\pi)^3B(k_1,k_2,k_3)\delta_\text{D}^3(\bm{k}_1+\bm{k}_2+\bm{k}_3)=\langle\delta(\bm{k}_1)\delta(\bm{k}_2)\delta(\bm{k}_3)\rangle.
\end{equation}
There is a host of information in the bispectrum. It provides tighter constraints on many cosmological parameters than with the power spectrum alone~\cite{Sefusatti_2006}, partially breaking the degeneracy between $\Omega_\text{M}$ and $\sigma_8$ in dark matter weak lensing surveys~\cite{Kilbinger_2005}, and breaking the degeneracy between galaxy bias and $\Omega_\text{M}$ in galaxy redshift surveys~\cite{Scoccimarro_1999}. However, the bispectrum is equivalent to $0$ for Gaussian $\delta(\bm{r})$. To probe parity violation, we would need four $\bm{k}$ vectors, as quadrilaterals in general do not respect mirror symmetry in three spacial dimensions, unlike triangles. This requires evaluating the four point function, the trispectrum~\cite{Cabass_2023,Coulton:2023oug}.

This Section outlines cosmological perturbation theory as one method of modelling the nonlinear matter content. Considering the complexity of this approach, we will then move on to simulation-based inference in Section~\ref{sec:simulation-based_inference}. In Chapter~\ref{ch:nf_random}, Chapter~\ref{ch:denoising}, and Chapter~\ref{ch:sr_diffusion}, a splitting between the Gaussian large length scales and non-Gaussian small lengths will ease the computation of our machine learning inference.



Perturbation theory in cosmology considers the time evolution of the matter density $\delta(\bm{r},t)$ and its velocity field $\bm{u}(\bm{r},t)$ as governed by the fluid continuity equation, the Euler equation, and the Poisson equation for the gravitational potential. To arrive at these three equations, we start with the phase space distribution of particles,
\begin{equation}
    f(\bm{r},\bm{p})=\sum_n\delta_\text{D}^3\left(\bm{r}-\bm{r}_n\right)\delta_\text{D}^3\left(\bm{p}-ma\bm{u}_n\right).
\end{equation}
The dark matter follows the collisionless Boltzmann equation, which states that the particles are conserved such that the full derivative of $f$ with respect to time is zero,
\begin{equation}
    \frac{\text{d}f(\bm{r},\bm{p},t)}{\text{d}t}=\frac{\partial f}{\partial t}+\frac{\bm{p}}{ma^2}\cdot\nabla f-m\sum_n\nabla\phi\cdot\nabla_{\bm{p}}f=0.
\end{equation}
The first three moments of $f$ give the matter density $\rho$, velocity $\bm{u}$, and a stress tensor $\bm{\sigma}$, as
\begin{gather}
    \rho(\bm{r},t)\equiv\frac{m}{a^3}\int\text{d}^3p\ f(\bm{r},\bm{p}),\\
    \rho(\bm{r},t)\bm{u}(\bm{r},t)\equiv\frac{1}{a^4}\int\text{d}^3p\ \bm{p}\ f(\bm{r},\bm{p}),\\
    \rho(\bm{r},t)\bm{u}(\bm{r},t)\bm{u}(\bm{r},t)-\bm{\sigma}(\bm{r},t)\equiv\frac{1}{ma^5}\int\text{d}^3p\ \bm{p}\bm{p}\ f(\bm{r},\bm{p}).
\end{gather}


The fluid equations are the zeroth moment and first moment of the Boltzmann equation,
\begin{gather}
    \int\text{d}^3p\ \frac{\text{d}f}{\text{d}t}=0,\\
    \int\text{d}^3p\ \bm{p}\ \frac{\text{d}f}{\text{d}t}=0.
\end{gather}
The zeroth moment gives the continuity equation, written in terms of the zero-mean overdensity $\delta$ as
\begin{equation}
\label{eq:pt_setup_1}
    \frac{\partial\delta}{\partial t}+\nabla\cdot\left((1+\delta)\bm{u}\right)=0.
\end{equation}
The first moment gives the Euler equation,
\begin{equation}
\label{eq:pt_setup_2}
    \frac{\partial\bm{u}}{\partial t}+\bm{u}\cdot\nabla\bm{u}+aH\bm{u}=-\nabla\Phi-\frac{1}{\rho}\nabla\cdot\bm{\tau}
\end{equation}
for an effective stress tensor $\bm{\tau}$. This $\bm{\tau}$ depends on $\bm{\sigma}$ as well as an integral over $\Phi$.
Along with the Poisson equation
\begin{equation}
\label{eq:pt_setup_3}
    \nabla^2\Phi=4\pi G\rho a^2\delta,
\end{equation}
Eqs.~\ref{eq:pt_setup_1},~\ref{eq:pt_setup_2}, and~\ref{eq:pt_setup_3} are complete when taking $\nabla\cdot\bm{\tau}$=0. These equations with $\nabla\cdot\bm{\tau}$=0 are solved using the ``standard" perturbation theory outlined below.

If we do not neglect the effective stress tensor $\bm{\tau}$, the $\nabla\Phi$ integral in $\nabla\bm{\tau}$ contains an infrared divergence, which can be understood in the Newtonian limit as a spacial integral over the inverse-square law of gravitation. Indeed, the source of our complexities in the late time matter distribution is due to gravitational collapse of matter at the smaller length scales. The common approach to handle $\bm{\tau}$ is to use an effective field theory by introducing a Fourier space cutoff $\Lambda$~\cite{Carrasco:2012cv}. In practice, this cutoff may be applied by taking a Gaussian smoothing of variance $\Lambda^2$ with any diverging observables. Above $\Lambda$ are highly nonlinear physics not suitable to perturbation theory. For example, at shell-crossing~\cite{Buchert:1997dr}, the dark matter particle trajectories begin to intersect, and therefore velocity field will become multi-valued at some locations in space.

Other than the stress tensor, Eqs.~\ref{eq:pt_setup_1},~\ref{eq:pt_setup_2}, and~\ref{eq:pt_setup_3} as written here have made some simplifying assumptions. A vorticity term $\nabla\times\bm{u}\propto1/a$ in the Euler equation has been taken to be zero, which is a good estimation at late times; N-body simulations give up to $(\nabla\times\bm{u})/(\nabla\cdot\bm{u})\sim0.01$~\cite{Jelic_Cizmek_2018} at linear scales. However, filamants and halo clusters are expected to exhibit vorticity, and filament vorticity has likely been recently observed~\cite{Wang_2021}. There is also no discrimination between baryons, photons, and neutrinos~\cite{Ma_1995}, instead treating all particle species together. The methods here could also be extended with the Zeldovich approximation~\cite{Munshi_1994} or a second-order Lagrangian perturbation theory approach~\cite{Crocce:2006ve}. Additionally, writing down the gravitational potential as $\Phi(\bm{r})$ in the first place is a Newtonian approximation.

The standard perturbation theory solves Eqs.~\ref{eq:pt_setup_1},~\ref{eq:pt_setup_2} with $\bm{\tau}=0$, and~\ref{eq:pt_setup_3} with the expansion in Fourier space
\begin{equation}
    \delta(\bm{k},a)=\sum_{n=1}^\infty D_+^n(a)\delta^{(n)}(\bm{k}),
\end{equation}
and a similar equation for $\nabla\cdot\bm{u}$. The $\delta^{(n)}$ are written as an expansion in the functions $F_n(\bm{k}_1,...,\bm{k}_n)$ as
\begin{equation}
    \delta^{(n)}(\bm{k},a)=D_+^n(a)\int\frac{\text{d}^3k_1}{(2\pi)^3}\ ...\int\frac{\text{d}^3k_n}{(2\pi)^3}\ \delta_\text{D}^3{\left(\bm{k}-\sum_{i=1}^n\bm{k}_i\right)}F_n(\bm{k}_1,...,\bm{k}_n)\prod_{i=1}^n\delta_1(\bm{k_i})
\end{equation}
Matching with linear order gives $F_1=1$. The next term is~\cite{Peebles:1980yev}
\begin{equation}
    F_2(\bm{k}_1,\bm{k}_2)=\frac{5}{7}+\frac{1}{2}\left(\frac{\bm{k}_1\cdot\bm{k}_2}{k_1k_2}\right)\left(\frac{k_1}{k_2}+\frac{k_2}{k_1}\right)+\frac{2}{7}\left(\frac{\bm{k}_1\cdot\bm{k}_2}{k_1k_2}\right)^2.
\end{equation}

\begin{figure}
    \centering
    \hspace{0.9cm}Power spectra at present day
    
    \includegraphics[width=0.495\textwidth]{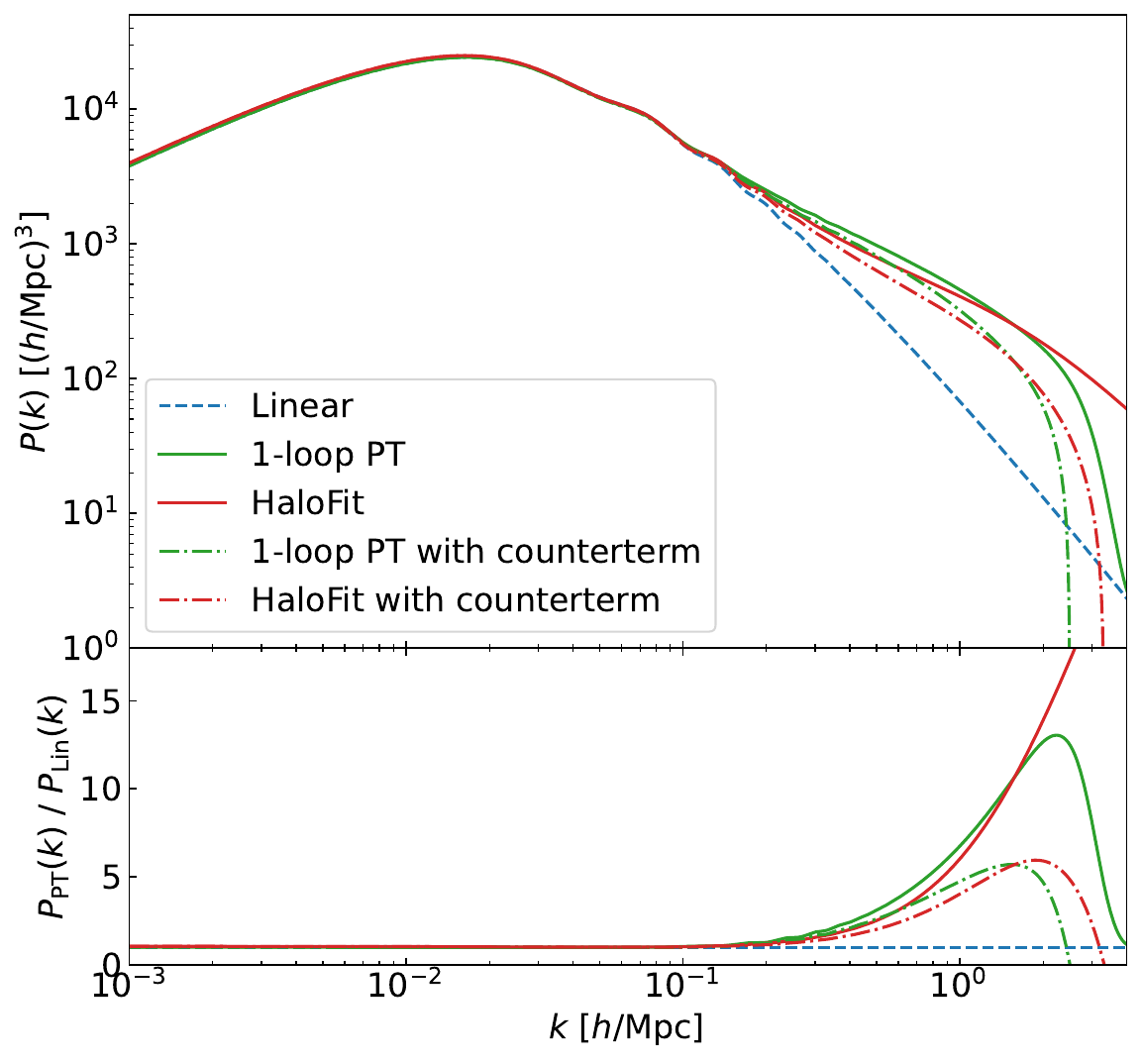}
    \caption[Matter power spectrum computed at linear and 1-loop order, and with HaloFit]{Matter power spectrum at $z=0$ comparing different levels of perturbation theory, along with a $c_s=1$ counterterm. Linear and 1-loop order are computed with \texttt{CLASS-PT}, and HaloFit is computed with \texttt{CLASS}.}
    \label{fig:ps_pt_class}
\end{figure}

The power spectrum with the perturbative expansion can be written as
\begin{equation}
    P(k,a)=P_\text{Lin}(k,a)+P_\text{1-loop}(k,a)+P_\text{Counter}(k,a) + \mathcal{O}\left(F_2F_3,F_4\right).
\end{equation}
The 1-loop term involves momentum integrals,
\begin{align}
    P_\text{1-loop}(k,a)=&+2\int\frac{\text{d}^3p}{(2\pi)^3}F_2^2(\bm{p},\bm{k}-\bm{p})P_\text{Lin}(p,a)P_\text{Lin}(|\bm{k}-\bm{p}|,a)\\
    &+6P_\text{Lin}(k,a)\int\frac{\text{d}^3p}{(2\pi)^3}F_3(\bm{p},-\bm{p},\bm{k})P_\text{Lin}(p,a),
\end{align}
where $F_3(\bm{k}_1,\bm{k}_2,\bm{k}_3)$ was solved for in~\cite{Fry:1983cj}. The 2-loop terms are similar integrals over $F_2F_3$ and $F_4$, where $F_4$ was solved for in~\cite{Goroff1986}. The counterterm can be written as
\begin{equation}
    P_\text{counter}(k,a)=-2c_s^2(a)k^2P_\text{Lin}(k,a)
\end{equation}
where $c_s$ is an effective speed of sound. It is approximately given by $c_s^2\propto\frac{c^2}{3\left(1+\frac{\partial\rho_\text{b}}{\partial\rho_\gamma}\right)}$, and is taken as a nuisance parameter with value $c_s=\mathcal{O}{\left(1\right)}\ \text{Mpc}/h$.
The present day power spectrum at 1-loop order is plotted along with a $c_s=1$ counterterm in Fig.~\ref{fig:ps_pt_class}, with comparison to the linear power spectrum. The 1-loop perturbation theory is calculated with \texttt{CLASS-PT}~\cite{Chudaykin_2020}. As a comparison with simulations, the HaloFit~\cite{Takahashi_2012} power spectrum is computed with \texttt{CLASS}~\cite{Diego_Blas_2011} in \texttt{nbodykit}~\cite{Hand_2018}. We see that at present day, there is a roughly order of magnitude error in the linear power spectrum at $k\gtrsim1\ h/\text{Mpc}$.

\begin{figure}
    \centering
    \begin{tabular}{cc}
        \hspace{0.8cm}Bispectra, $(k_1,k_2)=(0.15\frac{h}{\text{Mpc}},0.25\frac{h}{\text{Mpc}})$ & \hspace{0.8cm}Bispectra, $(k_1,k_2)=(0.6\frac{h}{\text{Mpc}},1.0\frac{h}{\text{Mpc}})$\\
        \includegraphics[width=0.474\textwidth]{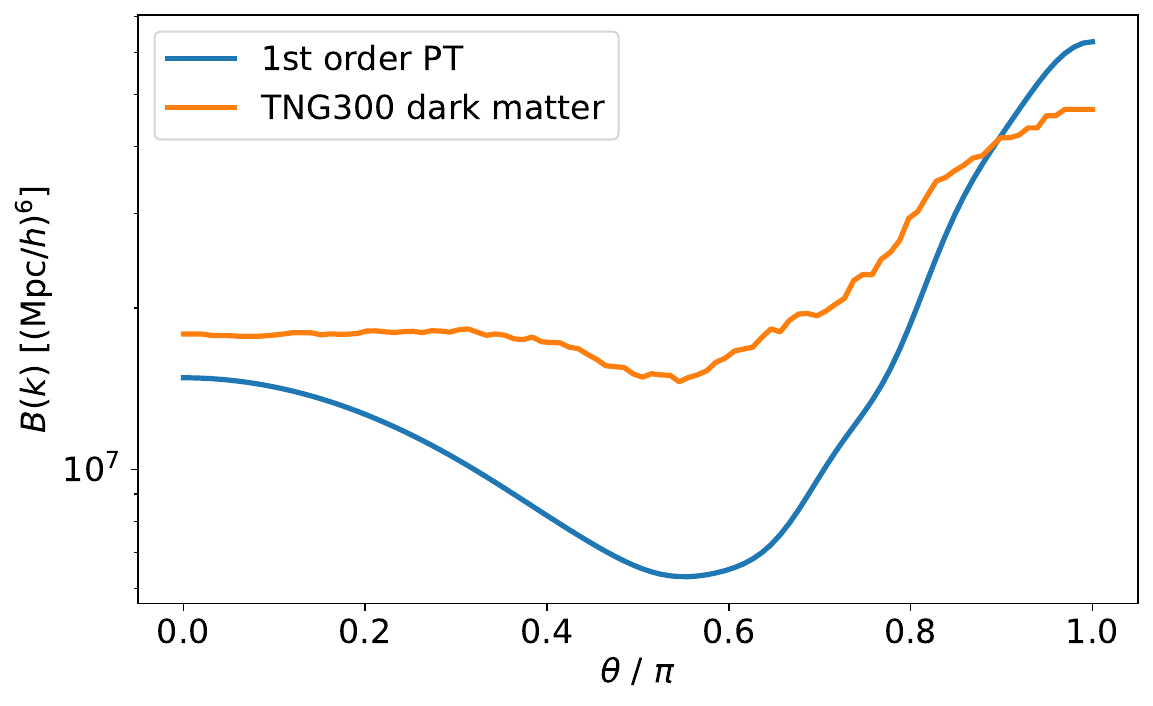} & \includegraphics[width=0.474\textwidth]{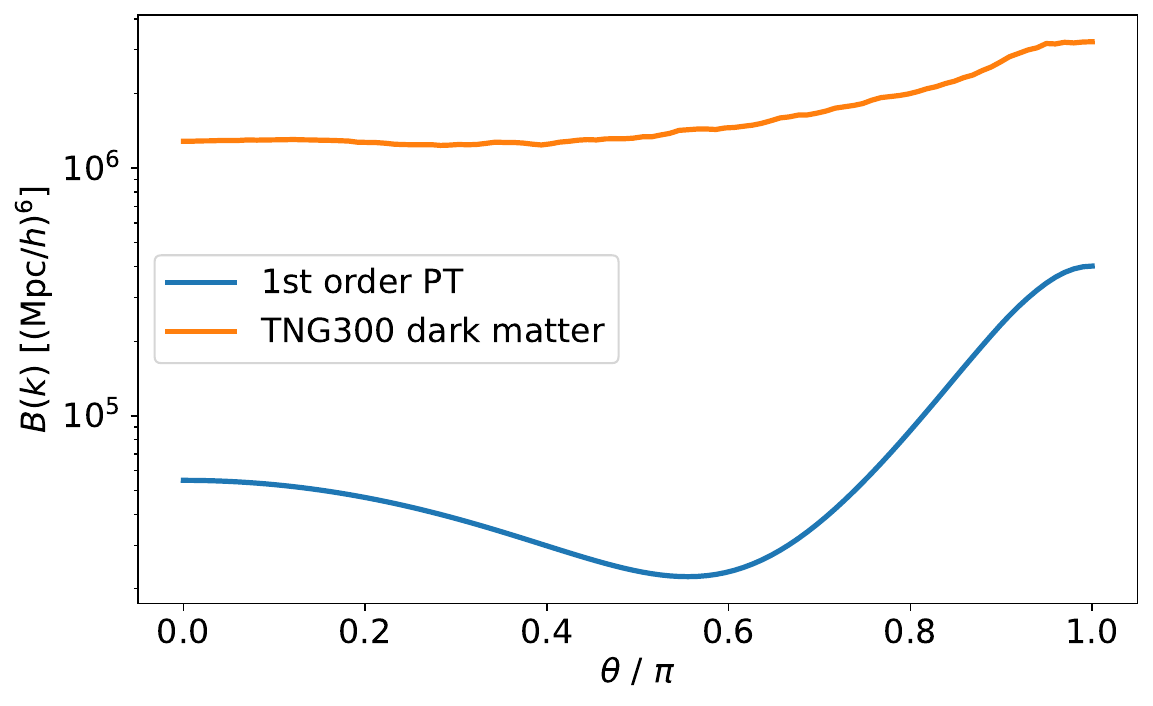}
    \end{tabular}
    \caption[Bispectrum computed an next-to-linear order, compared to TNG300-Dark]{Bispectrum at $z=0$ computed at next-to-linear order, and the full bispectrum calculated from TNG300-Dark, plotted as the angle $\theta$ between $k_1$ and $k_2$. (Left) $k$ values near the nonlinear regime. (Right) $k$ values in the nonlinear regime.}
    \label{fig:bs_f2}
\end{figure}

The bispectrum at next-to-linear order (NLO) is
\begin{align}
    B(k_1,k_2,k_3)=2\big(&F_2(\bm{k}_1,\bm{k}_2)P_\text{Lin}(k_1)P_\text{Lin}(k_2)\\
    +&F_2(\bm{k}_1,\bm{k}_3)P_\text{Lin}(k_1)P_\text{Lin}(k_3)\\
    +&F_2(\bm{k}_2,\bm{k}_3)P_\text{Lin}(k_2)P_\text{Lin}(k_3)\big).
\end{align}
This NLO bispectrum is between lowest order, where it is zero, and the one loop order, containing momentum integrals. The bispectrum is sensitive to $\Omega_\text{M}$, $\sigma_8$, and $n_s$ through the power spectrum, which in term depends on the transfer function. The NLO bispectrum is plotted in Fig.~\ref{fig:bs_f2} at $z=0$, with parameters used in IllustrisTNG: $\Omega_\text{M}=0.3089$, $\sigma_8=0.8159$, $n_s=0.9667$, and $h=0.6774$. It is compared with the Illustris TNG300-Dark~\cite{tng2017_1,tng2017_2, tng2017_3,tng2017_4,tng2017_5} simulation's bispectrum, computed at the field level with \texttt{Pylians}~\cite{pylians}. We will have more to say about the TNG300 simulation in Chapter~\ref{ch:sr_diffusion}, but for now we take TNG300-Dark to be a highly accurate representation of the dark matter density. Because the bispectrum is a function of three variables, for easier analysis it is plotted as a function of the angle $\theta$ between two vectors of magnitude $k_1$ and $k_2$. We see that for values $k_1=0.15\ h/\text{Mpc}$ and $k_2=0.25\ h/\text{Mpc}$ slightly into the nonlinear regime, the NLO bispectrum is quite inaccurate compared to TNG300-Dark, albeit still near the correct amplitude. However, for $k_1=0.6\ h/\text{Mpc}$, $k_2=1.0\ h/\text{Mpc}$, the NLO bispectrum has an order of magnitude error.

At one-loop order, the bispectrum is
\begin{equation}
    B_\text{1-loop}=B_{222}+B_{123}+B_{114}
\end{equation}
where $B_{222}$ is an integral over three factors of $F_2$, $B_{123}$ over $F_2F_3$, and $B_{114}$ over $F_4$. While these integrals have been evaluated for the approximations considered here~\cite{Scoccimarro:1997st}, we see that an perturbative analysis with the matter bispectrum at 1-loop order is already very difficult.

In principle, the statistical properties of $\delta(\bm{k})$ is fully described by every $n$-point correlation function
\begin{equation}
    \langle\delta(\bm{k}_1)...\delta(\bm{k}_n)\rangle.
\end{equation}
The time dependence of $\delta^{(n)}(k,a)$ being proportional to $D^n(a)\sim0.8^n$ at late times offers little help in truncating the expansion at higher order. Including vorticity, baryons, photons, massive neutrinos~\cite{Lesgourgues:2012uu}, alternative dark matter particles, or general relativistic corrections quickly exacerbates the complexity of going to higher order perturbation theory for higher order correlation functions. Including modified theories of gravity, even in the relatively simple case of $f(R)$ gravity, leads to questionable results in the perturbative regime~\cite{Nojiri_2011}. An approach that goes to the scales of individual galaxies is beyond the realm of perturbation theory.






\section{Simulation-based inference in cosmology}
\label{sec:simulation-based_inference}

This work solves three classes of problems with probabilistic machine learning models: parameter inference and data reconstruction with normalizing flows, and data generation with a denoising diffusion super-resolution emulator. These models are trained on various simulations of the large-scale structure with an approach known as likelihood-free inference. An explanation of the KL-divergence will be given, as this quantity will be used in the loss function training normalizing flows. As the following Chapters rely on simulated large-scale structure data, the methods of N-body simulations will also be outlined.

\subsection{Likelihood-free inference}

When a likelihood term can not be written down analytically, it could instead constructed by other means, such as with simulated data, with an approach broadly classified as likelihood-free inference. Despite its name, likelihood-free inference does not mean that a likelihood is completely removed from the inference pipeline. To illustrate in the case of parameter inference, a set of parameters $\bm{\phi}$ in a sample $\bm{x}$ has values taken from
\begin{equation}
    p(\bm{\phi}|\bm{x})\propto p(\bm{x}|\bm{\phi})\ p(\bm{\phi}).
\end{equation}
A simulator will forward model $\bm{x}$ from a set of parameters $\bm{\phi}$, generating a set of $N$ number of pairs $\{(\bm{x}_n,\bm{\phi}_n)\}_{n=1}^N$. However, for a simulator as complex as the ones we will use to model the large-scale structure, $p(\bm{x}|\bm{\phi})$ is intractable. The simulation has a large latent space with latent variables $\bm{z}$, and the likelihood is the marginalized probability over all possible trajectories through the latent space
\begin{equation}
\label{eq:likelihood_latent_1}
    p(\bm{x}|\bm{\phi})=\int p(\bm{x},\bm{z}|\bm{\phi})\ \text{d}\bm{z}.
\end{equation}
In an N-body simulation, $\bm{z}$ is the phase of the initial conditions along with every sequential time step of the simulation, for every particle interaction in each time step. There is clearly no way of evaluating Eq.~\ref{eq:likelihood_latent_1} for something as complicated as a simulation of the large-scale structure.


Similar to parameter inference, data reconstruction and conditional data generation can be written as a problem of obtaining a sample $\bm{y}$ from $p(\bm{y}|\bm{x})$. Here $\bm{x}$ could be data that is missing information, noisy, or low-resolution in some fashion. Sampling a reconstructed or high-resolution $\bm{y}$ conditional on some $\bm{x}$ is written as
\begin{equation}
    p{(\bm{y}|\bm{x})}\propto p{(\bm{x}|\bm{y})}\ p{(\bm{y})}.
\end{equation}
We may simulate many samples $\{\bm{y}_n\}_{n=1}^N$, where as in Eq.~\ref{eq:likelihood_latent_1}, $p({\bm{y}})$ is in principle an integral of $p({\bm{y},\bm{z}})$ over the simulator latent space $\bm{z}$. We require some method of representing either $p(\bm{y})$ if we are able to write down $p(\bm{x}|\bm{y})$, or $p(\bm{y}|\bm{x})$.


To overcome these issues, we learn an approximate form of the likelihood with machine learning models. We can simulate many sets of $\{(\bm{\phi}_n,\bm{x}_n)\}_{n=1}^N$ or $\{(\bm{y}_n,\bm{x}_n)\}_{n=1}^N$, and use this data to fit a high dimensional latent space to some representation of a PDF. For a machine learning model containing many parameters $\bm{\theta}$, our PDFs may be thought of as $p(\bm{x}|\bm{\phi};\bm{\theta})$, conditional on the model parameters. We will usually suppress writing out the model parameters $\bm{\theta}$.




\subsection{Comparing PDFs with the KL-divergence}
\label{sec:KL-div}

To fit one PDF to another, as will be required to train normalizing flows, a quantitative measure of the the similarity between the PDFs is required. A class of functions that measures the dissimilarity of probability distributions is the $f$-divergence~\cite{Renyi:1961}, defined as
\begin{equation}
    D_f\big(p(x)||q(x)\big)=\int q(x)f{\left(\frac{p(x)}{q(x)}\right)}\text{d}x.
\end{equation}
A few properties that make the $f$-divergence useful in fitting a distribution $q$ to a distribution $p$ are
\begin{itemize}
    \item $D_f\big(p||q\big)\geq0$ for all $p$, $q$,
    \item $D_f\big(p||p\big)=0$,
    \item $D_f\big(p||q\big)\neq0$ for $p\neq q$ unless $f(t)=c(t-1)$ for a constant $c$.
\end{itemize}
(The $f(t)=c(t-1)$ case has no interest here.)

The Kullback–Leibler (KL)-divergence~\cite{Kullback:1951} is defined as the $f$-divergence with $f(t)=t\ln{t}$; this gives
\begin{equation}
    D_{\text{KL}}\big(p(x)||q(x)\big)=\int p(x)\ln{\left(\frac{p(x)}{q(x)}\right)}\text{d}x.
\end{equation}
This choice of $f(t)=t\ln{t}$ can be motivated with the concept of entropy in information theory. For a discrete random variable $x$ with PDF $p(x)$, the entropy is defined as $-\sum_xp(x)\log{p(x)}$ for any base logarithm. The cross-entropy of $q(x)$ relative to $p(x)$ is defined as $-\sum_xp(x)\log{q(x)}$. The KL-divergence can be expressed in simpler terms as
\begin{equation}
    D_{\text{KL}}\big(p(x)||q(x)\big)=-\bigg(\text{Entropy of }p(x)\bigg)+\bigg(\text{Cross-entropy of }q(x)\text{ relative to }p(x)\bigg).
\end{equation}
In a sense, minimizing $D_{\text{KL}}\big(p(x)||q(x)\big)$ is equivalent to minimizing the cross-entropy of $q(x)$ relative to $p(x)$, while maximizing the entropy of $p(x)$.

As we will see, the goal of training a normalizing flow is to learn the transformation $T$ on $x$ that takes a distribution $q(x)$ to the distribution $p(x)$, and that is accomplished by minimizing the KL-divergence with respect to $q(x)$ given a $p(x)$. For such an optimization, the term $\int p(x)\ln{p(x)}\ \text{d}x$ is constant with respect to $q(x)$. Therefore, what is actually needed to minimize the KL-divergence is more simply
\begin{align}
    D_\text{KL}\big(p(x)||q(x)\big)\propto&-\int p(x)\ln{q(x)}\ \text{d}x\\
    =&-\int p(x)\bigg(\ln{(q_T(T(x)))}+\ln{|\det{J_T(T(x))}|}\bigg)\text{d}x
\end{align}

Of particular interest will be the KL-divergence of a Gaussian distribution with some other distribution. For $p(x)$ being a Gaussian distribution with mean $\mu_p$ and variance $\sigma_p^2$,
\begin{align}
    D_\text{KL}\big(p(x)||q(x)\big)&=\int p(x)\ln{(p(x))}\ \text{d}x-\int p(x)\ln{(q(x))}\ \text{d}x\\
    &=-\frac{1}{2}\left(1+\ln{\left(2\pi\sigma_p^2\right)}\right)-\frac{1}{\sqrt{2\pi\sigma_p^2}}\int e^{-\frac{1}{2}\left(\frac{x-\mu_p}{\sigma_p}\right)^2}\ln{(q(x))}\ \text{d}x
\end{align}
\sloppy To fit $q(x)$ to $p(x)$, gradients of KL-divergence need to be taken. The first term in $D_\text{KL}\big(p(x)||q(x)\big)$ is independent of $q(x)$ and is not needed to solve this problem.

To illustrate using the KL-divergence to measure and minimize the distance between two distributions, here is an example for two Gaussian distributions. The integral in the KL-divergence can be solved analytically if $p(x)$ and $q(x)$ are both Gaussian with (possibly) different means and variances. In such a case, in one-dimension, the KL-divergence evaluates to
\begin{equation}
    D_\text{KL}\big(p(x)||q(x)\big)=\ln{\frac{\sigma_q}{\sigma_p}}+\frac{(\mu_p-\mu_q)^2+\sigma_p^2}{2\sigma_q^2}-\frac{1}{2}.
\end{equation}
This can be seen to satisfy the properties of an f-divergence listed above, noticing that $D_\text{KL}\big(p(x)||q(x)\big)=0$ for $\mu_p=\mu_q$, $\sigma_p=\sigma_q$. In principle, solving the minimization problem requires taking the gradient of $D_\text{KL}\big(p(x)||q(x)\big)$. The gradient of this KL-divergence with respect to the $q$ parameters is
\begin{align}
    \frac{\partial}{\partial \mu_q}D_\text{KL}\big(p(x)||q(x)\big)&=\frac{\mu_q-\mu_p}{\sigma_q^2},\\
    \frac{\partial}{\partial\sigma_q}D_\text{KL}\big(p(x)||q(x)\big)&=\frac{-(\mu_p-\mu_q)^2+\sigma_q^2-\sigma_p^2}{\sigma_q^3},
\end{align}
and again it is apparent that the extrema is location at $\mu_p=\mu_q$, $\sigma_p=\sigma_q$.




\subsection{N-body simulations}

N-body simulations are used throughout this work as training data for machine learning models. Basic N-body simulations in cosmology compute the collisionless dynamics of dark matter particles under the influence of gravity in an expanding universe, and Chapters~\ref{ch:nf_random}-\ref{ch:denoising} use dark matter only simulations. All simulations used in this work have been run in a periodic box, and non-periodic data can be obtained by cropping the periodic boxes at the end of a simulation run. Our universe does not appear to be periodic, but our simulations are periodic simply to avoid a non-local gravitational collapse to the center of the volume. This will be relevant in Chapters~\ref{ch:denoising} and ~\ref{ch:sr_diffusion}, where machine learning models are trained on patches of a larger simulation box.

The Hamiltonian of an $N$ number of dark matter particles, each having mass $m_i$, in a gravitational potential $\Phi(\bm{r})$, is
\begin{equation}
    \mathcal{H}(\bm{p}_1,...,\bm{p}_N,\bm{r}_1,...,\bm{r}_N,t)=\sum_i\frac{(\bm{p}_i/a(t))^2}{2m_i}+\sum_{i,j}m_im_j\frac{\Phi(\bm{r}_i-\bm{r}_j)}{a(t)}.
\end{equation}
The value of $a(t)$ at any time step $t$ is obtained by integrating Eq.~\ref{eq:Friedmann_a}. Inter-particle interactions would necessitate a much more complicated Hamiltonian, as is the case for simulations containing baryons. We will work with a baryonic simulation in Chapter~\ref{ch:sr_diffusion}, leaving an explanation complicated of the methods of accurately running such baryonic simulations to the simulation's developer~\cite{Springel_2010}.

Throughout this work, dark matter particles within a simulation box have identical masses, and therefore the factors of $m_i$ can be absorbed. For a periodic box of length $L$, the peculiar gravitational potential is
\begin{equation}
    \nabla^2\Phi(\bm{r})=4\pi G\left(\sum_{\bm{n}}\delta(\bm{r}-\bm{n}L)-\frac{1}{L^3}\right).
\end{equation}
The $\bm{n}=(n_x,n_y,n_z)$ sum is taken over all combinations of $3$ integers.

The equations of motion are
\begin{align}
    \frac{\text{d}\bm{r}}{\text{d}t}&=\frac{\bm{v}(t)}{a(t)},\\
    \frac{\textbf{d}\bm{v}}{\text{d}t}&=-\frac{\text{d}a(t)/\text{d}t}{a(t)}\bm{v}(t)-\frac{1}{a^2(t)}\nabla\Phi(\bm{r})
\end{align}
The positions and velocities are obtained by integrating the equations of motion along a number of time steps. A discussion of efficient methods of solving the previous equations are beyond our scope. We rely on various different simulation software in this work which have been developed for speed, as with the particle mesh code \texttt{FastPM}~\cite{Feng:2016yqz}, or for accuracy, as with the TreePM codes \texttt{GADGET}~\cite{gadget2} and \texttt{Arepo}~\cite{Springel_2010}.

It is somewhat insightful to count the number of parameters needed to be specified in a dark matter simulation. The equations of motion depend on $\Phi(\bm{r})$ and $a(t)$. The potential $\Phi(\bm{r})$ explicitly depends on $L$. The scale factor $a(t)$ through Eq.~\ref{eq:Friedmann_a} with $\Omega_R=0$ depends on $\Omega_\text{M}$, $\Omega_\Lambda=1-\Omega_\text{M}$ (from a flat cosmology), and $H_0$. With some inspection of the previous equations, it is possible to absorb all dependency on $H_0$ into the positions and velocities; indeed, physical lengths, velocities, and Fourier modes in cosmology are often expressed with a factor of the reduced Hubble constant $h$. Therefore, given the initial conditions, the dark matter simulations we use depend on $2$ parameters, $\Omega_\text{M}$ and the simulation box length~$L$.





\section{Deterministic machine learning}
\label{sec:deterministic_ml}

This Section and Section~\ref{sec:deterministic_ml} will differentiate two broad approaches to machine learning models, which are referred to here as ``deterministic" and ``probabilistic" machine learning. These terms are not robustly defined, but it is illustrative to understand machine learning models as being part of these two classes for the purposes of this work.

\begin{figure}
\centering
\begin{tikzpicture}[node distance=0.65cm]
\footnotesize
\node[align=center, font=\bfseries] (title){Parameter inference};
\node (start) [cnn, below of=title] {Field $\bm{x}$: $c_\text{in}\times 128^2$};
\node (conv0) [cnn, below of=start, fill=azul!75] {$3\times3$ conv: $16\times126^2$};
\node (pool0) [cnn, below of=conv0, fill=red!15]  {$2\times2$ pool: $32\times63^2$};
\node (conv1) [cnn, below of=pool0, fill=azul!75] {$3\times3$ conv: $32\times61^2$};
\node (pool1) [cnn, below of=conv1, fill=red!15]  {$2\times2$ pool: $32\times30^2$};
\node (conv2) [cnn, below of=pool1, fill=azul!75] {$3\times3$ conv: $32\times28^2$};
\node (pool2) [cnn, below of=conv2, fill=red!15]  {$2\times2$ pool: $32\times14^2$};
\node (conv3) [cnn, below of=pool2, fill=azul!75] {$3\times3$ conv: $32\times12^2$};
\node (conv4) [cnn, below of=conv3, fill=azul!75]  {$3\times3$ conv: $32\times10^2$};
\node (conv5) [cnn, below of=conv4, fill=azul!75]  {$3\times3$ conv: $32\times8^2$};
\node (conv6) [cnn, below of=conv5, fill=azul!75]  {$3\times3$ conv: $32\times6^2$};
\node (conv7) [cnn, below of=conv6, fill=azul!75]  {$3\times3$ conv: $32\times4^2$};
\node (fc0)   [cnn, below of=conv7, fill=green!15] {Dense: $128$};
\node (fc1)   [cnn, below of=fc0,   fill=green!15] {Dense: $6$};
\node (end)   [cnn, below of=fc1] {$(\Omega_\text{M},\sigma_8,n_s,\sigma_{\Omega_\text{M}},\sigma_{\sigma_8},\sigma_{n_s})$};

\draw [arrow] (start.east) to [out=-30,in=30](conv0.east);
\draw [arrow] (conv0.east) to [out=-30,in=30](pool0.east) node[yshift=0.35cm] {\hspace{1.3cm}ReLU};
\draw [arrow] (pool0.east) to [out=-30,in=30](conv1.east);
\draw [arrow] (conv1.east) to [out=-30,in=30](pool1.east) node[yshift=0.35cm] {\hspace{1.3cm}ReLU};
\draw [arrow] (pool1.east) to [out=-30,in=30](conv2.east);
\draw [arrow] (conv2.east) to [out=-30,in=30](pool2.east) node[yshift=0.35cm] {\hspace{1.3cm}ReLU};
\draw [arrow] (pool2.east) to [out=-30,in=30](conv3.east);
\draw [arrow] (conv3.east) to [out=-30,in=30](conv4.east) node[yshift=0.35cm] {\hspace{1.3cm}ReLU};
\draw [arrow] (conv4.east) to [out=-30,in=30](conv5.east) node[yshift=0.35cm] {\hspace{1.3cm}ReLU};
\draw [arrow] (conv5.east) to [out=-30,in=30](conv6.east) node[yshift=0.35cm] {\hspace{1.3cm}ReLU};
\draw [arrow] (conv6.east) to [out=-30,in=30](conv7.east) node[yshift=0.35cm] {\hspace{1.3cm}ReLU};
\draw [arrow] (conv7.east) to [out=-30,in=30](fc0.east) node[yshift=0.35cm] {\hspace{1.5cm}Flatten};
\draw [arrow] (fc0.east)   to [out=-30,in=30](fc1.east) node[yshift=0.35cm] {\hspace{1.3cm}ReLU};
\draw [arrow] (fc1.east)   to [out=-30,in=30](end.east);
\footnotesize
\node[right of=conv1, xshift=7cm, font=\bfseries] (title){Data generation / data reconstruction};
\node (start) [cnn, below of=title, node distance=0.65cm] {Conditional field $\bm{x}$};
\node (model) [cnn, below of=start, fill=azul!75, node distance=1.95cm] {U-net $f(\bm{x};\bm{\theta})$};
\node (end) [cnn, below of=model, node distance=1.95cm] {Generated field $\bm{y}$};

\draw [arrow] (start.south) to (model.north);
\draw [arrow] (model.south) to (end.north);
\end{tikzpicture}
\caption[Examples of deterministic CNNs for data generation and parameter inference]{(Left) In a standard deterministic parameter inference model, a CNN is a mapping $f(\bm{x};\bm{\theta})=\bm{\phi}$ from data $\bm{x}$ to a list of parameters $\bm{\phi}$ with convolutions, pooling, and dense layers; group normalization (Appendix~\ref{sec:group_normalization}) and dropout (Appendix~\ref{sec:dropout}) could also be included in such a CNN. Shown are layer names, along with layer output channel $\times$ spacial dimensions. (Right) A deterministic model that generates data $\bm{y}$ conditional on an input $\bm{x}$ with a neural network (such as a U-net) $f(\bm{x};\bm{\theta})=\bm{y}$. A description of U-nets is given in Chapter~\ref{ch:sr_diffusion}.}
\label{fig:CNN_parameter_inference}
\end{figure}
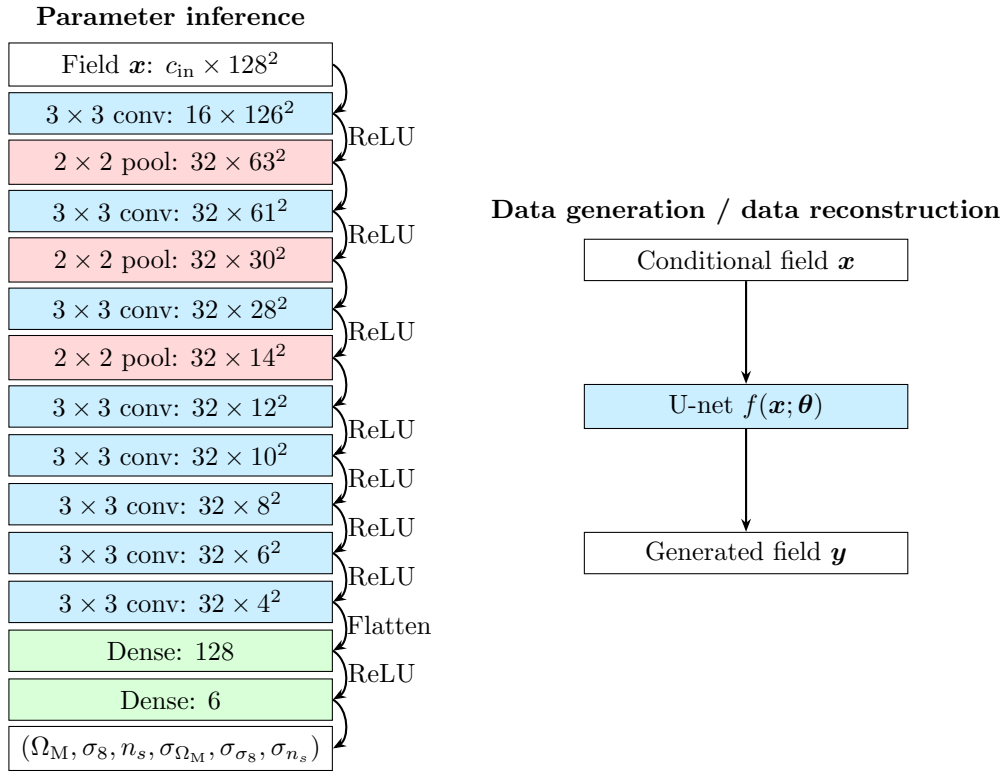

A standard, deterministic machine learning model can broadly be understood as a complicated mapping between sets of data. For example, parameter estimation or data classification of data $\bm{x}$ can be written as a mapping
\begin{equation}
    f{(\bm{x};\bm{\theta})}=\bm{\phi_\text{model}}
\end{equation}
for some set of parameters or classes $\bm{\phi}$. Considering cosmological data, $\bm{x}$ could be a list of numbers from a power spectrum, a 2-dimensional lensing survey image, or 3-dimensional large-scale structure data. A typical convolutional neural network (CNN) (see Appendix~\ref{sec:convolutions}) to map data to parameters is shown in Fig.~\ref{fig:CNN_parameter_inference} (left).

Considering data reconstruction and data generation, a deterministic machine learning approach would learn to generate fields $\bm{x}$ from fields $\bm{y}$ as a mapping
\begin{equation}
f(\bm{x};\bm{\theta})=\bm{y}.
\end{equation}
For this setup, $\bm{x}$ might be a low-resolution dark matter density field, and $\bm{y}$ a high-resolution dark matter field, or some other particle species. A powerful class of neural networks for $f$ to map between 2- or 3-dimensional fields would be the U-net~\cite{Ronneberger2015}. A diagram of a U-net built with many residual blocks~\cite{he2015deep} is shown in Fig.~\ref{fig:unet}.

Accurate parameter estimation of $(\Omega_\text{M},\sigma_8)$ has been done with basic CNN models on weak lensing maps\cite{Gupta:2018eev,Ribli:2018kwb,Fluri:2018hoy} and 3-dimensional dark matter fields \cite{ravanbakhsh2017estimating}. Additionally on 3-dimensional dark matter fields, inference with the three parameters $(\Omega_\text{M},\sigma_8,n_s)$ has been successful with a CNN~\cite{Mathuriya:2018luj,Lazanu_2021}. By definition, $\sigma_8$ can be determined from the amplitude of the linear power spectrum Eq.~\ref{eq:sigma_8_def}, and $n_s$ from the tilt of the linear power spectrum Eq.~\ref{eq:n_s_def}. However, it may be difficult to accurately obtain the linear power spectrum at small length scales with late time data. Indeed, there is currently a roughly $3\sigma$ tension in early vs. late time measurements of $\sigma_8$, with various different proposals for the source of this tension~\cite{Abdalla:2022yfr}.

The loss function that optimizes (Appendix~\ref{sec:sgd}) the model parameters $\bm{\theta}$ is often a mean squared error
\begin{equation}
    \mathcal{L}(\bm{x},\bm{\phi}_\text{true};\bm{\theta})=\sum_i\left(\phi_{i,\text{model}}-\phi_{i,\text{true}}\right)^2,
\end{equation}
where the sum is over the list of parameters ($\Omega_\text{M}$, etc.). It is also possible for the model to learn variances $\bm{\sigma}^2$ of the mean values $\bm{\phi}$, as simply a mapping from data $\bm{x}$ to $\bm{\phi}$ and $\bm{\sigma}^2$,
\begin{equation}
    f(\bm{x};\bm{\theta})=(\bm{\phi}_\text{model},\bm{\sigma}^2).
\end{equation}
Note that there is no a-priori ``true" $\bm{\sigma}_\text{true}$ to optimize the model with, hence the lack of a subscript on $\bm{\sigma}^2$. The model variance in $\bm{\phi}$ can be learned by including an additional term in the loss function. One way of directly estimating the variances $\bm{\sigma}^2$ is to extend the mean squared error loss as~\cite{Jeffrey:2020itg}
\begin{equation}
\label{eq:loss_sigma_noln}
    \mathcal{L}(\bm{x},\bm{\phi}_\text{true};\bm{\theta})=\sum_i\left(\phi_{i,\text{model}}-\phi_{i,\text{true}}\right)^2+\sum_i\left(\left(\phi_{i,\text{model}}-\phi_{i,\text{true}}\right)^2-\sigma_i^2\right)^2.
\end{equation}
This loss function Eq.~\ref{eq:loss_sigma_ln} is in the form of the Moment Network introduced in~\cite{Jeffrey:2020itg}, presenting loss functions containing a hierarchy of terms to estimate the posterior probability $p(\bm{\phi}|\bm{x})$. Another form of this loss function is~\cite{Villaescusa_Navarro_2022}
\begin{equation}
\label{eq:loss_sigma_ln}
    \mathcal{L}(\bm{x},\bm{\phi}_\text{true};\bm{\theta})=\ln{\left(\sum_i\left(\phi_{i,\text{model}}-\phi_{i,\text{true}}\right)^2\right)}+\ln{\left(\sum_i\left(\left(\phi_{i,\text{model}}-\phi_{i,\text{true}}\right)^2-\sigma_i^2\right)^2\right)}.
\end{equation}
This form of $\mathcal{L}$ with logarithms of each term is, in the eyes of gradient descent, much like multiplying the two terms together. In some cases where a specific $\phi_{i,\text{model}}$ is exceptionally difficult to fit to $\phi_{i,\text{true}}$, the loss function with logarithms may help in preventing such a parameter from getting ``left behind" and not properly extremized with the other parameters.

The mapping $f$ is still deterministic with the learned variance; inputting the same sample $\bm{x}$ into $f$ will return the same $(\bm{\phi},\bm{\sigma})$ every time, and the methods here do not actually obtain a representation of the full PDF $p(\bm{\phi}|\bm{x})$. However, learning the variance as in Eq.~\ref{eq:loss_sigma_noln} might be seen as a sort of first order correction to deterministic models, bringing us closer to a probabilistic machine learning model.

\section[Parameter inference with persistence images of dark matter halos]{Parameter inference with persistence images of\\dark matter halos}
\label{sec:persistence_images}

As an application of the deterministic machine learning methods discussed thus far, this Section maps topological data of dark matter halos to three cosmological parameters, $\Omega_\text{M}$, $\sigma_8$, and $n_s$. A slight aside from the ``field level" approach taken throughout the remainder of this work, this Section operates on persistence images, which are derived from the topology of point clouds. Persistence images are pixelated images in the same sense as particle data placed on a mesh, and so the methods here may be applied to any such matter density fields regardless. The results in this Section are part of an ongoing project~\cite{yipinprogress} on topological data analysis of the large-scale structure, and a continuation of the results in~\cite{yip2023learning}.

\subsection{Persistence images}

Persistent homology~\cite{Edelsbrunner2002,Zomorodian2004} is an area of topological data analysis that measures the topological features of point clouds across length scales. The point cloud is embedded into a simplical complex, containing simplices of points, edges, triangles, and tetrahedrons. The simplical complex may then be described by its homology groups, which there are $3$ of in $3$-dimensions: 0-cycles correspond to connected components or ``islands," 1-cycles correspond to loops, and 2-cycles correspond to voids. A filtration parameter controls simplices included in the simplical complex, where a small filtration has many islands and a large filtration is a fully-connected complex. 

\begin{figure}
    \centering
    \begin{tabular}{ccc}
        \includegraphics[width=0.31\textwidth]{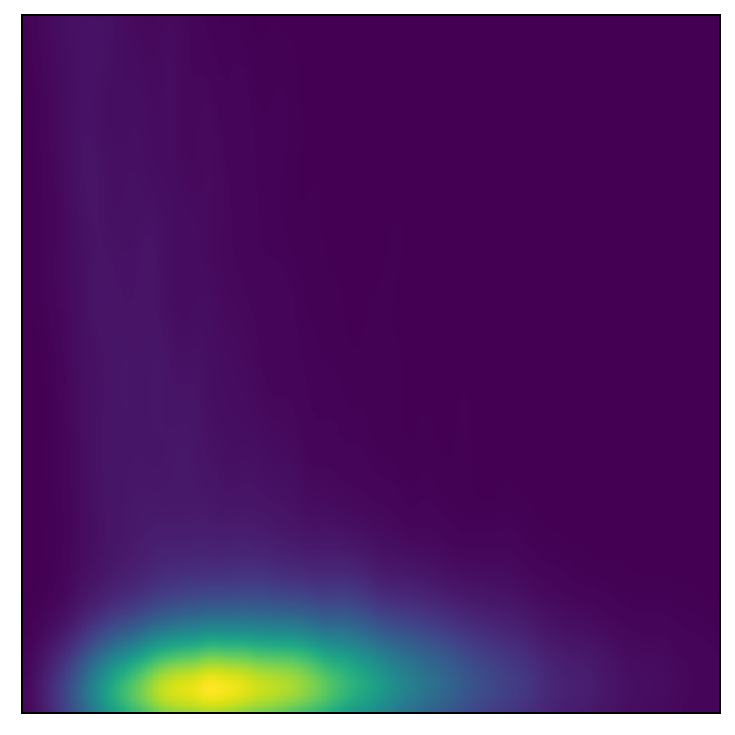} & \includegraphics[width=0.31\textwidth]{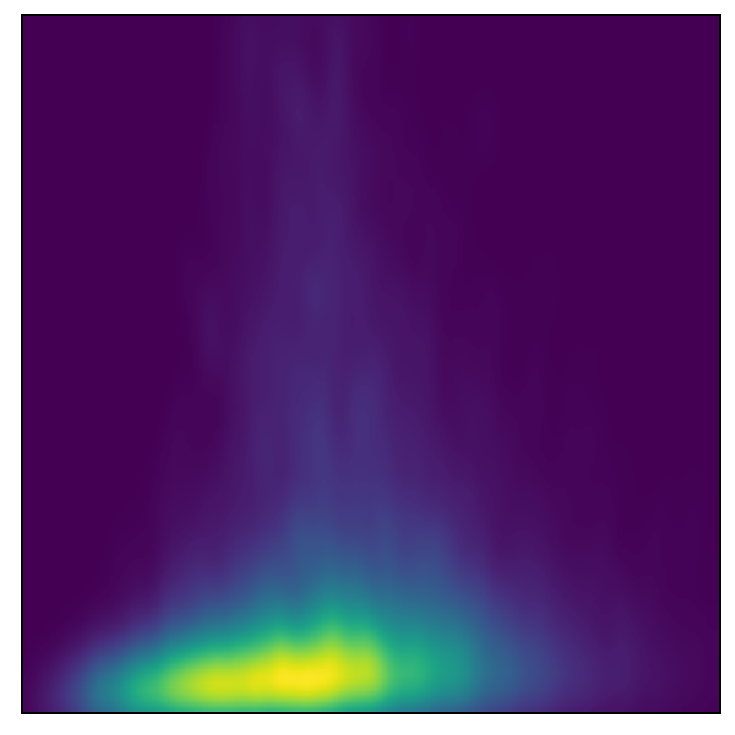} & \includegraphics[width=0.31\textwidth]{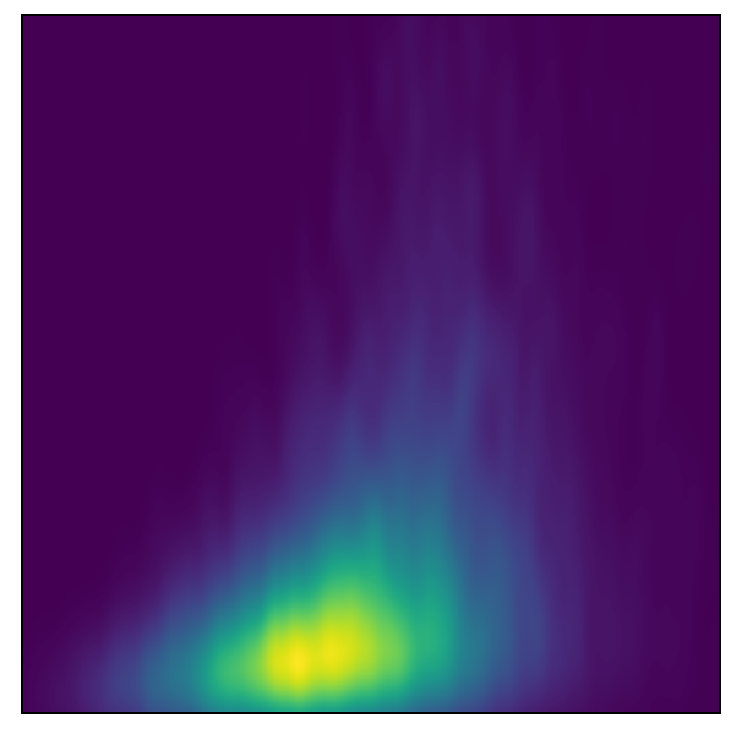}\\
        0-cycle & 1-cycle & 2-cycle
    \end{tabular}
    \caption[Persistence images of the large-scale structure]{Persistence images of dark matter halos.}
    \label{fig:persistence_images}
\end{figure}

Persistence diagrams are built from a growing filtration, keeping track of every $p$-cycle's birth and death. For each $p$-cycle, its corresponding persistence diagram is a 2-dimensional scatter plot of every birth and persistence in the simplical complex, where the persistence is the filtration size difference between the birth and death. Vectorizing the persistence diagrams gives persistence images, which may be used as an input to a CNN. Examples of persistence images are shown in Fig.~\ref{fig:persistence_images}, for 0-, 1-, and 2- cycles of dark matter halos.

With persistent homology, the topological features of point cloud data may be used for pattern recognition and parameter estimation~\cite{carlsson_2014}. Persistent homology has been used to measure primordial non-Gaussianity and place constraints on $f_\text{NL}^\text{loc}$ from dark matter simulations~\cite{Cole_2018,Biagetti_2021,Biagetti_2022}. In order to operate on persistence images, these works flattened the persistence images along their two dimensions by taking the sum along each of the two dimensions, allowing more straightforward inference taking a Gaussian likelihood. It was recently shown that to obtain optimal information of $\Omega_\text{M}$ and $\sigma_8$ from persistence images, the entire image should be part of the inference pipeline~\cite{yip2023learning}, the most straightforward approach being to use a CNN to map the persistence images to the cosmological parameters. A more detailed explanation of persistent homology may be found in any of the aforementioned references.

\begin{figure}
    \centering
    \includegraphics[width=\textwidth]{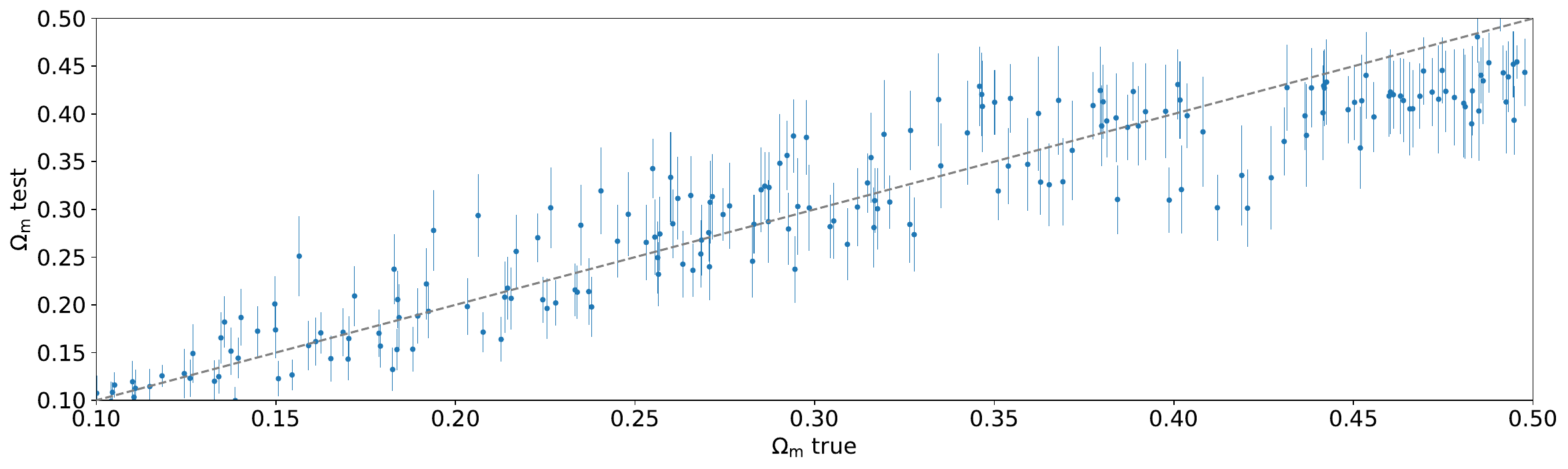}
    \includegraphics[width=\textwidth]{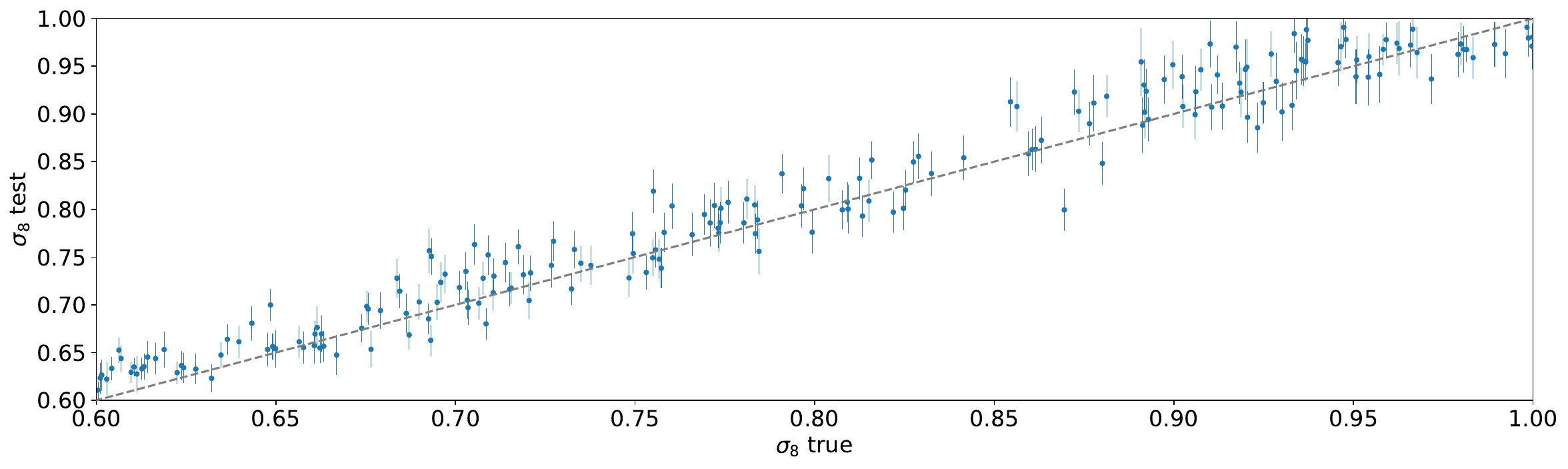}
    \includegraphics[width=\textwidth]{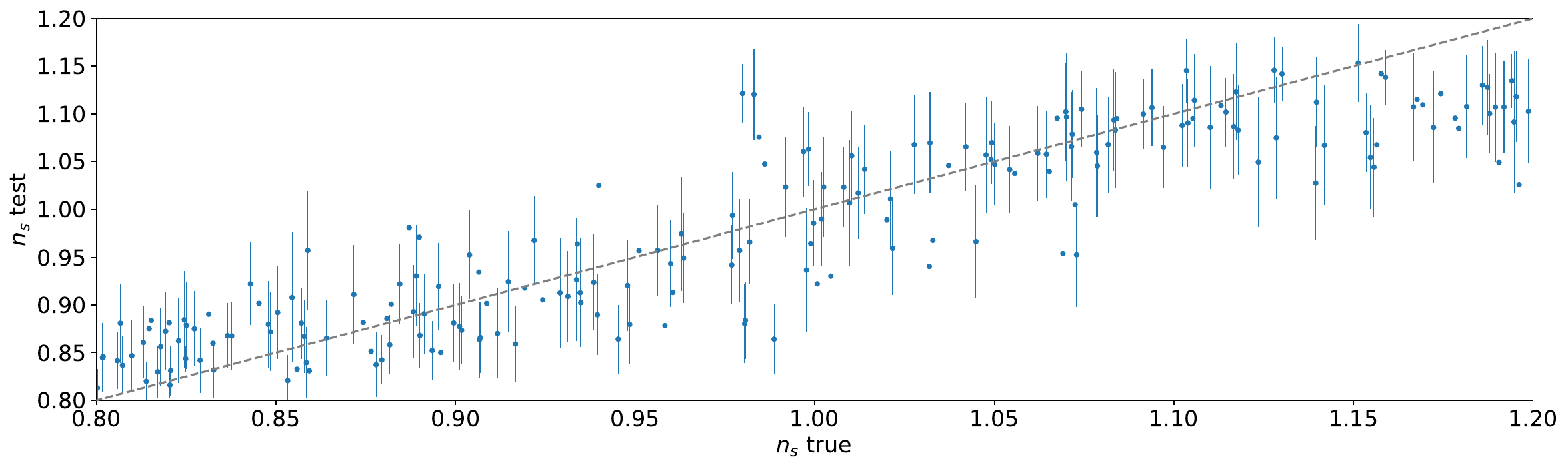}
    \caption[$\Omega_\text{M}$, $\sigma_8$, $n_s$ parameter inference with persistence images]{Parameter inference of persistence images with a CNN, predicting $\Omega_\text{M}$, $\sigma_8$, and $n_s$, along with learned $1\sigma$ error bars.}
    \label{fig:persistent_image_results}
\end{figure}

\subsection{Method for measuring cosmological parameters}

Our halo catalog to learn the topological features of the large-scale structure comes from the Quijote dark matter simulations~\cite{Villaescusa-Navarro:2019bje}, containing a Latin hypercube of $2000$ simulations varying the five parameters $\Omega_\text{M}$, $\sigma_8$, $n_s$, $\Omega_\text{B}$, and $h$; however, we will not consider $\Omega_\text{B}$ and $h$, and effectively marginalize our results over these two parameters. We use all three of the 0-, 1-, and 2-cycle persistence images as input to our model. We also introduce a hyperparameter $K$ of the filtration, to delay the inclusion of simplices to the simplicial complex~\cite{yip2024}. As a consequence, a persistent homology computation with a larger $K$ value would track topological features of larger scales, and a smaller $K$ vice versa. With a broad range of $6$ different $K$ cutoffs, we have a total of $18$ persistence images per halo field. The data is split as $1600$ training samples, $200$ validation samples, and $200$ test samples.

A basic CNN with pooling layers and ReLU activation functions is used to map persistence images to parameters, and a diagram of the model architecture is shown in Fig.~\ref{fig:CNN_parameter_inference} (left) with $c_\text{in}=18$ input channels. The output of our model is $6$ numbers, being model values for $\Omega_\text{M}$, $\sigma_8$, $n_s$, and their $1\sigma^2$ variances. This model has 129,174 parameters. We were unable to accurately infer $\Omega_\text{B}$ and $h$, but this is not surprising; dark matter simulations are not very sensitive to $\Omega_\text{B}$ and $h$, as information of these two parameters in the dark matter simulations is only found as small adjustments to the transfer function. The loss function used to learn the mapping from persistence images to parameters is Eq.~\ref{eq:loss_sigma_ln}. To help prevent overfitting, each image is given a random periodic shift, rotation, and chance of being mirrored. While these transformations break the symmetry of the data, they improve model performance with the somewhat small number of data samples. We train with a batch size of $128$, and training takes about $1$~hour on an RTX A4000 before the loss computed on the validation set shows overtraining.

\subsection{Results on $\Omega_\text{M}$, $\sigma_8$, and $n_s$}

Results are shown in Fig.~\ref{fig:persistent_image_results} for $\Omega_\text{M}$, $\sigma_8$, and $n_s$. We quantify the model's mean and $1\sigma$ results in several ways. Considering the mean results, the root-mean-squared error for $\Omega_\text{M}$ is $0.044$, $\sigma_8$ is $0.026$, and for $n_s$ is $0.053$. We may compare these values to the parameter ranges of $0.1<\Omega_\text{M}<0.5$, $0.6<\sigma_8<1.0$, and $0.8<n_s<1.2$. The coefficient of determination for a parameter $\phi_i$ is defined as
\begin{equation}
    R_i^2=1-\frac{\sum_i^N\left(\phi_{i,\text{model}}-\phi_{i,\text{true}}\right)^2}{\sum_i^N\left(\phi_{i,\text{model}}-\frac{1}{N}\sum_i^N\phi_{i,\text{model}}\right)^2}.
\end{equation}
Values of $R_i^2$ close to $1$ have accurately learned the mean values of the parameters. We get $R_{\Omega_\text{M}}^2=0.83$, $R_{\sigma_8}^2=0.95$, and $R_{n_s}^2=0.72$. Measurements of $\sigma_8$ are highly accurate, with measurements of $n_s$ being the least accurate of the three parameters considered.

Now we consider the accuracy of the learned $1\sigma$ error regions. We would expect $68.3\%$ of the model predictions to be within the $1\sigma$ error bars. The models predictions for $\Omega_\text{M}$ are $55.0\%$, $\sigma_8$ are $54.0\%$, and $n_s$ are $60.5\%$ within the learned $1\sigma$ error bars, indicating that the model is somewhat too confident with each of the parameters. Another measure of accuracy of the model's estimated errors is the reduced chi-squared, defined by
\begin{equation}
    \chi_i^2=\frac{1}{N}\sum_i^N\left(\frac{\phi_{i,\text{model}}-\phi_{i,\text{true}}}{\sigma_i}\right)^2.
\end{equation}
Values of $\chi_i^2$ below $1$ indicate under-confident error bars, while values above $1$ indicate the model is overly confident. We get $\chi^2_{\Omega_\text{M}}=1.40$, $\chi^2_{\sigma_8}=1.58$ and $\chi^2_{n_s}=1.44$. Again, the model is a bit too confident with the variance of each of the parameters. It is interesting to see that while $\sigma_8$ has the most accurate estimate of the mean, the model also gives $\sigma_8$ the most incorrect variance of the three parameters. This over-confidence may be the result of a small amount of overfitting. Training on more data samples, or carefully adjusting the model hyperparameters, may result in $\chi^2$ values closer to $1$.

\subsection{Discussion}

This Section has shown that the topological features of the large-scale structure contain information of $\Omega_\text{M}$, $\sigma_8$ and $n_s$, and these parameters may be accurately predicted from persistence images of dark matter halos. Considering the lack of robustness of machine learning models on out-of-distribution data, a persistent homology approach to parameter inference may be greatly aided in transitioning from simulated data to real data. For example, if a data set of dark matter halos contains errors from real data, such as a local volume of halo undercounting or a systematic scaling error, we would expect the resulting persistence image to still ``look" like a persistence image to the neural network. Considering the $K$ cutoff in the persistent homology, it would be interesting to measure the amount of information contained in the total $6$ values considered here against a fewer number of $K$ cutoffs. To make the model more robust at parameter inference on out-of-distribution data, we may also consider ``turning off" specific $K$ cutoffs depending on the test data. For the model trained on dark matter simulations, test data containing baryonic feedback will require a higher $K$ cutoff to block out the out-of-distribution small length modes. These direction are currently being explored~\cite{yipinprogress}.

In this analysis of persistence images, there is no mention of a PDF or any Bayesian statistical methods in the model architecture. Rather, the model learns the $1\sigma^2$ variance of each estimate of the mean as defined by the  loss function Eq.~\ref{eq:loss_sigma_ln}. We do not get covariances between the parameters, directional variances (for example, a tighter bound above the mean than below), or a $2\sigma^2$ variance (which could be different from twice the $1\sigma$ variance). Measuring the covariance between cosmological parameters crucial. For example, measurements of weak gravitational lensing of the CMB by the large-scale structure would expect a slight degeneracy between $\Omega_{M}$ and $\sigma_8$~\cite{Li_2023}. These features of the full PDF could be written into the loss function, but the model will likely have much greater difficulty in accurately learning a large number of parameters. An issue with having a vast number of model output parameters in the loss function is that between training steps, the model will bounce around the more difficult to predict values with great volatility as it balances optimizing the full range of parameters. Fixing the imprecision in the difficult loss function terms may require catering to each individual term separately by hand, by adding coefficients to terms in the loss function or training multiple models on subsets of the parameters. Instead, a model architecture that is designed to predict the actual PDF is a more elegant and less error-prone approach.

\section{Probabilistic approach to machine learning}
\label{sec:probabilistic_ml}


The following Chapters will use two types of probabalistic machine learning models, normalizing flows and a denoising diffusion probabilistic model. Normalizing flows will be used to learn the PDF $p(\bm{x})$ in Chapter~\ref{ch:nf_random}, and a PDF conditional on cosmological parameters $p(\bm{x}|\bm{\theta})$ in Chapter~\ref{ch:nf_parameters}. A denoising diffusion model is used in Chapter~\ref{ch:denoising} to sample high-resolution baryonic fields $\bm{y}$ from $p(\bm{y}|\bm{x})$, where $\bm{x}$ is a low-resolution dark matter field.

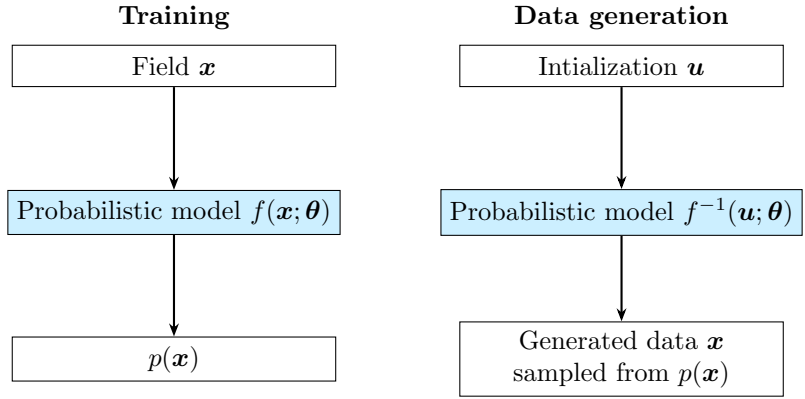
\begin{figure}
\centering
\begin{tikzpicture}[node distance=1.95cm]
\footnotesize
\node[align=center, font=\bfseries] (title){Training};
\node (start) [cnn, below of=title, node distance=0.65cm] {Field $\bm{x}$};
\node (model) [cnn, below of=start, fill=azul!75] {Probabilistic model $f(\bm{x};\bm{\theta})$};
\node (end) [cnn, below of=model] {$p(\bm{x})$};

\draw [arrow] (start.south) to (model.north);
\draw [arrow] (model.south) to (end.north);
\footnotesize
\node[right of=title, xshift=4cm, font=\bfseries] (title){Data generation};
\node (start) [cnn, below of=title, node distance=0.65cm] {Intialization $\bm{u}$};
\node (model) [cnn, below of=start, fill=azul!75, align=center] {Probabilistic model $f^{-1}(\bm{u};\bm{\theta})$};
\node (end) [cnn, below of=model, yshift=-0.cm, align=center] {Generated data $\bm{x}$\\sampled from $p(\bm{x})$};

\draw [arrow] (start.south) to (model.north);
\draw [arrow] (model.south) to (end.north);
\end{tikzpicture}
\caption[Schematic of a probabilistic machine learning model for data generation]{Schematic of probabilistic data generation, where the model $f(\bm{x};\bm{\theta})$ learns a PDF $p(\bm{x})$ from data $\bm{x}$.}
\label{fig:probabilistic_data_generation}
\end{figure}

\begin{figure}
\centering
\begin{tikzpicture}[node distance=1.95cm]
\footnotesize
\node[align=center, font=\bfseries] (title){Training};
\node (start) [cnn, below of=title, node distance=0.65cm] {Field $\bm{x}$ and parameters $\bm{\phi}$};
\node (model) [cnn, below of=start, fill=azul!75, align=center] {Probabilistic model\\$f(\bm{x},\bm{\phi};\bm{\theta})$};
\node (end) [cnn, below of=model] {$p(\bm{x}|\bm{\phi})$};

\draw [arrow] (start.south) to (model.north);
\draw [arrow] (model.south) to (end.north);
\end{tikzpicture}
\hspace{0.5cm}
\begin{tikzpicture}[node distance=1.95cm]
\footnotesize
\node[align=center, font=\bfseries] (title){Parameter inference};
\node (start) [cnn, below of=title, node distance=0.65cm] {Field $\bm{x}$};
\node (model0) [cnn, below of=start, xshift=-2.6cm, fill=azul!75, align=center] {Evaluate $\bm{\phi}$ domain in\\probabilistic model $f(\bm{x},\bm{\phi};\bm{\theta})$};
\node (model1) [cnn, below of=start, xshift=2.6cm, fill=azul!75, align=center] {Monte Carlo on $\bm{\phi}$ in\\probabilistic model $f(\bm{x},\bm{\phi};\bm{\theta})$};
\node (end0) [cnn, below of=model0] {$p(\bm{\phi}|\bm{x})$};
\node (end1) [cnn, below of=model1] {$p(\bm{\phi}|\bm{x})$};

\draw [arrow] (start.south) to (model0.north);
\draw [arrow] (start.south) to (model1.north);
\draw [arrow] (model0.south) to (end0.north);
\draw [arrow] (model1.south) to (end1.north);
\end{tikzpicture}
\caption[Schematic of a probabilistic machine learning model for parameter inference]{Schematic of probabilistic parameter inference, where the model $f(\bm{x},\bm{\phi})$ learns a PDF $p(\bm{x}|\bm{\phi})$ from data $\bm{x}$ having parameters $\bm{\phi}$.}
\label{fig:probabilistic_parameter_inference}
\end{figure}
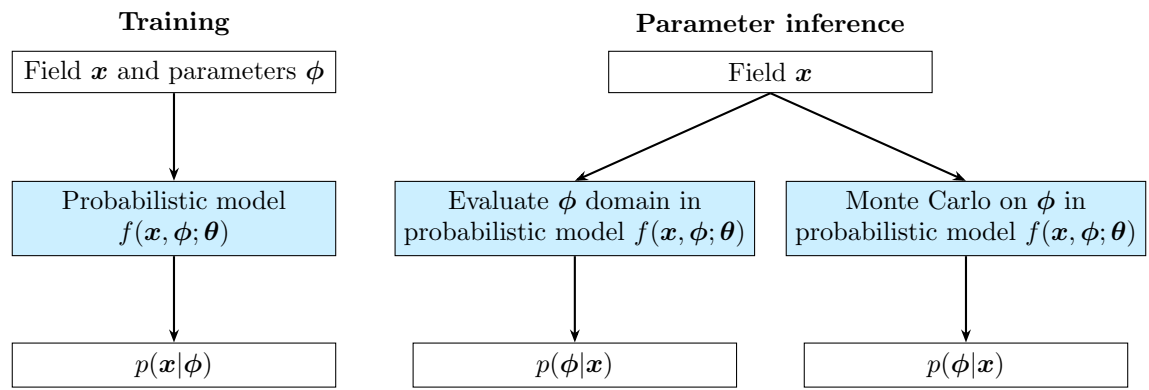

The probabilistic machine learning models used throughout this work have two separate ``phases" for training the model and doing inference with the model, as illustrated in Fig.~\ref{fig:probabilistic_data_generation} and Fig.~\ref{fig:probabilistic_parameter_inference}. The training phase learns the PDF, or some representation of a PDF which can be sampled from. In the inference phase, for the case of data generation, the PDF can be sampled from directly, or Monte Carlo methods may be applied using the PDF to stochastically sample from it. For parameter inference, standard Monte Carlo methods may be used with the PDF; however, as we will see in Chapter~\ref{ch:nf_parameters}, inference with normalizing flows is so fast that we may evaluate the PDF over an entire low-dimensional parameter space. As we will see in the case of normalizing flows and diffusion models, the inference phase is essentially the functional inverse of the training phase. This is quite different than the deterministic models described in the previous Section, which more simply mapped data points to data points.

Two other well-known probabilistic machine learning models not considered in this work include variational autoencoders~\cite{kingma2022autoencoding} and neural score matching~\cite{NEURIPS2019_3001ef25,song2020improved}. Variational autoencoders learn a low-dimensional laten space representation of some high-dimensional data. These latent variables $\bm{z}$ are used to sample from $p(\bm{x}|\bm{z})$, with a feature being that the $\bm{z}$ may be adjusted at inference time to smoothly interpolate between points in the model output data space. Neural score matching is a class of models that do not learn the PDF, but rather the score of the PDF, $\nabla_{\bm{x}}p(\bm{x})$. With these gradients, the maximum-a-posteriori solution to $p(\bm{x})$ may obtained; additionally, annealed Hamiltonian Monte Carlo~\cite{score_matching_2020} or annealed Langevin diffusion~\cite{NEURIPS2019_3001ef25} may be used to generate individual samples from $p(\bm{x})$.



\chapter[Normalizing Flows for Random Fields in Cosmology]{Normalizing Flows for\\Random Fields in Cosmology}
\label{ch:nf_random}
Normalizing flows are a powerful tool to create flexible probability distributions with a wide range of potential applications in cosmology. In this Chapter we study normalizing flows which represent cosmological observables at the field level, rather than at the level of summary statistics such as the power spectrum. We evaluate the performance of different normalizing flows for both density estimation and sampling of near-Gaussian random fields, and check the quality of samples with different statistics such as power spectrum and bispectrum estimators. We explore aspects of these flows that are specific to cosmology, such as flowing from a physically motivated base distribution and evaluating the density estimation results in the analytically tractable correlated Gaussian case.

\section{Introduction}
\label{sec:nf_introduction}

Normalizing flows are a major recent development in probabilistic machine learning (see~\cite{2019arXiv191202762P} for a review), providing a powerful tool to represent flexible and expressive probability densities. Many recent works have shown their usefulness for a wide range of applications, such as image generation~\cite{2018arXiv180703039K,2016arXiv160508803D}, variational inference~\cite{rezende2016variational}, and likelihood-free inference~\cite{2017arXiv170507057P,2019arXiv191101429C}. In cosmology, normalizing flows have recently been used to represent the posterior distribution of summary statistics such as the power spectrum or cosmological parameters~\cite{Alsing:2019xrx,DiazRivero:2020oai}. In the present Chapter, we use normalizing flows to represent the probability density of fields, such as the matter distribution, directly at the field level.

We start with a brief review of normalizing flows. A normalizing flow is a natural way to construct flexible probability distributions by transforming a simple base distribution (often Gaussian) into a complicated target distribution. This is done by applying a series of learned diffeomorphisms to the base distribution. Given a base distribution $p_u(\bm{u})$ of a random variable $\bm{u}$, the target distribution $p_x(\bm{x})$ is given by
\begin{equation}
    p_x(\bm{x})=p_u(\bm{u})\left|\det J_T(\bm{u})\right|^{-1}
\end{equation}
where $T$ is the transformation (or ``flow") $\bm{x}=T(\bm{u})$ and $J_T$ is its Jacobian. We can construct a more complex transformation $T$ by composing a number $K$ of simple transformations $T_k$,
\begin{align}
    T&=T_K(T_{K-1}(...(T_1(\bm{u},\bm{\theta}_1),...),\bm{\theta}_{K-1}),\bm{\theta}_K)\\
    &=T_K\circ\cdots\circ T_1.
\end{align}
These simple transformations must have the property that they have tractable inverses and a tractable Jacobian determinant. They depend on learned parameters and can be parametrized using neural networks. In this way very expressive densities can be constructed. Taking $\bm{z}_0=\bm{u}$ and $\bm{z}_K=\bm{x}$, the transformation at each step $k$ is
\begin{align}
    \bm{z}_{k}=T_k(\bm{z}_{k-1}),
\end{align}
and the Jacobian determinant is
\begin{equation}
    \ln\left|J_{T}(\bm{z})\right|=\ln\left|\prod_{k=1}^KJ_{T_{k}}(\bm{z}_{k-1})\right|=\sum_{k=1}^{K}\ln\left|J_{T_{k}}(\bm{z}_{k-1})\right|.
\end{equation}

\begin{figure}
    \centering
    \includegraphics[width=\textwidth]{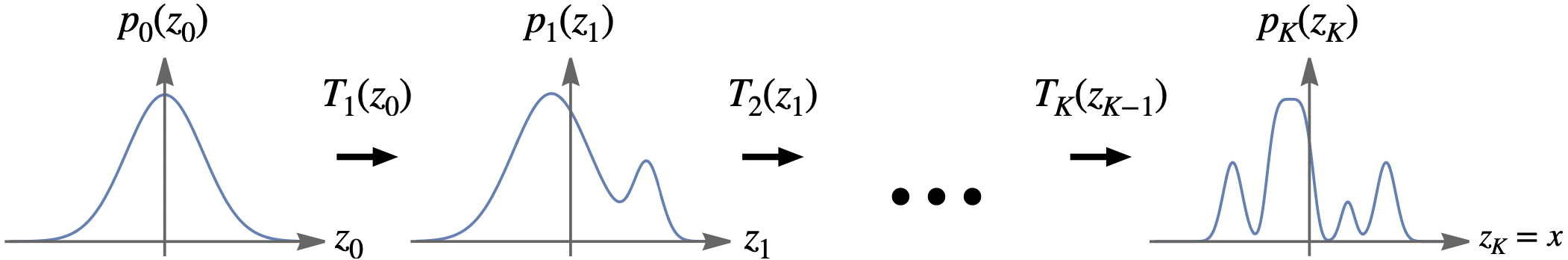}
    \caption[Illustration of a normalizing flow on a single variable]{Illustration of a normalizing flow on a single variable. The transformations $T_k$ transform a Gaussian base distribution $p_0(z_0)$ to a more complicated distribution $p_K(z_K)$.}
    \label{fig:nf_T}
\end{figure}

Once the flow is learned, two basic statistical operations can be performed efficiently: density evaluation (computing $p_x(\bm{x})$ given a sample $\bm{x}$) and sampling from $p_x(\bm{x})$. These operations can be used for statistical inference purposes. The difference between normalizing flows and ordinary neural network techniques are that the former are representing normalized probability densities rather than arbitrary, deterministic mappings from input to output, as explained in Chapter~\ref{ch:introduction} Section~\ref{sec:deterministic_ml}. A number of flow architectures that are applicable to 2-dimensional fields have been proposed~\cite{2019arXiv191202762P}, some of which are able to generate high quality images, as we will see. While normalizing flows are often not quite as expressive as Generative Adverserial Networks (GANs), Variational Autoencoders (VAE), or denoising diffusion models, they come with the statistical interpretation of a probability density function (PDF) and in particular offer exact density evaluation. 

It is a priori very plausible that normalizing flows could be strong at representing PDFs of fields in cosmology. Many fields in cosmology, such as the matter distribution, start as Gaussian fields in the far past. They then become progressively more non-Gaussian with time due to non-linear interactions. In the same way, a Gaussian field base distribution of a normalizing flow becomes progressively more non-Gaussian by the applications of more transformations $T_k$. Such a normalizing flow is therefore a natural candidate to represent somewhat (non-)Gaussian PDFs of matter fields at late times. In principle, normalizing flows are capable of representing any smooth, strictly positive PDF. A constructive proof using nonlinear independent component analysis was presented in~\cite{Hyvarinen:1999}, and written in the language of normalizing flows in~\cite{2019arXiv191202762P}.

Normalizing flows for fields in cosmology may be used for data analysis in several different ways. In particular, they are capable of representing non-Gaussian prior distributions of small-scale fields in cosmology for better posterior analysis. For example, the small-scale Cosmic Microwave Background (CMB) lensing potential~\cite{Lewis:2006fu} is not Gaussian and using a non-Gaussian prior $p(\bm{x})$, learned from simulations, in a MAP or HMC analysis would allow to increase the signal-to-noise under suitable conditions. Such an approach is taken is Chapter~\ref{ch:denoising}. In another direction, variational inference~\cite{2016arXiv160100670B} fits an approximate posterior distribution to samples from the unknown true posterior, and is usually more efficient than Monte Carlo methods. Normalizing flows are ideally suited for variational inference~\cite{rezende2016variational}. In our lensing potential example, this would mean to make a variational estimate of the lensing potential rather than a MAP as in~\cite{Carron:2017mqf} or an HMC sampling as in~\cite{Millea:2020iuw}. Hamiltonian Variational Inference~\cite{2014arXiv1410.6460S} may also be useful for this purpose. In addition, the trained flow can also be used in forward direction for sampling, for example to sample from foreground contamination in CMB data analysis, as recently shown in~\cite{Thorne:2021nux} using variational auto encoders rather than normalizing flows. Before we may use normalizing flows for these challenging practical applications, it is important to quantify their performance on simpler problems. This is the purpose of this study.

\section{Flow architecture and training process}

In this Section we review the architecture of some normalizing flows, explaining the form of the transformations $T_k$ and base distribution $p_u(\bm{u})$. We start with a basic real NVP flow, which we use in the main text of this paper, and then briefly discuss some extensions of it, one of which (Glow) we apply to N-body simulations in Section~\ref{sec:glow} and use in Chapter~\ref{ch:nf_parameters}. In general, the goal of constructing normalizing flows is to make them both flexible as well as computationally tractable, and the flows discussed here have both of these properties. For a more detailed review of the motivation behind these constructions we refer to the original papers.

\subsection{The real NVP flow}
\label{sec:realnvp}

The first flow with success in creating high quality images was the real-valued non-volume preserving (real NVP) flow~\cite{2016arXiv160508803D}. Compared to its predecessors~\cite{2014arXiv14108516D}, this flow is expressive and fast both for sampling and inference. This flow has also recently received some attention for its use to represent PDFs in lattice field theory~\cite{Albergo:2021vyo}. This is the basic flow we will be using in most of this study, and we will consider the more complicated Glow normalizing flow in Section~\ref{sec:glow}. A simplification of our application compared to most implementations of real NVP is that the latter was constructed for RGB images (3 channels) while we represent a scalar field (1 channel); however, a multi-scale architecture that mixes channels will be considered with Glow.

\subsubsection{Affine coupling layers}

The basis of this flow is the affine coupling layer, an operation that rescales and shifts a subset of the random variables depending on the value of the other random variables. First, the layer splits the set of all random variables (in our application all pixels of the random field) denoted as $\bm{x}$, into two subsets $\bm{x}_1$ and $\bm{x}_2$. Then their values are updated as
\begin{align}
    \bm{x}_1'&=e^{\bm{s}(\bm{x}_2)}\bm{x}_1+\bm{t}(\bm{x}_2),\\
    \bm{x}_2'&=\bm{x}_2
\end{align}
Here $\bm{s}$ and $\bm{t}$ are vector valued, and so each pixel in $\bm{x}_1$ is rescaled and shifted separately. This transformation guarantees invertibility as well as a triangular Jacobian which is computationally very easy to evaluate and invert. While the scaling transformation is simple, its flexibility comes from the free form of the functions $\bm{s}$ and $\bm{t}$ and stacking many such layers, with different partitions into subsets. The Jacobian matrix of this transformation $T_k$ is
\begin{equation}
    J_{ij}(\bm{x})=\frac{\partial T_k(x_i)}{\partial x_j}=
    \begin{pmatrix}
        \bm{I}_{\frac{d}{2}\times\frac{d}{2}} & \bm{0}_{\frac{d}{2}\times\frac{d}{2}}\\
        \frac{\partial T(x_k)}{\partial x_i} & \text{diag}{\left(e^{s(x_l)}\right)}.
    \end{pmatrix}
\end{equation}
As required, the determinant is easy to compute, being the product of the diagonal.

\subsubsection*{Checkerboard masking}

For images, an elegant choice of partition into subsets $\bm{x}_1$ and $\bm{x}_2$ is the checkerboard masking proposed in \cite{2016arXiv160508803D}. On the checkerboard, we either use the white pixels for $\bm{x}_1$ and the black pixels for $\bm{x}_2$ or vice versa. Between each consecutive affine coupling layer we switch the sets $\bm{x}_1$ and $\bm{x}_2$, allowing every pixel to be transformed.

\begin{figure}
    \centering
    \includegraphics[scale=0.5]{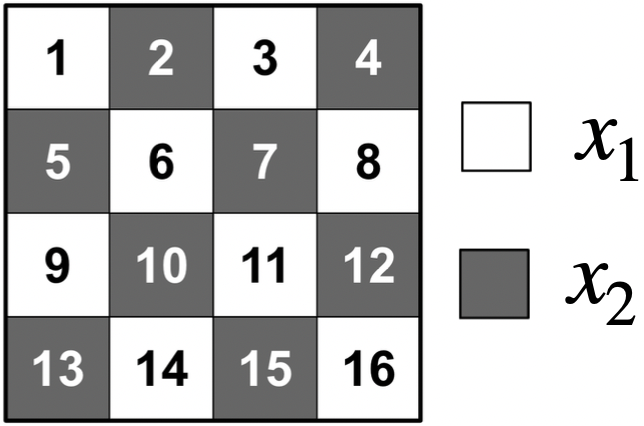}
    \caption[Checkerboard mask used in the real NVP normalizing flow]{Checkerboard mask used in the real NVP normalizing flow}
    \label{fig:checkerboard_mask}
\end{figure}

\subsubsection*{CNN to determine the affine parameters}

We now need to define the functions $\bm{s}(\bm{x}_2)$ and $\bm{t}(\bm{x}_2)$. As $\bm{x}_2$ are spatially organized, rather than arbitrary collections of random variables, it is natural to use a standard convolutional neural network for this purpose (as was done in \cite{2016arXiv160508803D}) which enforces translational symmetry. As the normalizing flow learns probabilities, the CNN needs to conserve dimensions, so we use stride 1 convolutions with and no pooling (Appendix~\ref{sec:convolutions}). To implement periodic boundary conditions we use the common approach of circular padding. As in~\cite{Albergo:2021vyo} we use 3 convolutional layers with kernel size 3 and leaky ReLU activation functions. The number of channels is in this order: 1 (the scalar PDF values), 16 (arbitrary number of feature maps), 16 (arbitrary number of feature maps), 2 (the output variables $\bm{s}$ and $\bm{t}$).

\subsubsection*{Stacking the layers}

As our default configuration, as in \cite{Albergo:2021vyo}, we stack $K=16$ affine coupling layers, each with their own CNN to parametrise the affine transformation $\bm{s}$ and $\bm{t}$.

\begin{figure}
\centering
\begin{tikzpicture}[node distance=0.65cm]
\footnotesize

\pgfdeclarelayer{bg}
\pgfsetlayers{bg, main}

\node[align=center, font=\bfseries] (title){Real NVP};
\node (start) [cnn, below of=title] {Training data};
\node (cb0) [cnn, below of=start, fill=black!15] {Checkerboard mask};
\node (cnn0) [cnn, below of=cb0, fill=azul!75] {$\bm{s},\bm{t}$ CNN};
\node (at0) [cnn, below of=cnn0, fill=orange!30] {Affine transformation};
\node (end) [cnn, below of=at0, fill=red!0] {Gaussian field};

\draw [arrow] (start.east) to [out=-30,in=30](cb0.east);
\draw [arrow] (cb0.east) to [out=-30,in=30](cnn0.east);
\draw [arrow] (cnn0.east) to [out=-30,in=30](at0.east);
\draw [arrow] (at0.west) to [out=135,in=-135](cb0.west) node[yshift=-0.68cm] {\hspace{-1.3cm}$\times L$};
\draw [arrow] (at0.east) to [out=-30,in=30](end.east);

\end{tikzpicture}
\hspace{1cm}
\begin{tikzpicture}[node distance=0.65cm]
\footnotesize

\pgfdeclarelayer{bg}
\pgfsetlayers{bg, main}

\node[align=center, font=\bfseries] (title){Reverse real NVP};
\node (start) [cnn, below of=title] {Gaussian field};
\node (cb0) [cnn, below of=start, fill=black!15] {Checkerboard mask};
\node (cnn0) [cnn, below of=cb0, fill=azul!75] {$\bm{s},\bm{t}$ CNN};
\node (at0) [cnn, below of=cnn0, fill=orange!30] {Inv. affine transformation};
\node (end) [cnn, below of=at0, fill=red!0] {Generated data};

\draw [arrow] (start.east) to [out=-30,in=30](cb0.east);
\draw [arrow] (cb0.east) to [out=-30,in=30](cnn0.east);
\draw [arrow] (cnn0.east) to [out=-30,in=30](at0.east);
\draw [arrow] (at0.west) to [out=135,in=-135](cb0.west) node[yshift=-0.68cm] {\hspace{-1.3cm}$\times L$};
\draw [arrow] (at0.east) to [out=-30,in=30](end.east);

\end{tikzpicture}

\ 

\begin{tikzpicture}[node distance=0.65cm]
\footnotesize

\node[align=center, font=\bfseries] (title){$\bm{s},\bm{t}$ CNN};
\node (conv0) [cnn, below of=title] {Image: $1\times n^2$};
\node (conv1) [cnn, below of=conv0, fill=azul!75] {$3\times3$ conv: $c\times n^2$};
\node (conv2) [cnn, below of=conv1, fill=azul!75] {$3\times3$ conv: $c\times n^2$};
\node (conv3) [cnn, below of=conv2, fill=azul!75] {$3\times3$ conv: $2\times n^2$};
\node (st) [cnn, below of=conv3] {$\bm{s},\bm{t}$: $2\times n^2$};

\draw [arrow] (conv0.east) to [out=-30,in=30](conv1.east);
\draw [arrow] (conv1.east) to [out=-30,in=30](conv2.east) node[yshift=0.35cm] {\hspace{2.4cm}Leaky ReLU};
\draw [arrow] (conv2.east) to [out=-30,in=30](conv3.east) node[yshift=0.35cm] {\hspace{2.4cm}Leaky ReLU};
\draw [arrow] (conv3.east) to [out=-30,in=30](st.east) node[yshift=0.35cm] {\hspace{1.3cm}Tanh};

\end{tikzpicture}

\caption[Real NVP normalizing flow and CNN architectures]{(Top left) The real NVP flow learns the Gaussian base distribution through a series of affine transformations. The checkerboard mask reverses parity after every layer $L$. (Top right) The reverse real NVP flow generates a complicated distribution from a Gaussian base distribution through a series of inverse affine transformations. (Bottom) The CNN used to find $\bm{s},\bm{t}$ in the (inverse) affine transformations. The CNN layers are listed with output channel $\times$ spacial dimensions.}

\label{fig:RNVP_tikz}
\end{figure}
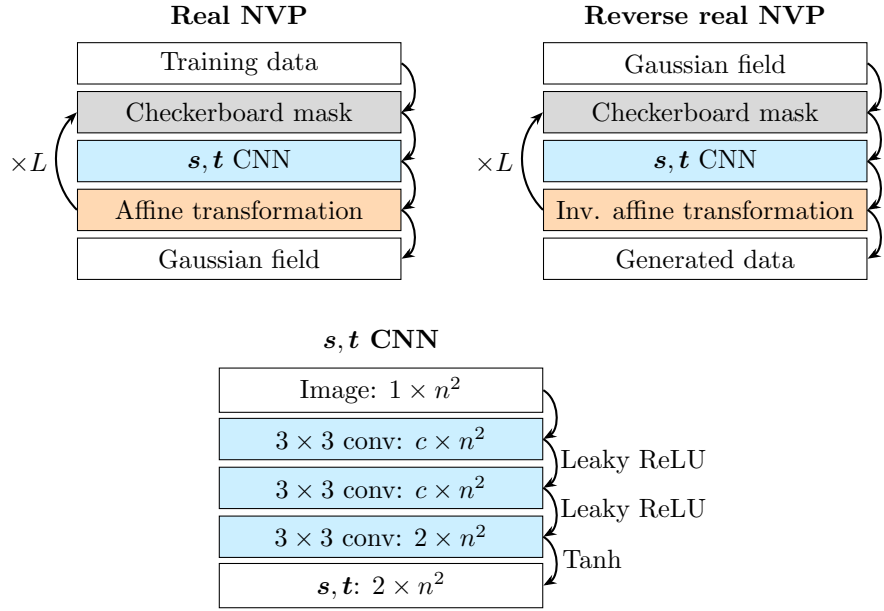

This architecture has a total of 44,320 trainable parameters. Our PyTorch~\cite{NEURIPS2019_9015} implementation of this normalizing flow is taken from~\cite{Albergo:2021vyo}, with minor modifications such as a different loss function and different base distributions (to be discussed Section~\ref{sec:correlated_prior}). An illustration of the architecture is shown in Fig.~\ref{fig:RNVP_tikz}.

\subsection{Other flows of interest}

\begin{figure}
    \centering
    \includegraphics[scale=0.4]{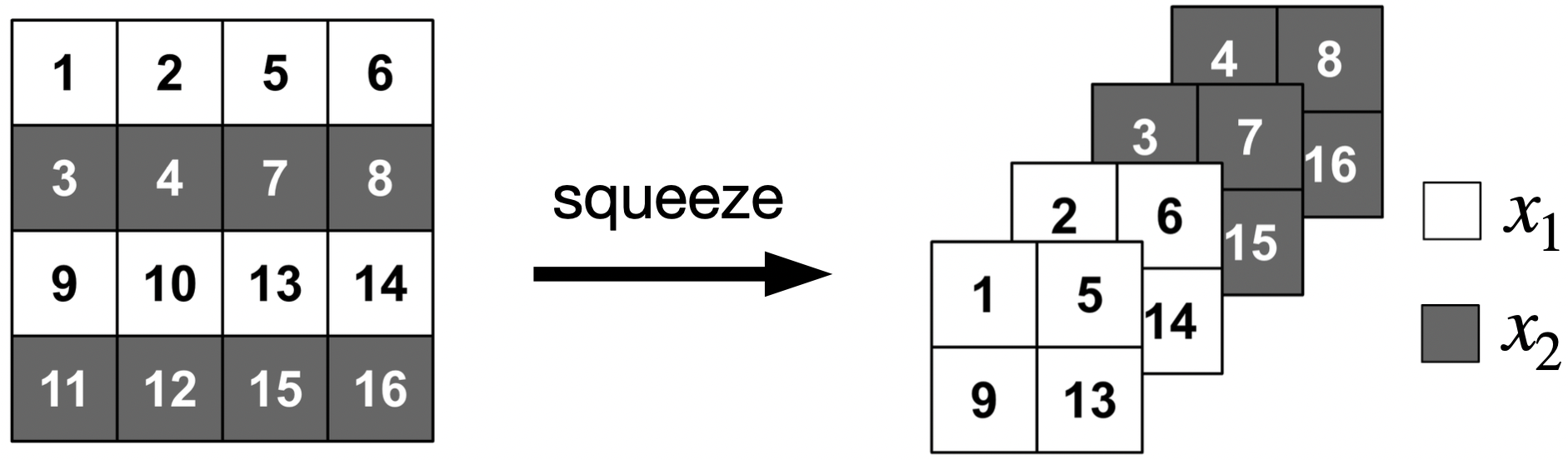}
    \caption[Multi-scale architecture used in the Glow normalizing flow]{The multi-scale architecture used in Glow is able to correlate distant pixels more easily than the checkerboard masking.}
    \label{fig:squeeze}
\end{figure}

There are a number of modifications to the real NVP flow just presented. Real NVP can have a multi-scale architecture in place of the checkerboard mask; the original real NVP paper~\cite{2016arXiv160508803D} used a multi-scale architecture to help with representing larger images. The idea is to reshape the random field to a smaller spatial size while proportionally increasing the number of channels, as shown in Fig.~\ref{fig:squeeze}, thus conserving the number of dimensions as required. One can then mix the channels in a number of ways.

A well-known extension of the real NVP flow with multi-scale architecture is the Glow~\cite{2018arXiv180703039K} normalizing flow. Glow mixes channels with invertible $1\times1$ convolutions and introduces a modified batch normalization layer. A more detailed explanation of Glow with results on N-body simulations is in Section~\ref{sec:glow}.

The normalizing flow transformations may also be done in Fourier space, where the invertible transformation on fields $\bm{x}$ is
\begin{equation}
    f=\Psi\left(\mathcal{F}^{-1}[\tilde{T}(k)\cdot\tilde{\bm{x}}(\bm{k})]\right).
\end{equation}
Here $\Psi$ is monotonically increasing as required for invertibility, $\tilde{T}$ is positive, and $\tilde{y}\equiv\mathcal{F}{(y)}$. The product between $\tilde{T}(k)$ and $\tilde{\bm{x}}$ is a simple element-wise multiplication. The absolute value of the Jacobian determinant for such a transformation is
\begin{equation}
    \left|\frac{\text{d}f}{\text{d}x}\right|=\prod_i^\text{pixels}\frac{\partial\Psi\left(\mathcal{F}^{-1}[\tilde{T}\cdot\tilde{\bm{x}}]\right)}{\partial\left(\mathcal{F}^{-1}[\tilde{T}\cdot\tilde{\bm{x}}]\right)}\prod_j^{k\ \text{modes}}\frac{\text{d}\left(\mathcal{F}^{-1}[\tilde{T}\cdot\tilde{\bm{x}}]\right)}{\text{d}\bm{x}}.
\end{equation}
The second set of factors simplifies as
\begin{align}
    \frac{\text{d}\left(\mathcal{F}^{-1}[\tilde{T}\cdot\tilde{\bm{x}}]\right)}{\text{d}x}&=\frac{\partial\left(\mathcal{F}^{-1}[\tilde{T}\cdot\tilde{\bm{x}}]\right)}{\partial(\tilde{T}\cdot\tilde{\bm{x}})}\frac{\partial(\tilde{T}\cdot\tilde{\bm{x}})}{\partial\tilde{x}}\frac{\text{d}\tilde{\bm{x}}}{\text{d}x}\\
    &=e^{i\bm{k}\cdot\bm{r}}\tilde{T}e^{-i\bm{k}\cdot\bm{r}}\\
    &=\tilde{T}.
\end{align}
Therefore, the Jacobian determinant is
\begin{equation}
    \left|\frac{\text{d}f}{\text{d}x}\right|=\prod_i^\text{pixels}\frac{\partial\Psi\left(\mathcal{F}^{-1}[\tilde{T}\cdot\tilde{\bm{x}}]\right)}{\partial\left(\mathcal{F}^{-1}[\tilde{T}\cdot\tilde{\bm{x}}]\right)}\prod_j^{k\ \text{modes}}\tilde{T}(k_j).
\end{equation}
Such a transformation was used in the Translationally and Rotationally Equivariant Flow (TRENF)~\cite{trenf}, showcased on dark matter fields.

\subsection{Flow training}

To train the flow, we minimize the forward KL divergence introduced in Chapter~\ref{ch:introduction} Section~\ref{sec:KL-div}, a measure of the relative entropy between the target distribution $p^*_{\bm{x}}(\bm{x})$ and the transformed base distribution $p_{\bm{x}}(\bm{x})$. In the notation of this Chapter, the forward KL divergence can be expressed as
\begin{align}
    \mathcal{L}(\bm{x},\bm{u};\bm{\theta})&=D_{\text{KL}}\big(p^*_{\bm{x}}(\bm{x})\ ||\ p_{\bm{x}}\left(\bm{x};\bm{\theta}\right)\big)\\
    &=-\mathbb{E}_{p^*_{\bm{x}}(\bm{x})}\big(\ln p_{\bm{x}}(\bm{x};\bm{\theta})-\ln p^*_{\bm{x}}(\bm{x};\bm{\theta})\big)\\
    &=-\mathbb{E}_{p^*_{\bm{x}}(\bm{x})}\big(\ln p_{\bm{u}}\left(T^{-1}\left(\bm{x};\bm{\theta}\right)\right)+\ln\left|\det{J_{T^{-1}}(\bm{x};\bm{\theta})}\right|\big)+\mathbb{E}_{p^*_{\bm{x}}(\bm{x})}\ln p^*_{\bm{x}}(\bm{x};\bm{\theta})
\end{align}
where $T$ is the flow transformation, with learned parameters $\bm{\theta}$. The final term here is a constant that we do not need to calculate. Here ``forward" denotes the order of $p^*_x$ and $p_x$ above. The expectation values are estimated as
\begin{equation}\label{eq:loss_kl}
    \mathcal{L}\left(\bm{\theta}\right)=-\frac{1}{N}\sum_{n=1}^N\left(\log p_u\left(T^{-1}\left(\bm{x}_n;\bm{\theta}\right)\right)+\log\left|\det J_{T^{-1}}(\bm{x}_n;\bm{\theta})\right|\right)+\text{constant},
\end{equation}
where $\bm{x}_n$ are the training samples from $p^*_{\bm{x}}(\bm{x})$. Minimizing the Monte Carlo approximation of the KL divergence is thus equivalent to fitting the flow model to the training samples by maximum likelihood estimation.

For the Sections that follow, we use the Adam optimizer (Appendix~\ref{sec:sgd}) to minimize the loss with respect to the parameters $\bm{\theta}$. We use a learning rate of $10^{-3}$ and batch size of $128$.

\section{Application to Gaussian fields}

As the first toy application, we show that we can flow from random uncorrelated Gaussian noise to a correlated Gaussian random field with a known power spectrum. We verify both the quality of the samples, as well as the quality of the density estimation operation. While in the Gaussian case we can sample efficiently and evaluate the density analytically and the flow treatment has thus no practical value, it is the basis for extensions to intractable non-Gaussian fields.

\subsection{Sample quality}

\begin{figure}[h!]
\centering
Base distribution samples

\includegraphics[width=\textwidth, trim=0 10 0 0.65cm, clip]{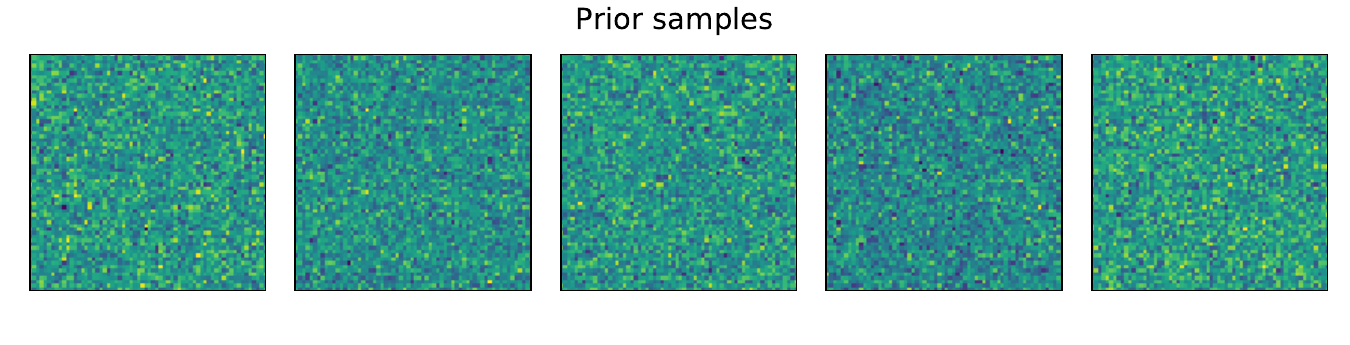}

Model samples

\includegraphics[width=\textwidth, trim=0 10 0 0.65cm, clip]{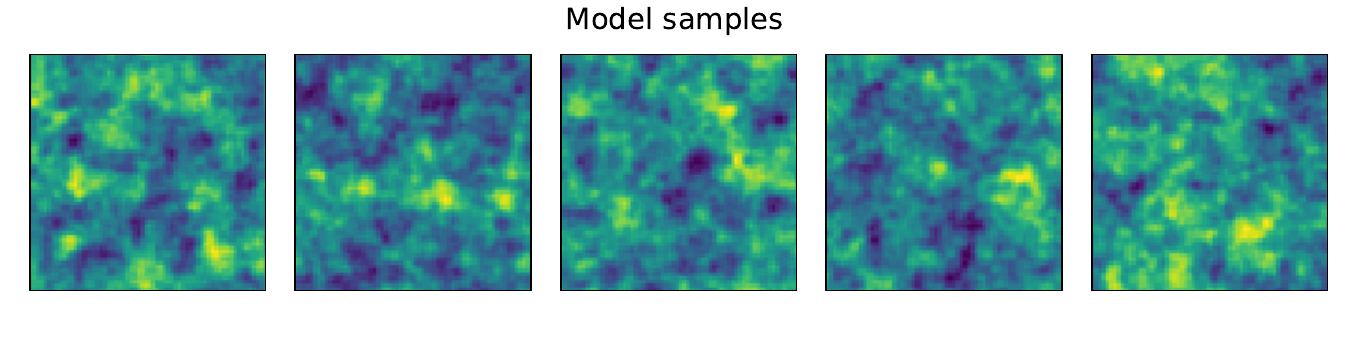}

Target samples

\includegraphics[width=\textwidth, trim=0 10 0 0.65cm, clip]{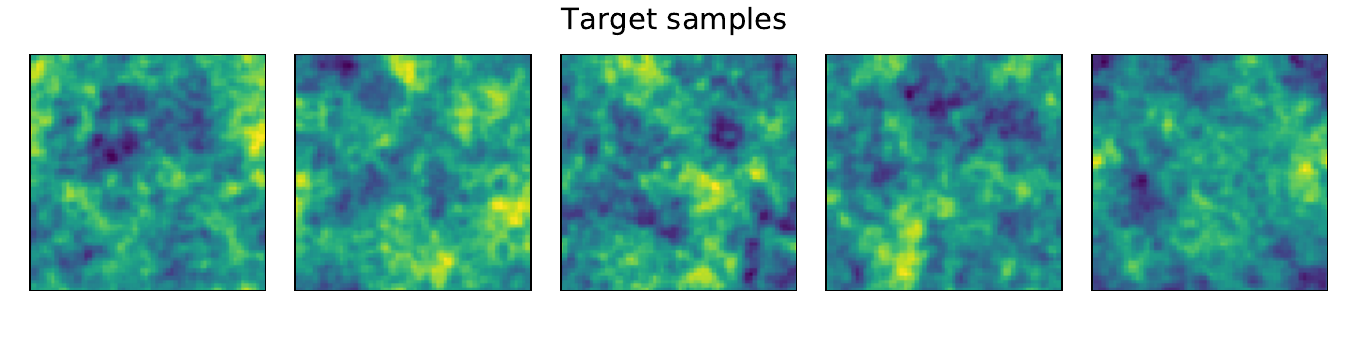}
\caption[Gaussian fields samples trained on real NVP]{Gaussian field samples with a CMB power spectrum. Random samples from the base distribution (top), corresponding flow samples (middle), and training samples (bottom). The training data is using a CMB power spectrum on a 4 degree sky patch on $64\times64$ pixels.}
\label{fig:gausssamples1}
\end{figure}

\begin{figure}[h!]
\centering
    \includegraphics[width=0.495\textwidth]{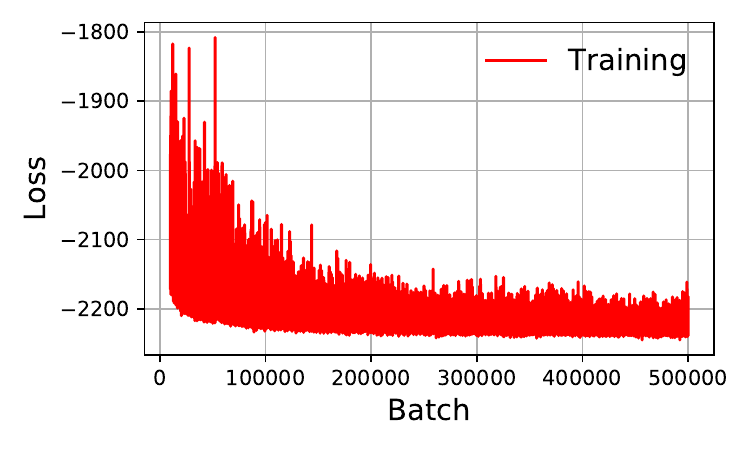}
    \includegraphics[width=0.495\textwidth]{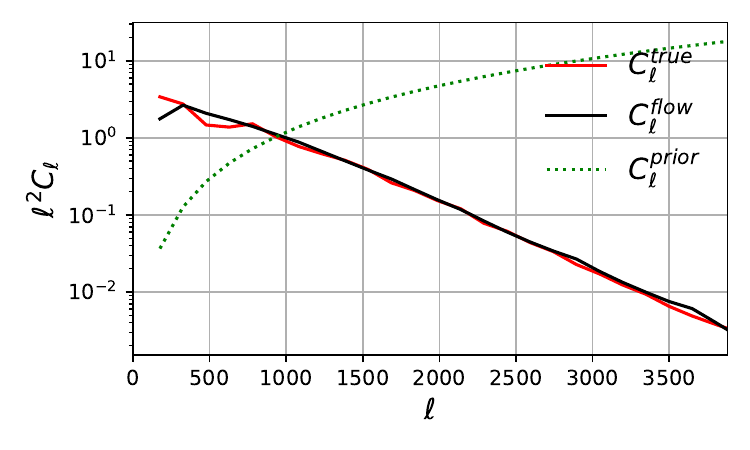}\\
    \includegraphics[width=0.495\textwidth]{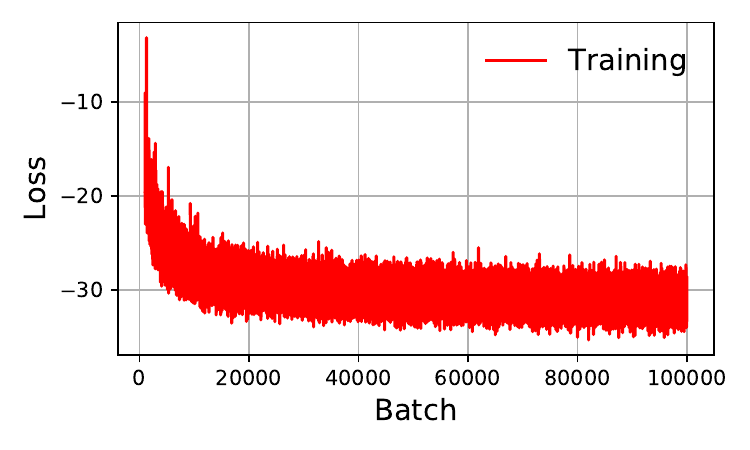}
    \includegraphics[width=0.495\textwidth]{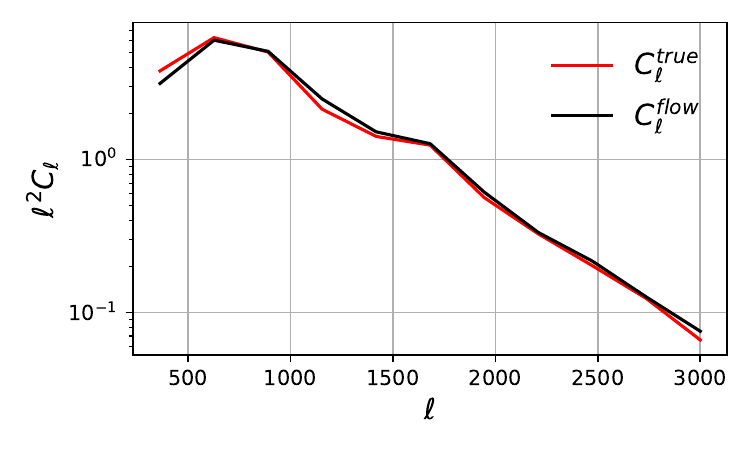}\\
    \caption[Loss convergence for Gaussian fields trained with real NVP]{Top left: Loss convergence of the $64$~px side length Gaussian field training (starting from batch 10,000). Training samples are generated on the fly, so there is no validation set. Top right: Comparison of the base, model, and training set power spectrum, averaged over 10,000 maps. The x-axis is the multipole expansion coefficient $\ell$. While the model power spectrum is generally correct, the $64$~px case (top row) does not resolve the BAO wiggle. On the bottom we show the same plots for $16$~px maps. Here the BAO wiggle is resolved better. We improve BAO and training convergence results below.}
    \label{fig:gaussps}
\end{figure}

We train the real NVP flow described in Section~\ref{sec:realnvp} on samples from a correlated Gaussian field, with a CMB temperature power spectrum. We use this as an example power spectrum from cosmology with baryon acoustic oscillations (BAO) features in it, but we are not interested in representing Gaussian primary CMB for practical applications. Here we flow from an uncorrelated (or noise) Gaussian base distribution. We use patches of $64\times64$ pixels covering a sky angle side length of 4 degrees. We have experimented with infinite ``on the fly" generated training samples as well as fixed size training samples with 10,000 or 1,000 samples. Sampling random correlated Gaussian fields with a given power spectrum is explained in Appendix~\ref{sec:gaussian_random_fields}, the computational expense being a single fast Fourier transform of the field. We show the infinite case here, but we only find a difference with limited training data for the density estimation task which we discuss below. Samples from the base, model, and target (training) distributions are shown in Fig.~\ref{fig:gausssamples1}. By eye, the model samples look like the training data. We also compare their power spectrum in Fig.~\ref{fig:gaussps} and find good agreement. The power spectrum has however not fully resolved the BAO feature. It is likely that with longer training and more network capacity this could be improved. However, we will fix this problem in the next Section more elegantly using a correlated base distribution, and in Section~\ref{sec:glow} using a different flow architecture. The total training time was about 40 hours on an RTX 3090 for the $64$~px maps, and uses 3.8 GB of GPU memory. We will also find much faster convergence with a correlated base distribution in the next section.

\subsection{Density evaluation}

For some of the data analysis applications discussed in the introduction, we need to use the flow in reverse direction for density evaluation. There are two different tasks we can consider here: in-distribution density evaluation (IID: independent, identical distribution) and out-of-distribution density evaluation. We focus here on IID density evaluation, leaving out-of-distribution to subsequent Chapters. 

\begin{figure}[h!]
\centering
\rotatebox{90}{\small\hspace{1.3cm}$\ln{p(\bm{x})}$}
\includegraphics[width=0.95\textwidth, trim={1cm 1cm 0 0}, clip]{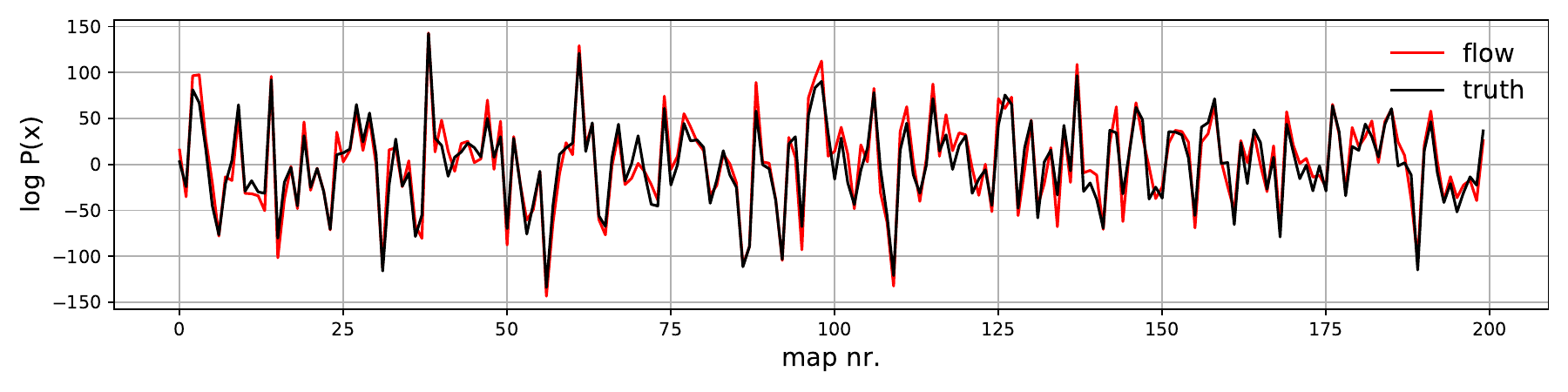}

\hspace{1.05cm}\small Map number
\caption[PDF evaluation for Gaussian random fields from the real NVP flow compared to the true probability of the sample]{Density evaluation $\ln p(x)$ result from the flow compared to the true log probability of the sample (mean shifted to zero), for 200 maps with $64\times64$ pixel resolution. The cross-correlation coefficient is 0.993 (estimated from 10,000 maps). The flow has thus learned density evaluation on new IID samples. Here we used on the fly created training data, see main text for results with limited training data.}
\label{fig:density_estimate_gauss}
\end{figure}

We would like to verify that IID samples, when run backwards through the flow (with uncorrelated Gaussian noise base distribution), are assigned a probability that corresponds to their true probability. Starting with a tractable Gaussian distribution allows us to compare the flow probability with the exact analytic probability. We sample 10,000 new IID samples $\bm{x}$ from the same distribution as the training data and reverse flow them to obtain $\ln{p_{\text{flow}}(\bm{x})}$. For the configuration above, we find that the cross correlation coefficient between $\ln{p_{\text{flow}}(\bm{x})}$ and $\ln{p_{\text{true}}(\bm{x})}$ is $r\approx0.993$. We plot these quantities for 200 random example maps in Fig.~\ref{fig:density_estimate_gauss}. We thus find that the flow can be used for density evaluation in reverse mode with high accuracy.

We examined whether precise density evaluation results were only possible due to the on the fly created training data, which ensured that the training data covered a large number of independent points in sample space. Indeed we found that by training on 1,000 Gaussian maps, the cross correlation coefficient dropped to about $r\approx0.97$, while with 10,000 Gaussian maps we found $r\approx0.98$. At the level of the power spectrum we found no difference with respect to the on the fly created training data, and we did not reach a domain of over-training with the limited training data. Generating non-Gaussian training data can be computationally expensive in practice. Depending on the size of the training data set, different sample augmentation techniques could be used to get the best possible density evaluation results after training.

\section{Local non-Gaussianity and correlated base distribution}
\label{sec:correlated_prior}

After verifying the flow performance in the simple case of an analytically tractable correlated Gaussian field, we now consider perhaps the simplest non-Gaussian random field in cosmology. Local non-Gaussianity \cite{Komatsu:2001rj} is generated by transforming a correlated Gaussian field $\bm{u}_\text{G}(x)$ as
\begin{align}
\label{eq:fnlmapmake}
    \bm{x}_\text{NG}(\bm{r})=\bm{u}_\text{G}(\bm{r})+\tilde{f}_{\text{NL}}^{\text{local}}\left(\bm{u}_\text{G}^2(\bm{r})-\langle\bm{u}_\text{G}^2(\bm{r})\rangle\right)
\end{align}
It is easy to draw samples from this non-Gaussian distribution $\bm{x}_\text{NG}(\bm{r})$, since all we need to do is to square the Gaussian field and add it to the original field with some amplitude. This form of non-Gaussianity is generated in cosmology for example by multi-field inflation~\cite{Byrnes:2010em}. We will discuss a second form of non-Gaussianity in Appendix~\ref{sec:nongaussestim}. Unlike in cosmology, here we normalize the random field so that $\bm{u}_\text{G}$ has variance $1$, so that $\tilde{f}_\text{NL}^{\text{local}} = 1$ indicates $\mathcal{O}(1)$ non-Gaussianity. To make this difference clear we have added the tilde to the notation. We are primarily interested in strongly non-Gaussian fields here, induced by gravitation and astrophysics, not in the tiny effects generated during inflation.

\begin{figure}[h!]
\centering
Base distribution samples

\includegraphics[width=\textwidth, trim=0 10 0 0.65cm, clip]{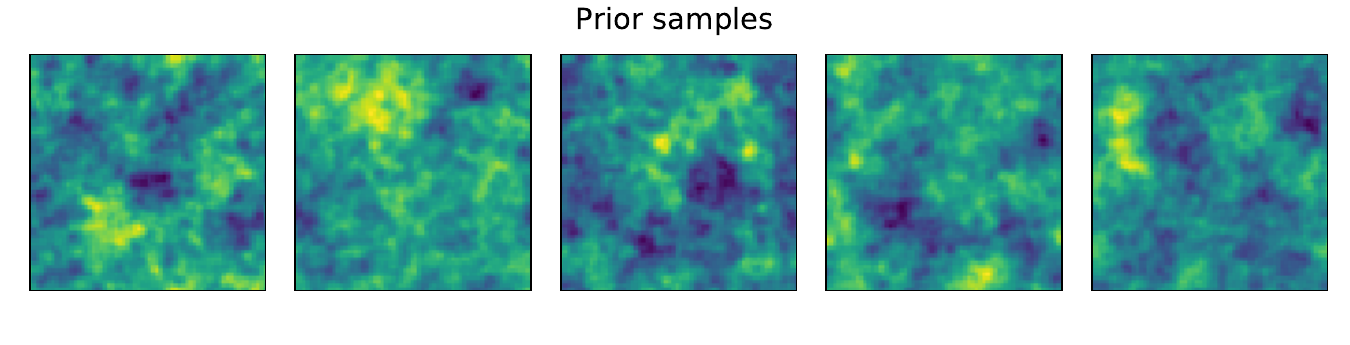}

Model samples

\includegraphics[width=\textwidth, trim=0 10 0 0.65cm, clip]{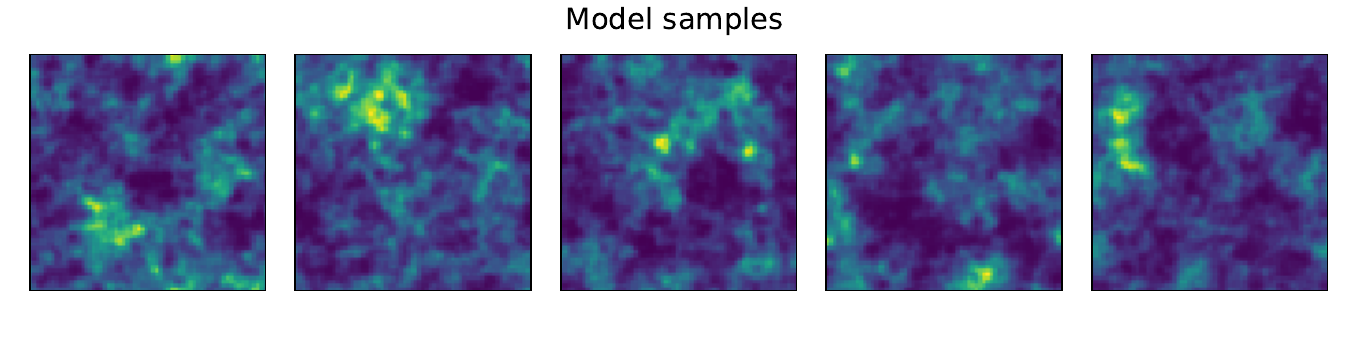}

Target samples

\includegraphics[width=\textwidth, trim=0 10 0 0.65cm, clip]{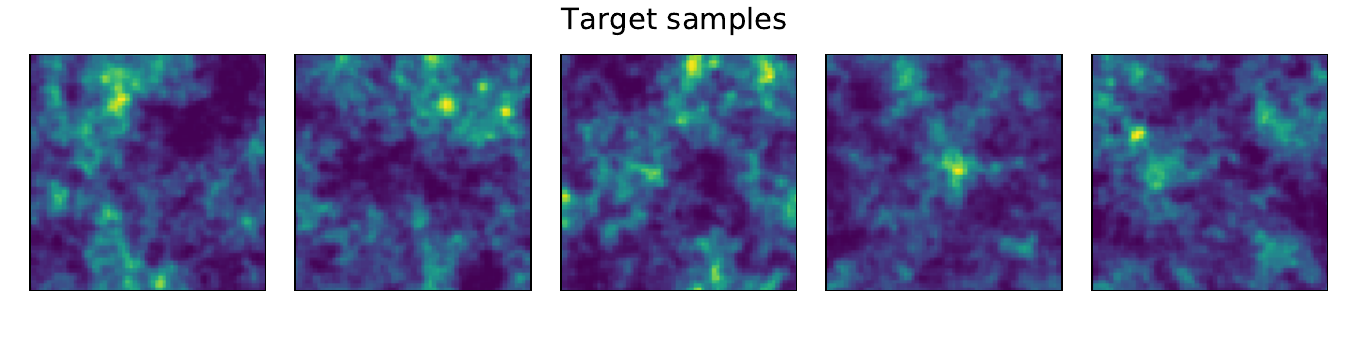}
\caption[Local non-Gaussianity field samples trained on real NVP]{Local non-Gaussianity field samples with $\tilde{f}_\text{NL}^{\text{local}} = 0.2$. Gaussian base distribution samples (top), flow samples (middle), and training samples (bottom). The training data is generated by drawing Gaussian maps from a CMB power spectrum on a 4 degree sky patch represented on $64\times64$ pixels and making them non-Gaussian with Eq.~\ref{eq:fnlmapmake}. Comparing the base and model samples, we find that the network makes the Gaussian base samples more non-Gaussian, with maxima and minima becoming more pronounced.}
\label{fig:fnlsamples1}
\end{figure}

\begin{figure}[h!]
\centering
\includegraphics[width=0.495\textwidth]{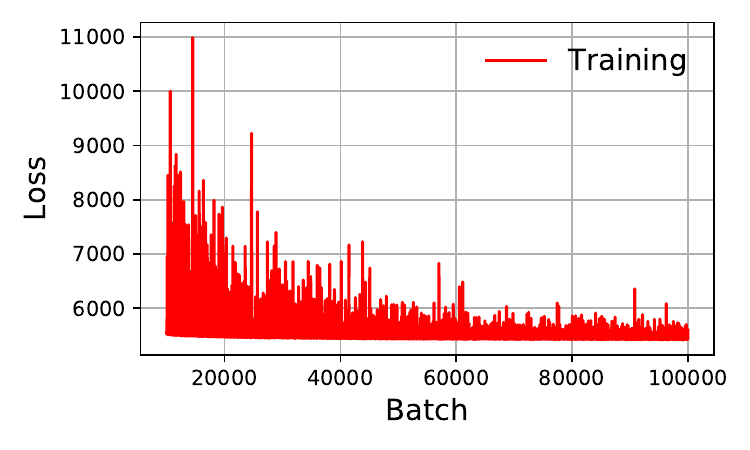}
\includegraphics[width=0.495\textwidth]{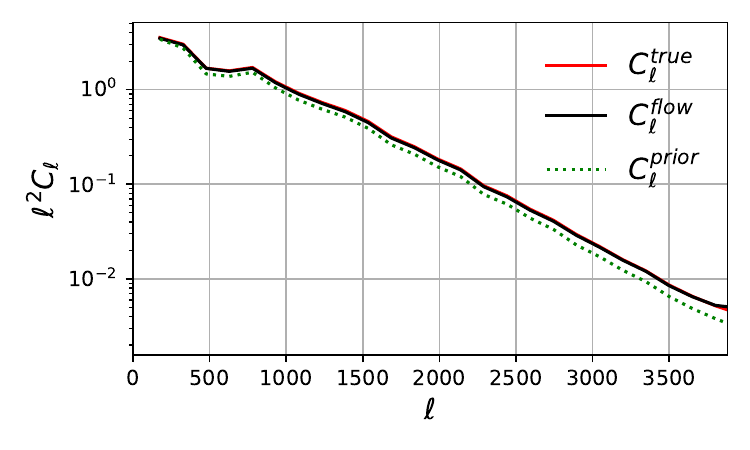}\\
\caption[Loss convergence for local non-Gaussian fields trained with real NVP]{Left: Loss during training (starting from batch 10,000) for the  $f_\text{NL}=0.2$ example with correlated base distribution. Convergence is about 10 times faster than with the random uncorrelated prior. Right: Comparison of the base, model, and training set power spectrum averaged over 10,000 maps, for the $f_\text{NL}=0.2$ example. With the correlated base, the BAO wiggle is resolved almost perfectly.}\label{fig:fnlps}
\end{figure}

We train the same flow model as in the previous section, with the same training parameters. However, we now flow from a correlated Gaussian field with the correct power spectrum. This makes the training much more efficient as the flow has a better starting point. In practice for the non-Gaussian fields of interest such as the matter field, it always makes sense to flow from a Gaussian field with the right power spectrum. We show samples from the base, model, and training distributions in Fig.~\ref{fig:fnlsamples1}. As for the Gaussian case, we find that the samples are indistinguishable by eye from the training data. We show the power spectrum of the samples in Fig.~\ref{fig:fnlps} (right), including the power spectrum of the base distribution that now matches the training data. The flow thus learns to induce the correct non-Gaussianity while keeping the power spectrum intact. The correlated base distribution also improves the training convergence. The flow trains roughly 10 times faster than with the uncorrelated base, as illustrated by the training loss in Fig.~\ref{fig:fnlps} (left).

We now measure the non-Gaussianity in the samples with the method reviewed in Appendix~\ref{sec:nongaussestim}. We estimate the amplitude of two non-Gaussian templates: the local and equilateral non-Gaussianity. In the present case the training data has local non-Gaussianity by definition, however the equilateral template has some overlap with the local template, and it is useful to measure both in the model samples whose non-Gaussianity can be non-local. We find the following mean estimated values, averaging over 10,000 training and flow samples with $64$~px resolution:

\begin{center}
\begin{tabular}{cccc}
    \hline
    $\tilde{f}_\text{NL}^{\text{local}}$ training & $\tilde{f}_\text{NL}^{\text{local}}$ flow & $\tilde{f}_\text{NL}^{\text{equi}}$ training & $\tilde{f}_\text{NL}^{\text{equi}}$ flow\\
    \hline
    0.2 & 0.187 & 0.150 & 0.146\\
    0.05 & 0.047 & 0.037 & 0.035\\
    \hline
\end{tabular}
\end{center}

We are also interested in the per sample variance of the estimated non-Gaussianity:

\begin{center}
\begin{tabular}{ccccc}
    \hline
    $\tilde{f}_\text{NL}^{\text{local}}$ training & $\sigma(\tilde{f}_\text{NL}^{\text{local}})$ training & $\sigma(\tilde{f}_\text{NL}^{\text{local}})$ flow & $\sigma(\tilde{f}_\text{NL}^{\text{equi}})$ training & $\sigma(\tilde{f}_\text{NL}^{\text{equi}})$ flow\\
    \hline
    0.2 & 0.074 & 0.085 & 0.044 & 0.040\\
    0.05 & 0.024 & 0.029 & 0.017 & 0.017\\
    \hline
\end{tabular}
\end{center}

We find that the flow is accurate to about $5\%$ in the mean and to about $10\%$ in the per sample variance. As discussed in more detail in Appendix~\ref{sec:nongaussestim} we have normalized the non-Gaussianity estimator from simulations, so that we recover the correct $\tilde{f}_\text{NL}^{\text{local}}$ value in the training data. We are probing the strongly non-Gaussian regime here, so the variance depends on the mean value, rather than being constant as in the weakly non-Gaussian case.

In this section we have used a correlated base distribution with a physical power spectrum. When implementing the correlated base distribution, it is important not to over-count degrees of freedom when evaluating $p(\bm{u})$. We are using the \texttt{rfftn} function in PyTorch, which uses a slightly degenerate data representation, meaning there is some symmetry in the array of Fourier coefficients. We carefully masked out degenerate information to calculate a correct density value (Appendix~\ref{sec:fourier_transforms}).

\section{Non-Gaussian fields from N-body simulations}
\label{sec:nbodyflow}

Finally we tackle the important use case of representing field PDFs from N-body simuations. We use 100 high-resolution simulations from the publicly available Quijote \cite{Villaescusa-Navarro:2019bje} suite of N-body simulation data to generate training patches of the projected matter field. The simulations were run in a box size of $1\ \text{Gpc}/h$ and use $1024^3$ dark matter particles. We use snapshots generated at $z=2$ to estimate the matter density in the 3-dimensional simulation volume by painting the dark matter particles on a 3-dimensional mesh of size $1024\times1024\times1024$ using the cloud-in-cell~\cite{1981csup.book.....H} mass assignment scheme as implemented in \texttt{nbodykit}~\cite{Hand:2017pqn}. With this particular choice for the mesh size and redshift, we are able to resolve scales of size $k\gtrsim1\ h/\text{Mpc}$, which are well into the non-linear regime. We subsequently divide the simulation volume into subboxes of size $128\times128\times128$ and project the density along a dimension to get a realization of a 2-dimensional field. We do not perform any data augmentation on these patches and generated a total of 51,200 independent patches.

\begin{figure}
\centering
Base distribution samples

\includegraphics[width=0.65\textwidth, trim=0 10 0 0.65cm, clip]{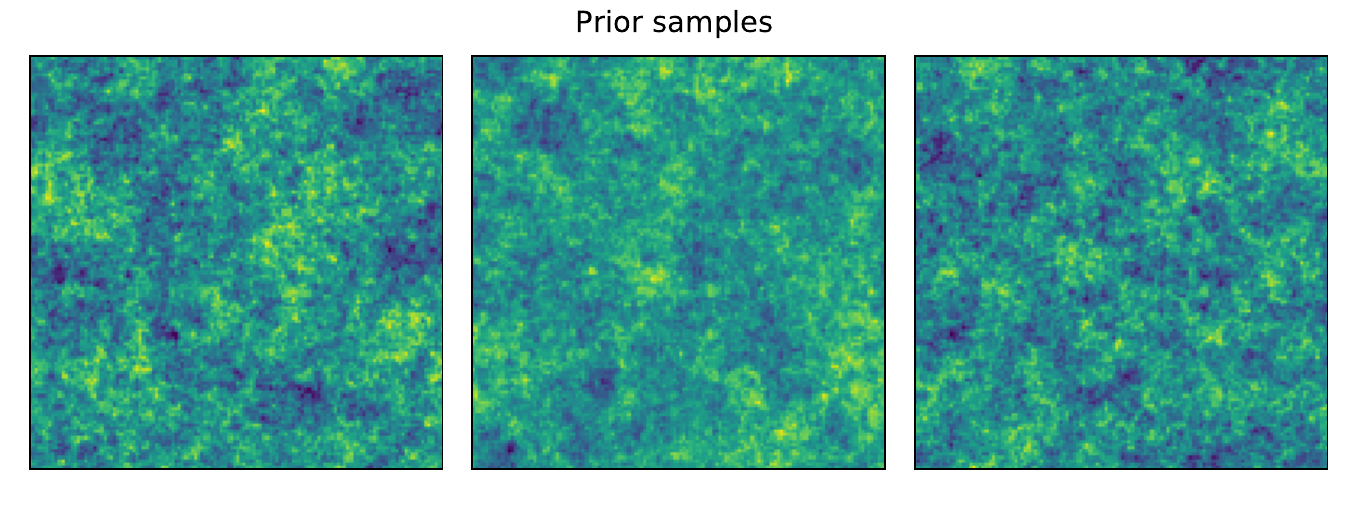}

Model samples

\includegraphics[width=0.65\textwidth, trim=0 10 0 0.65cm, clip]{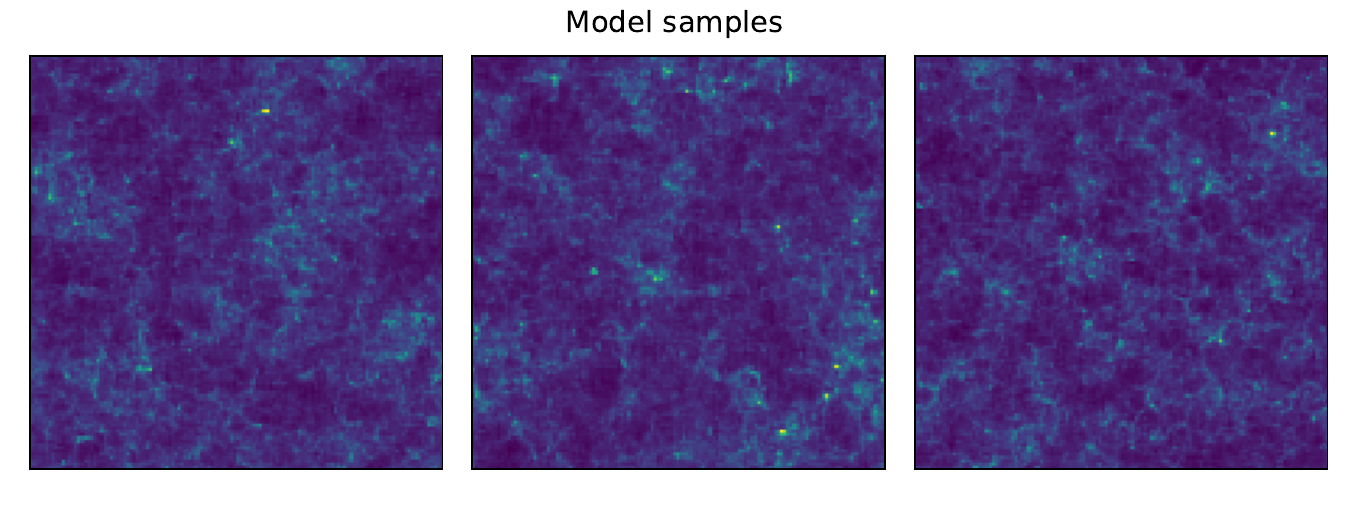}

Target samples

\includegraphics[width=0.65\textwidth, trim=0 10 0 0.65cm, clip]{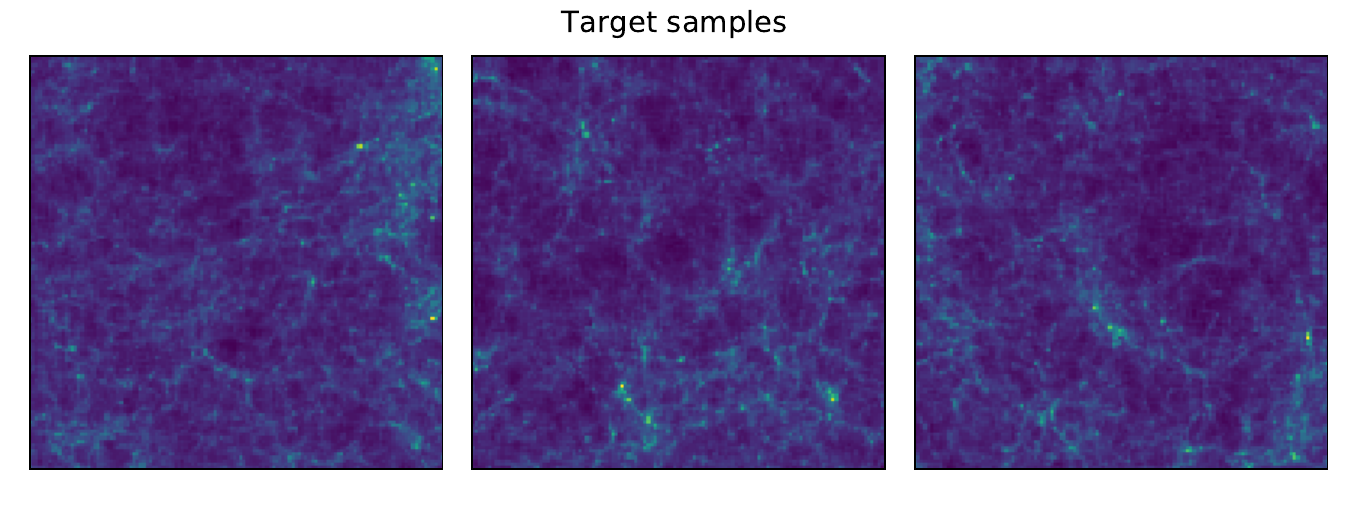}
\caption[N-body simulations with real NVP]{N-body simulation example trained on real NVP flow. Each sample is derived from a 3D volume of ($125 \ h^{-1}\ \text{Mpc})^3$ by projecting the matter field along an axis such that the final 2D map has a resolution of $128\times128$ pixels. (Top) Samples from the Gaussian base distribution with the same power spectrum as the samples. Middle: Model samples from the trained flow, derived from the base samples. It is visible that the flow amplifies the structures seeded in the Gaussian base. (Bottom) independent samples from the training data.}\label{fig:nbodysamples}
\end{figure}

\begin{figure}[h!]
\centering
\includegraphics[width=0.495\textwidth]{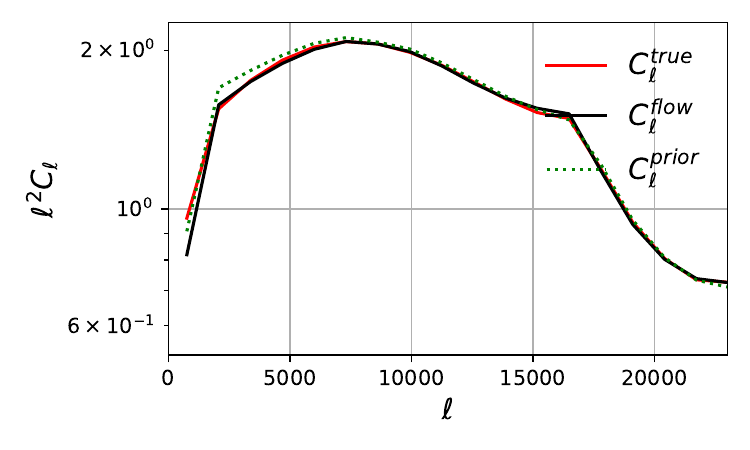}
\caption[Power spectrum of N-body simulations with real NVP]{Matter power spectrum of the projected N-body simulation samples (red), trained flow model (black), and correlated input Gaussian base (green), averaging over 10,000 maps such as those in Fig.~\ref{fig:nbodysamples}. For consistency with previous sections we again use the multipole scale $\ell$ on the x-axis, rather than the comoving wave vector $k$.}\label{fig:nbodyps}
\end{figure}

We trained the same real NVP flow as in the previous section. The training time for this $128\times128$ pixel data set was about 50 hours on an RTX 3090 and used about 11 GB of GPU memory. We found again that it is advantageous to flow from a correlated Gaussian field with the right power spectrum compared to flowing from Gaussian noise. This base distribution leads to faster convergence and better sample quality. We show random samples from the base, training and model distributions in Fig.~\ref{fig:nbodysamples}, finding visually excellent results. We note that while there is certainly some correlation between base and N-body initial conditions on large scales, we are not learning physical structure formation. The base distribution is only used as a good starting point to flow from.  In Fig.~\ref{fig:nbodyps} we find an almost perfect power spectrum of the flow, with the exception of the smallest $\ell$ bin where the flow power is about $10\%$ too large. We obtain the following non-Gaussianity measurements, sampling over each 10,000 training and model maps:

\begin{center}
\begin{tabular}{cccc}
    \hline
    $\tilde{f}_\text{NL}^{\text{local}}$ training & $\tilde{f}_\text{NL}^{\text{local}}$ flow & $\tilde{f}_\text{NL}^{\text{equi}}$ training & $\tilde{f}_\text{NL}^{\text{equi}}$ flow\\
    \hline
    0.85 & 0.78 & 0.61 & 0.62\\
    \hline
\end{tabular}
\end{center}

\begin{center}
\begin{tabular}{cccc}
    \hline
    $\sigma(\tilde{f}_\text{NL}^{\text{local}})$ training & $\sigma(\tilde{f}_\text{NL}^{\text{local}})$ flow & $\sigma(\tilde{f}_\text{NL}^{\text{equi}})$ training & $\sigma(\tilde{f}_\text{NL}^{\text{equi}})$ flow\\
    \hline
    0.41 & 0.29 & 0.24 & 0.24\\
    \hline
\end{tabular}
\end{center}

We find that the flow is accurate to $5$ to $10\%$. While this accuracy is sufficient for some applications (such as providing a better prior on non-linear scales than a Gaussian prior), we found that the more complicated Glow normalizing flow gives even better results on N-body simulations. We discuss this flow and its training results in Section~\ref{sec:glow}.

The present non-Gaussianity results are not correctly normalized since we did not account for non-Gaussian estimator variance and masking effects. This is not important here since our goal is only to establish equivalence between the training and flow samples, rather than providing physically exact measurements. In fact, the N-body training patches are not periodic (unlike the data in the previous sections), so for both power spectrum and bispectrum estimators we should apodize the mask and numerically estimate the masking bias. Alternatively we could generate periodic training data rather than cutting a larger simulation into patches. For simplicity we have done neither of these two options here. This is justified because our goal is only to compare the N-body and flow results, not to estimate physical parameters.

\section{Glow applied to N-body simulations}
\label{sec:glow}

A highly expressive modification of the real NVP flow is Glow~\cite{2018arXiv180703039K}. Glow has been shown to generate higher quality RGB images than real NVP, and we thus examine if it also improves results on cosmological data. On the other hand, the flow architecture is more complicated; in particular, it does not flow from a spatially organized base distribution, so we cannot straightforwardly flow from a correlated Gaussian base as we did for real NVP. It would also be more difficult to enforce periodicity, which however for the present application is not required.

\subsection{Flow architecture}

Glow uses a multi-scale architecture with a series of $K$ steps of flow, each step consisting of an activation normalization (actnorm) layer, an invertible $1\times 1$ convolution, and an affine coupling layer. The actnorm layer, introduced in~\cite{2018arXiv180703039K}, is a generalization of batch normalization, initializing trainable scale and bias parameters on each batch such that each channel has zero mean and unit variance. The learnable $1\times1$ convolution mixes channels, a generalization of permuting the channels or alternating the subsets $\bm{x}_1$ and $\bm{x}_2$ in the checkerboard masking as in real NVP. The affine coupling layers have learned functions $\bm{s}$ and $\bm{t}$ as described in Section~\ref{sec:nf_introduction}. We use 3 convolutional layers with kernel sizes $3$, then $1$, then $3$. The two hidden layers have ReLU activation functions and 128 channels each. The final convolution of each CNN is initialized with zeros as recommended in \cite{2018arXiv180703039K}.

Prior to the $K$ steps of flow, a squeeze operation reshapes each $2\times2\times c$ region of the input tensor to $1\times1\times 4c$. The squeeze operation thus reshapes the input tensors from $n\times n\times c$ to $\frac{n}{2}\times\frac{n}{2}\times4c$.

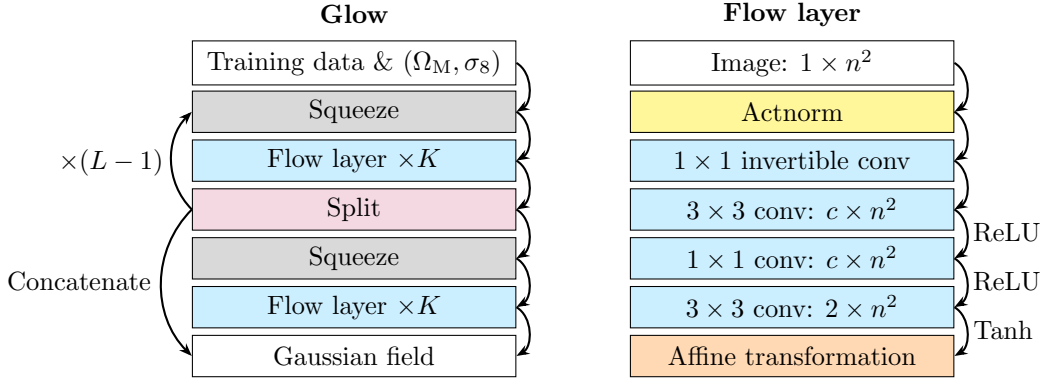
\begin{figure}
\centering

\begin{tikzpicture}[node distance=0.65cm]
\footnotesize

\pgfdeclarelayer{bg}
\pgfsetlayers{bg, main}

\node[align=center, font=\bfseries] (title){Glow};
\node (start) [cnn, below of=title] {Training data \& $(\Omega_\text{M},\sigma_8)$};
\node (sqz0) [cnn, below of=start, fill=black!15] {Squeeze};
\node (flow0) [cnn, below of=sqz0, fill=azul!75]  {Flow layer $\times K$};
\node (spl) [cnn, below of=flow0, fill=purple!15] {Split};
\node (sqz1) [cnn, below of=spl, fill=black!15] {Squeeze};
\node (flow1) [cnn, below of=sqz1, fill=azul!75]  {Flow layer $\times K$};
\node (end) [cnn, below of=flow1, fill=red!0] {Gaussian field};

\draw [arrow] (start.east) to [out=-30,in=30](sqz0.east);
\draw [arrow] (sqz0.east) to [out=-30,in=30](flow0.east);
\draw [arrow] (flow0.east) to [out=-30,in=30](spl.east);
\draw [arrow] (spl.west) to [out=135,in=-135](sqz0.west) node[yshift=-0.68cm] {\hspace{-2.15cm}$\times(L-1)$};
\draw [arrow] (spl.west) to [out=225,in=-225]( end.west) node[yshift=1cm] {\hspace{-3cm}Concatenate};
\draw [arrow] (spl.east) to [out=-30,in=30](sqz1.east);
\draw [arrow] (sqz1.east) to [out=-30,in=30](flow1.east);
\draw [arrow] (flow1.east) to [out=-30,in=30](end.east);

\end{tikzpicture}
\hspace{1cm}
\begin{tikzpicture}[node distance=0.65cm]
\footnotesize

\node[align=center, font=\bfseries] (title){Flow layer};
\node (start) [cnn, below of=title] {Image: $1\times n^2$};
\node (actn0) [cnn, below of=start, fill=yellow!50] {Actnorm};
\node (1x1)   [cnn, below of=actn0, fill=azul!75] {$1\times1$ invertible conv};
\node (conv0) [cnn, below of=1x1,   fill=azul!75] {$3\times3$ conv: $c\times n^2$};
\node (conv1) [cnn, below of=conv0, fill=azul!75] {$1\times1$ conv: $c\times n^2$};
\node (conv2) [cnn, below of=conv1, fill=azul!75] {$3\times3$ conv: $2\times n^2$};
\node (at0)   [cnn, below of=conv2, fill=orange!30] {Affine transformation};

\draw [arrow] (start.east) to [out=-30,in=30](actn0.east);
\draw [arrow] (actn0.east) to [out=-30,in=30](1x1.east);
\draw [arrow] (1x1.east)   to [out=-30,in=30](conv0.east);
\draw [arrow] (conv0.east) to [out=-30,in=30](conv1.east) node[yshift=0.35cm] {\hspace{1.4cm}ReLU};
\draw [arrow] (conv1.east) to [out=-30,in=30](conv2.east) node[yshift=0.35cm] {\hspace{1.4cm}ReLU};
\draw [arrow] (conv2.east) to [out=-30,in=30](at0.east) node[yshift=0.35cm] {\hspace{1.3cm}Tanh};

\end{tikzpicture}

\caption[The Glow normalizing and CNN flow architectures]{(Left) The Glow normalizing flow uses the multi-scale architecture. (Right) The flow layers, along with layer output channel $\times$ spacial dimensions. Glow introduces activation-normalization and $1\times1$ invertible convolutional layers.}

\label{fig:glow_architecture}
\end{figure}

After $K$ steps of flow, the tensor is split into halves along the channel dimension. The squeeze, flow steps, and split form an $L-1$ number of levels, followed by a final level not containing a split function. A diagram of the Glow architecture is shown in Fig.~\ref{fig:glow_architecture}.

It is interesting to note that this implementation of Glow, unlike our real NVP flow, does not have a straightforward way to enforce periodicity due to its multi-scale architecture. We make no changes to Glow between periodic and non-periodic data, as we do not find any obvious issues in expressibility with Glow between the two. A periodic implementation of Glow has been developed with invertible $d\times d$ in place of the invertible $1\times 1$ convolutions~\cite{pmlr-v97-hoogeboom19a}. This periodic Glow was shown to have improved performance on images of galaxies, the argument being that most of the pixels at the edges of the images have nearly the same values of black.

We use a PyTorch implementation of Glow taken from \cite{GlowGithub} that has been slightly modified to include the loss function Eq.~\ref{eq:loss_kl} and use the Adam optimizer.

\subsection{Application to N-body simulations}

\begin{figure}[h!]
\centering
\includegraphics[width=0.495\textwidth]{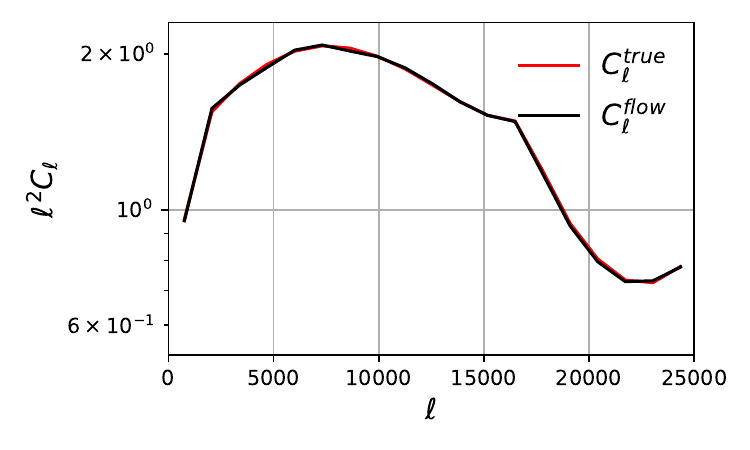}
\caption[Power spectrum for N-body fields trained with real NVP]{The power spectrum of the training data and the samples generated with Glow agree very closely.}\label{fig:glow_nbody_ps}
\end{figure}

\begin{figure}
\centering
Model samples

\includegraphics[width=0.65\textwidth, trim=0 10 0 0.65cm, clip]{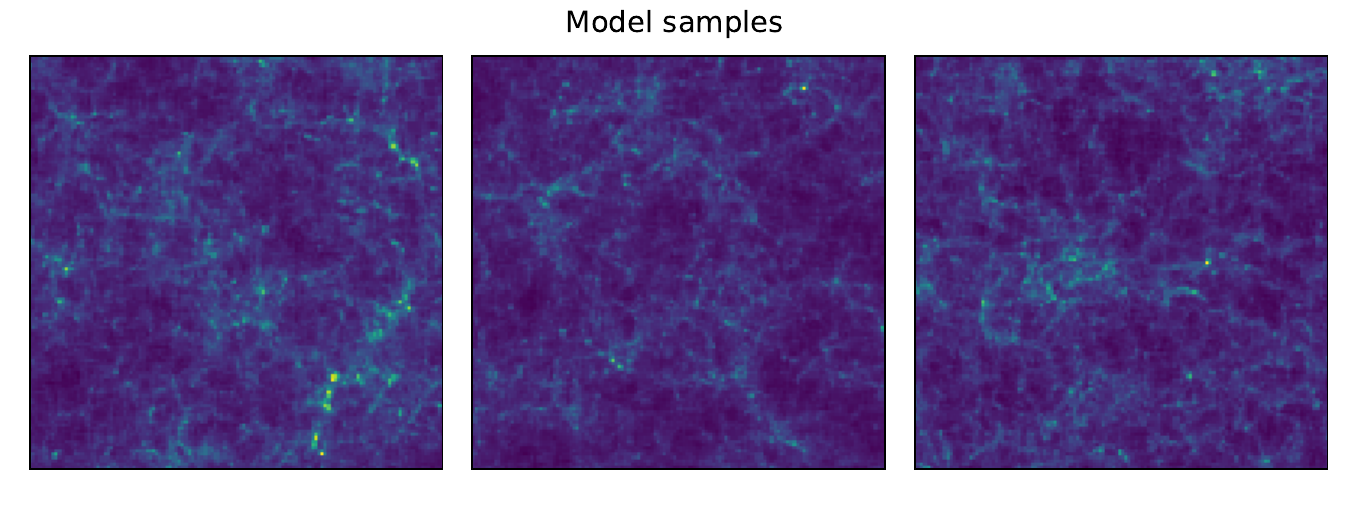}

Target samples

\includegraphics[width=0.65\textwidth, trim=0 10 0 0.65cm, clip]{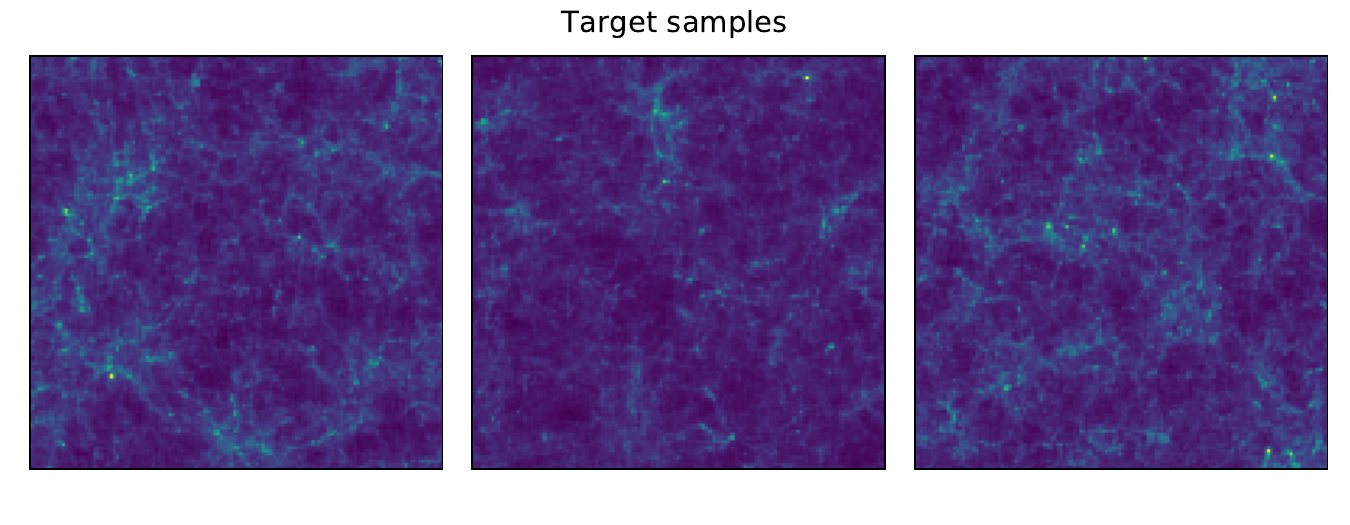}
\caption[N-body simulations trained on Glow]{N-body simulation example from Section~\ref{sec:nbodyflow} trained with Glow. Top: Model samples from the trained flow. Bottom: Samples from the training data. Training and model samples cannot be distinguished visually.}\label{fig:nbodysamples_glow}
\end{figure}

The performance of Glow on cosmological data is demonstrated here by training on the N-body patches introduced in Section~\ref{sec:nbodyflow}. We train with a depth of flow $K=16$ and $L=3$ levels; this gives a total of $1.6$~million trainable parameters. We use Adam optimization with a learning rate of $5\times10^{-4}$ and batch size of 32, using about 11 GB of GPU memory. The training converges in about 2 hours, much more quickly quickly than with real NVP on the same N-body data. Glow samples from the N-body training are shown in Fig.~\ref{fig:nbodysamples_glow}, and the power spectra in Fig.~\ref{fig:glow_nbody_ps} (right). The Glow flow has a near perfect power spectrum, including in the smallest $l$ bin (largest spatial scale) where the basic checkerboard real NVP flow in Section~\ref{sec:nbodyflow} was suboptimal. We obtain the following non-Gaussianity measurements, averaging over 10,000 training and model samples:

\begin{center}
\begin{tabular}{cccc}
    \hline
    $\tilde{f}_\text{NL}^{\text{local}}$ training & $\tilde{f}_\text{NL}^{\text{local}}$ flow & $\tilde{f}_\text{NL}^{\text{equi}}$ training & $\tilde{f}_\text{NL}^{\text{equi}}$ flow\\ 
    \hline
    0.85 & 0.85 & 0.61 & 0.63\\
    \hline
\end{tabular}
\end{center}

\begin{center}
\begin{tabular}{cccc}
    \hline
    $\sigma(\tilde{f}_\text{NL}^{\text{local}})$ training & $\sigma(\tilde{f}_\text{NL}^{\text{local}})$ flow & $\sigma(\tilde{f}_\text{NL}^{\text{equi}})$ training & $\sigma(\tilde{f}_\text{NL}^{\text{equi}})$ flow\\
    \hline
    0.40 & 0.37 & 0.24 & 0.30\\
    \hline
\end{tabular}
\end{center}

We find that the Glow flow is accurate at the percent level in $\tilde{f}_\text{NL}^{\text{local}}$, an improvement over the real NVP flow in Section~\ref{sec:nbodyflow}, and has similar performance in $\tilde{f}_\text{NL}^{\text{equi}}$. As in Section~\ref{sec:nbodyflow} we have not corrected for the masking bias as our goal was only to show equivalence between training and model distribution.

While results presented here used 11 GB of memory for 128 channels in the two hidden layers, we find almost the same performance reducing to 32 channels in each hidden layer, which reduces the GPU memory size to 4.1 GB. The 32 channel version has a significant error only in the highest $l$ bin, being about 10\% in the power spectrum. When extending to 3-dimensional data, using a memory efficient architecture will be important.

\section{Conclusion}

The goal of this study was to evaluate how good certain normalizing flows are at representing PDFs of the near-Gaussian random fields that appear in cosmology. Our results are encouraging. We find that we can train flows whose samples have power spectra and non-Gaussianity accurate to a few percent. Beyond sample generation, we have tested the inverse operation of density evaluation in the analytically tractable Gaussian case and again found percent level accuracy in the PDF. This opens up the possibility of using flows for inference of cosmological random fields in various ways.

Unlike typical applications of normalizing flows such as image generation, our physical setup allows us to use a better base distribution than the commonly used independent Gaussian random variables. We have confirmed that starting with a Gaussian field with a physical power spectrum facilitates learning of the target distribution.

We have measured the non-Gaussianity of the flow distribution and compared it to the target distribution, using the bispectrum estimation formalism of cosmology. Exactly measuring three-point correlation functions in statistical machine learning may be useful more generally to qualify the performance of different normalizing flows in representing PDFs.

The main text of the paper was developed with the real NVP flow, which is conceptually simple. However, we found that in the case of N-body simulations a more complicated architecture with Glow converges faster and gives higher quality results than the real NVP flow with a correlated base distribution. Perhaps the Glow architecture could be combined with a correlated base to be even more efficient.

We have considered 2-dimensional maps here, up to a side length of $128$~px. In practice, analyzing a small but very deep sky patch with HMC such as for CMB lensing in~\cite{Millea:2020iuw} could benefit from a non-Gaussian prior distribution of the lensing potential represented with a normalizing flow. We are in particular interested in applying flows to kinetic Sunyaev-Zeldovich tomography~\cite{Deutsch:2017ybc,Smith:2018bpn}, where the scales that generate the kSZ signal are highly non-linear. For this purpose, we would have to train a 3-dimensional flow of the matter distribution and develop a method to efficiently patch together many boxes of the small-scale matter distribution into a large-scale CMB map. As discussed in the introduction, the present normalizing flows are also very suitable for variational inference at field level, such as variational inference of the initial conditions of the matter distribution of the universe. Larger maps and 3-dimensional flows will require the use of multiple GPUs and memory optimized flow architectures. 

For some applications it would be useful to make the flow PDF dependent on astrophysical or cosmological parameters. Glow is made to be conditional on cosmological parameters in Chapter~\ref{ch:nf_parameters} by extending the affine transformation to be dependent upon these parameters. Additionally, flows can depend not only on the parameters of the transformations $T_k$ but also on learned parameters of the base distribution. We have not made use of this possibility here, however in particular the correlated Gaussian base could easily be extended to depend on parameters that modify the power spectrum.

Since the research in this Chapter was conducted, we have made several attempts at extending the real NVP and Glow normalizing flows to 3 spacial dimensions. Real NVP appears to be far from expressive enough, as it fails at learning the non-Gaussianity of 3-dimensional late time matter fields. A 3-dimensional implementation of Glow constantly sends pixels to infinity. Unfortunately, it seems that normalizing flows are not up to the task of learning very high dimensional distributions. Normalizing flows are theoretically capable of representing any distribution, but this is much more difficult in practice. In Chapter~\ref{ch:sr_diffusion}, a more expressive denoising diffusion model will be used on 3-dimensional fields.

\chapter[Field Level Cosmological Parameter Inference with a\\Conditional Normalizing Flow]{Field Level Cosmological Parameter Inference with a Conditional Normalizing Flow}
\label{ch:nf_parameters}
Inference on high-dimensional data, such as images and volumetric simulations, has become much more accessible in recent years from advances in machine learning technology. This Chapter uses a normalizing flow to learn the PDF of field level dark matter simulations conditional on cosmological parameters. We perform conditional data generation and parameter inference directly from the learned conditional PDF. The results for $\Omega_\text{M}$ and $\sigma_8$ from the field level data are shown to be more accurate than with the power spectrum. 


\section{Introduction}

As explained in Chapter~\ref{ch:nf_random}, a normalizing flow learns the PDF $p(\bm{x})$ from a series of transformations $T$ on a base distribution $p(\bm{u})$. In this Chapter, we demonstrate how the PDF can be made to be conditional on parameters $\phi$ to obtain $p(\bm{x}|\bm{\phi})$. Learning the conditional PDF $p(\bm{x}|\bm{\phi})$ enables conditional data generation and parameter inference. With the conditional PDF, from Bayes' theorem, the parameters $\bm{\phi}$ may be found as
\begin{equation}
    p(\bm{\phi}|\bm{x})\propto p(\bm{x}|\bm{\phi})\ p(\bm{\phi}),
\end{equation}
We may easily find the proportionality constant by normalizing the probabilities to unity.

In this Chapter, two parameters are considered, $\bm{\phi}\in\{\Omega_\text{M},\sigma_8\}$. There is much difficulty in accurately predicting the baryon density $\Omega_\text{b}$ and Hubble constant $h$ from large-scale structure N-body simulations, as these parameters only show up as small effects in transfer function. Further, by using two conditional parameters, the entire $\bm{\phi}$ manifold of $p(\bm{x}|\bm{\phi})$ can be more easily illustrated than for a higher-dimensional parameter space. For some applications, it may be useful to extend the methods of this Chapter to include $n_s$, which is sensitive to N-body simulations.

In practice, there are several methods to design a normalizing flow as conditional on parameters $\bm{\phi}$ supplied to the model. One method is to use a hypernetwork~\cite{hypernetworks} by training the flow on the fields $\bm{x}$, and then training a separate network to adjust the flow's weights conditional on $\bm{\phi}$. Such a method was employed with the TRENF normalizing flow~\cite{trenf}. Another method would be to concatenate the $\bm{\phi}$ into new channels of data in the flow layers. Such a concatenation could be done before any of the convolutions, while still defining the final convolution to output $2$ channels for the $\bm{s},\bm{t}$ affine parameters. The concatenation method is not ideal, as it requires copying each parameter in $\bm{\phi}$ into $n^2$ numbers ($n^2$ being the number of pixels in the data) to match the dimensions of the fields, and this unnecessarily adds extra data to the model. Furthermore, the multiscale architecture of Glow is designed to move spacial dimensions into channel dimensions; this is physically motivated by allowing the flow to more easily learn the spacial correlations in the data. Including the parameters $\bm{\phi}$ as extra concatenated channels may increase the difficulty in the flow learning spacial correlations.

For this Chapter, the architecture uses a parameter embedding into the CNN to allow the flow to be conditional on $\bm{\phi}$. In each flow step, a learned representation of $\bm{\phi}$ is arithmetically added between two of the CNN layers that decide $\bm{s},\bm{t}$. In this way, the CNN weights are partially determined by $\bm{\phi}$ (and therefore $\bm{s},\bm{t}$) and thus the every step of the flow is conditional on $\bm{\phi}$.

We use an idealized setup to demonstrate our method of field level parameter inference with a normalizing flow. The resolution of the simulated fields we operate on goes to highly nonlinear modes of $k_\text{max}\sim1 h/\text{Mpc}$ at redshift $z=0$. We compare our field level parameter inference results to the power spectrum with the same range of modes, finding that the power spectrum leaves out a significant amount of information. We note that this $k_\text{max}$ is larger than would be considered by standard galaxy survey length scales at low redshift~\cite{SDSS:2008tqn}, as an EFT analysis that can marginalize over biases would optimistically go to $\sim0.3h/\text{Mpc}$. However, weak lensing surveys at larger redshifts closer to $z\sim1$~\cite{DES:2021wwk} are less nonlinear, may do a power spectrum analysis up to $k_\text{max}\sim1 h/\text{Mpc}$. A more detailed discussion of applying our models to real data in general is given in Chapter~\ref{ch:conclusion}.

\section{Conditional Glow for parameter inference}

In this Section, we describe the normalizing flow model used to learn the conditional PDF $p(\bm{x}|\Omega_\text{M},\sigma_8)$, and we also describe the simulated data that the model trains on.

\subsection{Model details}

The Glow normalizing flow is made to be conditional on cosmological parameters $\Omega_\text{M}$ and $\sigma_8$, with a diagram in~Fig.~\ref{fig:glow_cond}. The actnorm and convolutional layers prior to the parameter embedding are able to learn how to optimally accept the parameter embedding. The Glow architecture used in this Chapter has $L=3$ resolution levels, and $K=8$ conditional flow steps per level. The convolutions are set up as in the previous Chapter, and shown in Fig.~\ref{fig:glow_cond}; there are three convolutional layers with kernel sizes $3$, then $1$, then $3$, and $c=32$ channels per convolution.

The parameter embedding to make the flow conditional goes through a linear layer with $c=32$ parameters, and is then arithmetically added to the output of the first $3\times3$ convolution each flow layer. Another possible configuration for the parameter embedding is to add the output from the linear layer to the output of the $1\times1$ invertible convolution; some experimenting showed similar performance between these two configurations.

\begin{figure}
\centering
\textbf{Conditional Glow for parameter inference}

\ 

\hspace{4cm}\begin{tikzpicture}[node distance=0.65cm]
\footnotesize

\pgfdeclarelayer{bg}
\pgfsetlayers{bg, main}

\node[align=center, font=\bfseries] (title){Conditional Glow};
\node (start) [cnn, below of=title] {Training field};
\node (param) [cnn, right of=start, xshift=4cm] {Parameters};
\node (sqz0) [cnn, below of=start, fill=black!15] {Squeeze};
\node (flow0) [cnn, below of=sqz0, fill=azul!75]  {Conditional flow layer $\times K$};
\node (spl) [cnn, below of=flow0, fill=purple!15] {Split};
\node (sqz1) [cnn, below of=spl, fill=black!15] {Squeeze};
\node (flow1) [cnn, below of=sqz1, fill=azul!75]  {Flow layer $\times K$};
\node (end) [cnn, below of=flow1, fill=red!0] {Gaussian field};

\draw [arrow] (start.east) to [out=-30,in=30](sqz0.east);
\draw [arrow] (param.south) to [out=-90,in=0](flow0.east);
\draw [arrow] (sqz0.east) to [out=-30,in=30](flow0.east);
\draw [arrow] (flow0.east) to [out=-30,in=30](spl.east);
\draw [arrow] (spl.west) to [out=135,in=-135](sqz0.west) node[yshift=-0.68cm] {\hspace{-2.15cm}$\times(L-1)$};
\draw [arrow] (spl.west) to [out=225,in=-225]( end.west) node[yshift=1cm] {\hspace{-3cm}Concatenate};
\draw [arrow] (spl.east) to [out=-30,in=30](sqz1.east);
\draw [arrow] (sqz1.east) to [out=-30,in=30](flow1.east);
\draw [arrow] (flow1.east) to [out=-30,in=30](end.east);

\end{tikzpicture}

\ 

\ 

\begin{tikzpicture}[node distance=0.65cm]
\footnotesize

\node[align=center, font=\bfseries] (title){Conditional flow layer};
\node (start) [cnn, below of=title, xshift=-2.5cm] {Image: $1\times n^2$};
\node (actn0) [cnn, below of=start, fill=yellow!50] {Actnorm};
\node (1x1)   [cnn, below of=actn0, fill=azul!75] {$1\times1$ invertible conv};
\node (conv0) [cnn, below of=1x1,   fill=azul!75] {$3\times3$ conv: $c\times n^2$};
\node (plus) [circ, below of=conv0, xshift=2.5cm, yshift=-.5cm] {$\pmb{+}$};
\node (conv1) [cnn, below of=plus, yshift=-0.5cm, fill=azul!75] {$1\times1$ conv: $c\times n^2$};
\node (conv2) [cnn, below of=conv1, fill=azul!75] {$3\times3$ conv: $2\times n^2$};
\node (at0)   [cnn, below of=conv2, fill=orange!30] {Affine transformation};

\node (cond0) [cnn, right of=start, xshift=4.25cm] {Parameters};
\node (cond1) [cnn, below of=cond0, fill=green!15] {Linear: $c$};

\draw [arrow] (start.east)  to [out=-30,in=30](actn0.east);
\draw [arrow] (actn0.east)  to [out=-30,in=30](1x1.east);
\draw [arrow] (1x1.east)    to [out=-30,in=30](conv0.east);
\draw [arrow] (conv0.south) to (plus) node[yshift=0.35cm] {\hspace{-4.cm}ReLU};
\draw [arrow] (plus)        to (conv1);
\draw [arrow] (conv1.east)  to [out=-30,in=30](conv2.east) node[yshift=0.35cm] {\hspace{1.4cm}ReLU};
\draw [arrow] (conv2.east)  to [out=-30,in=30](at0.east) node[yshift=0.35cm] {\hspace{1.7cm}Sigmoid};

\draw [arrow] (cond0.east)  to [out=-30,in=30](cond1.east) node[yshift=0.35cm] {\hspace{1.4cm}ReLU};
\draw [arrow] (cond1.south) to (plus) node[yshift=1.1cm] {\hspace{3.8cm}ReLU};

\end{tikzpicture}

\caption[Conditional Glow normalizing flow architecture with parameter embedding]{The Glow normalizing flow is made conditional with parameter embedding in each conditional flow layer. In the experiment presented in this Chapter, two cosmological parameters $(\Omega_\text{M},\sigma_8)$ are expanded to $c$ channels with a fully connected network, and are arithmetically added between image convolutions.}

\label{fig:glow_cond}
\end{figure}

\subsection{Data}

The data for this Chapter is simulated with \texttt{FastPM}~\cite{Feng:2016yqz}, a fast particle mesh N-body solver. We simulate 5,000 boxes with physical length $512\ \text{Mpc}/h$, containing $128^3$ particlces in a $128^3$~px meshgrid. All simulated volumes are given random parameters uniformly distributed in $\Omega_\text{M}\in(0.1,0.5)$ and $\sigma_8\in(0.6,1.0)$. There are $10$ time steps computed from scale factor $a_i=0.1$ to $a_f=0$. The simulations are then projected to $2$ dimensions with depth $128\ \text{Mpc}/h$ twice, with a $128\ \text{Mpc}/h$ gap between the two projections. This projection is done in each of the 3 dimensions, and so 6 projections are made from each cube for a total of 30,000 images. The gap between 2-dimensional projections allows multiple images taken from the same simulation volume to be nearly uncorrelated.

\subsection{Model training}

The data is given an 80-10-10 split as training, validation, and test data. We train with a batch size of 64 on an A4000 GPU, the model trains in about 2 hours. The validation set shows no overtraining.

\section{Results}
\label{sec:results_cond_glow}

\begin{figure}
    \centering
    
    \hspace{0.35cm}\textbf{Model samples for varied $\Omega_\text{M}$ and $\sigma_8$}
    \vspace{-.1cm}
    
    \footnotesize
    \rotatebox{90}{
    \begin{tabular}{ccccc}
        \vspace{-.23cm}\hspace{0.05cm}$\Omega_\text{M}=0.1$ & \hspace{.17cm}$\Omega_\text{M}=0.2$ & \hspace{.17cm}$\Omega_\text{M}=0.3$ & \hspace{.17cm}$\Omega_\text{M}=0.4$ & \hspace{.17cm}$\Omega_\text{M}=0.5$
    \end{tabular}
    }
    \includegraphics[width=.625\textwidth]{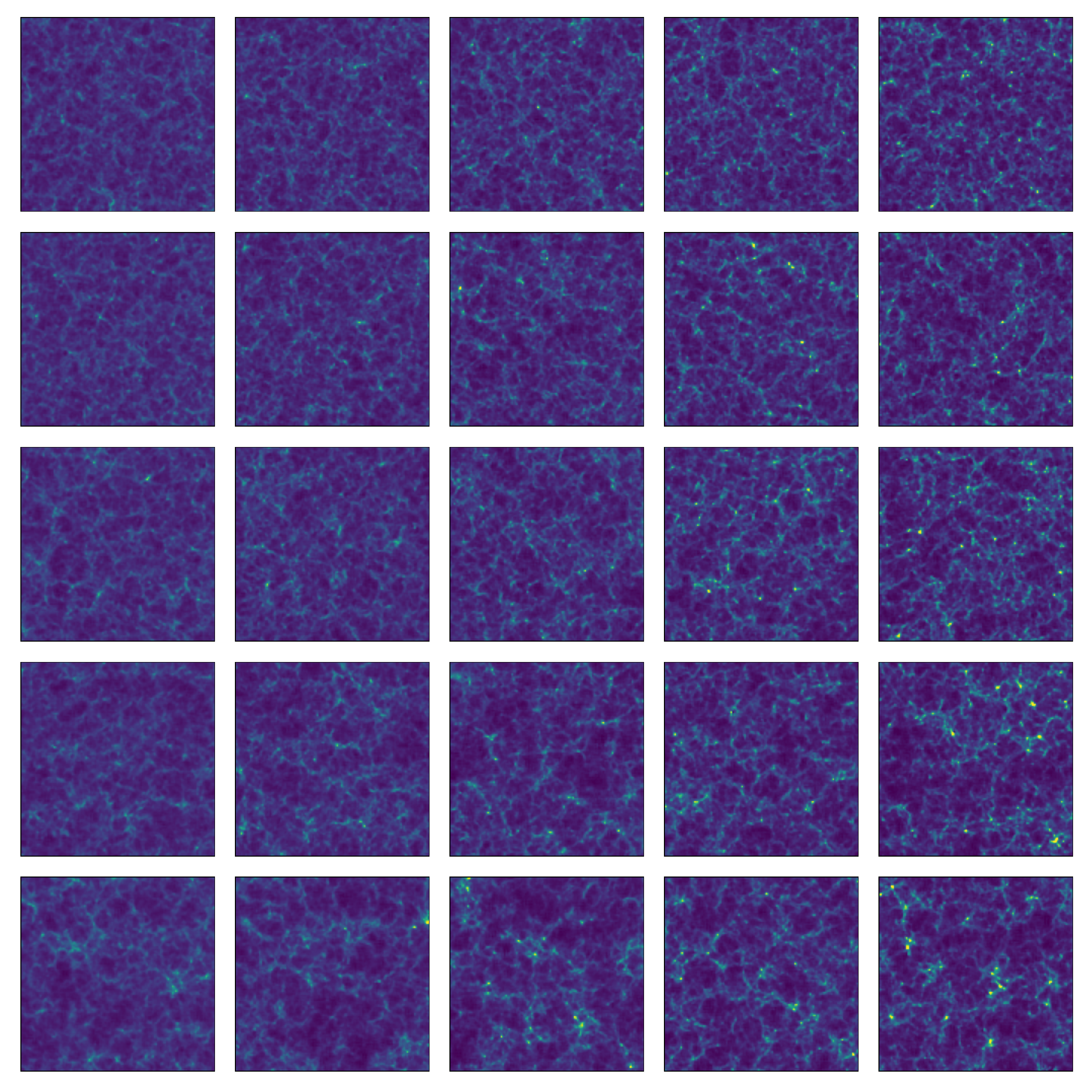}

    \vspace{-.23cm}
    \begin{tabular}{ccccc}
        \hspace{.33cm}$\sigma_8=0.6$ & \hspace{.33cm}$\sigma_8=0.7$ & \hspace{.33cm}$\sigma_8=0.8$ & \hspace{.33cm}$\sigma_8=0.9$ & \hspace{.33cm}$\sigma_8=1.0$
    \end{tabular}

    \normalsize
    \hspace{0.35cm}\textbf{Truth samples for varied $\Omega_\text{M}$ and $\sigma_8$}
    \vspace{-.1cm}
    
    \footnotesize
    \rotatebox{90}{
    \begin{tabular}{ccccc}
        \vspace{-.23cm}\hspace{0.05cm}$\Omega_\text{M}=0.1$ & \hspace{.17cm}$\Omega_\text{M}=0.2$ & \hspace{.17cm}$\Omega_\text{M}=0.3$ & \hspace{.17cm}$\Omega_\text{M}=0.4$ & \hspace{.17cm}$\Omega_\text{M}=0.5$
    \end{tabular}
    }
    \includegraphics[width=.625\textwidth]{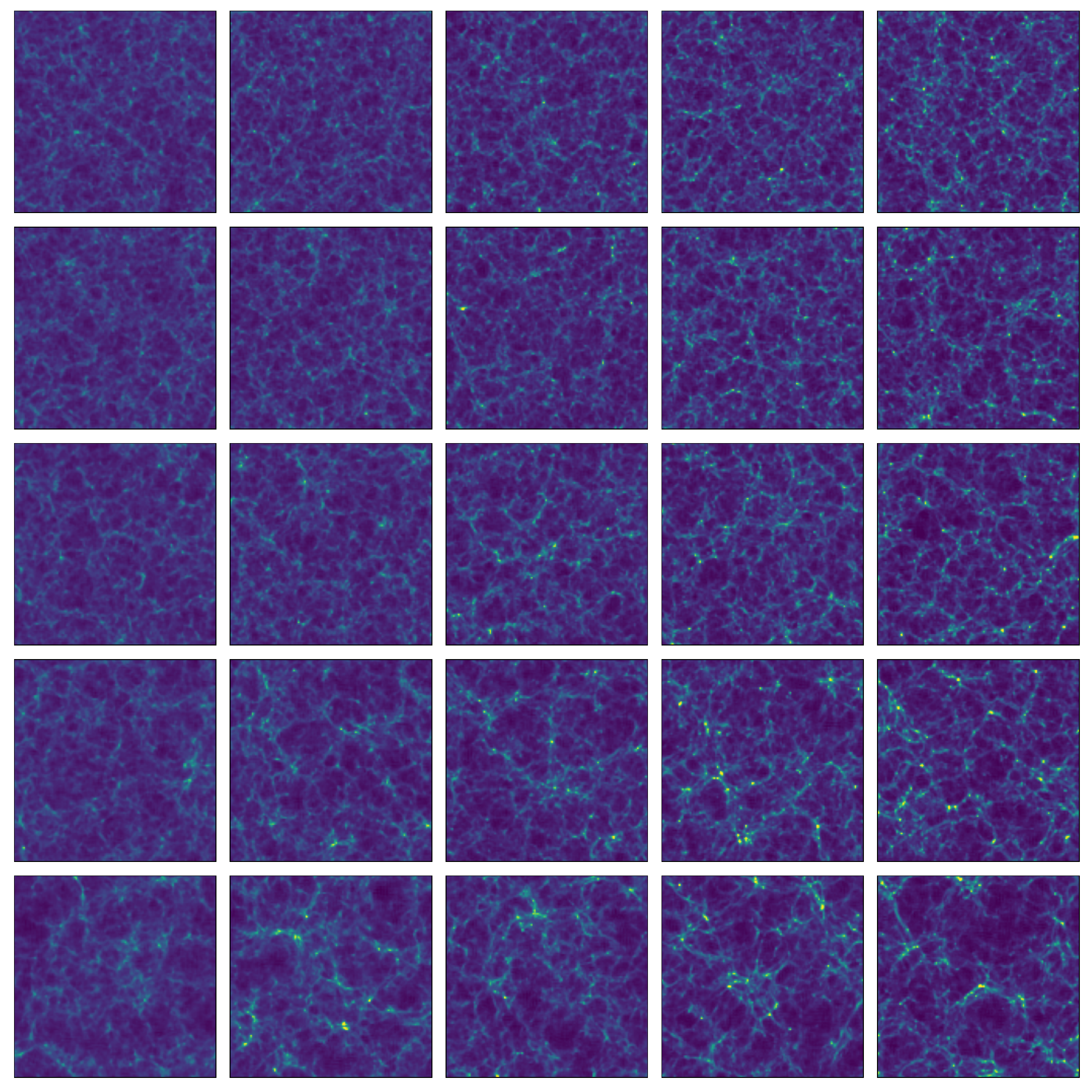}

    \vspace{-0.23cm}
    \begin{tabular}{ccccc}
        \hspace{.33cm}$\sigma_8=0.6$ & \hspace{.33cm}$\sigma_8=0.7$ & \hspace{.33cm}$\sigma_8=0.8$ & \hspace{0.33cm}$\sigma_8=0.9$ & \hspace{.33cm}$\sigma_8=1.0$
    \end{tabular}
    
    \caption[Conditional Glow random samples for varied $\Omega_\text{M}$ and $\sigma_8$]{Model samples (top) and training samples (bottom) for varied $\Omega_\text{M}$ and $\sigma_8$. Each sample is generated from a random Gaussian base distribution $\bm{u}$ and the conditional flow $T_{\bm{\phi}}$.}
    \label{fig:cond_Glow_visual_om_s8}
\end{figure}

\sloppy Visual results for random samples generated from the conditional flow model are shown in Fig.~\ref{fig:cond_Glow_visual_om_s8}, for conditional parameters with $\Omega_\text{M}\in\{0.1,0.2,0.3,0.4,0.5\}$ and $\sigma_8\in\{0.6,0.7,0.8,0.9,1.0\}$. These samples are generated from the model by supplying the respective $(\Omega_\text{M},\sigma_8)$ along with a random Gaussian field for each sample. By visually comparing the model output samples and training samples, they follow the same patterns across the varied parameters. Increasing $\Omega_\text{M}$ decreases the void sizes, and increasing $\sigma_8$ increases the amplitude of the overdense features.

\begin{figure}
    \centering
    
    \hspace{0.3cm}\textbf{Power spectra for varied $\Omega_\text{M}$, $\sigma_8$}
    \vspace{-.1cm}
    
    \footnotesize
    \rotatebox{90}{
    \begin{tabular}{ccccc}
        \vspace{-.23cm}\hspace{1.05cm}$\Omega_\text{M}=0.1$ & \hspace{1.23cm}$\Omega_\text{M}=0.2$ & \hspace{1.23cm}$\Omega_\text{M}=0.3$ & \hspace{1.23cm}$\Omega_\text{M}=0.4$ & \hspace{1.23cm}$\Omega_\text{M}=0.5$
    \end{tabular}
    }
    \includegraphics[width=0.98\textwidth]{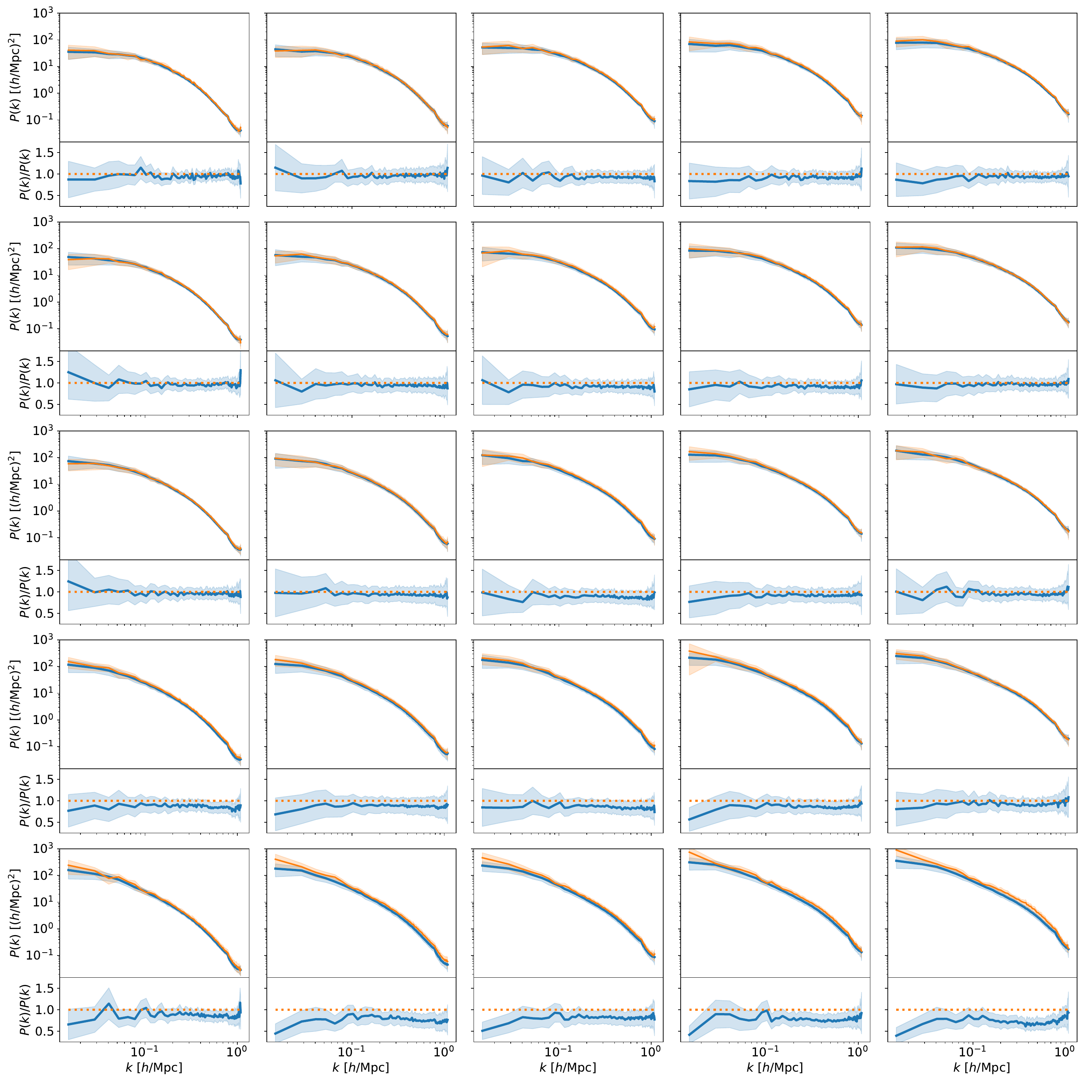}
    
    \vspace{-.23cm}
    
    \begin{tabular}{ccccc}
        \hspace{0.95cm}$\sigma_8=0.6$ & \hspace{1.37cm}$\sigma_8=0.7$ & \hspace{1.37cm}$\sigma_8=0.8$ & \hspace{1.37cm}$\sigma_8=0.9$ & \hspace{1.37cm}$\sigma_8=1.0$
    \end{tabular}
    
    \caption[Conditional Glow power spectra for varying $\Omega_\text{M}$ and $\sigma_8$]{Conditional Glow power spectra for varying $\Omega_\text{M}$ and $\sigma_8$, comparison between the model output (blue) and training data (orange). Plots are averages of 100 samples with $1\sigma$ regions.}
    \label{fig:cond_Glow_ps_om_s8}
\end{figure}

Power spectra are shown in Fig.~\ref{fig:cond_Glow_ps_om_s8} for the varied $\Omega_\text{M}$ and $\sigma_8$, comparing the model output with the training data, each having $1\sigma$ bands. The model samples are accurate to the truth for most of the parameter space; however, the model output becomes slightly inaccurate at the $\Omega_\text{M}=0.1$ extreme for all length scales across all $\sigma_8$, with truth samples just outside of the model samples' $1\sigma$ region.

\begin{figure}
    \centering
    
    \hspace{0.35cm}\textbf{Model samples from the same base distribution, example 1}
    \vspace{-.1cm}
    
    \footnotesize
    \rotatebox{90}{
    \begin{tabular}{ccccc}
        \vspace{-.23cm}\hspace{.05cm}$\Omega_\text{M}=0.1$ & \hspace{.17cm}$\Omega_\text{M}=0.2$ & \hspace{.17cm}$\Omega_\text{M}=0.3$ & \hspace{.17cm}$\Omega_\text{M}=0.4$ & \hspace{.17cm}$\Omega_\text{M}=0.5$
    \end{tabular}
    }
    \includegraphics[width=.625\textwidth]{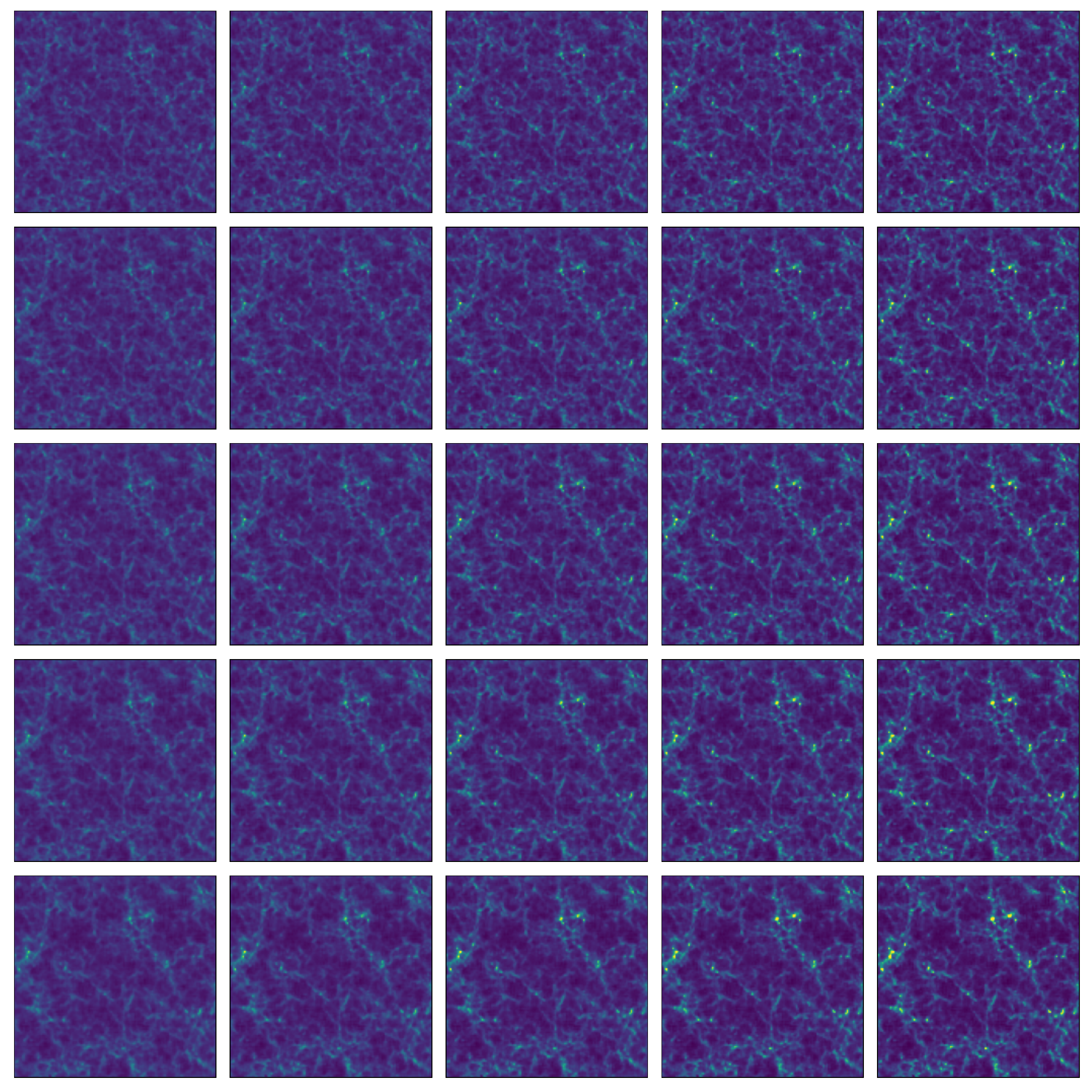}

    \vspace{-.23cm}
    \begin{tabular}{ccccc}
        \hspace{.3cm}$\sigma_8=0.6$ & \hspace{.3cm}$\sigma_8=0.7$ & \hspace{.3cm}$\sigma_8=0.8$ & \hspace{.3cm}$\sigma_8=0.9$ & \hspace{.3cm}$\sigma_8=1.0$
    \end{tabular}

    \normalsize
    \hspace{0.35cm}\textbf{Model samples from the same base distribution, example 2}
    \vspace{-.1cm}
    
    \footnotesize
    \rotatebox{90}{
    \begin{tabular}{ccccc}
        \vspace{-.23cm}\hspace{.05cm}$\Omega_\text{M}=0.1$ & \hspace{.17cm}$\Omega_\text{M}=0.2$ & \hspace{.17cm}$\Omega_\text{M}=0.3$ & \hspace{.17cm}$\Omega_\text{M}=0.4$ & \hspace{.17cm}$\Omega_\text{M}=0.5$
    \end{tabular}
    }
    \includegraphics[width=.625\textwidth]{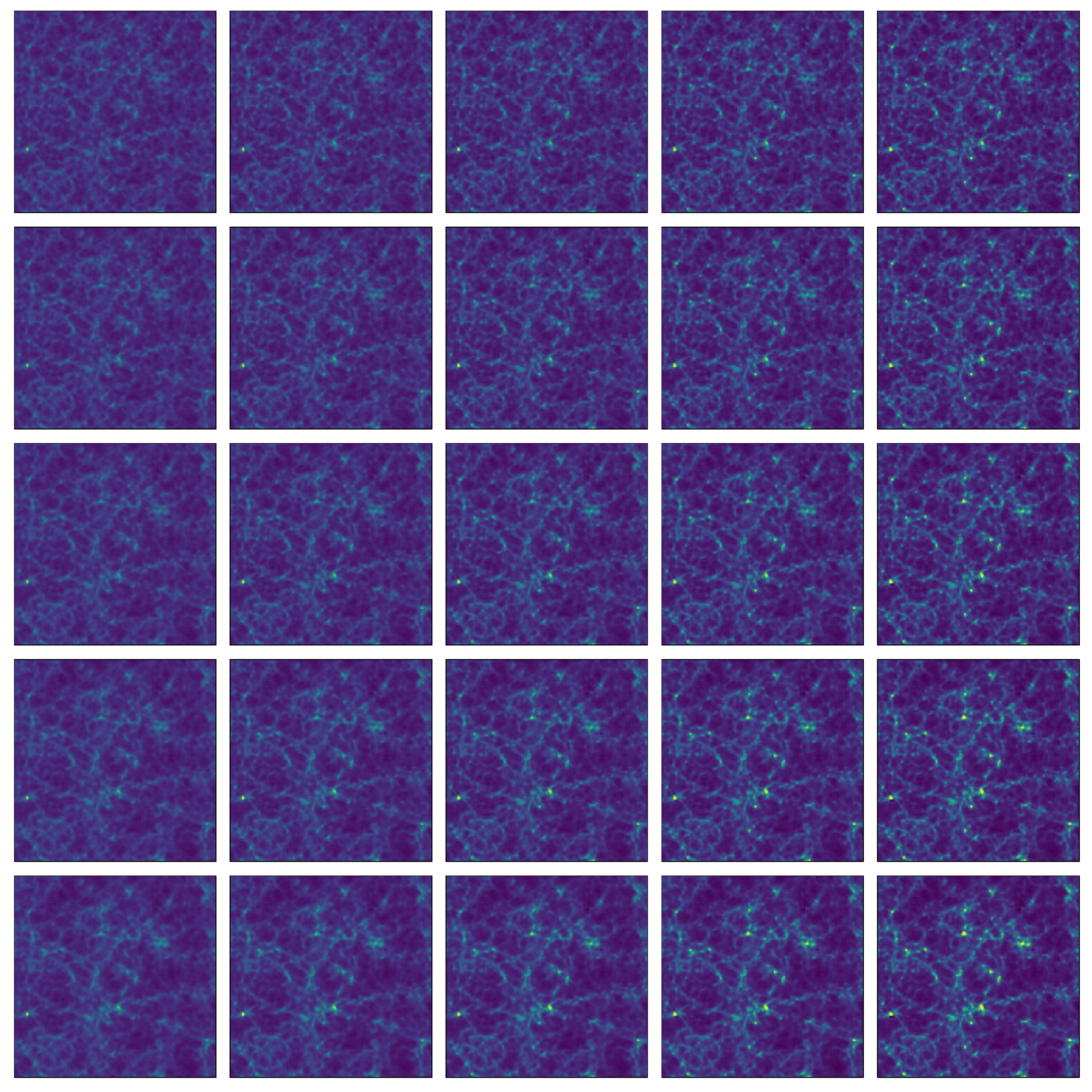}

    \vspace{-0.23cm}
    \begin{tabular}{ccccc}
        \hspace{.3cm}$\sigma_8=0.6$ & \hspace{.3cm}$\sigma_8=0.7$ & \hspace{.3cm}$\sigma_8=0.8$ & \hspace{0.3cm}$\sigma_8=0.9$ & \hspace{.3cm}$\sigma_8=1.0$
    \end{tabular}
    
    \caption[Model samples from the same base distribution for varied $\Omega_\text{M}$ and $\sigma_8$]{Model samples across the $(\Omega_\text{M},\sigma_8)$ manifold generated from the same base distribution. Two sets of results are shown.}
    \label{fig:cond_Glow_visual_om_s8_samebase}
\end{figure}

Many various model samples can be generated from the same base distribution, allowing samples to be generated across the $(\Omega_\text{M},\sigma_8)$ manifold with the same phases. Results are shown in Fig.~\ref{fig:cond_Glow_visual_om_s8_samebase} for two instances of varied $(\Omega_\text{M},\sigma_8)$ samples generated from the same base distribution. Visually inspecting the parameter space, decreasing $\Omega_\text{M}$ increases the size of voids, while overdensity features gain larger amplitudes for increasing $\sigma_8$. Meanwhile, the phases of each set of samples are apparently very similar, with each halo, filament, and void having nearly the same location in each sample across the $(\Omega_\text{M},\sigma_8)$ plane.

\begin{figure}
    \centering
    \begin{tabular}{cc}
        \hspace{.8cm}Power spectra varied $\Omega_\text{M}$ with $\sigma_8=0.8$ & \hspace{.8cm}Power spectra varied $\sigma_8$ with $\Omega_\text{M}=0.3$\\
        \includegraphics[width=0.49\textwidth]{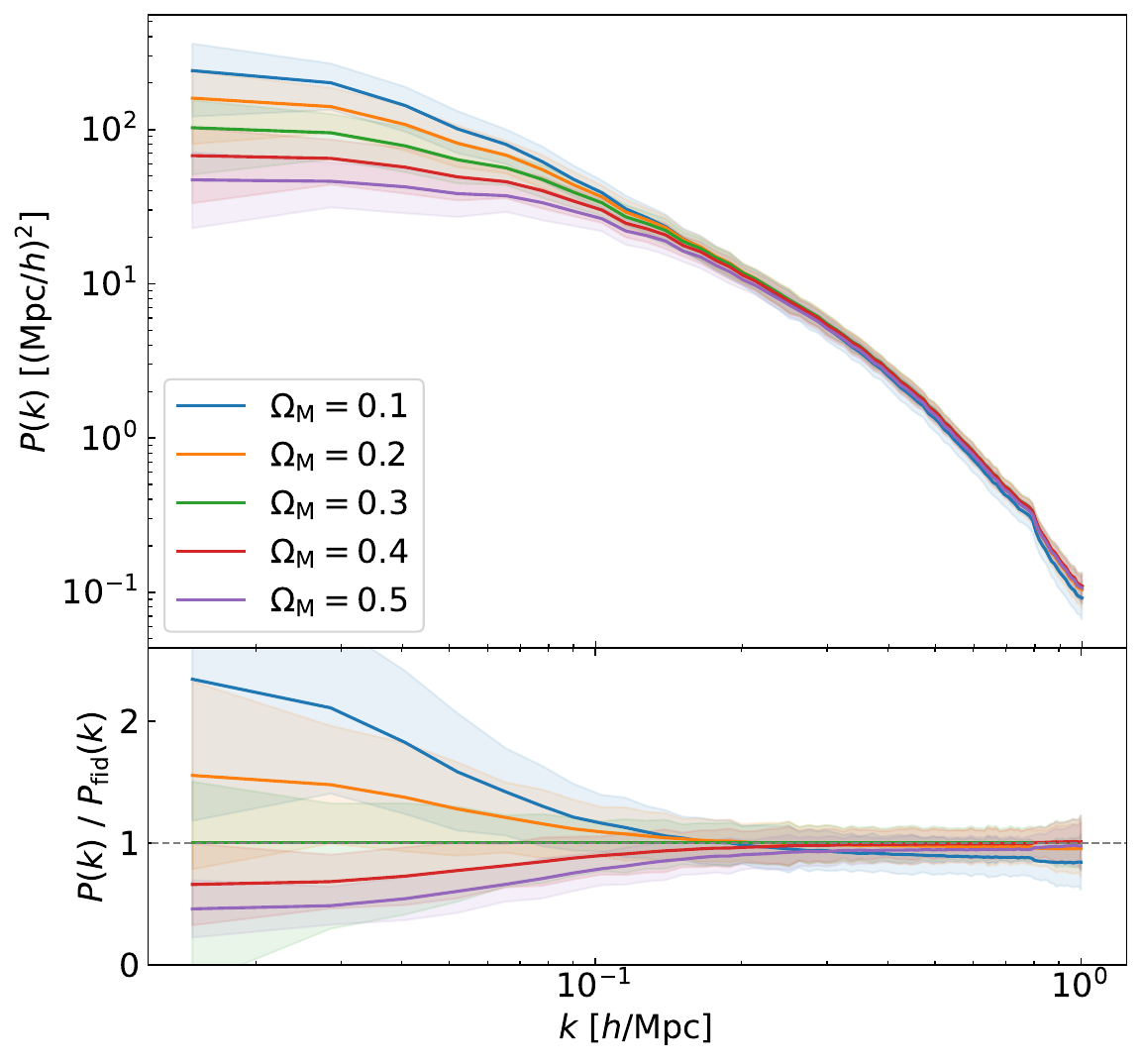} & \includegraphics[width=0.49\textwidth]{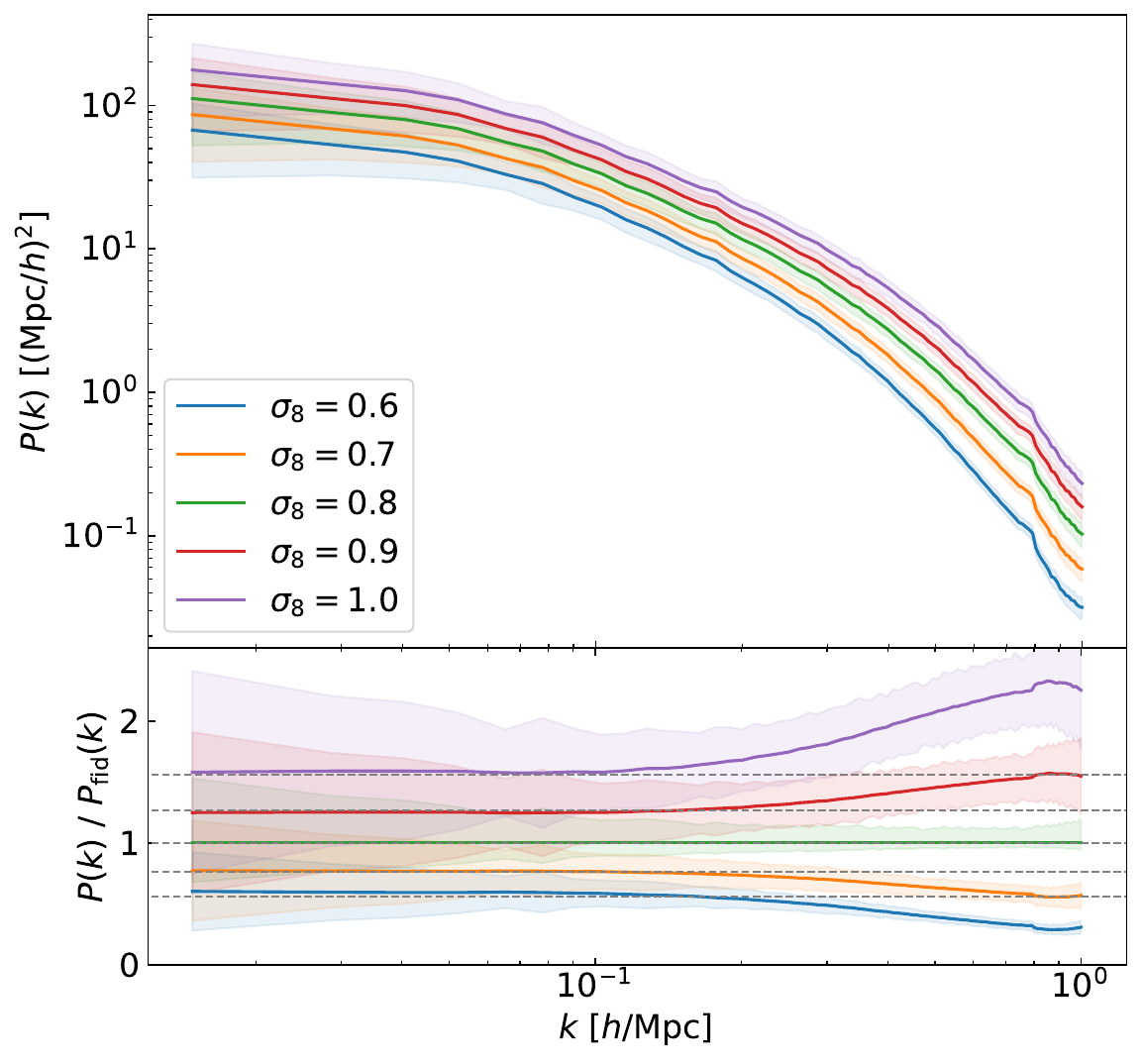}
    \end{tabular}
    \caption[The power spectra for varied $\Omega_\text{M}$, $\sigma_8=0.8$ from the same base distribution]{Power spectra for varied $\Omega_\text{M}$ with $\sigma_8=0.8$ (left) and varied $\sigma_8$ with $\Omega_\text{M}=0.3$ (right). Shown are the mean of $100$ samples, with each sample varied in $(\Omega_\text{M},\sigma_8)$ from the same base distribution $\bm{u}$. The $P(k)$ ratio plot for varied $\sigma_8$ has lines with values $(\sigma_8/0.8)^2$, for which the nonlinear side of $P(k)$ accurately matches.}
    \label{fig:ps_same_base}
\end{figure}

\sloppy The power spectra for samples generated from the same base distribution are shown in Fig.~\ref{fig:ps_same_base}. The results are shown for 100 samples, with each sample generated in $\Omega_\text{M}\in\{0.1,0.2,0.3,0.4,0.5\}$ and $\sigma_8\in\{0.6,0.7,0.8,0.9,1.0\}$. Although quantifying the accuracy of varying $\Omega_\text{M}$ is a bit difficult, we expect the amplitude of the linear power spectrum to be proportional to $\sigma_8^2$~(Eq.~\ref{eq:sigma_8_def}). At nonlinear scales and late times, an increased matter amplitude causes stronger gravitational interactions, and thus increased power. This is clearly seen in the ratio of $P(k)/P_\text{fid}(k)$, where the powers diverge from each starting at the nonlinear scale of $k\sim0.2h/\text{Mpc}$.

\section{$\Omega_\text{M}$, $\sigma_8$ parameter inference}

In the previous section, the normalizing flow learned the PDF $p(\bm{x}|\Omega_\text{M},\sigma_8)$. This Section now performs field level parameter inference of $p(\Omega_\text{M},\sigma_8|\bm{x})$. Because $\Omega_\text{M}$ and $\sigma_8$ were chosen in a uniform distribution, $p(\Omega_\text{M},\sigma_8|\bm{x})$ is proportional to $p(\bm{x}|\Omega_\text{M},\sigma_8)$; the proportionality constant is found by normalizing probabilities to unity.

Results for $p(\Omega_\text{M},\sigma_8|\bm{x})$ are shown in Fig.~\ref{fig:cond_glow_param_inference} for 16 test samples $\bm{x}$ spanning the space of $(\Omega_\text{M},\sigma_8)$. All results for $\Omega_\text{M}$ and $\sigma_8$ shown here are within $3\sigma$ of the true values, where the model gives about $\sigma_{\Omega_\text{M}}\sim0.04$ and $\sigma_{\sigma_8}\sim0.01$ for each sample. The model also gives a slight degeneracy between $\Omega_\text{M}$ and $\sigma_8$, which is usually found in large-scale structure surveys~\cite{Li_2023}.

The model PDF is compared to a PDF found from the power spectra of the same fields. This PDF from the power spectra is obtained with a masked autoregressive flow (MAF)~\cite{2017arXiv170507057P}, a type of normalizing flow that stacks several Masked Autoencoder for Distribution Estimation~\cite{germain2015made} blocks. The power spectrum data on the $128$~px length fields contains $90$ bins, and $2$ extra bins containing the respective values of $\Omega_\text{M}$ and $\sigma_8$ are concatenated to each power spectrum. These lists of $92$ numbers are trained on with the MAF to learn the conditional $p(\bm{x}|\bm{\phi})$. The results presented here may be compared to similar approaches of cosmological parameter inference with various other normalizing flows in~\cite{Hassan:2021ymv,trenf,Dai:2023lcb}.


\tikzstyle{square} = [rectangle, fill=white, line width=0cm,
                      minimum height=1.0cm, text width=1.0cm, text centered]

\begin{figure}
\centering
\begin{tikzpicture}
\node (x) {\includegraphics[width=0.99\textwidth]{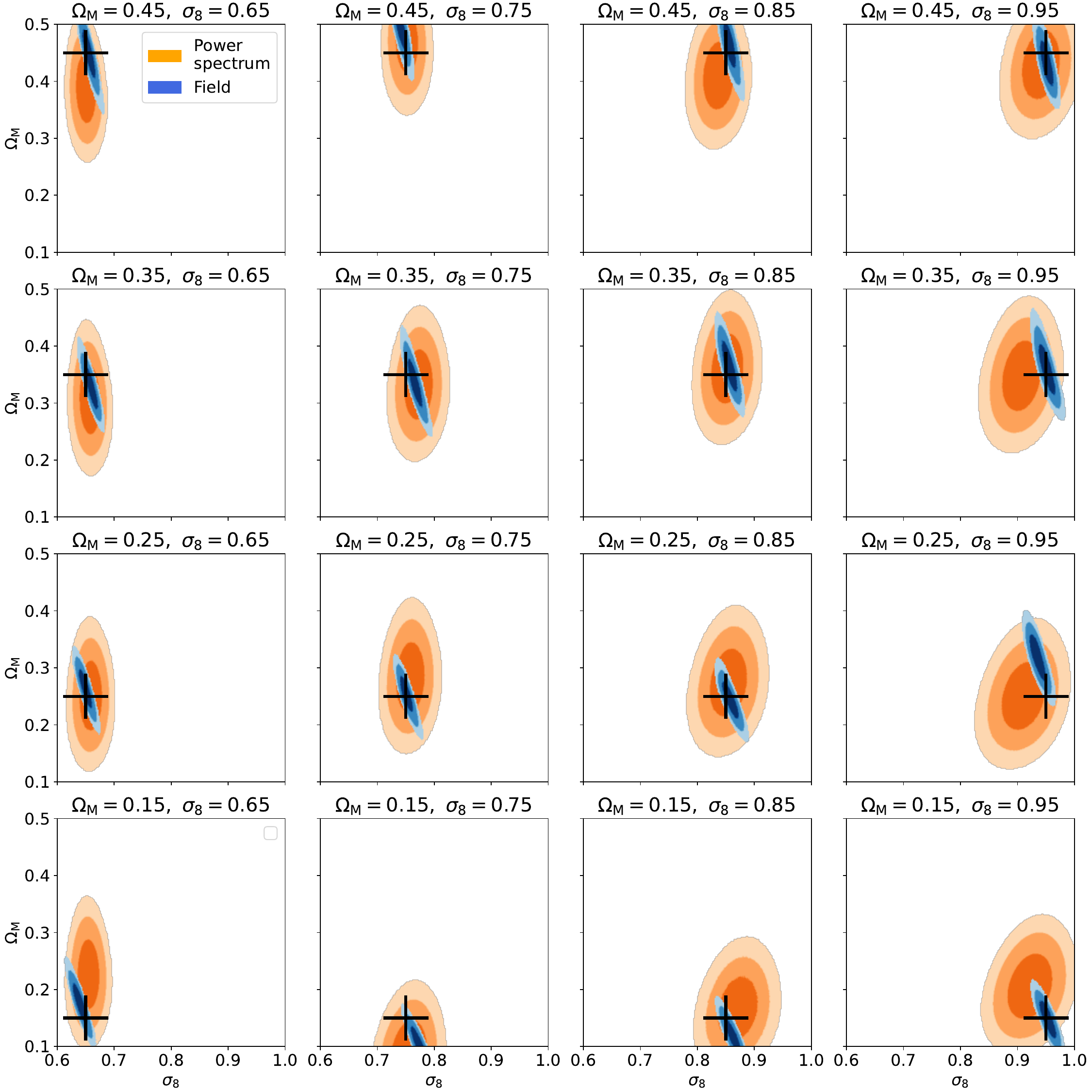}};

\node[square, xshift=-4.65cm, yshift=-4.65cm] (U) {};
\end{tikzpicture}
\caption[$\Omega_\text{M}$ and $\sigma_8$ parameter inference with conditional Glow]{Parameter inference with conditional Glow for test samples across the space of $(\Omega_\text{M},\sigma_8)$. Shown are $1\sigma$, $2\sigma$, and $3\sigma$ regions from the field-level data in blue, with the true value listed above each plot and marked in black. In comparison is the PDF from the power spectra in orange, obtained from a masked autoregressive flow}
\label{fig:cond_glow_param_inference}
\end{figure}

\begin{figure}
    \centering
    \includegraphics[width=0.99\textwidth]{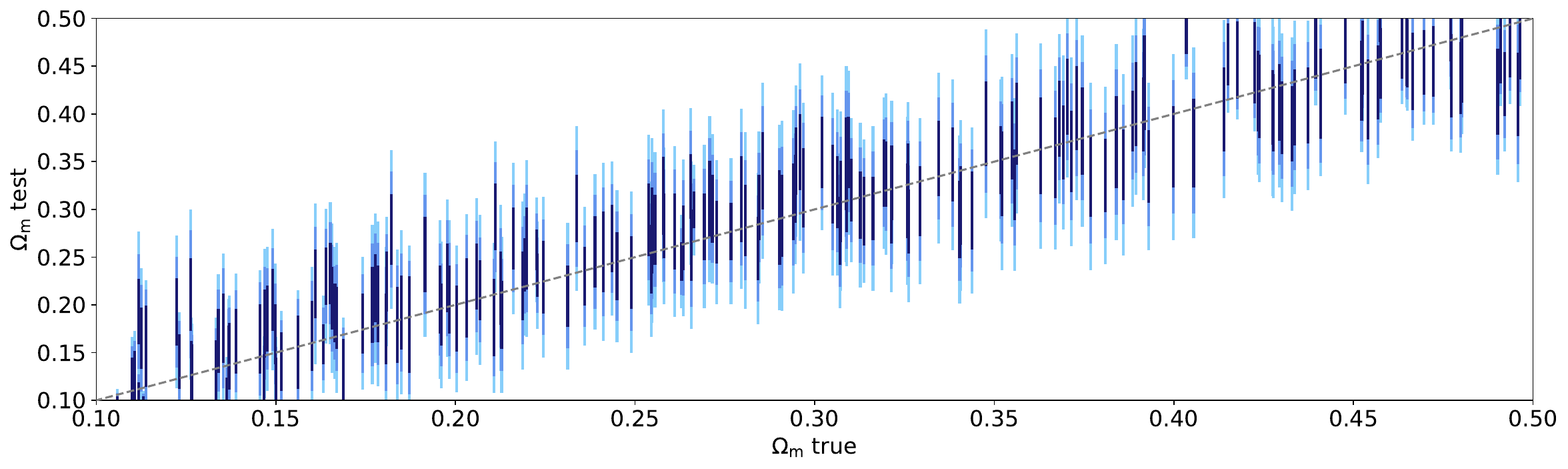}
    \includegraphics[width=0.99\textwidth]{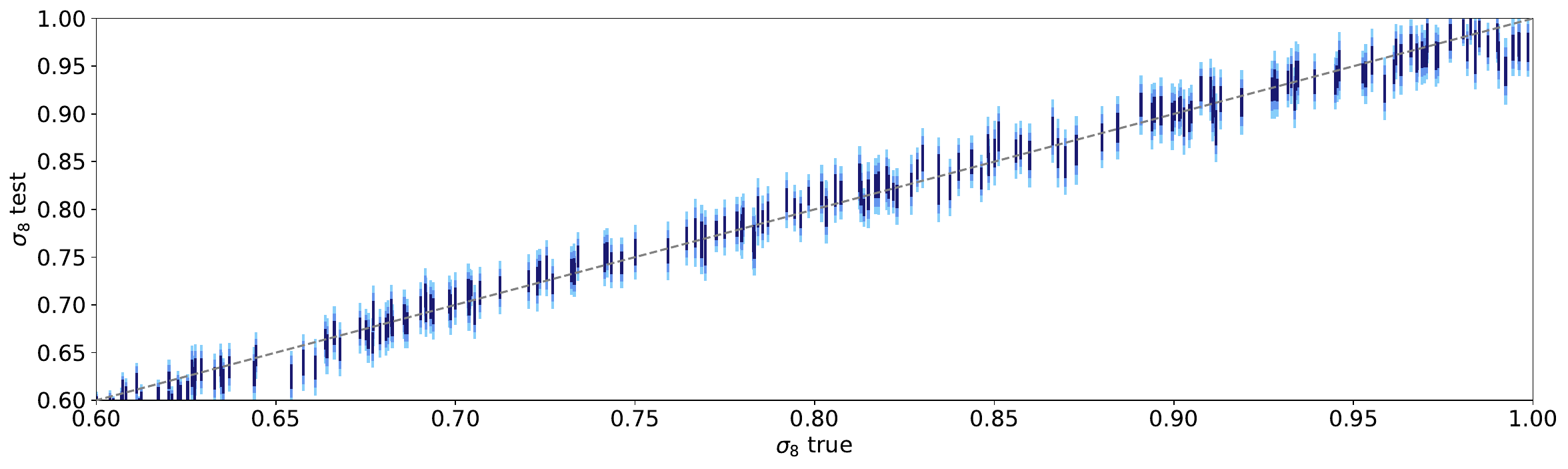}
    \caption[Marginalized $\Omega_\text{M}$ and $\sigma_8$ parameter inference with conditional Glow]{Parameter inference with conditional Glow for 200 random test samples, marginalized for $\Omega_\text{M}$ (top) and $\sigma_8$ (bottom). Shown are $1\sigma$, $2\sigma$, and $3\sigma$ regions.}
    \label{fig:cond_glow_param_inference_separate}
\end{figure}

Marginalized probabilities for $\Omega_\text{M}$ and $\sigma_8$ are separately plotted for 200 random samples from a test set in Fig.~\ref{fig:cond_glow_param_inference_separate}, again for $1\sigma$, $2\sigma$, and $3\sigma$ error bars. With these marginalized probabilities, we quantify the accuracy of our model in several ways. The root-mean-squared error for $\Omega_\text{M}$ is $0.032$, and for $\sigma_8$ is $0.012$. The coefficient of determination is defined as
\begin{equation}
    R_i^2=1-\frac{\sum_i^N\left(\phi_{i,\text{model}}-\phi_{i,\text{true}}\right)^2}{\sum_i^N\left(\phi_{i,\text{model}}-\frac{1}{N}\sum_i^N\phi_{i,\text{model}}\right)^2}.
\end{equation}
We get $R_{\Omega_\text{M}}^2=0.90$ and $R_{\sigma_8}^2=0.99$, indicating accurate estimates of the parameters across their domains, with more accurate inference of $\sigma_8$.

Now considering the size of the error bars obtained from the PDF, we would expect $68.3\%$ of the model predictions to be within the $1\sigma$ error bars, $95.5\%$ within $2\sigma$, and $99.7\%$ within $3\sigma$. The model predictions for $\Omega_\text{M}$ are $74.5\%$, $91.5\%$, and $98.0\%$ within the learned $1\sigma$, $2\sigma$, and $3\sigma$ regions, respectively; for $\sigma_8$, we get predictions $71.0\%$, $90.5\%$, and $96.5\%$ within the $1\sigma$, $2\sigma$, and $3\sigma$ regions. The model is slightly under-confident with the $1\sigma$ regions, while the $2\sigma$ and $3\sigma$ regions are slightly over-confident. Another measure of accuracy of the model's estimated $1\sigma$ errors is the reduced chi-squared, defined by
\begin{equation}
    \chi_i^2=\frac{1}{N}\sum_i^N\left(\frac{\phi_{i,\text{model}}-\phi_{i,\text{true}}}{\sigma_i}\right)^2.
\end{equation}
We get $\chi_{\Omega_\text{M}}^2=0.763$ and $\chi_{\sigma_8}^2=0.744$, again indicating under-confidence at the marginalized $1\sigma$ level. The model is more accurate in its mean value predictions of $\sigma_8$ than $\Omega_\text{M}$, however, the much tighter error bars on $\sigma_8$ over $\Omega_\text{M}$ give similar chi-squared values between the two parameters.

\label{sec:parameter_transfer}

\section{Conclusion}

Parameter inference at the field level has become greatly aided recently by various machine learning approaches. This Chapter has used a normalizing flow conditional on $\Omega_\text{M}$ and $\sigma_8$ to learn the PDF $p(\bm{x}|\Omega_\text{M},\sigma_8)$. With the PDF in hand, parameter inference $p(\Omega_\text{M},\sigma_8|\bm{x})$ and data generation can be performed. We find that the flow based PDF has significantly tighter parameter constraints than those obtained from the power spectra of the same fields.

A conceptually straightforward further direction would be to include additional parameters in the inference scheme. As N-body simulations are sensitive to only a few parameters as well as the initial conditions, the most obvious candidate would be $n_s$. However, the computational workload of going from 2 to 3 parameters would be significant with the methods presented in this Chapter, sampling every grid point on the entire parameter space. Moving from this grid point evaluation to standard Monte Carlo methods would be necessary for a higher dimensional parameter space.

Using the latent space mapping from $\bm{x}$ to $\bm{u}$, an interesting application of the normalizing flow would be to adjust the parameters of some new (possibly slightly out-of-distribution) field $\bm{x}+\delta\bm{x}$. In this way, a high quality simulation box $\bm{x}+\delta\bm{x}$ can be run a single time, and its cosmological parameters may be moved along the space of $\bm{\phi}$ that the flow has learned.

This Chapter has considered dark matter N-body simulations, a simplification from the significantly more computationally intensive baryonic physics. Applying the methods of this Chapter to real data must consider the effects of baryons~\cite{Huang_2019}. It would be straightforward to train a conditional normalizing flow on the CAMELS~\cite{camels2021} simulation suite, which does model baryons. CAMELS contains simulations in a latin hypercube of 6 parameters, $\Omega_\text{M}$, $\sigma_8$, and 4 astrophysical parameters. However, significant adjustments must be made for the model to accurately obtain PDFs from real data. Real data may have survey masks, systematic errors, and a different overall scaling that the model has not accounted for. A seemingly small discrepancy may adversely affect the model. For example, the value of $\sigma_8$ is related to the amplitude of the linear power spectrum. If real data is not scaled to exactly what the model trained on, the model's confident measurement of $\sigma_8$ may be wrong by many standard deviations. Additionally, applying our model to real survey data would require a more careful consideration of the length scales we model in our simulations, and a method to marginalize over simulation uncertainties. A longer discussion of applying out-of-distribution data to machine learning models is in Chapter~\ref{ch:conclusion}.

\chapter{Denoising Non-Gaussian Fields in Cosmology with Normalizing Flows}
\label{ch:denoising}
Fields in cosmology, such as the matter distribution, are observed by experiments up to experimental noise. The first step in cosmological data analysis is usually to denoise the observed field using an analytic or simulation driven prior. On large enough scales, such fields are Gaussian, and the denoising step is known as Wiener filtering. However, on smaller scales probed by upcoming experiments, a Gaussian prior is substantially sub-optimal because the true field distribution is highly non-Gaussian. Using normalizing flows, it is possible to learn the non-Gaussian prior from simulations (or from more high-resolution observations), and use this knowledge to denoise the data more effectively. We show that we can train a flow to represent the matter distribution of the universe, and evaluate how much signal-to-noise can be gained as a function of the experimental noise under idealized conditions. We also introduce a patching method to reconstruct fields on arbitrarily large images by dividing them up into small maps (where we reconstruct non-Gaussian features), and patching the small posterior maps together on large scales (where the field is Gaussian).

\section{Introduction}
\label{sec:introduction_denoising}

Normalizing flows~\cite{2019arXiv191202762P} have been shown to be very effective at learning high-dimensional probability distribution functions (PDFs), in particular when the random variables are spatially organized as in an image. This has led to a lot of recent work where PDFs in physics have been parameterized with flows, in particular in the domain of lattice QCD~\cite{Albergo:2021vyo}. In our precursor work~\cite{rouhiainen2021} presented in Chapter~\ref{ch:nf_random}, we evaluated how well various flows can learn sample generation and density estimation of cosmological fields. In the present Chapter, we use the learned flow for a practical application, finding the maximum a posteriori (MAP) value of a noisy observation. The potential gain of the method is that observations from galaxy survey telescopes or intensity mapping could be denoised using a normalizing flow to ultimately reach better cosmological constraints. While a simulation driven prior comes with some baryonic uncertainty, the learned prior only needs to be closer to reality than the usual Gaussian assumption to successfully improve denoising.

In cosmology, apart from our previous work~\cite{rouhiainen2021}, flows have recently been used to represent the matter distribution of the universe in~\cite{trenf}. This paper designed a rotation equivariant flow, TRENF, specifically for cosmology, while here we used a classical real NVP flow~\cite{2016arXiv160508803D} which is only translationally symmetric. The TRENF paper is using flows to measure cosmological parameters, while here we use it to reconstruct a field from a noisy observation. The closest existing works which we are aware of are~\cite{score_matching_2020} and~\cite{score_matching_2022}, which also aim to improve the posterior of a noisy observation of a Gaussian field, by using a learned prior. However in their case, a score matching approach was used which learns gradients, rather than a normalizing flow that gives the complete normalized PDF. Normalizing flows have the advantage that they are straightforward to use both for sampling and for inference, which makes them easy to interpret and visualize. In the present Chapter we consider the somewhat idealized case of an observation of the matter distribution corrupted by Gaussian noise, and study how much signal to noise can be gained depending on the noise in the experiment. We will use the simulated matter distribution as a proxy for observable non-Gaussian fields that depend on the matter distribution, including the smoothed galaxy field in a galaxy survey, the convergence map of galaxy surveys, or the secondary anisotropies of a high-resolution CMB survey (such as kSZ and CMB lensing)~\cite{Carlstrom_2002,Lewis:2006fu}. Applications to realistic observations should be presented in future work.

\begin{figure}
    \centering
    \begin{tabular}{ccc}
        \textbf{Type of data} & \textbf{Field $\bm{x}$} & \textbf{Base distribution $\bm{u}$}\\
        \begin{tabular}{l}In-distribution\end{tabular} & \begin{tabular}{l}\includegraphics[width=0.23\textwidth]{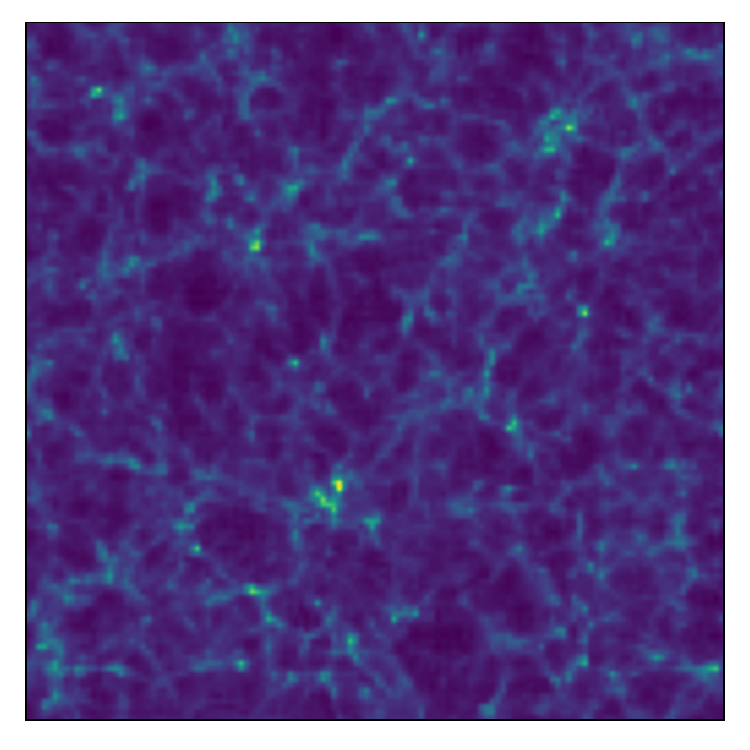}\end{tabular} & \begin{tabular}{l}\includegraphics[width=0.23\textwidth]{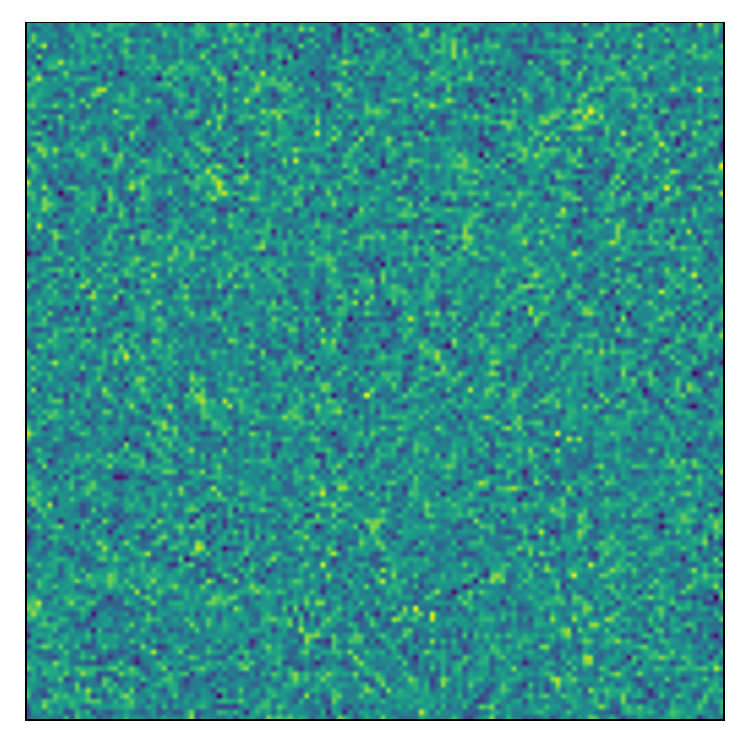}\end{tabular}\\
        \begin{tabular}{c}Out-of-distribution\\Foreground removed\end{tabular} & \begin{tabular}{l}\includegraphics[width=0.23\textwidth]{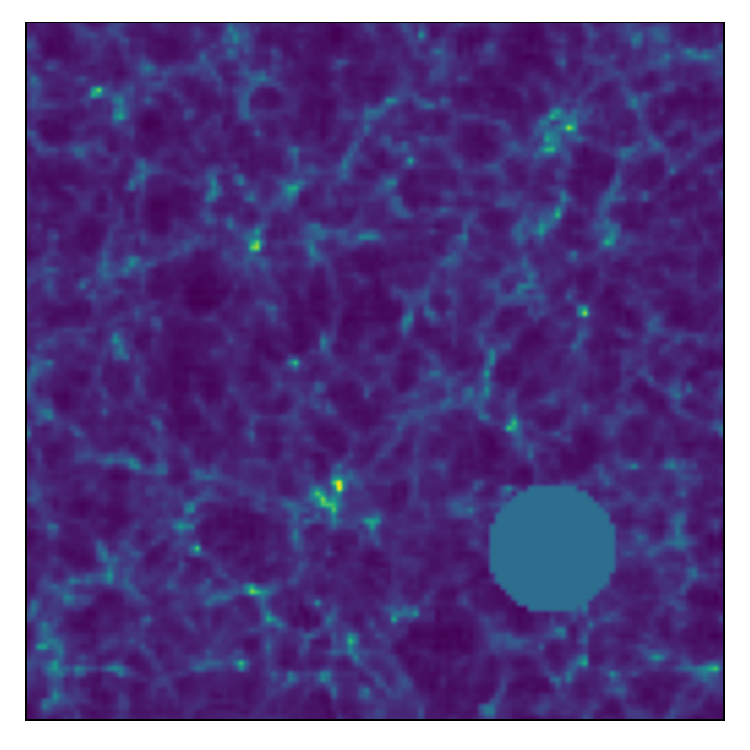}\end{tabular} & \begin{tabular}{c}\includegraphics[width=0.23\textwidth]{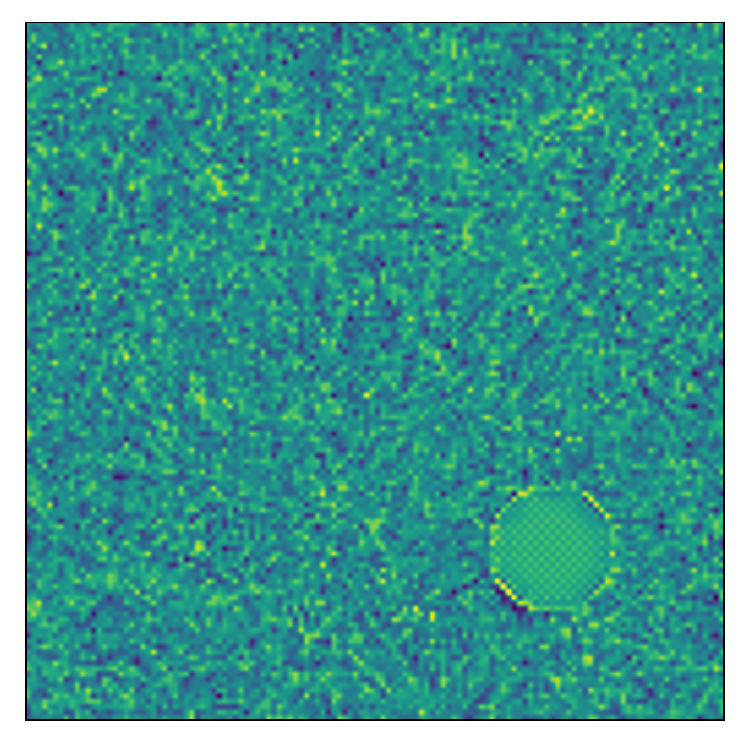}\end{tabular}\\
        \begin{tabular}{c}Out-of-distribution\\Noisy\end{tabular} & \begin{tabular}{l}\includegraphics[width=0.23\textwidth]{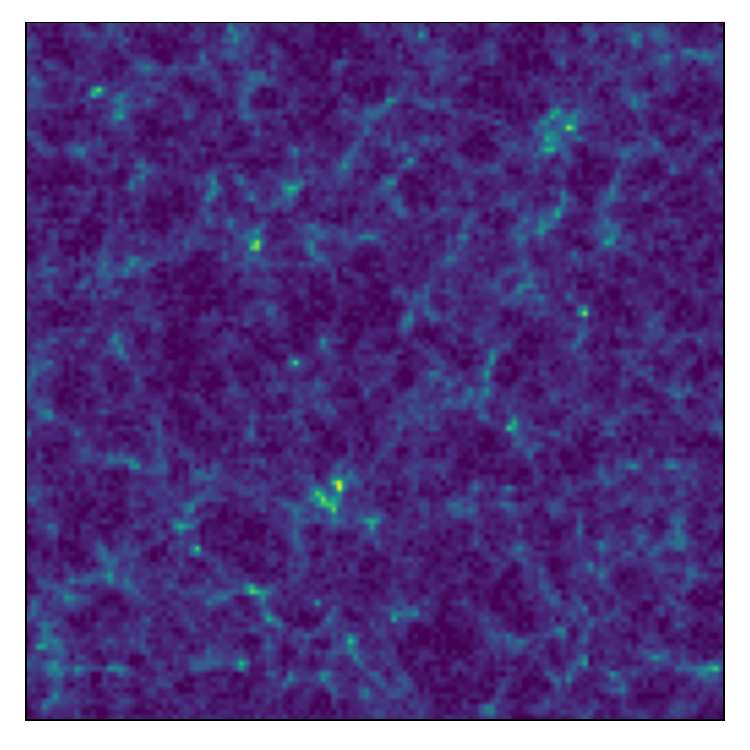}\end{tabular} & \begin{tabular}{l}\includegraphics[width=0.23\textwidth]{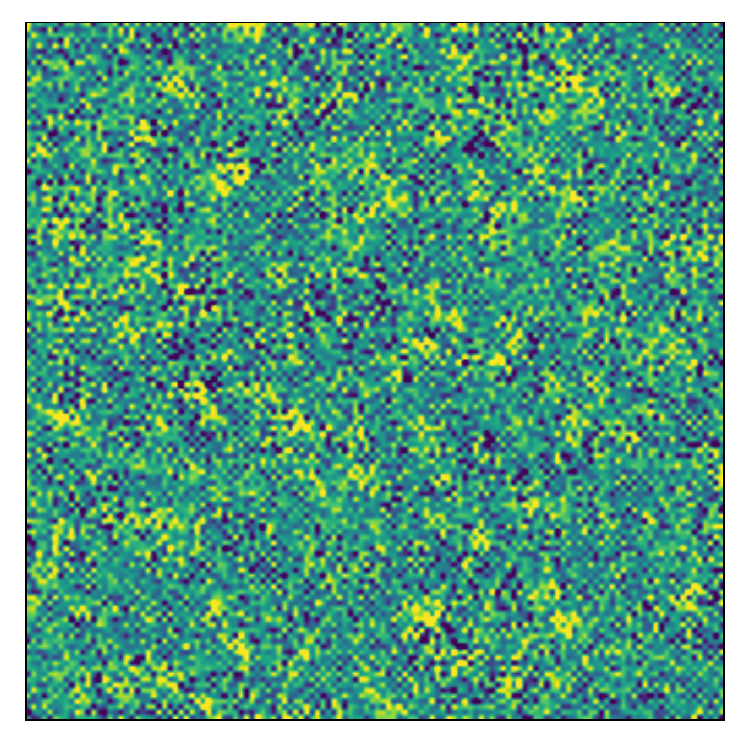}\end{tabular}\\
        \begin{tabular}{c}Out-of-distribution\\Wrong scale\end{tabular} & \begin{tabular}{l}\includegraphics[width=0.23\textwidth]{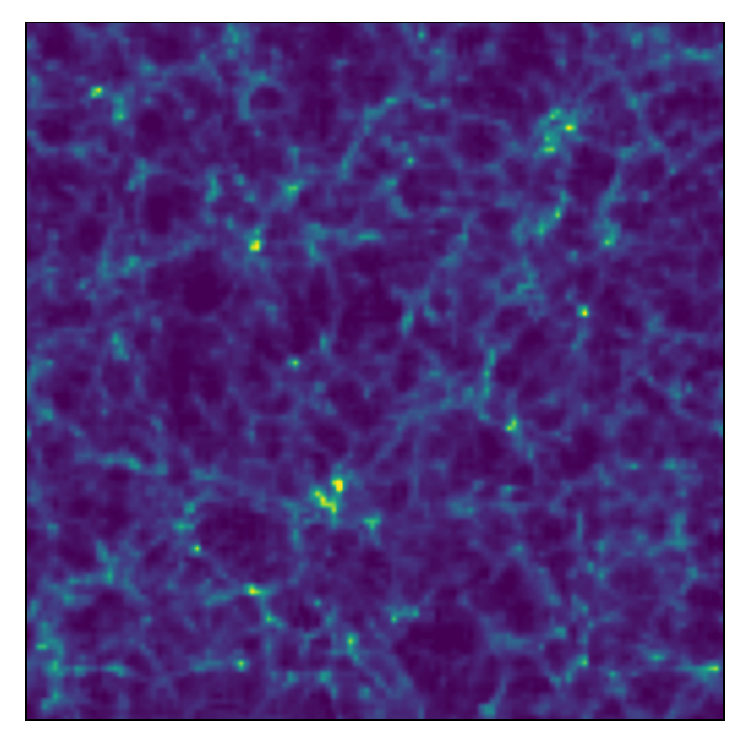}\end{tabular} & \begin{tabular}{l}\includegraphics[width=0.23\textwidth]{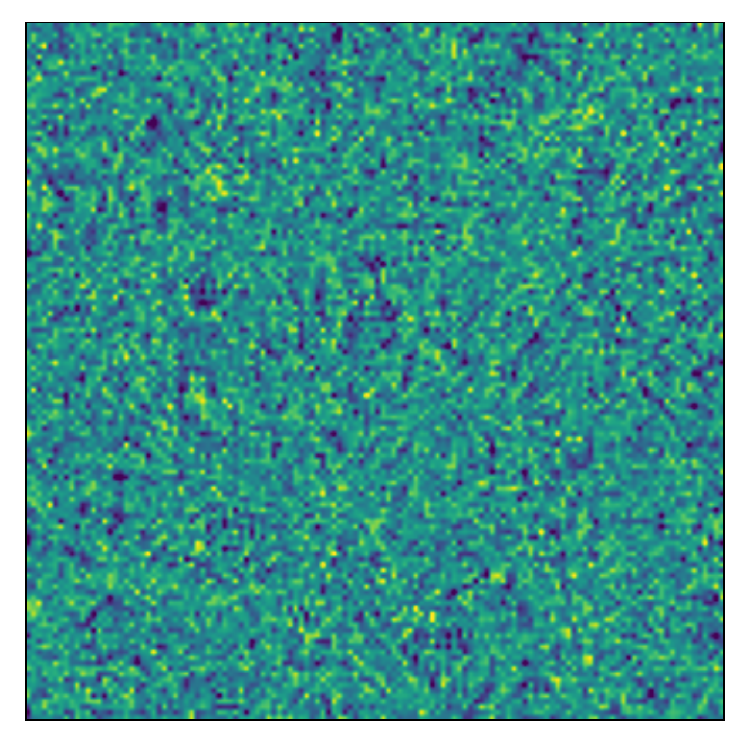}\end{tabular}\\
        \begin{tabular}{c}Out-of-distribution\\Downscaled\end{tabular} & \begin{tabular}{l}\includegraphics[width=0.23\textwidth]{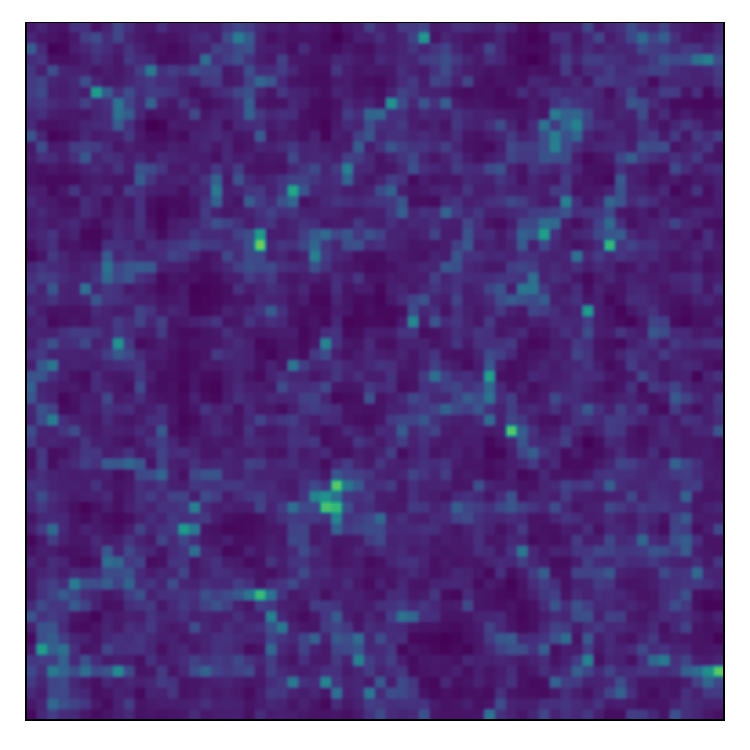}\end{tabular} & \begin{tabular}{l}\includegraphics[width=0.23\textwidth]{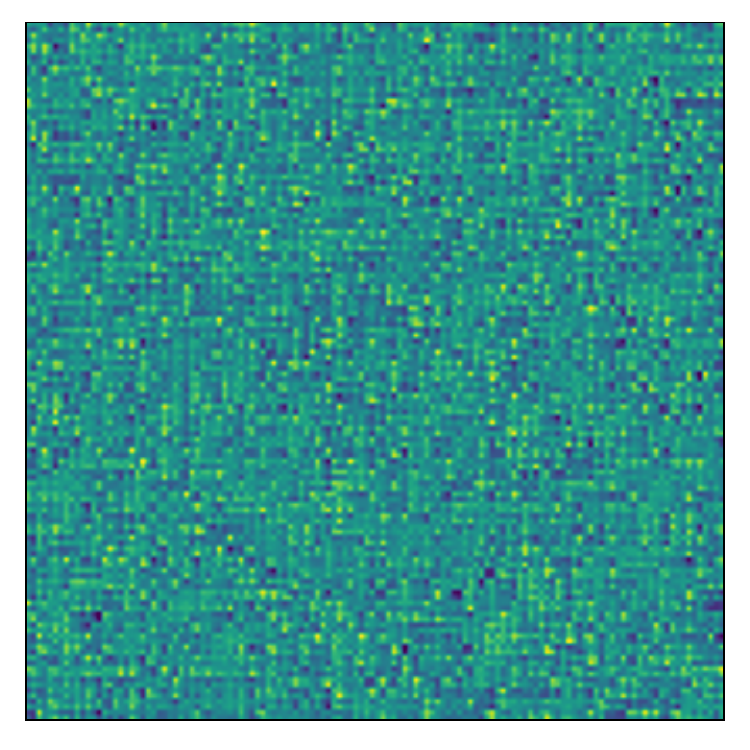}\end{tabular}
    \end{tabular}
    \caption[Normalizing flow detecting out-of-distribution data]{Normalizing flows may detect out-of-distribution data by flowing from fields $\bm{x}$ to the base distribution $\bm{u}$. For each in-distribution and out-of-distribution field $\bm{x}$ shown, the base distribution $\bm{u}$ is computed with a real NVP flow that has been trained on the in-distribution data. Only the in-distribution $\bm{x}$ correctly flows to a Gaussian field.}
    \label{fig:recon_illustration}
\end{figure}

\begin{figure}
    \centering
    \includegraphics[width=0.75\textwidth]{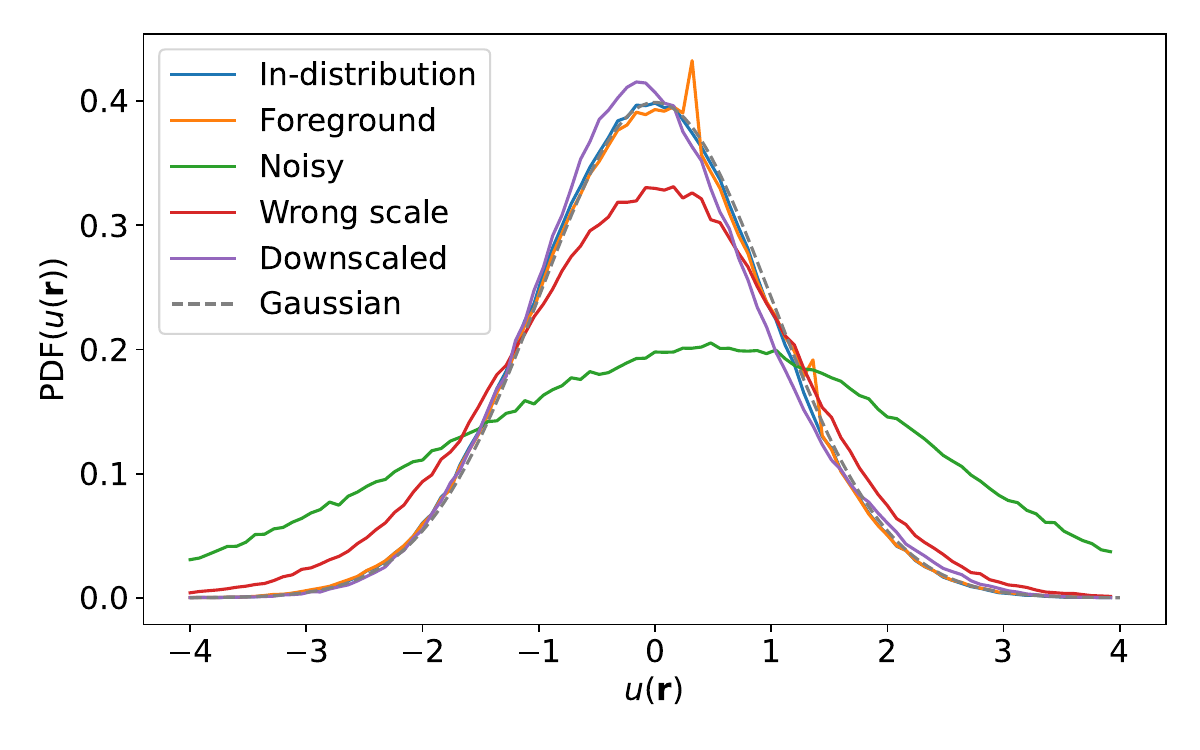} 
    \caption[Base distribution PDF for out-of-distribution data]{PDFs of in-distribution and out-of-distribution data, measured with their real NVP flow base distribution $\bm{u}$. The in-distribution $\bm{u}$ closely matches a Gaussian distribution, while each of the out-of-distribution examples are inaccurate. Each curve is averaged over $50$ fields.}
    \label{fig:u_hist}
\end{figure}

An illustration of how a normalizing flow detects out-of-distribution data is shown in Fig.~\ref{fig:recon_illustration}. A real NVP flow trained on the ``in-distribution" data flows to a Gaussian base distribution $\bm{u}$. However, for each of the out-of-distribution examples, $\bm{u}$ is not a Gaussian distribution. Fig.~\ref{fig:u_hist} plots the PDF of each $\bm{u}$, where is it apparent that the in-distribution $\bm{u}$ fits to a Gaussian curve, while the out-of-distribution $\bm{u}$ do not. Recalling that there are two terms in the loss function, the PDF of $\bm{u}$ matching Gaussian noise, and a second term related to the Jacobian determinant of every flow transformation $T$, the problem is actually worse than depicted. The transformations $T$ have never seen the out-of-distribution data and will produce Jacobian determinants reflecting that.

We outline our normalizing flow architecture, and describe how to find posterior reconstructed maps using a learned prior distribution, in Section~\ref{method}. We show our MAP results in Section~\ref{MAP_results} for reconstructing noisy N-body simulation data, comparing the flow results with Wiener filtering. In Section~\ref{HMC}, Hamiltonian Monte Carlo (HMC) is used to explore the parameter space of solutions to the denoising problem. Finally in Section~\ref{patching}, we demonstrate a method of seamlessly patching together many $128$~px length posterior maps to reconstruct a much larger $1024$~px map that might otherwise be too computationally difficult to address without patching.

\section{Method}
\label{method}

In this Section we describe the normalizing flow as well as the optimization procedure we use to denoise the observed matter field.

\subsection{Flow architecture}

The real NVP flow \cite{2016arXiv160508803D}, used in this Chapter to learn the matter distribution from simulations, is widely used and is expressive and fast for both sampling and inference. Here we outline the specifics of the network we use, with a more thorough discussion of normalizing flows in Chapter~\ref{ch:nf_random}. Our implementation resembles \cite{Albergo:2021vyo}, stacking $16$ affine coupling layers \cite{2014arXiv14108516D}, each with their own CNN. We use 3 convolutional layers with kernel size 3 and leaky ReLU activation functions, and we use $12$ hidden feature maps after each of the first and second convolutions. This architecture has a receptive field (Appendix~\ref{sec:convolutions}) of $97$~px, which on our data corresponds to a Fourier mode of $k=1.6\times10^{-2}\ h/\text{Mpc}$. In practice however, the real NVP network is learning the larger $k$ modes more accurately \cite{rouhiainen2021}. This is acceptable for our purposes, as we will reconstruct the small $k$ modes with Wiener filtering. Testing different sets of hyperparameters, we found improved generalization to out-of-distribution data with this setup of a relatively small 26,336 parameters, at the cost of having a smaller receptive field.

The small number of parameters can be motivated with the Akaike information criterion. For a statistical model with a number of $N_{\bm{\theta}}$ parameters, the relative quality of the model is~\cite{Akaike:1974}
\begin{equation}
    2N_{\bm{\theta}}-2\ln{\left(p(\bm{s}|\bm{d})\right)}.
\end{equation}
The Akaike information basically means that the preferred model has the least number of parameters while still being accurate.

The real NVP flow can be trained on either periodic or non-periodic data simply by setting the padding mode of the network convolutions to either periodic or zero padding. We use the same flow architecture for periodic data (Section~\ref{MAP_results} and Section~\ref{HMC}) and non-periodic data (Section~\ref{patching}) in this Chapter, only changing the convolution padding mode.

We decided against using a more complex flow architecture such as the Glow normalizing flow~\cite{2018arXiv180703039K}, which normally uses more training parameters than real NVP with the benefit of correlating longer length scales with pixel squeezing and channel mixing operations. We also experimented with the rotationally equivariant TRENF flow \cite{trenf}, which uses $\mathcal{O}\left(100\right)$ number of parameters in a typical setup; however with our implementation we were not able match the real NVP flow results.

\subsection{Finding the posterior with an optimizer}

In a cosmological experiment, such as a survey of the galaxy distribution, one can often assume that the vector of data $\bm{d}=\bm{s}+\bm{n}$ is a sum of mutually uncorrelated signal $\bm{s}$ and noise $\bm{n}$. The first step in cosmological data analysis is often to find the MAP $\hat{\bm{s}}$ of the signal given the data $\bm{d}$. The MAP is given by maximizing the posterior
\begin{align}
    \ln{p(\bm{s}|\bm{d})}&\propto \ln{p(\bm{d}|\bm{s})}+\ln{p(\bm{s})}\\
    &=-\frac{1}{2}(\bm{s}-\bm{d})^\text{T}\bm{N}^{-1}(\bm{s}-\bm{d})+\ln{p(\bm{s})}
\end{align}
with respect to the signal $\bm{s}$ to find the MAP $\hat{\bm{s}}$. Here we assumed that the noise of the experiment, which appears in the likelihood, is Gaussian with covariance $\bm{N}=\langle\bm{n}\bm{n}^\text{T}\rangle$, which is usually the case in practice. If we also assume that the signal is a Gaussian field with covariance $\bm{S}=\langle \bm{s}\bm{s}^\text{T}\rangle$, the prior is 
\begin{align}
    \ln{p(\bm{s})} 
    &\propto-\frac{1}{2}\bm{s}^\text{T}\bm{S}^{-1}\bm{s},
\end{align}
and there is an analytic solution to the maximization called Wiener filtering, given by
\begin{equation}
    \hat{\bm{s}}_\text{WF}=\frac{\bm{S}}{\bm{S}+\bm{N}}\bm{d}.
\end{equation}
The division here is taken element-wise.

Wiener filtering is very common in cosmology, see for example \cite{Carron:2017mqf}, \cite{Munchmeyer:2019kng}. However, upcoming surveys in cosmology such as Rubin Observatory \cite{LSSTSciBook} or Simons Observatory \cite{simons_observatory} probe the matter distribution with such high resolution, that scales are being measured where the Gaussianity assumption of the signal prior does not hold at all. Until recently, it would have been difficult to improve upon this assumption, because no good analytic expressions for the matter distribution~$p(\bm{s})$ at non-Gaussian scales exist. In this Chapter, we introduce the reconstruction of non-Gaussian signal maps $\bm{s}$ where the prior $\ln{p(\bm{s})}$ is learned with a normalizing flow. 

Based on the learned differentiable prior represented by the normalizing flow, we can either find the MAP solution to the posterior $\ln{p(\bm{s}|\bm{d})}$, or perform HMC to make probabilistic instances of the posterior $\ln{p(\bm{s}|\bm{d})}$. Extremizing the prior $p{(\bm{s})}$ to find the MAP solution is possible due to the gradients $\nabla_{\bm{s}}p{(\bm{s})}$ being immediately accessible with the flow. In this Chapter we demonstrate both MAP and HMC solutions, but first focus on the MAP for simplicity. A benefit of using a learned prior over directly training a neural network for denoising is that the noise matrix $N$ only appears in the likelihood term $\ln{p(\bm{d}|\bm{s})}$, which is easy to compute for Gaussian noise. Therefore a single trained flow used as the prior $\ln{p(\bm{s})}$ may be used to denoise any amount of noise $N$.

\subsection{Data}
\label{sec:denoising_data}

We use the particle mesh code \texttt{FastPM} \cite{Feng:2016yqz} to generate an ensemble of simulations of the matter distribution of patches of the universe, and project them to 2 dimensions for computational simplicity. Using cosmological parameters $\Omega_\text{M}=0.315$ and $\sigma_8=0.811$, we simulate $128^3$ particles in a $512\ \text{Mpc}/h$ side-length periodic box, and run 10 steps from scale factor $a=10$ to $a=1$. The particles are then fitted to mesh, creating 3-dimensional arrays of $128^3$~px. We make four 2-dimensional projections per 3-dimensional box by projecting 2 dimensions by a quarter of the box length along the third dimension. We make a total of 48,000 periodic matter density maps, split 80-10-10 as training, validation, and test sets.

\subsection{Model training}
\label{sec:denoising_training}

Our real NVP flow is trained on the $128^2$~px 2-dimensional projections of the simulations with an RTX A4000, using a batch size of 96 with a random rotation and flip given to each map. We minimize the Kullback–Leibler divergence \cite{2019arXiv191202762P} between the flow mapping of a Gaussian noise base distribution and our simulation target distribution with an Adam optimizer of learning rate $10^{-3}$, reduced by half on when the loss plateaus. The loss converges in $\sim10^6$ training cycles, in about 10~hours.

\begin{figure}
    \centering
    \includegraphics[width=0.65\textwidth]{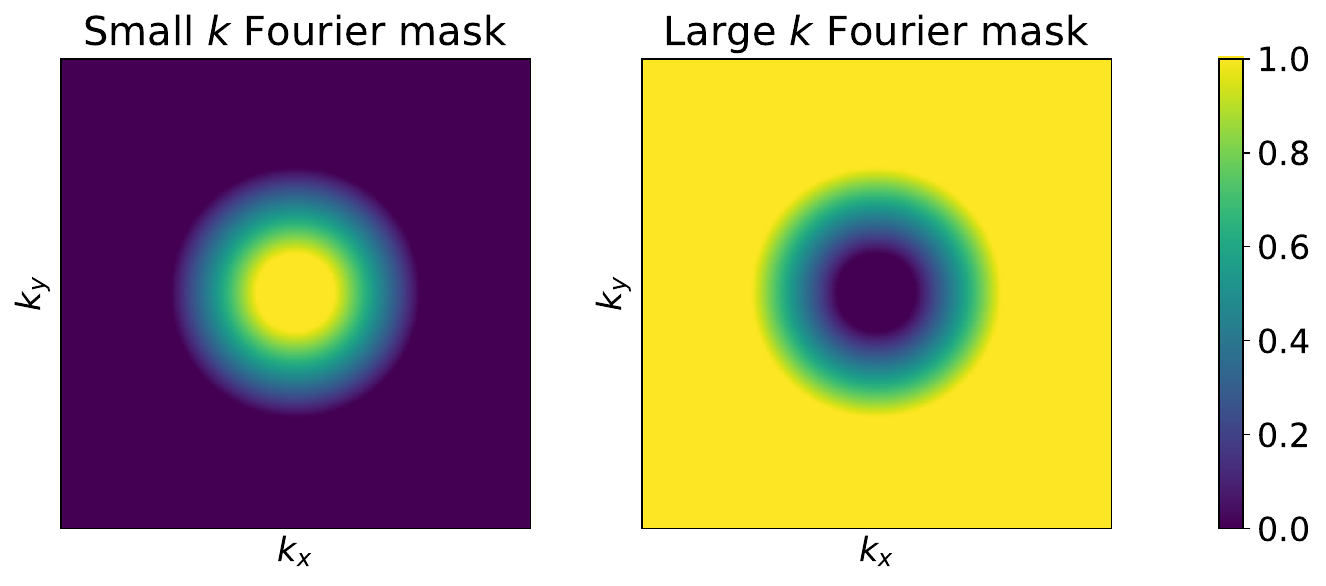}
    \caption[Fourier masks to split linear and nonlinear modes]{Wiener filtering is accurate on small $k$ modes, and we train a flow to be accurate on large $k$ modes. We apply a small $k$ Fourier mask (left) to Wiener filtered maps, and a large $k$ mask (right) to the flow maps. Our final reconstructed map is the sum in Fourier space of these two results. We found slightly more accurate summary statistics by using a smooth cutoff.}
    \label{fig:fourier_masks}
\end{figure}

After the flow is trained, we make simulated noisy data maps, by adding pixel-wise shot noise to the independent test data, and mask it to mimic a typical survey geometry. On this simulated data, we run an Adam optimizer to find MAP maps with our flow prior by extremizing $\ln{p(\bm{s}|\bm{d})}$. We found that we obtain the best results by using the flow prior only on small non-Gaussian scales, while optimizing the large linear scales with ordinary Wiener filtering (where it is optimal). Thus we make a smooth Fourier cutoff at about $k=0.2\ h/\text{Mpc}$, taking the small $k$ modes from Wiener filtering and the large $k$ modes from the flow posterior. In all results that follow, the ``flow" results are calculated after doing this Fourier splitting. An illustration of our Fourier masks is in Fig.~\ref{fig:fourier_masks}.

By splitting off the linear modes, we assume no large-scale to small-scale coupling. This assumption is not exactly correct in cosmology, and our reconstruction is thus not optimal. We will explore modelling such large-scale to small-scale coupling in the future using conditional normalizing flows, potentially improving the reconstruction further by conditioning the flow on the large-scale environment. However, our current implicit factorization assumption of the PDF in Fourier space is sufficient to improve the reconstruction as we shall now see.

\section{Results}
\label{MAP_results}

Examples demonstrating how our flow posterior reconstructs information on noisy, masked maps are shown in Fig.~\ref{fig:map_maps}. The noise in these examples has st. dev. $0.5\tilde{\sigma}$ and $1.0\tilde{\sigma}$, where we call $\tilde{\sigma}$ the pixel-wise st.~dev. of our training data.

\newcommand\fourimwidth{0.248}
\newcommand\fourhspace{-2.mm}
\newcommand\fourvspace{-6.5mm}

\begin{figure}[h]
    \centering
    {\Large$0.5\tilde{\sigma}$ noise}
    
    \medskip
    \begin{subfigure}{\fourimwidth\textwidth}
        \includegraphics[width=\textwidth]{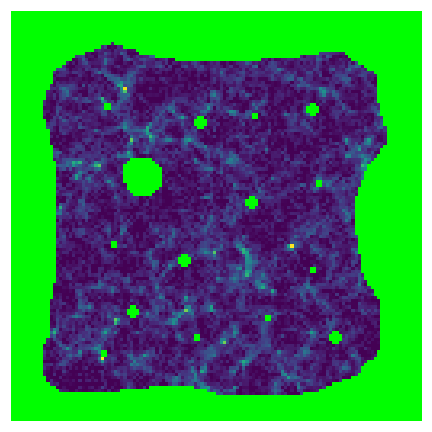}
        \vspace*{\fourvspace}
        \caption*{Observed}
    \end{subfigure}
    \hspace*{\fourhspace}
    \begin{subfigure}{\fourimwidth\textwidth}
        \includegraphics[width=\textwidth]{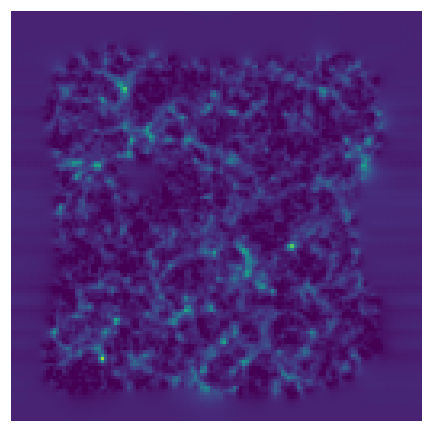}
        \vspace*{\fourvspace}
        \caption*{Wiener filtered}
    \end{subfigure}
    \hspace*{\fourhspace}
    \begin{subfigure}{\fourimwidth\textwidth}
        \includegraphics[width=\textwidth]{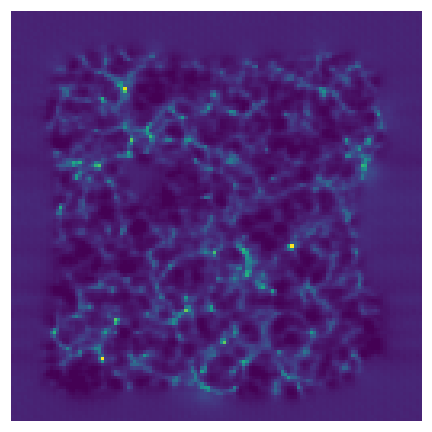}
        \vspace*{\fourvspace}
        \caption*{Flow posterior}
    \end{subfigure}
    \hspace*{\fourhspace}
    \begin{subfigure}{\fourimwidth\textwidth}
        \includegraphics[width=\textwidth]{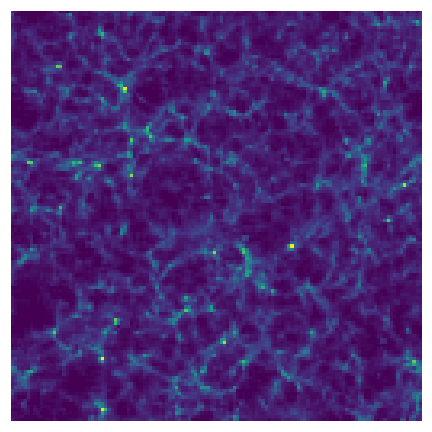}
        \vspace*{\fourvspace}
        \caption*{Truth}
    \end{subfigure}
    
    \medskip
    
    {\Large$1.0\tilde{\sigma}$ noise}
    
    \medskip
    
    \centering
    \begin{subfigure}{\fourimwidth\textwidth}
        \includegraphics[width=\textwidth]{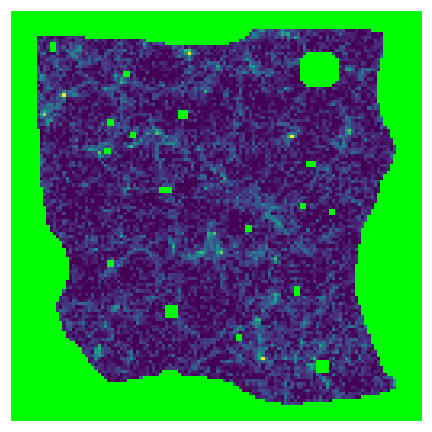}
        \vspace*{\fourvspace}
        \caption*{Observed}
    \end{subfigure}
    \hspace*{\fourhspace}
    \begin{subfigure}{\fourimwidth\textwidth}
        \includegraphics[width=\textwidth]{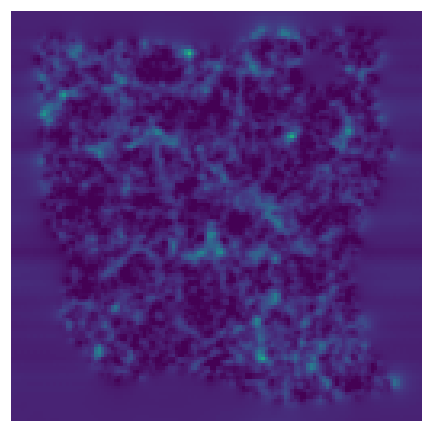}
        \vspace*{\fourvspace}
        \caption*{Wiener filtered}
    \end{subfigure}
    \hspace*{\fourhspace}
    \begin{subfigure}{\fourimwidth\textwidth}
        \includegraphics[width=\textwidth]{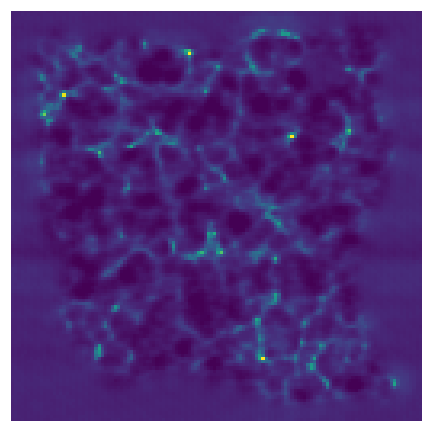}
        \vspace*{\fourvspace}
        \caption*{Flow posterior}
    \end{subfigure}
    \hspace*{\fourhspace}
    \begin{subfigure}{\fourimwidth\textwidth}
        \includegraphics[width=\textwidth]{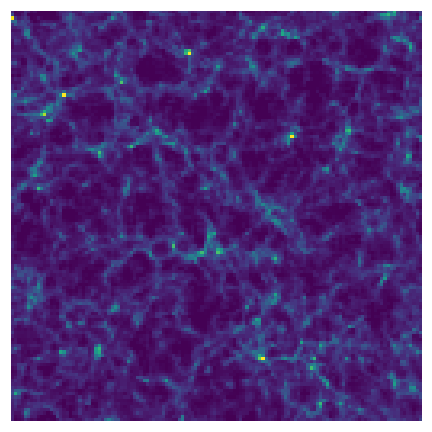}
        \vspace*{\fourvspace}
        \caption*{Truth}
    \end{subfigure}
    
    \caption[Observed, reconstructed, and truth maps for noisy and masked N-body simulations]{Observed, reconstructed posterior, and truth maps for $0.5\tilde{\sigma}$ and $1.0\tilde{\sigma}$ noise and a mask (where $\tilde{\sigma}$ is the st. dev. of the training data). Wiener filtering reduces noise on large and moderate length scales at the cost of over-smoothing small length scales. The flow posterior maps correctly retain the high frequency modes. The maps have length $128$~px, with physical length $512\ \text{Mpc}/h$ and a projected depth of $\Delta z=128\ \text{Mpc}/h$.}
    \label{fig:map_maps}
\end{figure}

We measure the quality of our posterior maps in several ways. The power spectra on discrete data may be computed as
\begin{equation}
    P(k)=\sum_{k_i-\frac{\Delta k}{2}<k<k_i+\frac{\Delta k}{2}}
    \frac{\left|\delta(\bm{k})\right|^2}{N_i},
\end{equation}
We also record the MSE per pixel, and the reconstruction noise defined by
\begin{equation}
    N^\text{rec}(k)=\langle(\bm{\varepsilon}^\text{rec})^\dagger\bm{\varepsilon}^\text{rec}\rangle
\end{equation}
where
$\bm{\varepsilon}^\text{rec}=\bm{s}^\text{rec}-\bm{s}^\text{truth}$. We also measure the accuracy of the reconstruction with the Fourier mode cross-correlation coefficient:
\begin{equation}
    r(k)=\frac{P^\text{true,rec}(k)}{\sqrt{P^\text{true}(k)P^\text{rec}(k)}}.
\end{equation}
where $P^\text{true,rec}(k)$ is the cross power spectrum. By ``$\text{rec}$," we mean either the reconstructed flow posterior or Wiener filtered map. Similar to the power spectrum, the cross power spectrum is defined by
\begin{equation}
    (2\pi)^2P^\text{true,rec}(k)\delta_\text{D}(\bm{k}+\bm{k}')=\langle\delta^\text{true}(\bm{k})\delta^\text{rec}(\bm{k}')\rangle,
\end{equation}
or for discrete data,
\begin{equation}
    P^\text{true,rec}(k)=\sum_{k_i-\frac{\Delta k}{2}<k<k_i+\frac{\Delta k}{2}}
    \frac{\left|\delta^\text{true}(\bm{k})\delta^\text{rec}(\bm{k})\right|}{N_i},
\end{equation}
where $N_i$ is the number of modes in the $k$-bin.

We present results for reconstructing 100 maps in our test set for the $0.5\tilde{\sigma}$ and $1.0\tilde{\sigma}$ noise setups. Fig.~\ref{fig:map_ps} shows plots of the power spectrum of the posterior maps and $N(k)$ (left), and cross-correlation~(right). We find improvement with the flow MAP against Wiener filtering on all scales above the nonlinear scale $k\sim0.2\ h/\text{Mpc}$, with an improvement of up to a factor of 2 in this setup for large~$k$. For $0.5\tilde{\sigma}$ noise, the flow posterior MSE is 29\% lower than the Wiener filtering MSE, and for $1.0\tilde{\sigma}$ noise the flow MSE is 22\% lower.

The improvement factor also depends strongly on the non-Gaussianity of the field. For example if we reduce the depth range (given by the radial distance $z$) of our 2-dimensional maps from $\Delta z=128\ \text{Mpc}/h$ to $\Delta z=32\ \text{Mpc}/h$, we find an improvement on the smallest scales by a factor of about $3.5$ in $r$. It would be interesting to study the possible improvement with volumetric data using a normalizing flow in 3 spacial dimensions, if the expressibility of normalizing flows could handle such high-dimensional data.

\newcommand\twoimwidth{0.4999}

\begin{figure}
    \centering
    {\Large$0.5\tilde{\sigma}$ noise}
    
    \medskip
    
    \includegraphics[scale=\twoimwidth]{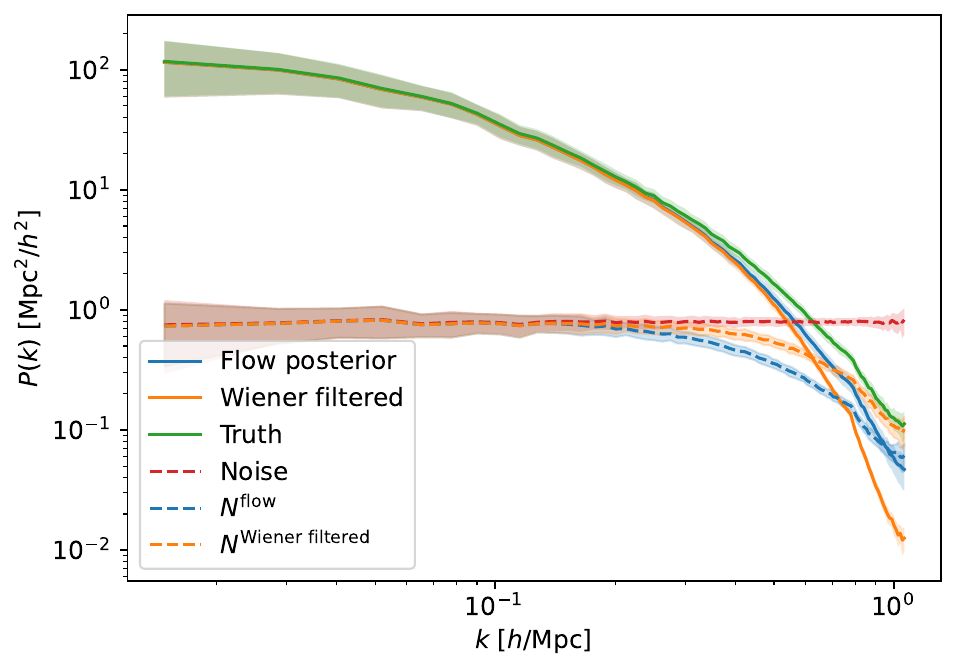}
    \includegraphics[scale=\twoimwidth]{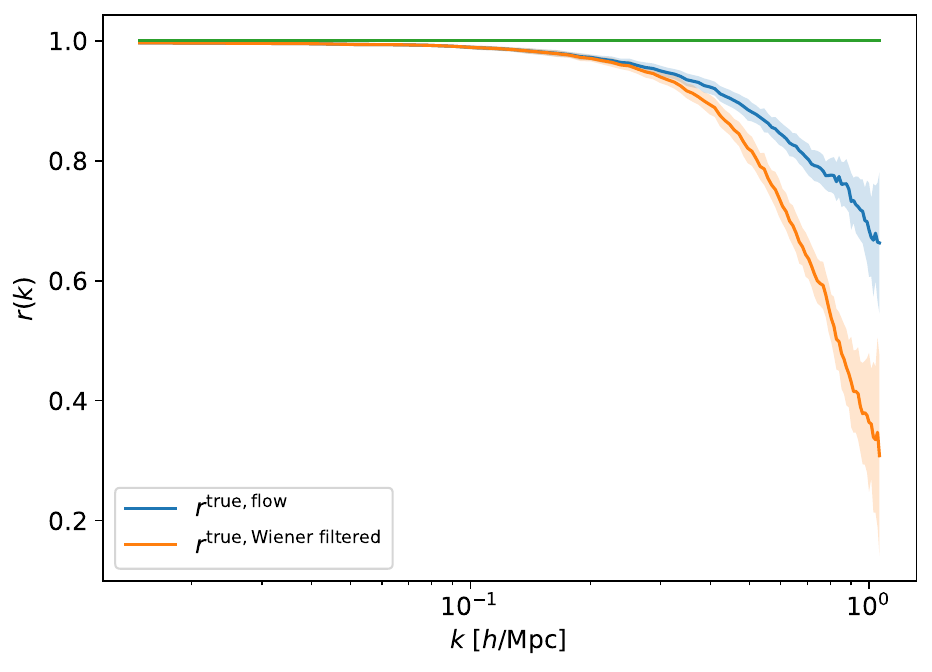}
    
    \medskip
    
    {\Large$1.0\tilde{\sigma}$ noise}
    
    \medskip
    
    \includegraphics[scale=\twoimwidth]{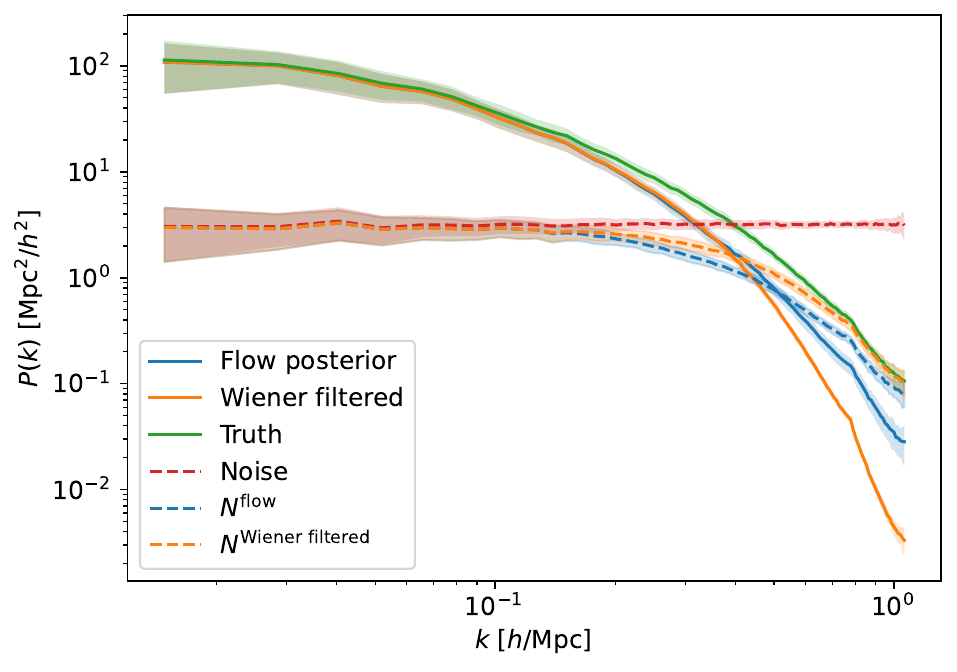}
    \includegraphics[scale=\twoimwidth]{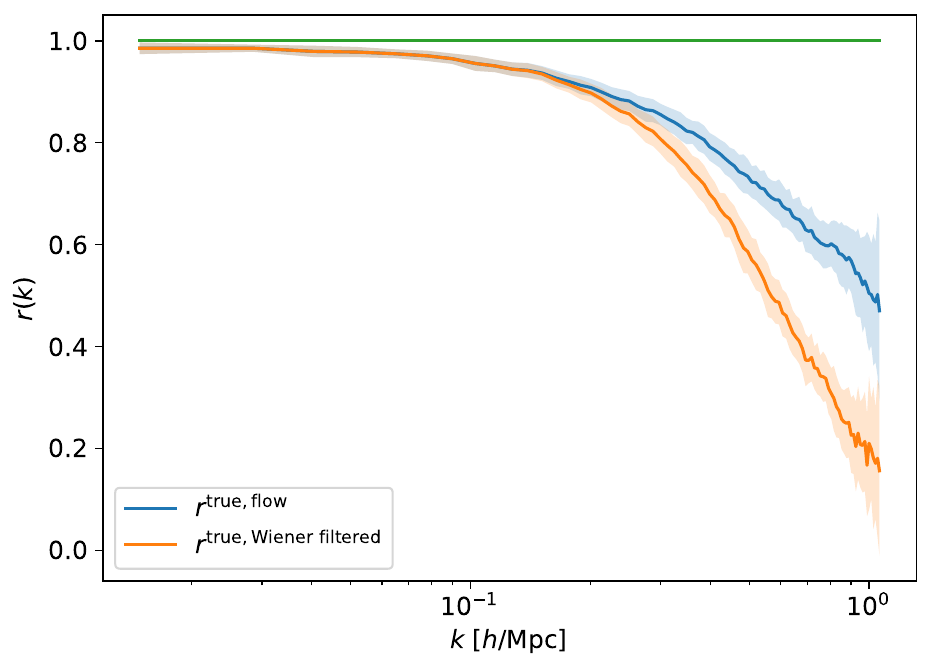}
    \caption[Summary statistics for denoising N-body simulations]{Power spectra (left) and cross-correlation (right) comparing our flow posterior with Wiener filtering for denoising $0.5\tilde{\sigma}$ and $1.0\tilde{\sigma}$ noise, averaged over 100 maps, along with $1$ st. dev. confidence intervals, computed without a mask. We find improvement with the flow over Wiener filtering on all modes above the nonlinear scale $k\sim0.2\ h/\text{Mpc}$: the flow power spectrum is closer to the truth, $N^\text{flow}$ is lower than $N^\text{Wiener\ filtered}$, and $r^\text{true,\ flow}$ has up to a factor of $2$ improvement at larger $k$ modes. In this example the projected depth of the map is $\Delta z=128\ \text{Mpc}/h$ which is relatively large; a smaller depth leads to more non-Gaussianity and thus even larger improvements.}
    \label{fig:map_ps}
\end{figure}

\begin{figure}
    \centering
    {\includegraphics[scale=\twoimwidth]{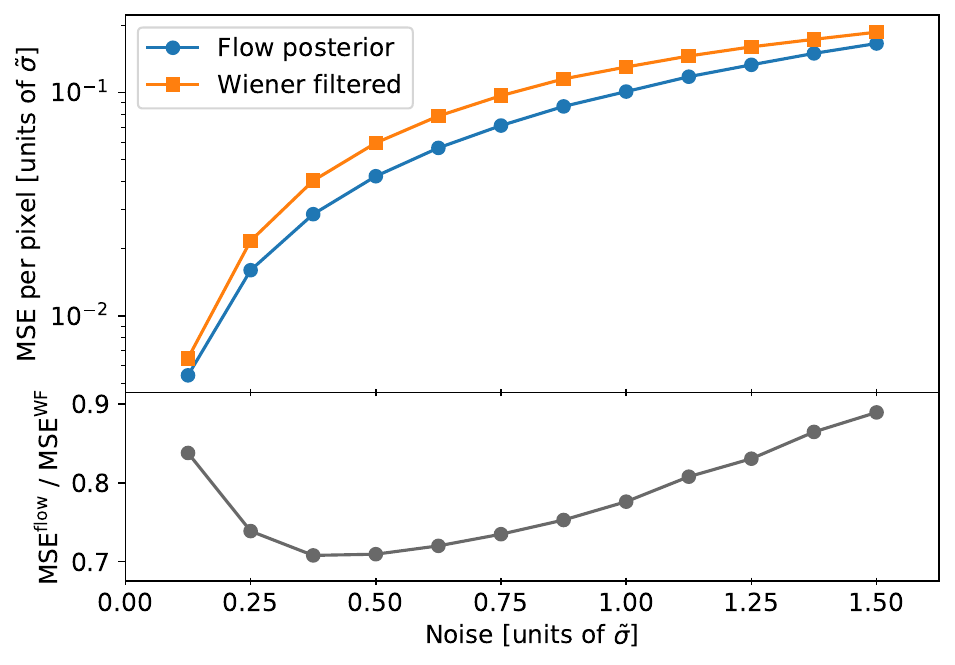}}
    \caption[MSE per pixel across a range of noise levels]{MSE per pixel between posterior and truth maps for a range of noise levels, calculated on 100 maps at each point. The flow improves denoising at all noise levels relative to Wiener filtering, with the largest improvement at a noise half the signal for a 30\% MSE reduction over Wiener filtering.}
    \label{fig:mse}
\end{figure}

We also examined how the improvement in the reconstruction depends on the noise in the map (without a mask). Fig.~\ref{fig:mse} shows the MSE per pixel calculated as a function of noise level (relative to~$\tilde{\sigma}$), comparing the flow posterior with Wiener filtering. We find a lower MSE with the flow at all noise levels. The greatest reduction in the flow's MSE over Wiener filtering is for the noise around half of the signal ($0.5\tilde{\sigma}$), giving about a $30\%$ improvement, with lesser improvements at either the low or high noise limit. Intuitively, if the noise is small, the gains by a better prior will be small since the prior matters less. On the other hand, for very large noise, the prior will begin to dominate over the likelihood. A difficulty, as we have found, is that the normalizing flow is trained on IID data, but the posterior optimization enters domains of the configuration space that are out-of-distribution and thus the flow may not generalize well to such cases \cite{nf_ood}. We found that it is advantageous for generalization to use a flow with relatively few training parameters, and that splitting the $k$ modes as explained above helps at large noise.

\section{Generating posterior samples with Hamiltonian Monte Carlo}
\label{HMC}

In the previous section we found the MAP solution to the denoising problem. We can also generate many diverse instances of posterior maps with the HMC algorithm. HMC obtains samples from the parameter space of a probability density $U\left(\bm{s}\right)$ containing parameters $\bm{s}$, and introduces auxiliary momentum variables $\bm{p}$. The momentum adds a term $\frac{1}{2}\bm{p}^\text{T}\bm{M}^{-1}\bm{p}$ to the Hamiltonian (where the mass matrix $\bm{M}$ is often taken to be diagonal), allowing the use of Hamiltonian dynamics to explore the space of posterior solutions for $\bm{s}$. The Hamiltonian is
\begin{align}
    H(\bm{s},\bm{p})&=U(\bm{s})+\frac{1}{2}\bm{p}^\text{T}\bm{M}^{-1}\bm{p}\\
    &=-\ln{p\left(\bm{s}|\bm{d}\right)}+\frac{1}{2}\bm{p}^\text{T}\bm{M}^{-1}\bm{p},
\end{align}
which corresponds to the familiar equations of motion,
\begin{align}
    \frac{\text{d}\bm{s}}{\text{d}t}&= \frac{\partial H}{\partial\bm{p}},\\
    \frac{\text{d}\bm{p}}{\text{d}t}&=-\frac{\partial H}{\partial\bm{s}}.
\end{align}
The position $\bm{s}$ and momentum $\bm{p}$ are evolved under Hamiltonian dynamics for a small time step by numerically integrating Hamilton's equations of motion. Every HMC sample containing the set of parameters $\bm{s}$ will be correlated to its previous sample in the Monte Carlo chain; the goal is to run a long enough chain to produce a number of uncorrelated samples.

We generate HMC samples with \texttt{hamiltorch} \cite{hamiltorch}, and use the no U-turn sampler (NUTS) \cite{NUTSHMC}. A main feature of NUTS is that it requires less fine-tuning of the step size. Each chain has a burn-in set of $500$ samples to set a step size, where there are $100$ steps per sample. The adapted step size from NUTS is about $0.02\tilde{\sigma}$ with a target acceptance rate of 0.8. For a single noisy map, we generate $2000$ HMC samples in this way, taking about $2$ hours to run each chain.

\begin{figure}
    \centering
    \begin{subfigure}{\fourimwidth\textwidth}
        \scalebox{1}[1]{\includegraphics[width=\textwidth]{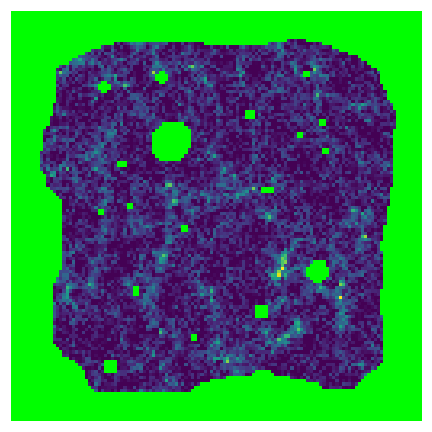}}
        \vspace*{\fourvspace}
        \caption*{Observed}
    \end{subfigure}
    \hspace*{\fourhspace}
    \begin{subfigure}{\fourimwidth\textwidth}
        \scalebox{1}[1]{\includegraphics[width=\textwidth]{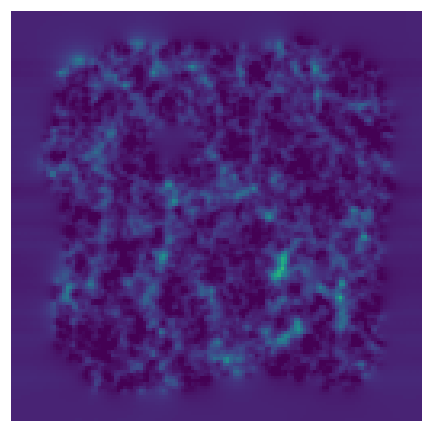}}
        \vspace*{\fourvspace}
        \caption*{Wiener filtered}
    \end{subfigure}
    \hspace*{\fourhspace}
    \begin{subfigure}{\fourimwidth\textwidth}
        \scalebox{1}[1]{\includegraphics[width=\textwidth]{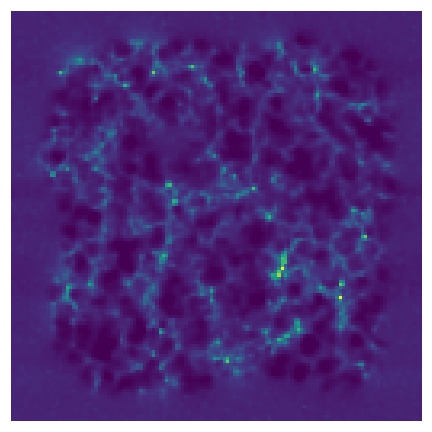}}
        \vspace*{\fourvspace}
        \caption*{Mean of 100 HMC}
    \end{subfigure}
    \hspace*{\fourhspace}
    \begin{subfigure}{\fourimwidth\textwidth}
        \scalebox{1}[1]{\includegraphics[width=\textwidth]{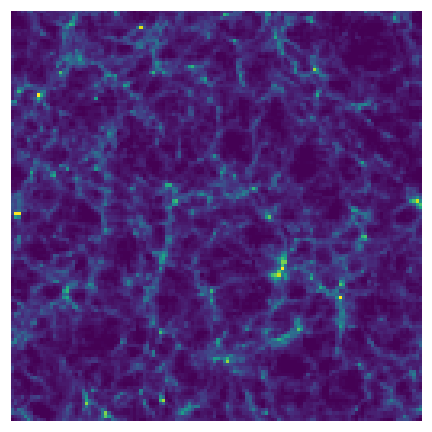}}
        \vspace*{\fourvspace}
        \caption*{Truth}
    \end{subfigure}
    
    \medskip
    
    \begin{subfigure}{\fourimwidth\textwidth}
        \scalebox{1}[1]{\includegraphics[width=\textwidth]{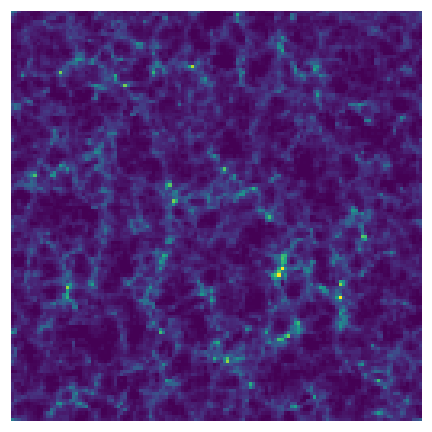}}
        \vspace*{\fourvspace}
        \caption*{HMC sample}
    \end{subfigure}
    \hspace*{\fourhspace}
    \begin{subfigure}{\fourimwidth\textwidth}
        \scalebox{1}[1]{\includegraphics[width=\textwidth]{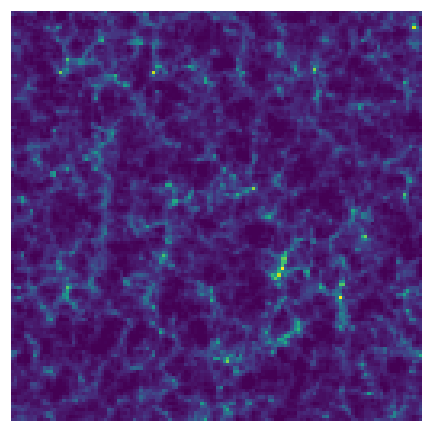}}
        \vspace*{\fourvspace}
        \caption*{HMC sample}
    \end{subfigure}
    \hspace*{\fourhspace}
    \begin{subfigure}{\fourimwidth\textwidth}
        \scalebox{1}[1]{\includegraphics[width=\textwidth]{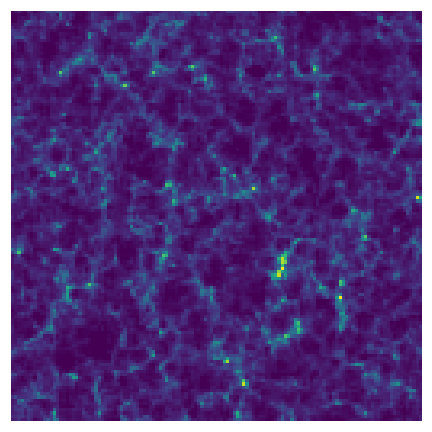}}
        \vspace*{\fourvspace}
        \caption*{HMC sample}
    \end{subfigure}
    \hspace*{\fourhspace}
    \begin{subfigure}{\fourimwidth\textwidth}
        \scalebox{1}[1]{\includegraphics[width=\textwidth]{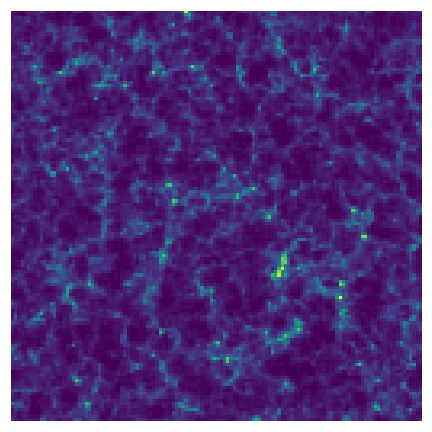}}
        \vspace*{\fourvspace}
        \caption*{HMC sample}
    \end{subfigure}
    
    \caption[Observed, reconstructed HMC posterior, and truth maps for noisy and masked N-body fields]{(Top row) Observed, reconstructed posterior mean of 100 HMC samples, and truth maps for $1.0\tilde{\sigma}$ noise and a mask. Wiener filtering reduces noise on large and moderate length scales at the cost of over-smoothing small length scales. The HMC posterior mean retains the high frequency modes. (Bottom row) Four HMC samples, showing the variety of possible posterior solutions.}
    \label{fig:hmc_result_128_map}
\end{figure}

Our results for reconstructing a masked map with $1.0\tilde{\sigma}$ noise is in Fig.~\ref{fig:hmc_result_128_map}. We selected $100$ samples evenly spread out along the HMC chain to take a pixel-wise mean as our final posterior mean map. The HMC posterior mean is of similar quality to the MAP solution in the previous section. We also show several HMC samples in Fig.~\ref{fig:hmc_result_128_map}, demonstrating the variety of possible solutions to the posterior reconstruction problem.

To obtain the posterior mean (rather than individual samples), we again Wiener filter the large scales and combine them with the HMC samples as described in the previous section. The computational cost of producing uncorrelated HMC samples is lowered with the Fourier splitting, as only the small-scale modes need to be de-correlated in the HMC chain. As is shown in \cite{reconstructVBS}, the autocorrelation length of HMC samples is about $2$ times larger on large length scales.

\begin{figure}
    \centering
    \includegraphics[scale=\twoimwidth]{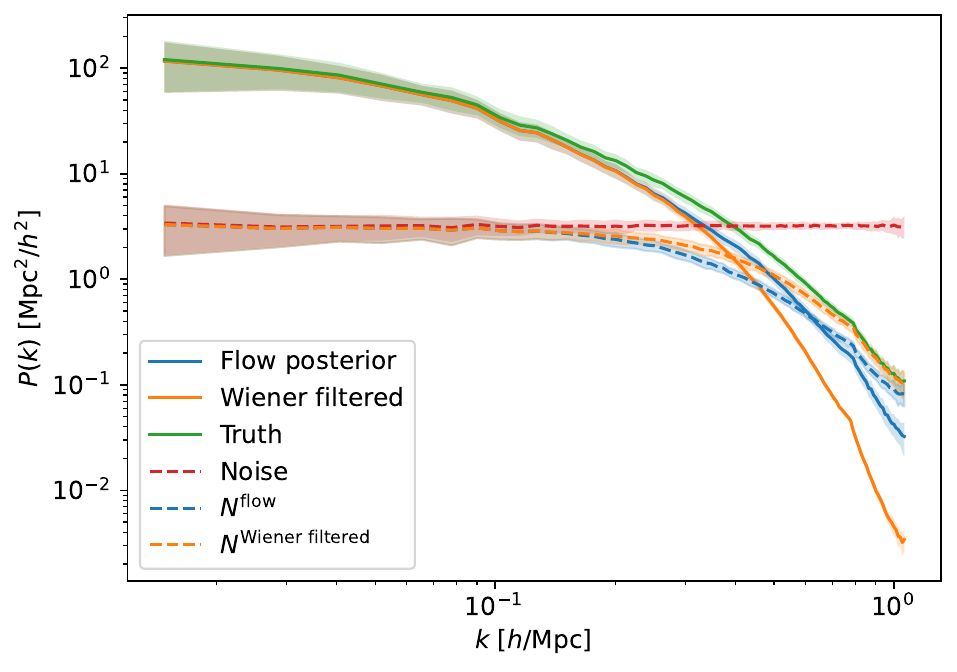}
    \includegraphics[scale=\twoimwidth]{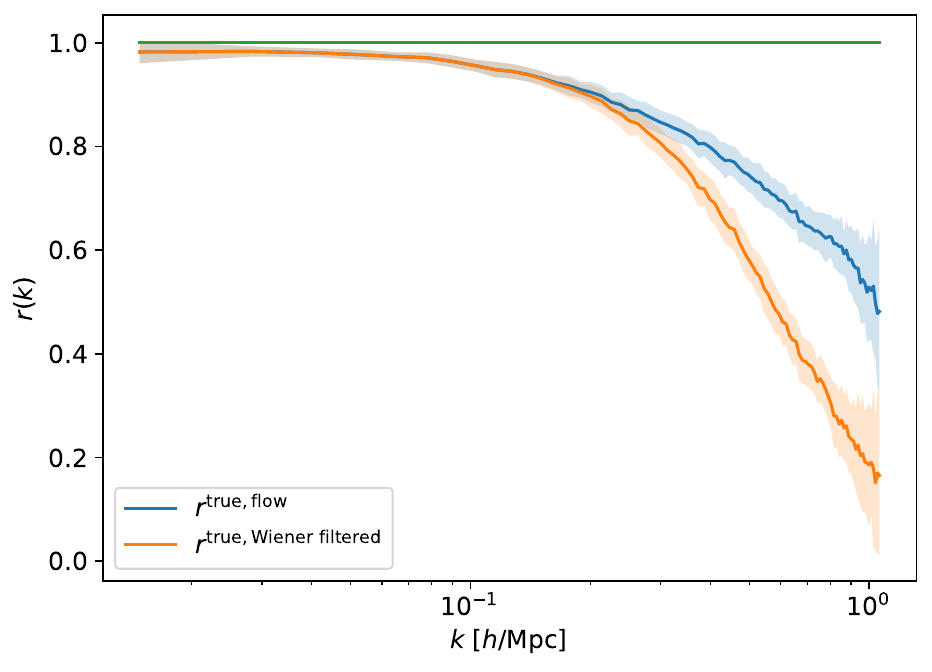}
    \caption[Summary statistics for HMC flow posterior mean]{Power spectra (left) and cross-correlation (right) comparing our HMC flow posterior mean with Wiener filtering for denoising $1.0\tilde{\sigma}$ noise. The posterior mean results are similar to the MAP results in Fig.~\ref{fig:map_ps}. Results are averaged over 100 posterior mean maps, each of which is computed from $100$ HMC samples.}
    \label{fig:hmc_ps_128}
\end{figure}

We also show summary statistics for running HMC chains on 100 different $1.0\tilde{\sigma}$ noisy maps (without a mask now). In Fig.~\ref{fig:hmc_ps_128} are the power spectra (left) and cross-correlation (right), averaged over the 100 different posterior mean maps (where each posterior mean is averaged over 100 HMC samples). The summary statistics here are similar to the previous section, except that the power spectrum of the posterior mean is a bit closer to the truth than the MAP solution.

While the posterior mean HMC maps lose power at large $k$ due to averaging over diverse large $k$ features between samples, the individual HMC samples do have the correct power spectrum. As we showed in \cite{rouhiainen2021}, normalizing flows are capable of generating matter distributions that match their training set power spectrum and non-Gaussianity nearly identically. Making samples from the posterior is required for a fully Bayesian analysis such as, for example, the CMB lensing reconstruction in \cite{Millea:2020iuw}.

\section{Patching maps together to reconstruct large maps}
\label{patching}

Above we discussed that we use an ordinary Wiener filter on large, linear scales, and only use the flow on small nonlinear scales. This Fourier splitting also allows us to reconstruct very large maps with a patching procedure, without the need to increase the flow dimension. Wiener filtering is computationally more tractable, so we may Wiener filter an entire large map, while the computationally more difficult flow posterior is computed on smaller patches in parallel. By patching the smaller flow filtered maps onto the large Wiener filtered map, we avoid having to train an extremely large flow.

As an example, we reconstruct a large map of length $n_\text{L}=1024$~px with a flow trained on non-periodic small maps of length $n_\text{S}=128$~px, but we can patch together arbitrarily large maps in principle. We use the same network architecture as described in Section~\ref{sec:introduction_denoising}, with zeros for the convolution padding to respect the non-periodicity. Our training data is as described in Section~\ref{MAP_results}, except our $128$~px 2-dimensional projections are cut out of $384$~px 3D simulations to get non-periodicity.

The reconstruction and patching procedure is as follows.
\begin{enumerate}
    \item Divide the large, $n_\text{L}$ length periodic map into a number of $(2n_\text{L}/n_\text{S})^2$ evenly spaced small maps of length $n_\text{S}$. These small maps have an $n_\text{S}\times n_\text{S}/2$ overlapping region with each of their four neighboring maps.
    \item Reconstruct the small-scale modes in these non-periodic small maps with a trained flow.
    \item Reconstruct the large-scale modes by applying Wiener filtering to the entire large map.
    \item To avoid discontinuities near the edges of the small maps, add only the the large $k$ modes from only the center $n_\text{S}/2$ length square region of the flow reconstructed maps to the Wiener filtered large map.
\end{enumerate}

The critical step that allows the smaller maps to be patched together without discontinuities is the Wiener filtering applied to the entire large map. This large-scale Wiener filtering is well within our computational constraints for very large maps encountered in cosmology, while training a flow on very large maps may be computationally infeasible. As explained above, by Wiener filtering large scales, we implicitly assume a factorization of the PDF in Fourier space. Here, by patching, we also assume a factorization in real space on the scale of the patches. Neither of these factorizations hold exactly true in cosmology, but they are sufficient approximations for our goal to to improve the reconstruction.

\begin{figure}[H]
    \centering
    \begin{subfigure}{0.63\textwidth}
        \includegraphics[trim={2mm 2mm 2mm 2mm}, clip, width=\textwidth]{masked_3_huge.png}
        \vspace*{\fourvspace}
        \caption*{Observed}
    \end{subfigure}
    \begin{subfigure}{0.63\textwidth}
        \includegraphics[trim={2mm 2mm 2mm 2mm}, clip, width=\textwidth]{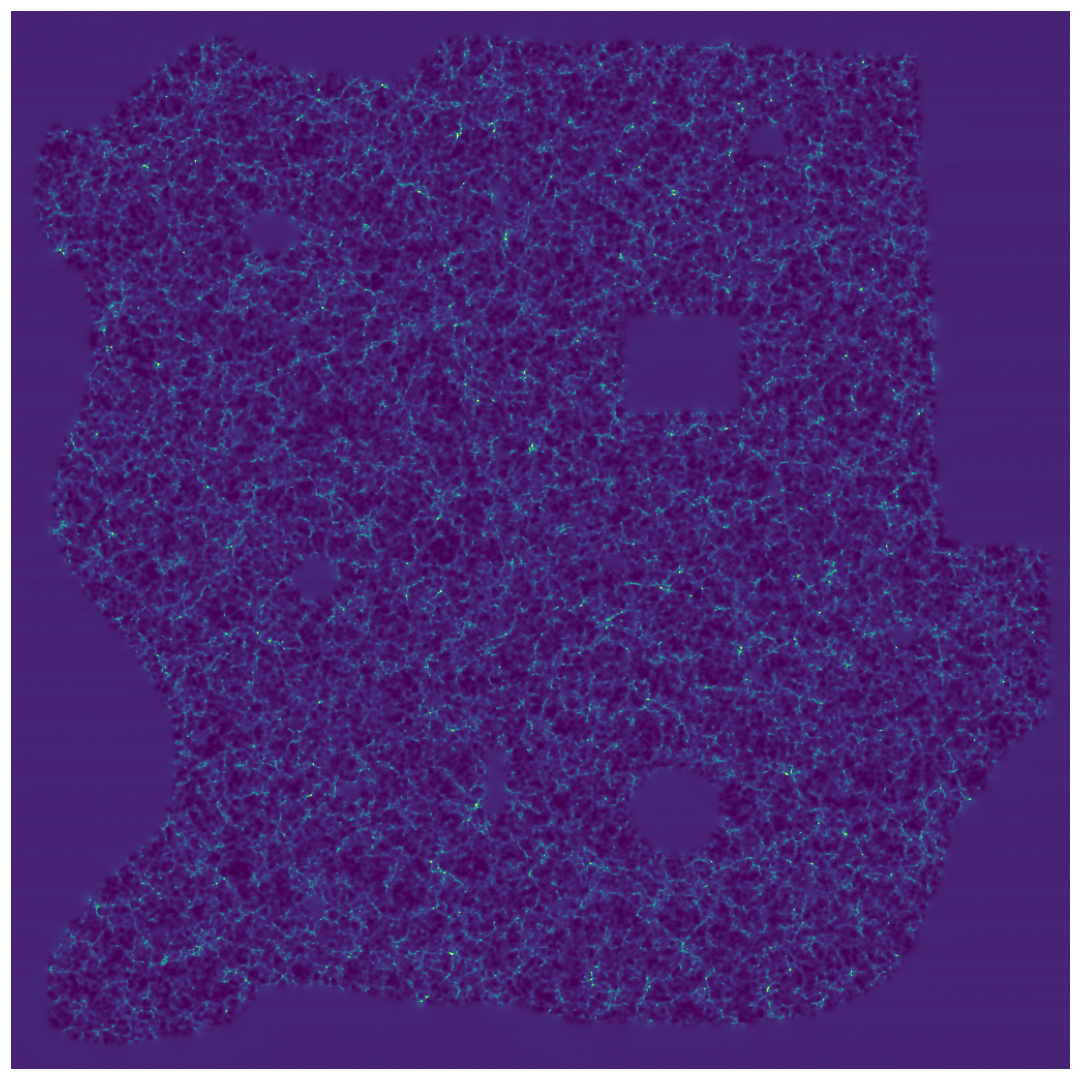}
        \vspace*{\fourvspace}
        \caption*{Flow posterior}
    \end{subfigure}
    
    \caption[Observed and reconstructed large maps patched together with many smaller maps]{Observed and reconstructed maps of length $1024$~px, and physical length $4096\ \text{Mpc}/h$ with projected depth $\Delta z=128\ \text{Mpc}/h$. The flow posterior map was patched together with 256 posterior maps of length $128$~px as described in Section~\ref{patching}.}
    \label{fig:result_bigmap}
\end{figure}

\begin{figure}[H]
    \centering
    \begin{subfigure}{0.32\textwidth}
        \includegraphics[trim={45cm 45cm 45cm 45cm}, clip, width=\textwidth]{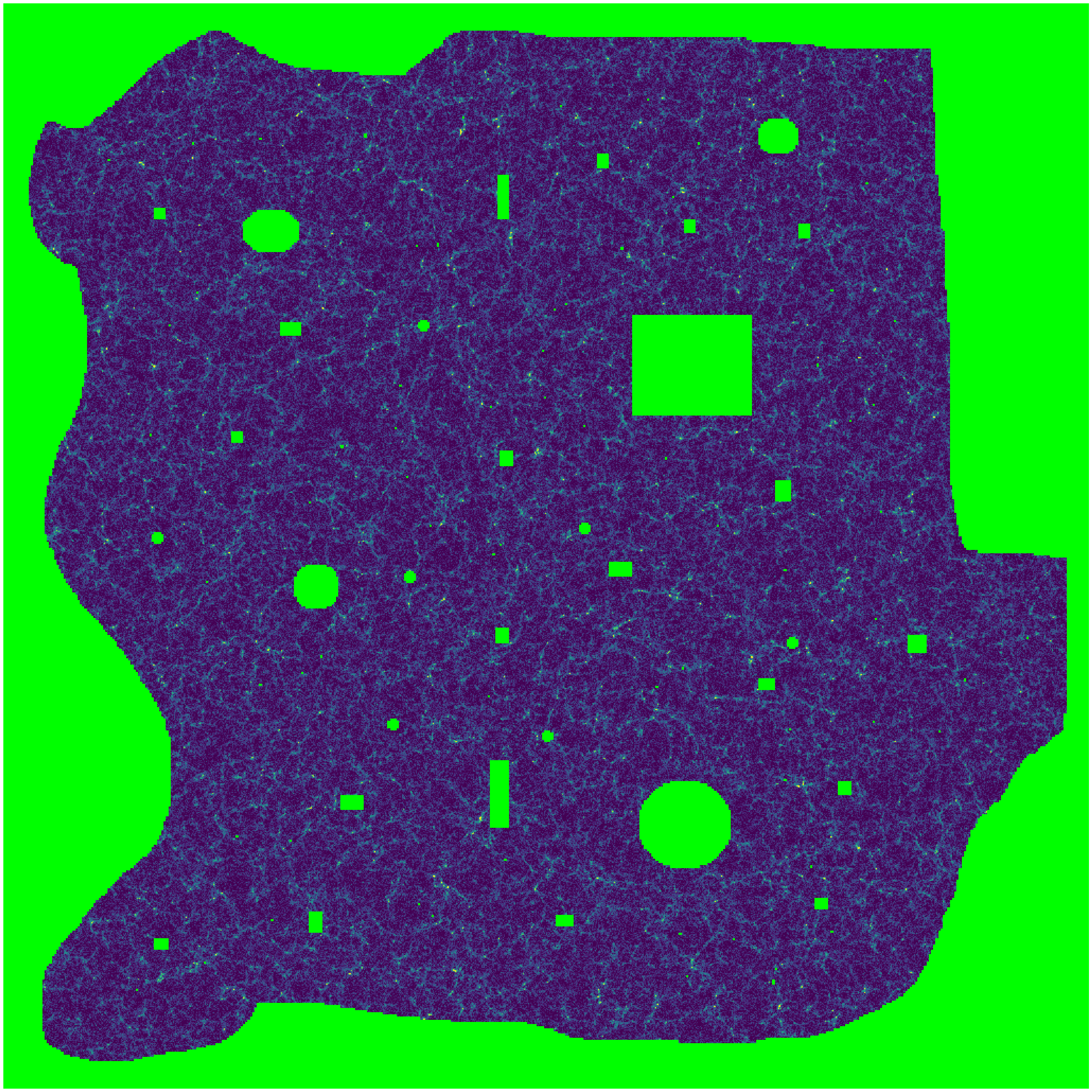}
        \caption*{Observed}
    \end{subfigure}
    \begin{subfigure}{0.32\textwidth}
        \includegraphics[trim={13.5cm 13.5cm 13.5cm 13.5cm}, clip, width=\textwidth]{flow_1p0_bigmap.png}
    \caption*{Flow posterior}
    \end{subfigure}
    \begin{subfigure}{0.32\textwidth}
        \includegraphics[trim={13.5cm 13.5cm 13.5cm 13.5cm}, clip, width=\textwidth]{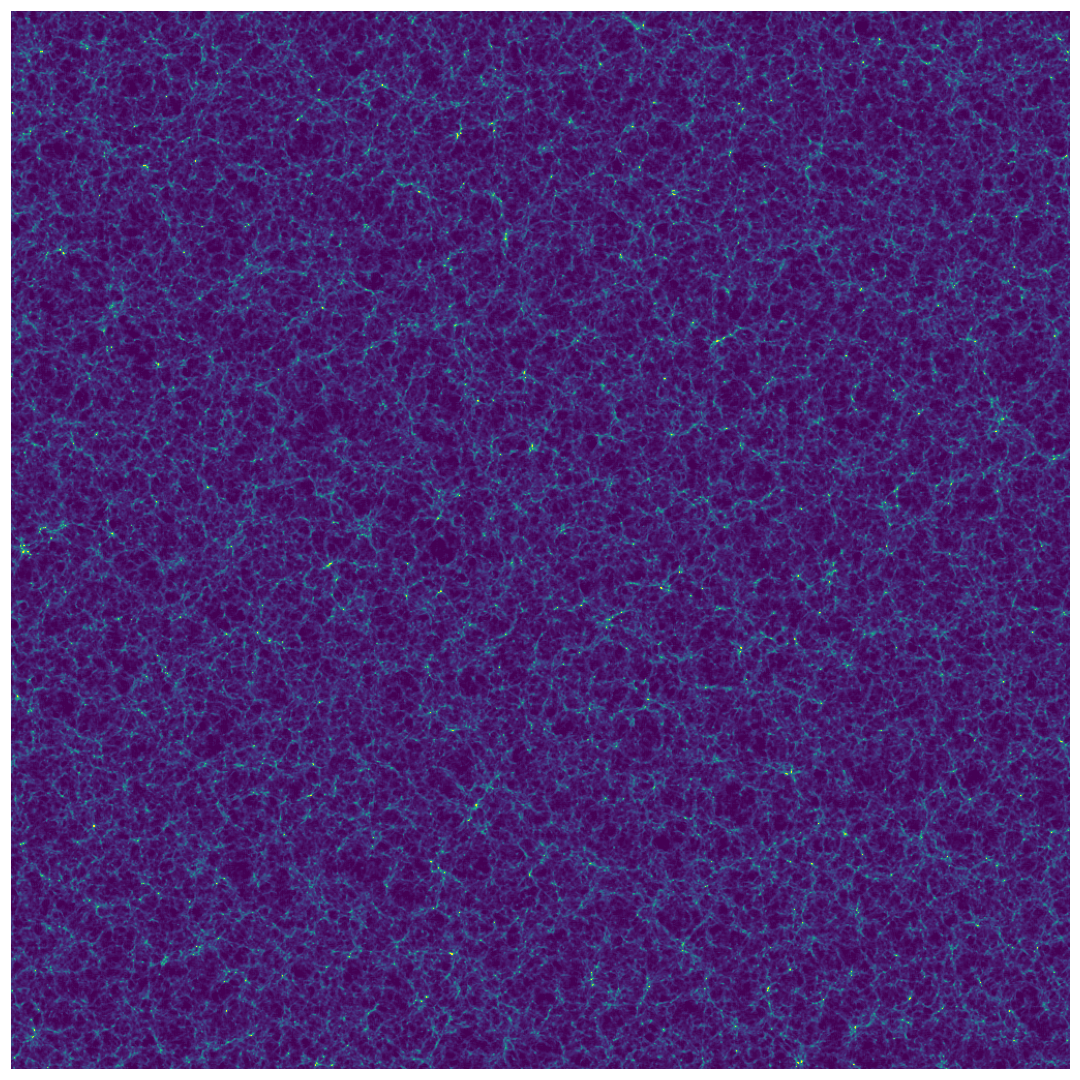}
    \caption*{Truth}
    \end{subfigure}
    
    \caption[Zoom of large maps patched together with many smaller maps]{Zoomed in view of the large observed, reconstructed, and truth maps of Fig.~\ref{fig:result_bigmap}.}
    \label{fig:result_bigmap_zoom}
\end{figure}

\begin{figure}[H]
    \centering
    \includegraphics[scale=\twoimwidth]{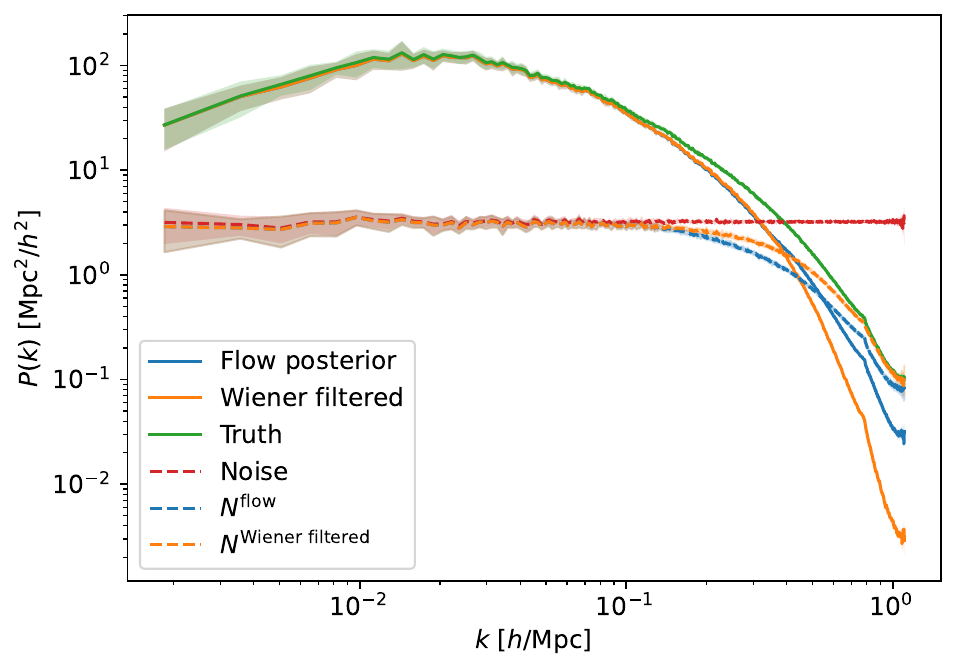}
    \includegraphics[scale=\twoimwidth]{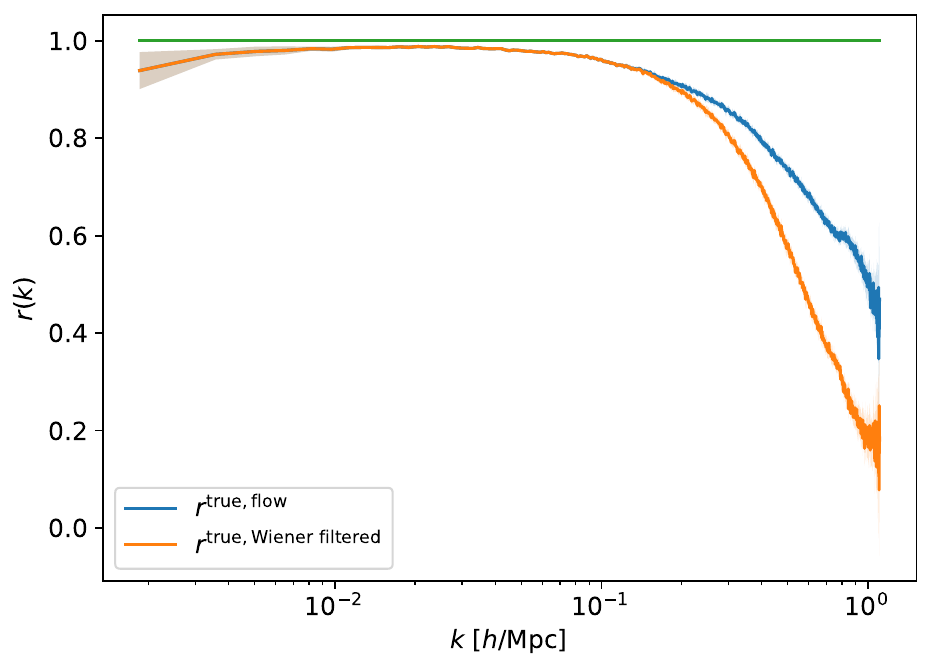}
    \caption[Summary statistics for denoising large fields with patching]{Power spectra (left) and cross-correlation (right) for denoising $1.0\tilde{\sigma}$ noise $1024$~px maps. We see no evidence of patching in the form of resonances or discontinuities in any of the summary statistics. Results are averaged over 10 different $1024$~px maps, computed without a mask.}
    \label{fig:bigmap_ps_128}
\end{figure}

High-resolution $1.0\tilde{\sigma}$ noise observed and reconstructed $1024$~px maps are shown in Fig.~\ref{fig:result_bigmap}, with zoomed in views in Fig.~\ref{fig:result_bigmap_zoom}, where the posterior map is reconstructed from 256 small $128$~px maps. There is no visible remnant of a grid where maps were patched together. Additionally, the summary statistics shown in Fig.~\ref{fig:bigmap_ps_128}, which have been averaged over 10 different $1024$~px maps, have no resonances or other oddities giving evidence of the patching.

Our patching method can also aid in generating large HMC samples. The number of steps to reach nearly independent HMC samples of $N$ parameters grows as $\mathcal{O}\left(N^{5/4}\right)$ \cite{neal_mcmc}, so there is a benefit to breaking up a large map and computing HMC samples on individual small maps in parallel, however at the cost of the approximations we just explained. It would be interesting to generalize our approach to a conditional patching to relax these approximations.

\section{Conclusion}

Normalizing flows are a powerful tool to deal with the non-Gaussianity of high-resolution cosmological surveys. The key feature of flows is that their exact likelihoods are tractable, and here we have made use of this feature to denoise simulated cosmological data. In our example setup, at non-linear scales, we gain up to 30\% reduction in MSE, and a factor of 2 increase in the cross-correlation coefficient, relative to Wiener filtering. The possible improvement depends strongly on the non-Gaussianity of the field. For example, reducing the projected depth from $\Delta z=128\ \text{Mpc}/h$ to $\Delta z=32\ \text{Mpc}/h$, we find an improvement on the smallest scales by a factor of $3.5$ in the cross-correlation. We also demonstrate a method of patching together many small posterior maps to denoise large maps that would be computationally difficult to address without patching. Our results were obtained by projected 2-dimensional maps. The generalization to 3 dimensions is mathematically straightforward but poses computational challenges for normalizing flows.

There are several other interesting follow-up projects. First, we have not included large-scale to small-scale coupling. We may train a conditional normalizing flow to sample small-scale structure conditioned on the large-scale environment, with an application to super-resolution emulation. Running a conditional flow in inference mode may allow us to improve our reconstruction further. We have also assumed that noise is Gaussian; this approximates instrument noise as well as shot noise, but is not always a good description, considering low mass galaxies with a high number density~\cite{2009PhRvL.103i1303S}. 

Furthermore, we would like to use our method to improve constraints on cosmological or astrophysical parameters. As we have seen, our method improves the reconstruction on non-linear scales, starting roughly at the scales where one usually cuts off a cosmological N-point function analysis, such as the power spectrum, due to theoretically uncontrolled non-linearities. To use the flow reconstruction for N-point function analysis would require a reliable simulation modelling of of these scales. Marginalizing over baryonic feedback is in principle possible with flows by conditioning them on unknown feedback parameters \cite{trenf}, but generating reliable training data is difficult. There are however also situations where a small-scale reconstruction can be used reliably for inference of cosmological parameters. This is the case in particular when small-scale modes can be used to reconstruct large-scale modes. For example, a better reconstruction of the mass field from a galaxy survey leads to an improved template for CMB lensing \cite{Schmittfull:2017ffw} or kSZ cross-correlation \cite{Munchmeyer:2018eey,Smith:2018bpn}. Such cross-correlation can be used for example to probe primordial non-Gaussianity \cite{Munchmeyer:2018eey,Schmittfull:2017ffw}. Unknown small-scale physics then appears in terms of biases on large scales which we can marginalize over. Future work should examine such applications.  


\chapter[Super-Resolution Emulation of Large Cosmological Fields with a\\3D Conditional Diffusion Model]{Super-Resolution Emulation of Large Cosmological Fields with a 3D Conditional Diffusion Model}
\label{ch:sr_diffusion}
High-resolution (HR) simulations in cosmology, in particular when including baryons, can take millions of CPU hours. On the other hand, low-resolution (LR) dark matter simulations of the same cosmological volume use minimal computing resources. We develop a denoising diffusion super-resolution emulator for large cosmological simulation volumes. Our approach is based on the image-to-image Palette diffusion model, which we modify to 3 dimensions. Our super-resolution emulator is trained to perform outpainting, and can thus upgrade very large cosmological volumes from LR to HR using an iterative outpainting procedure. As an application, we generate a simulation box with 8 times the volume of the Illustris TNG300 training data, constructed with over 9000 outpainting iterations, and quantify its accuracy using various summary statistics.

\section{Introduction}

Cosmological simulations of dark matter and baryonic matter are crucial for modern cosmology. They are required for theoretical studies, statistical method development, and parameter inference on real data. Unfortunately, high-resolution (HR) simulations with baryonic matter, even on moderate cosmological volumes, can require millions of CPU hours. On the other hand, low-resolution (LR) simulations of dark matter can easily be generated using much smaller computational resources. Our approach is to train a 3-dimensional diffusion model to upgrade LR simulations to super-resolution (SR) simulations, which emulate the HR simulations at our targeted resolution. Because the small scales of the HR simulation are not contained in the LR simulation, an SR emulator should be probabilistic, able to generate many SR simulations consistent with the same LR simulation. We thus require a stochastic generative model, which learns probabilistic small-scale physics from HR training data. In this work, we choose a denoising diffusion probabilistic model~\cite{sohl2015, ho2020denoising, palette}, because they are currently the best performing models for image generation, generally outperforming GANs on various image tasks ~\cite{dhariwal2021diffusion}, and are much more expressive than normalizing flows in higher dimensions~\cite{2019arXiv191202762P,kong2020expressive}.

A common feature of generative diffusion models on images is ``inpainting" and ``outpainting" (or ``outcropping"), whereby image data is stochastically generated in a masked region of the image. An outpainting task generates data outside the main body of the image, and can be performed iteratively to generate large images, as has been demonstrated in~\cite{palette}. Currently, diffusion models are often trained on $256$~px length images. In this work, computational limitations arising by moving to 3 dimensions leads us to training on $48$~px length cubes.

This work develops a stochastic treatment of super-resolution emulation combined with a conditional outpainting procedure that can make large volumes without boundary effects. Our outpainting method generates new SR volumes conditioned both on the LR simulation and on previously generated neighboring SR volumes. By repeatedly outpainting smaller, local volumes many times over, we are able to generate simulation volumes larger than the entire HR training data volume. This is possible due to the locality of the underlying physics of structure formation, where on large scales, LR and HR simulations with the same initial conditions give the same final result. Therefore, we are able to emulate features with length scales larger than the outpainting window so long as such features are in the LR conditional.

There have been various previous approaches to super-resolution of cosmological data. In \cite{kodiramanah2020}, the authors trained a Wasserstein GAN as a super-resolution emulator at field level, in a similar setup to our work. However, their approach did not probabilistically model the SR conditioned on the LR simulation. Instead, they generate a deterministic SR solution, conditioned on the HR initial conditions field. This is an interesting and efficient approach but does not allow to explore the probabilistic space of consistent SR simulations given an LR simulation. In a different series of works \cite{li2021_1, li2021_2, li2023}, a GAN is used to generate SR simulations stochastically at particle level, increasing the particle count in the simulation boxes by several orders of magnitude. This is a powerful alternative approach that is closer to N-body simulations, which naturally work with particles. We believe that both particle and field-level approaches can be useful. Cosmological survey analysis is usually done on a regular grid (for example, to take FFTs), so that particle positions are not necessarily required. Further, the work \cite{li2021_1} make large simulations by sewing together smaller SR volumes. However, they do not condition neighboring SR volumes on each other, which necessarily means that there will be boundary effects in the SR field. In our work, we explicitly condition the local SR volumes on their neighbors. Recently, \cite{schanz2023stochastic} used a diffusion model to perform super-resolution on 2-dimensional dark matter fields. Their work is technically most similar to ours, with their approach including a Fourier filter on the large-scale structure data to boost the importance of learning the nonlinear scales.

Diffusion models have been used in astrophysics for other purposes than super-resolution emulation. They have been used to generate astrophysical fields projected to 2 dimensions \cite{Mudur:2022gfq, Zhao:2023giv}, and reconstruct strong gravitational lensing images \cite{Karchev:2022ycy}. Further, a 3-dimensional diffusion model was recently used to infer initial conditions from present day dark matter simulations, using data meshed onto $128^3$ voxels~\cite{10.1093/mnrasl/slad152}. In the broader physical sciences, diffusion models have been used to generate high energy physics particle clouds \cite{Leigh:2023toe} and point clouds \cite{Mikuni:2023dvk}, map simulated Compact Muon Solenoid data to theoretical quantities \cite{Shmakov:2023kjj}, generate calorimeter showers \cite{Mikuni:2023tqg}, and generate atomic physics data \cite{Imani:2023blb}.

We briefly review the theory behind denoising diffusion models in Section~\ref{sec:diffusion_theory}. In Section~\ref{sec:data}, we describe the HR and LR training data, and LR test set data, used in this project. In Section~\ref{sec:diffusion_model}, we describe the diffusion model architecture with our iterative outpainting method. In Section~\ref{sec:results}, we show our results for a generated field larger than the entire training data volume. In Section~\ref{sec:results_variety}, we use the stochastic nature of diffusion models to generate a variety of SR fields conditional on a single LR field. In Section~\ref{sec:BAO_response}, we demonstrate SR emulation of baryon acoustic oscillations, which have length scales larger than our model's outpainting window. In Section~\ref{sec:variable_cosmos}, we test the robustness of the diffusion mode to slightly out-of-distribution data by varying the cosmological parameters of the LR field, and generating new SR fields.

\section{Diffusion model theoretical background}
\label{sec:diffusion_theory}

The image-to-image diffusion model is a probabilistic generative model that samples $\bm{y}$ from $p{(\bm{y}|\bm{x})}$, where the (usually high-resolution) generated image $\bm{y}$ is conditional on a (usually low-resolution) image $\bm{x}$. Applications of image-to-image diffusion include super-resolution, colorization, denoising, JPEG resoration, inpainting, outpainting, and other types image reconstruction. Here we briefly describe the theory behind the image-to-image diffusion model, with more details available in~\cite{ho2020denoising}. The equations here are valid in any number of spacial dimensions.

\begin{figure}[h!]
\centering
\textbf{\large Denoising diffusion training}

\begin{tikzpicture}
\node (x) {\includegraphics[width=.15\textwidth]{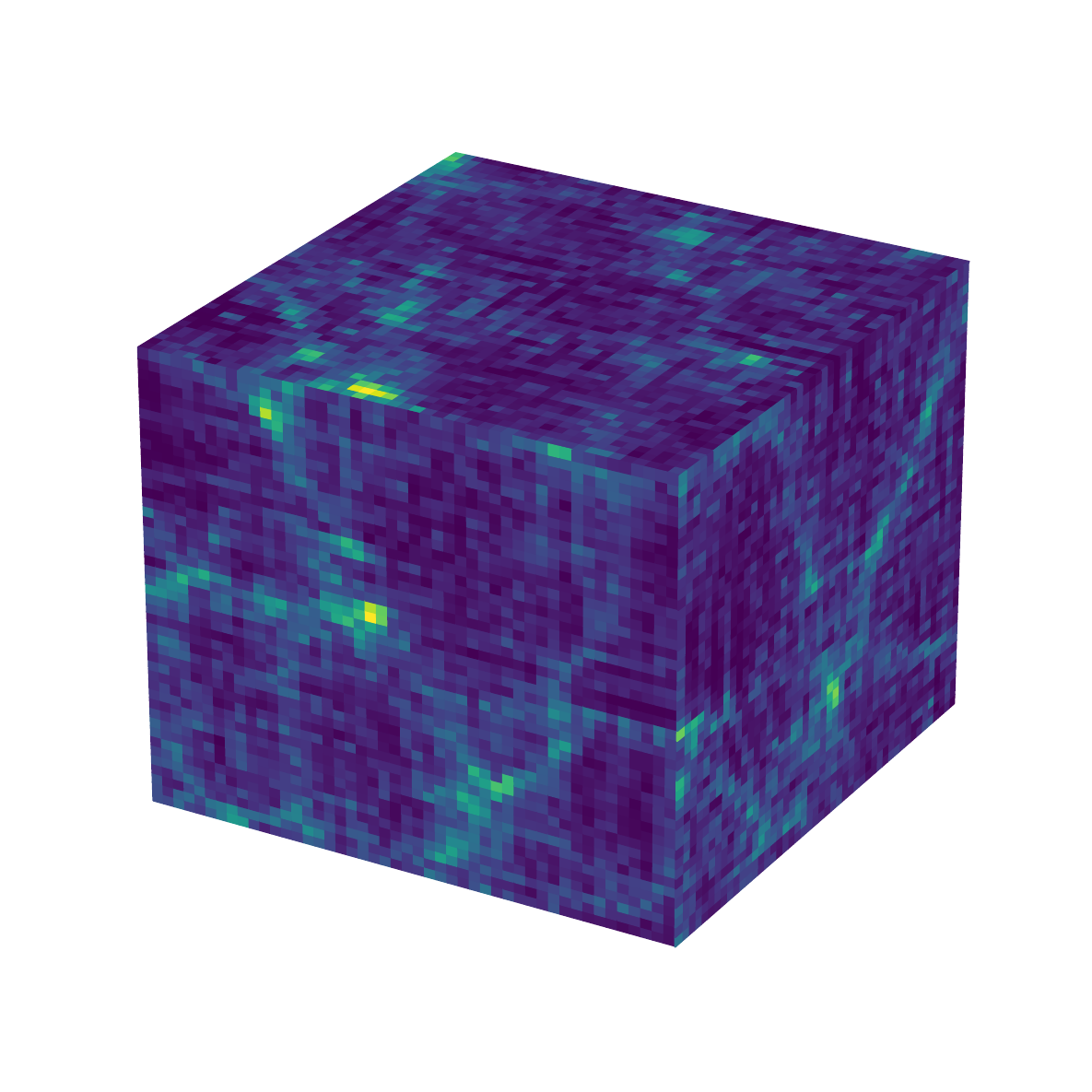}};
\node [below of=x, yshift=-.6cm] {Conditional $\bm{x}$};
\node (y_0) [above of=x, yshift=2.5cm] {\includegraphics[width=.15\textwidth]{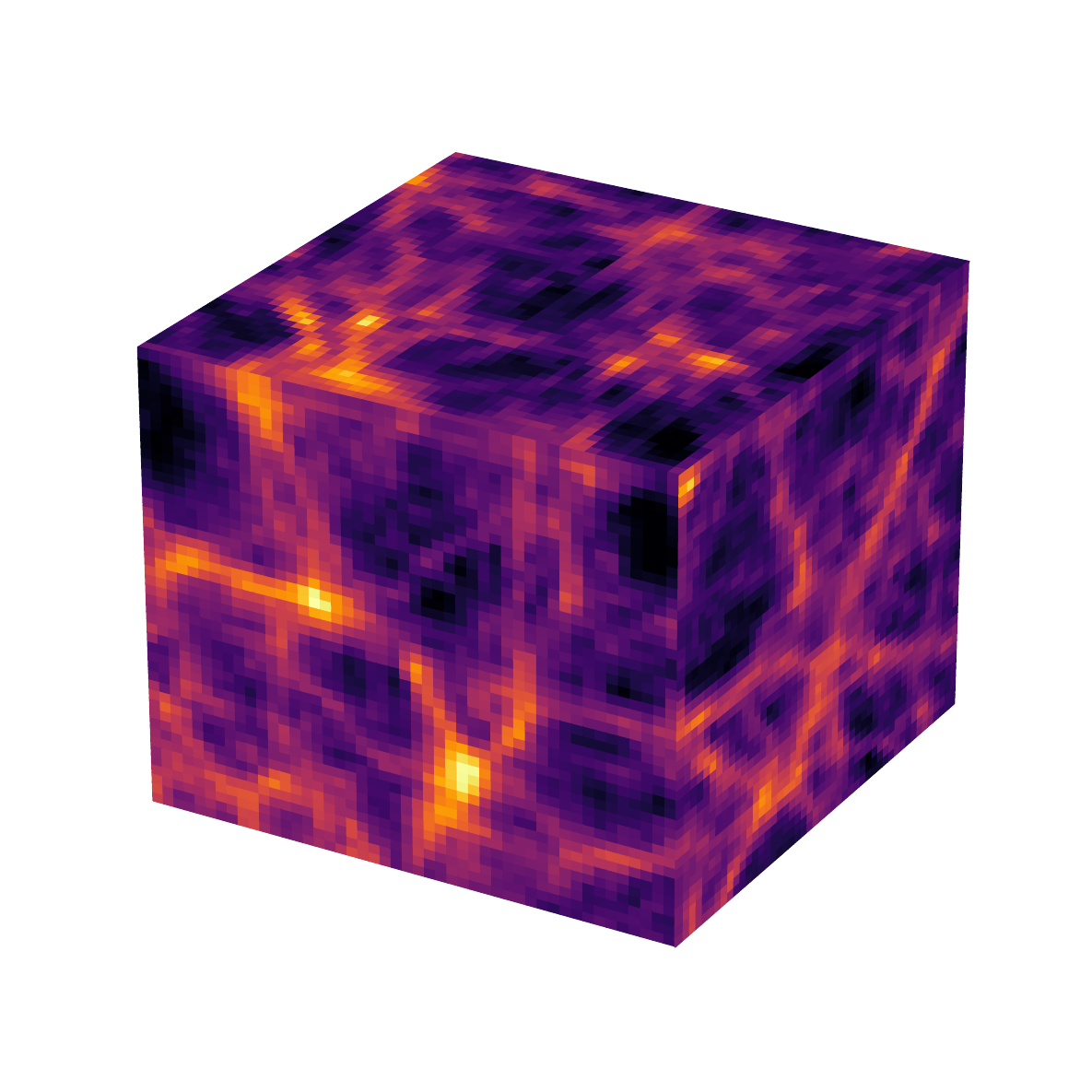}};
\node [below of=y_0, yshift=-.6cm] {Training data $\bm{y}_0$};
\node (y_t) [right of=y_0, xshift=4cm] {\includegraphics[width=.15\textwidth]{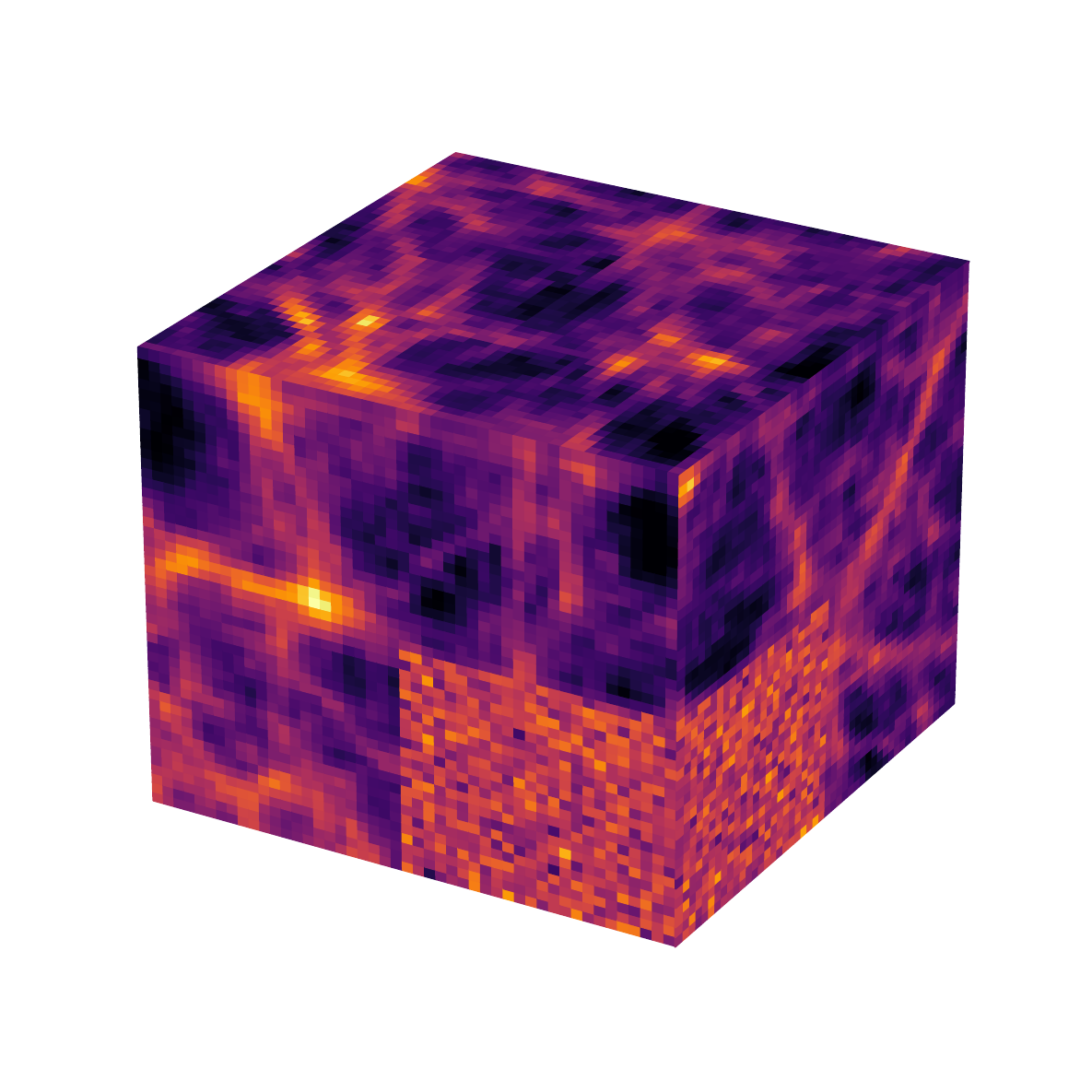}};
\node [below of=y_t, yshift=-.6cm] {$\bm{y}_t$};
\node (y_tm1) [right of=y_t, xshift=4cm] {\includegraphics[width=.15\textwidth]{images/sr_diffusion/images_outpainting/sr_cube_noise.pdf}};
\node (y_tm1_name) [below of=y_tm1, yshift=-.6cm] {$\bm{y}_{t-1}$};
\node (eps) [above of=y_0, yshift=2.5cm] {\includegraphics[width=.15\textwidth]{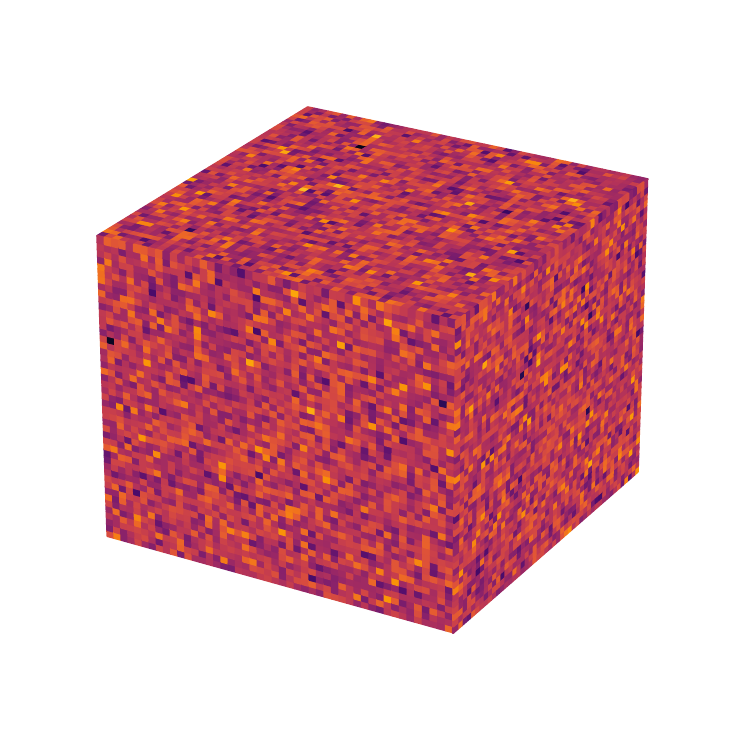}};
\node [below of=eps, yshift=-.6cm] {Random noise $\bm{\epsilon}$};
\node (loss) [above of=y_tm1, yshift=2.47cm, align=center, rectangle, text centered, draw=black, thick] {Loss\\function};

\draw [arrow] (x.east) to [out=0,in=-135](y_tm1.west);
\draw [arrow] (y_0.east) -- (y_t.west) node[midway, above, xshift=-0.25cm] {$\times\sqrt{\gamma_t}$};
\draw [arrow] (y_t.east) -- (y_tm1.west) node[midway, above, yshift=0.1cm] {$f_{\bm{\theta}}(\bm{x},\bm{y}_t,\gamma_t)$};
\draw [arrow] (y_tm1.north) -- (loss);
\draw [arrow] (eps.east) -- (y_t.west) node[midway, above, xshift=0.8cm, yshift=-0.1cm] {$\times\sqrt{1-\gamma_t}$};
\draw [arrow] (eps.east) -- ++(7.75cm, 0);
\end{tikzpicture}

\ 

\textbf{\large Denoising diffusion sample generation}

\vspace{0.25cm}
\begin{tikzpicture}
\node (x) {\includegraphics[width=.15\textwidth]{images/sr_diffusion/images_outpainting/lr_cube.pdf}};
\node [below of=x, yshift=-.6cm] {Conditional $\bm{x}$};
\node (y_T) [above of=x, yshift=2.5cm] {\includegraphics[width=.15\textwidth]{images/sr_diffusion/images_outpainting/sr_cube_noise.pdf}};
\node [below of=y_T, yshift=-.6cm] {Random noise mask $\bm{y}_T$};
\node (y_Tm1) [right of=y_T, xshift=4cm] {\includegraphics[width=.15\textwidth]{images/sr_diffusion/images_outpainting/sr_cube_noise.pdf}};
\node [below of=y_Tm1, yshift=-.6cm] {$\bm{y}_{T-1}$};
\node (y_0) [right of=y_Tm1, xshift=6cm] {\includegraphics[width=.15\textwidth]{images/sr_diffusion/images_outpainting/sr_cube.pdf}};
\node [below of=y_0, yshift=-.6cm] {Model output $\bm{y}_0$};
\node (ellipsis) [right of=y_Tm1, xshift=2.5cm, font=\bfseries] {\Large ...};

\draw [arrow] (x.east) -- (y_Tm1.west);
\draw [arrow] (x.east) to [out=0,in=-105](ellipsis.south);
\draw [arrow] (y_T.east) -- (y_Tm1.west) node[midway, above, yshift=0.1cm] {$f_{\bm{\theta}}(\bm{x},\bm{y}_T,\gamma_T)$};
\draw [arrow] (y_Tm1.east) -- ++(1cm, 0);
\draw [revarrow] (y_0.west) -- ++(-1cm, 0);
\path (y_Tm1.east) -- (y_0.west) node[midway, above, yshift=0.1cm] {Repeat $f_{\bm{\theta}}(\bm{x},\bm{y}_t,\gamma_t)$};
\end{tikzpicture}
\caption[Denoising diffusion model illustration]{Diffusion model training and generation. (Top) The training loop applies noise for a random step $t$ to the training data $\bm{y}_0$. The network $f_{\bm\theta}(\bm{x},\bm{y},\gamma_t)$ learns $\bm{\epsilon}$. The loss is computed between the network output $f_{\bm\theta}$ and truth $\bm{\epsilon}$. (Bottom) Data generation with a diffusion model is repeatedly applying $f_{\bm{\theta}}$ to a random noise field $\bm{y}_T$.}
\label{fig:model_illustration}
\end{figure}

The diffusion model has a forward process for training, and a reverse process for generation. In the forward diffusion step, a field $\bm{y}_t$ gets added noise from a field $\bm{y}_{t-1}$ as
\begin{equation}
    p(\bm{y}_t|\bm{y}_{t-1})=\mathcal{N}{(\bm{y}_t|\sqrt{1-\beta_t}\ \bm{y}_{t-1},\beta_t\bm{I})}.
\end{equation}
where $\beta_t$ control the amount of noise reduction between steps. By iterating this forward diffusion process, the field $\bm{y}_t$ at some step $t$ can be written in terms of the original field $\bm{y}_0$ as
\begin{align}
\label{eq:forward_diffusion_gamma}
    p(\bm{y}_t|\bm{y}_0)&=\prod_{i=1}^tp{(\bm{y}_i|\bm{y}_{i-1})}\\
    &=\mathcal{N}{(\bm{y}_t|\sqrt{\gamma_t}\ \bm{y}_0,(1-\gamma_t)\bm{I})}
\end{align}
where the noise parameters are condensed by defining
\begin{equation}
    \gamma_t\equiv\prod_{i=1}^t(1-\beta_i).
\end{equation}
A total of $T$ steps are taken. The $\gamma_t$ are chosen as model hyperparameters in such a way that by the end of the diffusion chain, $\sqrt{\gamma_T}\ \bm{y}_0$ is small relative to $\bm{y}_T$, and thus the noisy data has lost almost all resemblance to $\bm{y}_T$. Optimal values of $T$ and $\gamma_t$ are motivated in Section~\ref{sec:noise_schedule}

The reverse diffusion step given $\bm{y}_0$ is
\begin{equation}
\label{eq:reverse_step_0}
    p(\bm{y}_{t-1}|\bm{y}_0,\bm{y}_t)=\mathcal{N}{(\bm{y}_{t-1}|\bm{\mu},\sigma^2\bm{I})},
\end{equation}
having mean and variance
\begin{align}
    \bm{\mu}&=\frac{\sqrt{\gamma_{t-1}}\beta_t}{1-\gamma_t}\bm{y}_0+\frac{\sqrt{1-\beta_t}(1-\gamma_{t-1})}{1-\gamma_t}\bm{y}_t,\\
    \sigma^2&=\frac{(1-\gamma_{t-1})\beta_t}{1-\gamma_t}.
\end{align}
This reverse diffusion step depends on $\bm{y}_0$, which is precisely the data we wish to generate. The purpose of diffusion models is to generate fields in the distribution $\bm{y}_0$ without knowing the true value of $\bm{y}_0$, and so we need an equation for $\bm{y}_{t-1}$ in terms of $\bm{y}_t$ without $\bm{y}_0$. To achieve this, we fit a function $f_{\bm{\theta}}{(\bm{x},\bm{y}_t,\gamma_t)}$ to the distribution $\bm{\epsilon}\sim\mathcal{N}(\bm{0},\bm{I})$, by minimizing the loss function
\begin{equation}
    \label{eq:loss}
    \mathbb{E}_{(\bm{x},\bm{y})}\mathbb{E}_{\bm{\epsilon},\gamma}\left\Vert f_{\bm{\theta}}(\bm{x},\sqrt{\gamma}\ \bm{y}_0+\sqrt{1-\gamma}\ \bm{\epsilon},\gamma)-\bm{\epsilon}\right\Vert_2^2
\end{equation}
where $\left\Vert...\right\Vert_2^2$ is the squared $L_2$ norm. In practice, the function $f_{\bm{\theta}}{(\bm{x},\bm{y}_t,\gamma_t)}$ is a neural network with parameters $\bm{\theta}$, and the details of our network are described in Section~\ref{sec:network_architecture}.

There are multiple ways to solve for the reverse diffusion step~\cite{song2022denoising,lu2022dpmsolver}, and we have a brief discussion of these various approaches in Section~\ref{sec:conclusion}. In this work, we use the ``denoising diffusion probabilistic model" (DDPM) results of~\cite{ho2020denoising,palette} to solve the reverse diffusion problem. A lowest order estimate for the $\bm{y}_0$ in terms of $\bm{y}_t$ is to invert Eq.~\ref{eq:forward_diffusion_gamma} as
\begin{equation}
    \hat{\bm{y}}_0=\frac{1}{\sqrt{\gamma}_t}\left(\bm{y}_t-\sqrt{1-\gamma_t}\ f_{\bm{\theta}}{(\bm{x},\bm{y}_t,\gamma_t)}\right).
\end{equation}
The goal is not to simply remove some noise, but rather sample images from the distribution that $\bm{y}_0$ belongs to. The estimate $\hat{\bm{y}}_0$ is now inserted as $\bm{y}_0$ into Eq.~\ref{eq:reverse_step_0}, moving the mean and variance to~\cite{ho2020denoising}
\begin{align}
    \mu_{\bm{\theta}}(\bm{x},\bm{y}_t,\gamma_t)&=\frac{1}{\sqrt{1-\beta_t}}\left(\bm{y}_t-\frac{\beta_t}{\sqrt{1-\gamma_t}}f_{\bm{\theta}}{(\bm{x},\bm{y}_t,\gamma_t)}\right),\\
    \sigma_{\bm{\theta}}^2(\gamma_t)&=\beta_t.
\end{align}
Finally, the reverse diffusion step is
\begin{equation}
    \bm{y}_{t-1}=\frac{1}{\sqrt{1-\beta_t}}\left(\bm{y}_t-\frac{\beta_t}{\sqrt{1-\gamma_t}}f_{\bm{\theta}}{(\bm{x},\bm{y}_t,\gamma_t)}\right)+\sqrt{\beta_t}\ \bm{\epsilon}_t.
\end{equation}
This reverse diffusion step is iterated to obtain a sample $\bm{y}_0$ from $\bm{y}_T$. Because the starting field $\bm{y}_T$ is a random Gaussian field, we can obtain a variety of samples $\bm{y}_0$ without changing the network parameters $\bm{\theta}$.

\subsection{Choice of diffusion noise schedule}
\label{sec:noise_schedule}

The diffusion step sizes $\beta_t$ should be chosen carefully. Removing evidence of the initial field $\bm{y}_0$ from the final noise field $\bm{y}_T$ needs to be balanced with a computationally feasible number of steps $T$ to get there. Taking too few steps $T$ will leave traces of $\bm{y}_0$ in $\bm{y}_T$, while also making it overly difficult for the model to denoise large noise differences between adjacent steps.

Explicitly written out, $\bm{y}_t$ built from $\bm{y}_0$ is
\begin{align}
    \bm{y}_t&=\sqrt{\beta_t}\ \bm{y}_t-\sqrt{\beta_t}\sqrt{1-\beta_t}\ \bm{y}_{t-1}\\
    &\ \ \ +\sqrt{1-\beta_t}(\sqrt{\beta_{t-1}}\ \bm{y}_{t-1}-\sqrt{\beta_{t-1}}\sqrt{1-\beta_{t-1}}\ \bm{y}_{t-2}\\
    &\ \ \ +\sqrt{1-\beta_{t-1}}(...+\sqrt{1-\beta_0}\ \bm{y}_0)...)
\end{align}
Rewriting to isolate the true field $\bm{y}_0$ gives
\begin{equation}
    \bm{y}_t=\sqrt{\gamma_t}\ \bm{y}_0+\sum_{k=1}^t\sqrt{\frac{\gamma_t}{\gamma_k}}\ (\bm{y}_k-\sqrt{1-\beta}\ \bm{y}_{k-1}).
\end{equation}
The terms $(\bm{y}_k-\sqrt{1-\beta}\ \bm{y}_{k-1})$ are noise by construction, and therefore $\bm{y}_t$ is also noise in the limit $\sqrt{\gamma_t}\rightarrow0$. In practice, it is required that $\sqrt{\gamma_t}$ be much smaller than the overall scale of values in $\bm{y}_0$ to remove the information in $\bm{y}_0$.

\begin{figure}
    \centering
    \hspace{1.35cm}$\sqrt{\gamma_T}$ for linearly increasing $\beta_t$ and $\beta_0=10^{-6}$
    
    \includegraphics[width=0.7\textwidth]{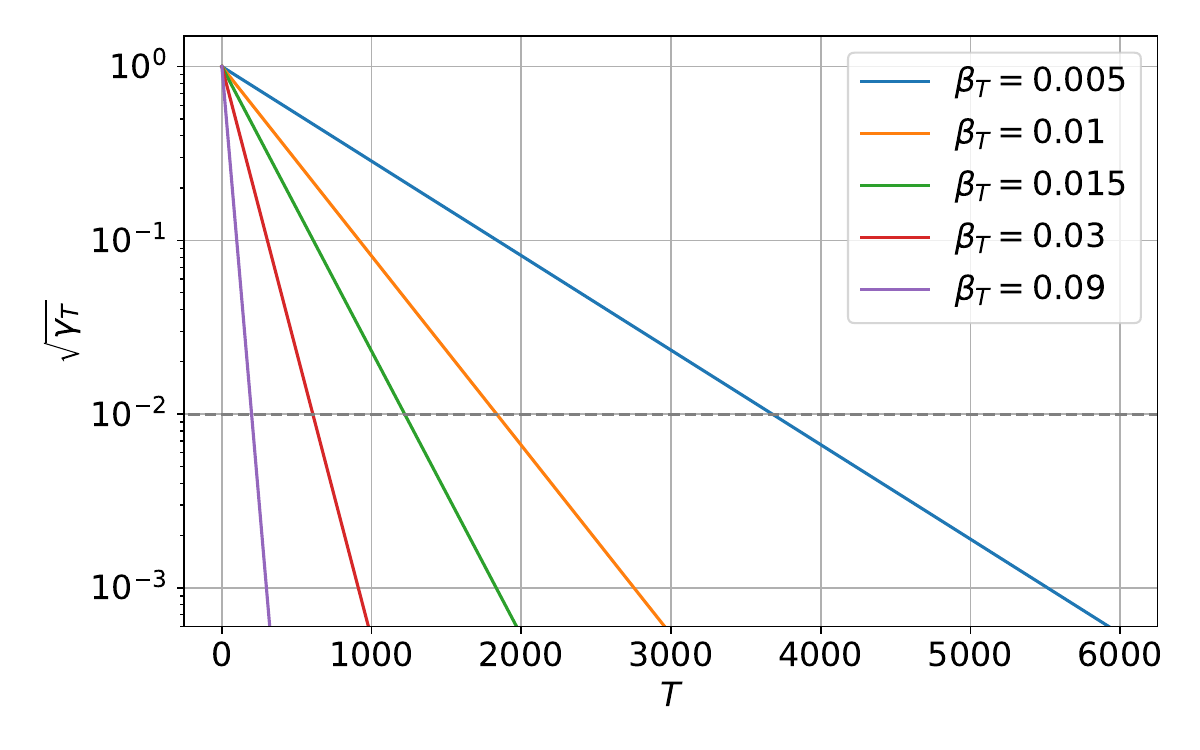}
    \caption[Determining the number of diffusion steps]{The error of the diffusion process as a function of the total number of diffusion steps $T$, for various values of the final step size $\beta_T$, with $\beta_t$ linearly increasing from a small $\beta_0=10^{-6}$. To get below one percent error (highlighted in gray) for a final $\beta_T=0.01$, more than $T=1850$ diffusion steps are required.}
    \label{fig:diffusion_betas}
\end{figure}

To demonstrate how to decide on a proper noise schedule, total number of diffusion steps $T$ required to achieve a certain accuracy in terms of $\sqrt{\gamma_t}$ is plotted in Fig.~\ref{fig:diffusion_betas}. We choose $\beta_t$ to be linearly increasing from a small initial $\beta_0=10^{-6}$. To get below one percent error for a final $\beta_T=0.01$, more than $T=1850$ diffusion steps are required; for $\beta_T=0.015$, the number of steps can be reduced to $T=1250$. With the DDIM approach used here, we cannot increase $\beta_T$ substantially above the percent level because larger denoising steps become more difficult to learn; other denoising algorithms with less steps and larger denoising per step are discussed in Section~\ref{sec:conclusion}. More complex noise schedules have been researched, such as a cosine noise schedule introduced in~\cite{improved_diffusion}, which help prevent overtraining by make the noise schedule more difficult than linear.

\section{Data}
\label{sec:data}

Here we describe the HR TNG300 training data and its accompanying LR conditional training data. We also describe our the LR conditional test data, which has different initial conditions (or phases), and a larger volume for Section~\ref{sec:results}.

\subsection{High-resolution data from TNG300}



The HR data to learn the baryonic physics of the large-scale structure comes from IllustrisTNG~\cite{tng2017_1,tng2017_2, tng2017_3,tng2017_4,tng2017_5}, a set of three gravo-magnetohydrodynamical simulations run on \texttt{Arepo}~\cite{Springel_2010}. We picked TNG300 because it covers a large enough volume to include both linear and nonlinear physics, and to demonstrate that a single HR simulation can be enough to train a model that generates larger volumes. The TNG300 run  simulated $2500^3$ matter and $2500^3$ dark matter particles in a $205\ \text{Mpc}/h\approx300\ \text{Mpc}$ length cube, with cosmological parameters $\Omega_\Lambda=0.6911$, $\Omega_\text{M}=0.3089$, $\sigma_8=0.8159$, $n_\text{s}=0.9667$, and $h=0.6774$. The simulation has an initial field at redshift of $z=127$ generated with \texttt{NGenIC}, using the Zel'dovich approximation. TNG300 computed over $10$ million time steps down to $z=0$, and we use the $z=0.01$ snapshot. We use the cloud-in-cell mass assignment scheme to place the TNG300 baryons onto a $264^3$~px cubic mesh. We will comment on increasing this resolution in Section \ref{sec:conclusion}. Although we only use one channel of data in this work, the baryon particle density, we can in principle include more channels in our HR data to learn the temperature, electron density, etc. Generating more than one channel, possibly with an increased pixel resolution, is left to future work, with a discussion also in Section~\ref{sec:conclusion}.



\subsection{Low-resolution conditional data}
\label{sec:lr_data}

Our goal is to quickly construct new HR matter fields from LR dark matter fields as a conditional for our diffusion model. With \texttt{Arepo}, we simulate dark matter fields to create the LR conditional training and LR test data. Our conditional fields are LR compared to TNG300 in three ways: they are dark matter only simulations with no baryonic physics, they contain far fewer particles, and they run across fewer time steps.

To simulate the LR training data, we use the same \texttt{NGenIC} initial seed, box size, and cosmological parameters as TNG300 described above, but now with $128^3$ dark matter particles, computing about $\sim2000$ time steps from $z=127$ to $z=0.01$. This simulation runs in about $10$ CPU hours. The HR-LR training pairs are constructed by randomly cropping $16000$ cubes of size $48^3$~px $(17.1\ \text{Mpc}/h)^3$ out of the full $264^3$~px fields, where 18\% of the volume has been set aside as a validation set. Respecting the statistics of the fields, we also give each pair of cubes a random $\pi/2$ rotation and random chance of being mirrored. While the amount of training data is somewhat limited, we do not find evidence of over-training or ``remembering of phases" in our tests of sample diversity in Section \ref{sec:results_variety}. We also checked that our model did not perform better on the LR training data then on LR test data with various statistics.



For the LR test data, we run this simulation again, but now with a different \texttt{NGenIC} initial seed. We make two sets of LR test data. For the main result of this paper, presented in Section~\ref{sec:results}, the LR field is a $410\ \text{Mpc}/h$ length box, having 8 times the volume of TNG300; therefore, we proportionally increase to $256^3$ dark matter particles as to have the same particle density as the LR training data. For Section~\ref{sec:results_variety} and Appendix~\ref{sec:variable_cosmos}, we simulate LR fields at $205\ \text{Mpc}/h$. There is no HR truth for our LR test data, and so we will use summary statistics from the TNG300 training data as a truth comparison.


\section{Conditional diffusion model approach}
\label{sec:diffusion_model}

Our approach is based on the the image-to-image Palette diffusion model \cite{ho2020denoising,palette}, which is capable of performing outpainting to generate larger images. We modify the model to 3 dimensions and train it to be conditional on our 3-dimensional LR simulation.

\subsection{Iterative outpainting to generate large fields}
\label{sec:iterative_outpainting}

\begin{figure}
    \centering
    \includegraphics[width=0.75\textwidth, trim={6.5cm, 5.5cm, 3.5cm, 5.5cm}, clip]{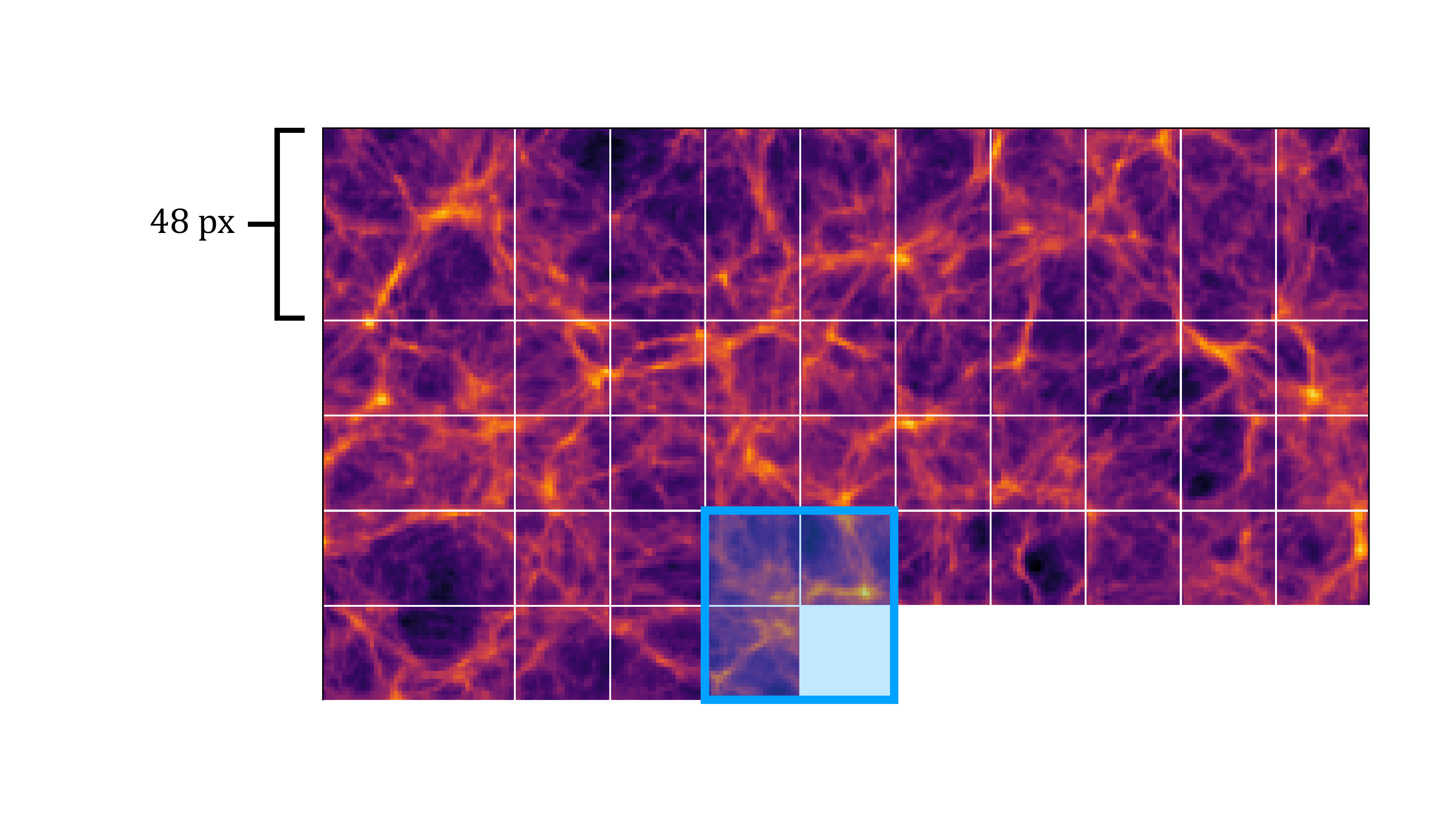}
    \caption[Iterative outpainting illustration]{Iterative outpainting illustration, 2-dimensional projection. The blue cube contains the previously generated SR fields adjacent to the subsequent volume ready to be outpainted. After a new volume is generated, the blue cube then moves in row-major order for the next SR volume to be generated. After a plane of small volumes is generated to SR, the outpaintings continue in the third dimension (perpendicular to the page), and the next plane is generated, with each SR volume also conditional on the previous plane.}
    \label{fig:iterative_outpainting}
\end{figure}

This work uses massively iterative outpainting that can generate cosmological volumes much larger than the training data. Given a large LR field, an SR field is generated sequentially, patch by patch, with each new patch generated conditional on both the LR field and adjacent SR fields. The LR conditional guides the large length scales of the SR field, while the adjacent SR conditional ensures that newly generated SR fields are smooth and physically consistent with the larger SR volume. Training the model to outpaint requires masking sub-volumes of the HR data in the HR-LR pairs, as shown in~Fig.\ref{fig:model_illustration}, and thus masked sub-volumes at inference can be generated given surrounding SR data. An illustration of the iterative outpainting is shown in Fig.~\ref{fig:iterative_outpainting}.

Our iterative outpainting method is motivated by the nature of mode coupling in the large-scale structure. On large scales, for $k\lesssim0.1\ h/\text{Mpc}$ today, the evolution of physical perturbations is linear, as modes evolve independently. On intermediate scales, $0.1\ h/\text{Mpc}\lesssim k\lesssim 0.5\ h/\text{Mpc}$ modes evolve nonlinearly but should be accurately captured by the LR dark matter N-body simulation. On smaller scales, $k\gtrsim 0.5\ h/\text{Mpc}$ we want the diffusion model to model nonlinearity and baryonic feedback. This informs us about the minimum physical size required for the outpainting volumes, and the diffusion model we describe here has a fundamental mode of $k_{24\text{px}}=0.34\ h/\text{Mpc}$. The diffusion model can modify the results of the LR simulation on physical scales smaller than this scale, and can thus take into account mode coupling on these scales. Mode coupling is included by the LR simulation on large scales, but cannot be modified by the diffusion model due to its outpainting window size, and we thus assume that the LR simulation is correct on these scales. In position space, the $24$~px window corresponds to a physical length of $18.6\ \text{Mpc}/h$. This can be compared to the typical particle displacement of $5\  \text{Mpc}/h$, with an upper limit of $\sim20\ \text{Mpc}/h$~\cite{kodiramanah2020}. Most of this displacement is already included in the LR simulation, so our window size is sufficient.

In detail, the iterative outpainting procedure works as follows. First, a $48^3$~px SR cube is generated conditional on a $48^3$~px LR cube; this first cube is shown in the top left of Fig.~\ref{fig:iterative_outpainting}. We move in row-major order, outpainting $24$~px at a time. The second SR volume generated is thus conditional on both its underlying $48^3$~px LR cube, as well as the $24$~px length right half of the first SR cube. After an entire plane is generated, the outpaintings move to the third dimension, with every subsequent plane conditional on the previously generated plane. The outpaintings continue in this way until the entire LR volume is generated to SR. We never break conditionality on previous adjacent SR regions, even after moving into the third dimension. Thus throughout the full SR volume, every locally outpainted volume is conditional on all adjacent previously generated volumes.

In some small regions at the outpainting boundaries, a slight discontinuity develops in the SR model output. To remedy this, we apply a linear interpolation in the $2$~px wide strip at the outpainting boundaries. This interpolation has negligible affect on the summary statistics, while making the results visually appear slightly more accurate.

A potential issue we have found is the generated data progressively slipping out of the distribution that the diffusion model was trained on. Suppose that at some early point $i$ in the outpainting chain, the diffusion model generates an out-of-distribution $\bm{y}_i+\bm{\delta}\bm{y}_i$ as a sample from $p(\bm{y}_i|\bm{x},\bm{y}_{<i})$. The next sample would be generated from $p(\bm{y}_{i+1}|\bm{x},\bm{y}_{<i},\bm{y}_i+\bm{\delta}\bm{y}_i)$, and because the model has not trained with any $\bm{y}_i+\bm{\delta}\bm{y}_i$, the out-of-distribution problem may compound. This problem arose for training the model using apparently too few model parameters. In this case, the first SR cube in the outpainting chain was too Gaussian, albeit still resembling the conditional field. However, after several more outpainting iterations, the model output quality collapsed, and appeared as very smooth blob. We overcame this problem by sufficiently increasing the number of model parameters (described below) as to never slip out of the model's learned distribution.

\begin{figure}
    \centering
    
    \begin{tabular}{cccc}
    \vspace{-0.4cm}\includegraphics[width=0.22\textwidth]{images/sr_diffusion/images_outpainting/sr_cube_noise_1.pdf} & \includegraphics[width=0.22\textwidth]{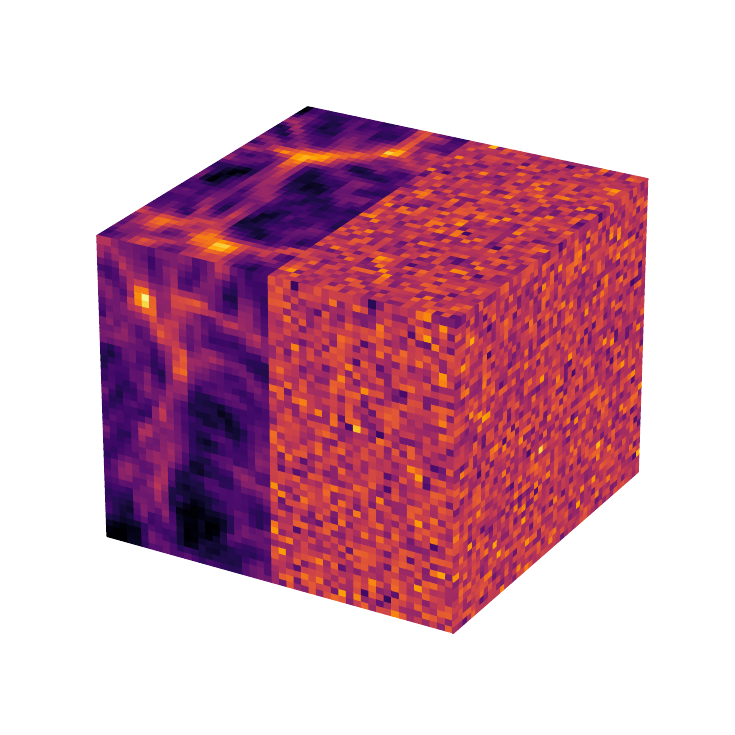} & \includegraphics[width=0.22\textwidth]{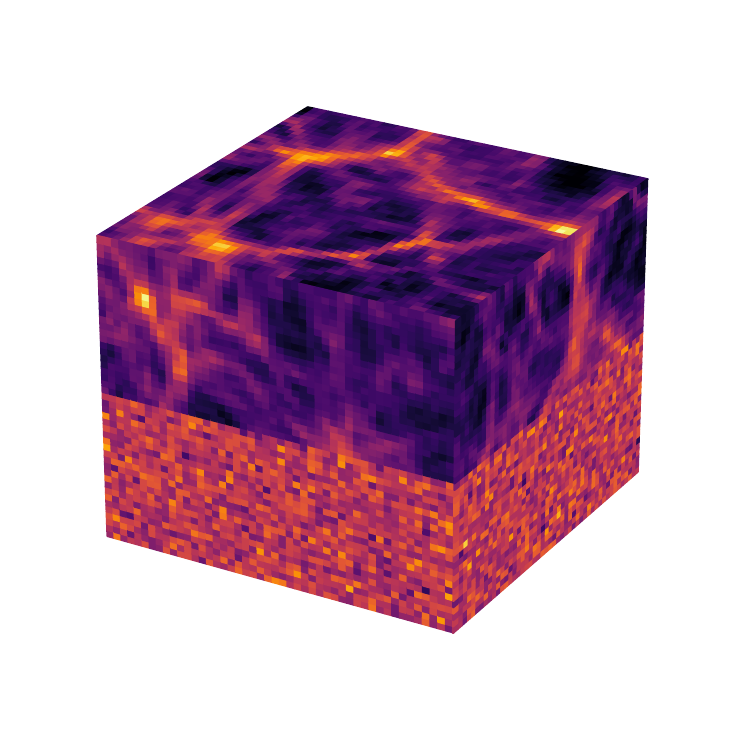} & \includegraphics[width=0.22\textwidth]{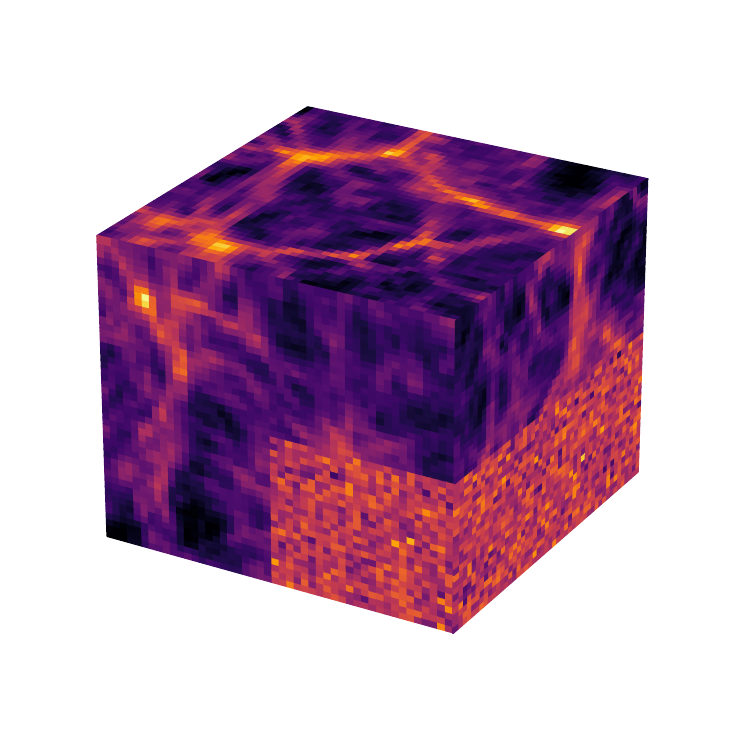}\\
    \vspace{-0.1cm}First cube & First row in & First column & First plane's\\
     & first plane & first plane & main volume\\
    \vspace{-0.4cm}\includegraphics[width=0.22\textwidth]{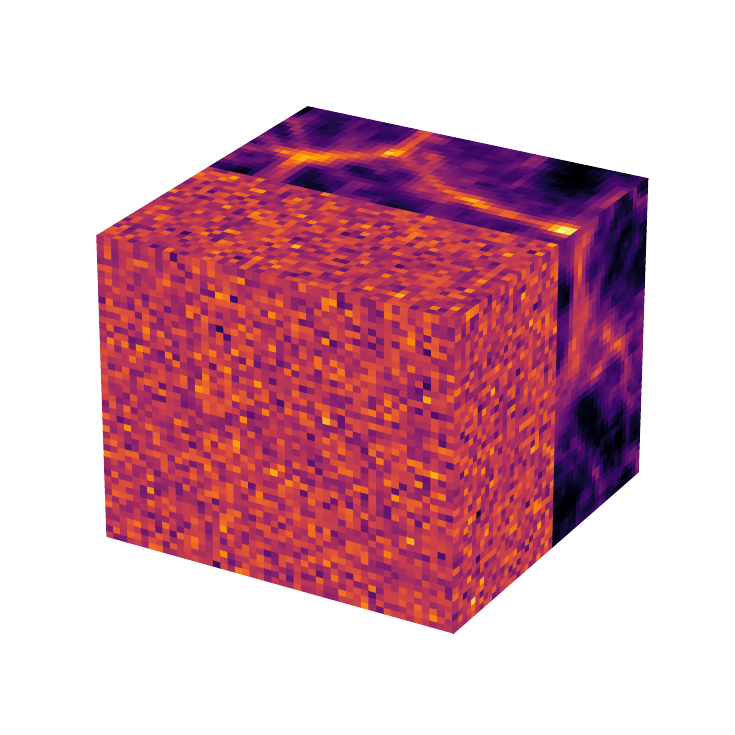} & \includegraphics[width=0.22\textwidth]{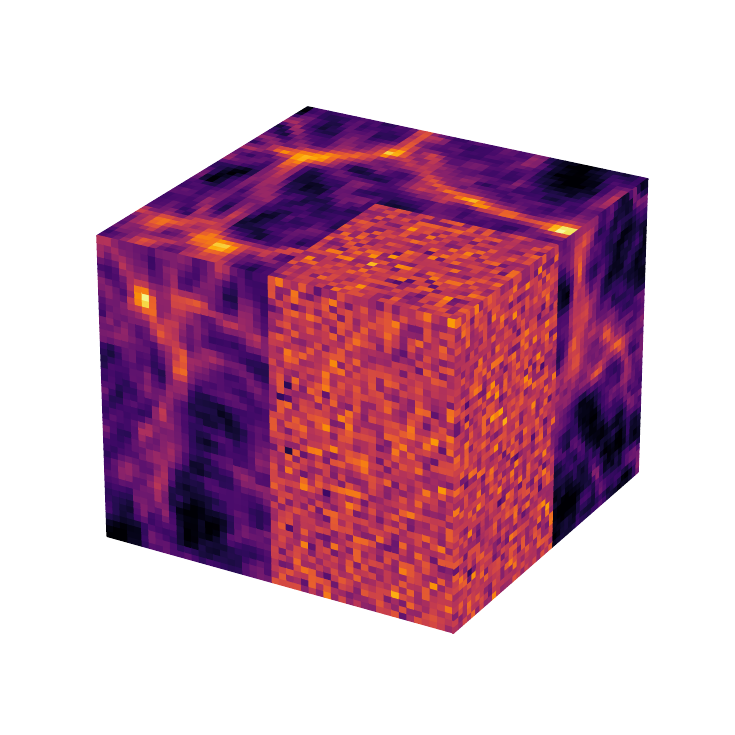} & \includegraphics[width=0.22\textwidth]{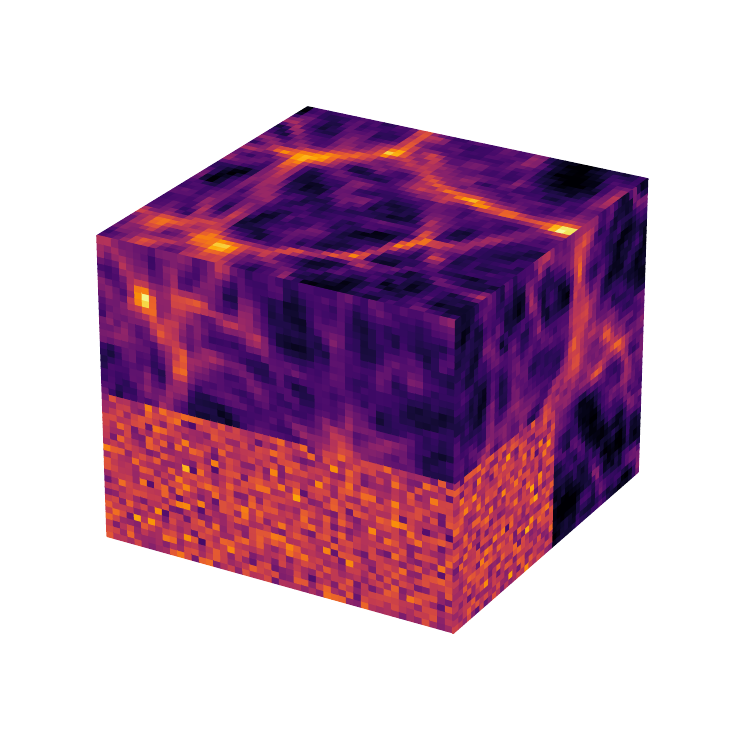} & \includegraphics[width=0.22\textwidth]{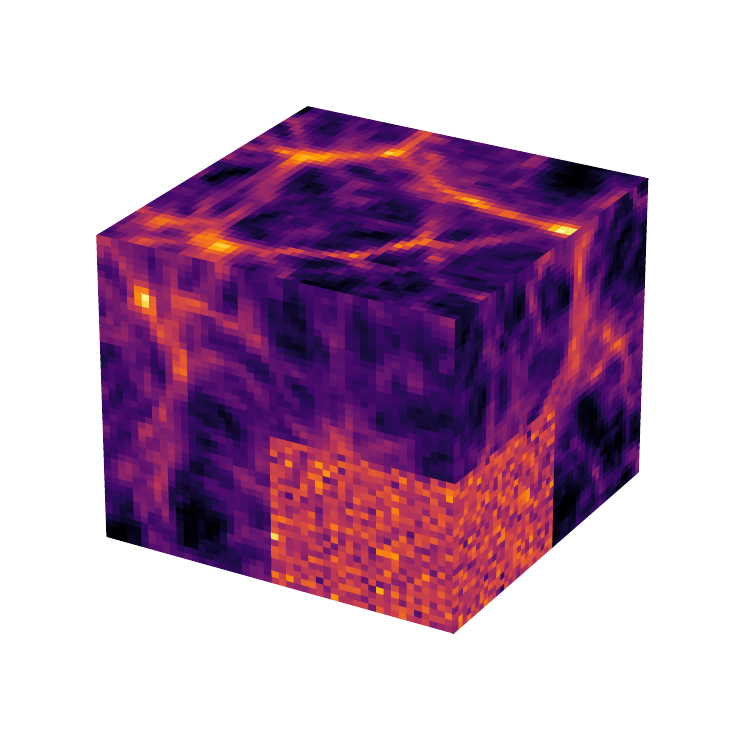}\\
    First volume in & First row in & First column in & Main volume\\
    subsequent planes & subsequent planes & subsequent planes &
    \end{tabular}
    
    \caption[Mask types to outpaint a 3-dimensional volume]{There are 8 mask types (4 considering rotational symmetry) required to iteratively outpaint the full 3-dimensional volume. Each mask type is labelled as to where in the full volume it is used for outpainting. The top row of masks are used to outpaint the full first plane of cubes. The three ``subsequent planes" masks on the bottom row are for outpainting the first row and column of every plane of cubes after the first plane. The vast majority of the outpaintings belong to the ``main volume" mask (bottom right) containing noise in a single octant, used for volumes that are not located in the first plane of cubes, nor the first row or first column of any planes.}
    \label{fig:masking_demo}
\end{figure}

\subsection{Model details}
\label{sec:network_architecture}

\begin{figure}
\begin{center}
\begin{tikzpicture}[node distance=0.75cm]
\def\unetwidth{2cm}
\def\unetheight{0.45cm}
\def\xs{5.6em}
\def\resfill{yellow!40}
\def\bgfill{orange!30}
\def\catfill{white}
\tiny

\normalsize
\node[align=center, font=\bfseries] (titleunet){U-net};
\tiny

\node (input) [cnnr, left of=titleunet, xshift=-.5cm, yshift=2.3cm] {Field: $2\times n^3$};

\node (cnn0) [cnnr, below of=input, fill=azul!75] {Conv: $64\times n^3$};
\node (resd0) [cnnr, left of=cnn0, xshift=-\xs, fill=\resfill] {Res: $64\times n^3$};
\node (resd1) [cnnr, left of=resd0, xshift=-\xs, fill=\bgfill] {BG: $64\times\frac{n}{2}^3$};
\node (resd2) [cnnr, left of=resd1, xshift=-\xs, fill=\resfill] {Res: $128\times\frac{n}{2}^3$};
\node (resd3) [cnnr, left of=resd2, xshift=-\xs, fill=\bgfill] {BG: $128\times\frac{n}{4}^3$};
\node (resd4) [cnnr, left of=resd3, xshift=-\xs, fill=\resfill] {Res: $256\times\frac{n}{4}^3$};

\node (resm0) [cnnr, left of=resd4, xshift=-\xs, fill=\resfill] {Res: $256\times\frac{n}{4}^3$};
\node (att) [cnnr, below of=resm0, fill=blue!15] {Att: $256\times\frac{n}{4}^3$};
\node (resm1) [cnnr, below of=att, fill=\resfill] {Res: $256\times\frac{n}{4}^3$};

\node (resu0) [cnnr, right of=resm1, xshift=\xs, fill=\resfill] {Res: $256\times\frac{n}{4}^3$};
\node (resu1) [cnnr, right of=resu0, xshift=\xs, fill=\resfill] {Res: $256\times\frac{n}{4}^3$};
\node (resu2) [cnnr, below of=resu1, fill=\bgfill] {BG: $256\times\frac{n}{2}^3$};
\node (resu3) [cnnr, right of=resu2, xshift=\xs, fill=\resfill] {Res: $128\times\frac{n}{2}^3$};
\node (resu4) [cnnr, right of=resu3, xshift=\xs, fill=\resfill] {Res: $128\times\frac{n}{2}^3$};
\node (resu5) [cnnr, below of=resu4, fill=\bgfill] {BG: $128\times n^3$};
\node (resu6) [cnnr, right of=resu5, xshift=\xs, fill=\resfill] {Res: $64\times n^3$};
\node (resu7) [cnnr, right of=resu6, xshift=\xs, fill=\resfill] {Res: $64\times n^3$};

\node (out0) [cnnr, below of=resu7, fill=red!15] {GN: $64\times n^3$};
\node (out1) [cnnr, below of=out0, fill=azul!75] {Conv: $1\times n^3$};

\draw [arrow] (input) to (cnn0);

\draw [arrow] (cnn0) to (resd0);
\draw [arrow] (resd0) to (resd1);
\draw [arrow] (resd1) to (resd2);
\draw [arrow] (resd2) to (resd3);
\draw [arrow] (resd3) to (resd4);
\draw [arrow] (resd4) to (resm0);

\draw [arrow] (resm0) to (att);
\draw [arrow] (att) to (resm1);

\draw [arrow] (resm1) to (resu0);
\draw [arrow] (resu0) to (resu1);
\draw [arrow] (resu1) to (resu2);
\draw [arrow] (resu2) to (resu3);
\draw [arrow] (resu3) to (resu4);
\draw [arrow] (resu4) to (resu5);
\draw [arrow] (resu5) to (resu6);
\draw [arrow] (resu6) to (resu7);

\draw [arrow, dashed] (cnn0) to (resu7);
\draw [arrow, dashed] (resd0) to (resu6);
\draw [arrow, dashed] (resd1) to (resu4);
\draw [arrow, dashed] (resd2) to (resu3);
\draw [arrow, dashed] (resd3) to (resu1);
\draw [arrow, dashed] (resd4) to (resu0);

\draw [arrow] (resu7) to (out0);
\draw [arrow] (out0) to (out1) node[xshift=-0.38cm, rotate=90] {\hspace{.85cm}SiLU};

\normalsize
\node[align=center, font=\bfseries, right of=titleunet, xshift=6cm] (titleres){\hspace{-.2cm}Residual block};
\tiny

\node (cnn0) [cnn, below of=titleres, xshift=-1.3cm, yshift=.15cm, minimum width=\unetwidth, minimum height=\unetheight] {Field}; 
\node (cnn1) [cnn, below of=cnn0, fill=red!15, minimum width=\unetwidth, minimum height=\unetheight] {GN}; 
\node (cnn2) [cnn, below of=cnn1, fill=azul!75, minimum width=\unetwidth, minimum height=\unetheight] {Conv}; 

\node (emb0) [cnn, right of=cnn0, xshift=1.75cm, minimum width=\unetwidth, minimum height=\unetheight] {Embedding}; 
\node (emb1) [cnn, below of=emb0, fill=green!15, minimum width=\unetwidth, minimum height=\unetheight] {Linear}; 

\node (plus0) [circ, below of=cnn2, xshift=5em, yshift=-.65em] {$\bm{+}$};
\node (cnn3) [cnn, below of=plus0, yshift=-.65em, fill=red!15, minimum width=\unetwidth, minimum height=\unetheight] {GN}; 
\node (cnn4) [cnn, below of=cnn3, fill=azul!75, minimum width=\unetwidth, minimum height=\unetheight] {Conv}; 
\node (plus1) [circ, below of=cnn4, yshift=-.65em] {$\bm{+}$};
\node (cnn5) [cnn, below of=plus1, yshift=-.65em, minimum width=\unetwidth, minimum height=\unetheight, draw=none] {};

\draw [arrow] (cnn0) to (cnn1);
\draw [arrow] (cnn1) to (cnn2) node[yshift=0.37cm] {\hspace{.85cm}SiLU};

\draw [arrow] (emb0) to (emb1) node[yshift=0.37cm] {\hspace{.85cm}SiLU};

\draw [arrow] (cnn2.south) to (plus0);
\draw [arrow] (emb1.south) to (plus0);

\draw [arrow] (plus0) to (cnn3);
\draw [arrow] (cnn3) to (cnn4) node[yshift=0.35cm] {\hspace{1.9cm}SiLU, Dropout};
\draw [arrow] (cnn4) to (plus1);

\coordinate (topleft) at ([xshift=-1.3cm]cnn0);
\coordinate (bottomleft) at ([xshift=-2.59cm]plus1);
\draw [arrow] (cnn0) to (topleft) to (bottomleft) to (plus1);

\draw [arrow] (plus1) to (cnn5);

\normalsize
\node[align=center, font=\bfseries, below of=plus1, yshift=-.3cm] (titlebg){\hspace{-.2cm}BigGAN residual block};
\tiny

\node (cnn0) [cnn, below of=titlebg, xshift=-1.3cm, yshift=.15cm, minimum width=\unetwidth, minimum height=\unetheight] {Field};
\node (cnn1) [cnn, below of=cnn0, fill=red!15, minimum width=\unetwidth, minimum height=\unetheight] {GN};
\node (cnn2) [cnn, below of=cnn1, fill=azul!75, minimum width=\unetwidth, minimum height=\unetheight] {Conv};
\node (cnn2) [cnn, below of=cnn1, fill=azul!75, minimum width=\unetwidth, minimum height=\unetheight] {Conv};
\node (pool) [cnn, below of=cnn2, fill=orange!35, minimum width=\unetwidth, minimum height=\unetheight] {Down/Up};

\node (emb0) [cnn, right of=cnn0, xshift=1.75cm, minimum width=\unetwidth, minimum height=\unetheight] {Embedding};
\node (emb1) [cnn, below of=emb0, fill=green!15, minimum width=\unetwidth, minimum height=\unetheight] {Linear};

\node (plus0) [circ, below of=pool, xshift=5em, yshift=-.65em] {$\bm{+}$};
\node (cnn3) [cnn, below of=plus0, yshift=-.65em, fill=red!15, minimum width=\unetwidth, minimum height=\unetheight] {GN};
\node (cnn4) [cnn, below of=cnn3, fill=azul!75, minimum width=\unetwidth, minimum height=\unetheight] {Conv};
\node (plus1) [circ, below of=cnn4, yshift=-.65em] {$\bm{+}$};
\node (cnn5) [cnn, below of=plus1, yshift=-.65em, draw=none] {};

\draw [arrow] (cnn0) to (cnn1);
\draw [arrow] (cnn1) to (cnn2) node[yshift=0.37cm] {\hspace{.85cm}SiLU};
\draw [arrow] (cnn2) to (pool);

\draw [arrow] (emb0) to (emb1) node[yshift=0.35cm] {\hspace{.85cm}SiLU};

\draw [arrow] (pool.south) to (plus0);
\draw [arrow] (emb1.south) to (plus0);

\draw [arrow] (plus0) to (cnn3);
\draw [arrow] (cnn3) to (cnn4) node[yshift=0.37cm] {\hspace{1.9cm}SiLU, Dropout};   
\draw [arrow] (cnn4) to (plus1);

\coordinate (topleft) at ([xshift=-1.3cm]cnn0);
\coordinate (bottomleft) at ([xshift=-2.59cm]plus1);
\draw [arrow] (cnn0) to (topleft) to (bottomleft) to (plus1);

\draw [arrow] (plus1) to (cnn5);

\small
\node[align=center, font=\bfseries, below of=plus1, xshift=-3cm, yshift=0cm] (titleleg){Legend};
\tiny

\node (convleg) [cnn, below of=titleleg, xshift=-1cm, yshift=.2cm, fill=azul!75, minimum width=\unetwidth, minimum width=\unetwidth, minimum height=\unetheight] {Conv};
\node[align=left, font=\bfseries, right of=convleg, xshift=2cm] {$3\times3\times3$ convolution};
\node (resleg) [cnn, below of=convleg, yshift=.15cm, fill=\resfill, minimum width=\unetwidth, minimum height=\unetheight] {Res};
\node[align=left, font=\bfseries, right of=resleg, xshift=2cm] {Residual block};
\node (bgleg) [cnn, below of=resleg, yshift=.15cm, fill=\bgfill, minimum width=\unetwidth, minimum height=\unetheight] {BG};
\node[align=left, font=\bfseries, right of=bgleg, xshift=2cm] {BigGAN residual block};
\node (attleg) [cnn, below of=bgleg, yshift=.15cm, fill=blue!15, minimum width=\unetwidth, minimum height=\unetheight] {Att};
\node[align=left, font=\bfseries, right of=attleg, xshift=2cm] {QKV attention};
\node (gnleg) [cnn, below of=attleg, yshift=.15cm, fill=red!15, minimum width=\unetwidth, minimum height=\unetheight] {GN};
\node[align=left, font=\bfseries, right of=gnleg, xshift=2cm] {$32$ group normalization};
\node (duleg) [cnn, below of=gnleg, yshift=.15cm, fill=orange!35, minimum width=\unetwidth, minimum height=\unetheight] {Down/Up};
\node (dulegdescrip) [align=left, font=\bfseries, right of=duleg, xshift=2cm] {$2\times$ down/up sampling};
\coordinate (concatleft)  at ([xshift=-.9cm, yshift=-.52cm]duleg);
\coordinate (concatright) at ([xshift= .9cm, yshift=-.52cm]duleg);
\draw [arrow, dashed] (concatleft) to (concatright);
\node[align=left, font=\bfseries, below of=dulegdescrip, yshift=.15cm] {Channel concatenation};

\draw [arrow] (plus1) to (cnn5);

\end{tikzpicture}
\caption[U-net used to learn each denoising diffusion step]{The U-net used in this work (left) is made of residual blocks (top right), with BigGAN residual blocks (center right) used to downsample and upsample the fields. Shown after every layer name in the U-net is the layer's output channels $\times$ volume.}
\label{fig:unet}
\end{center}
\end{figure}
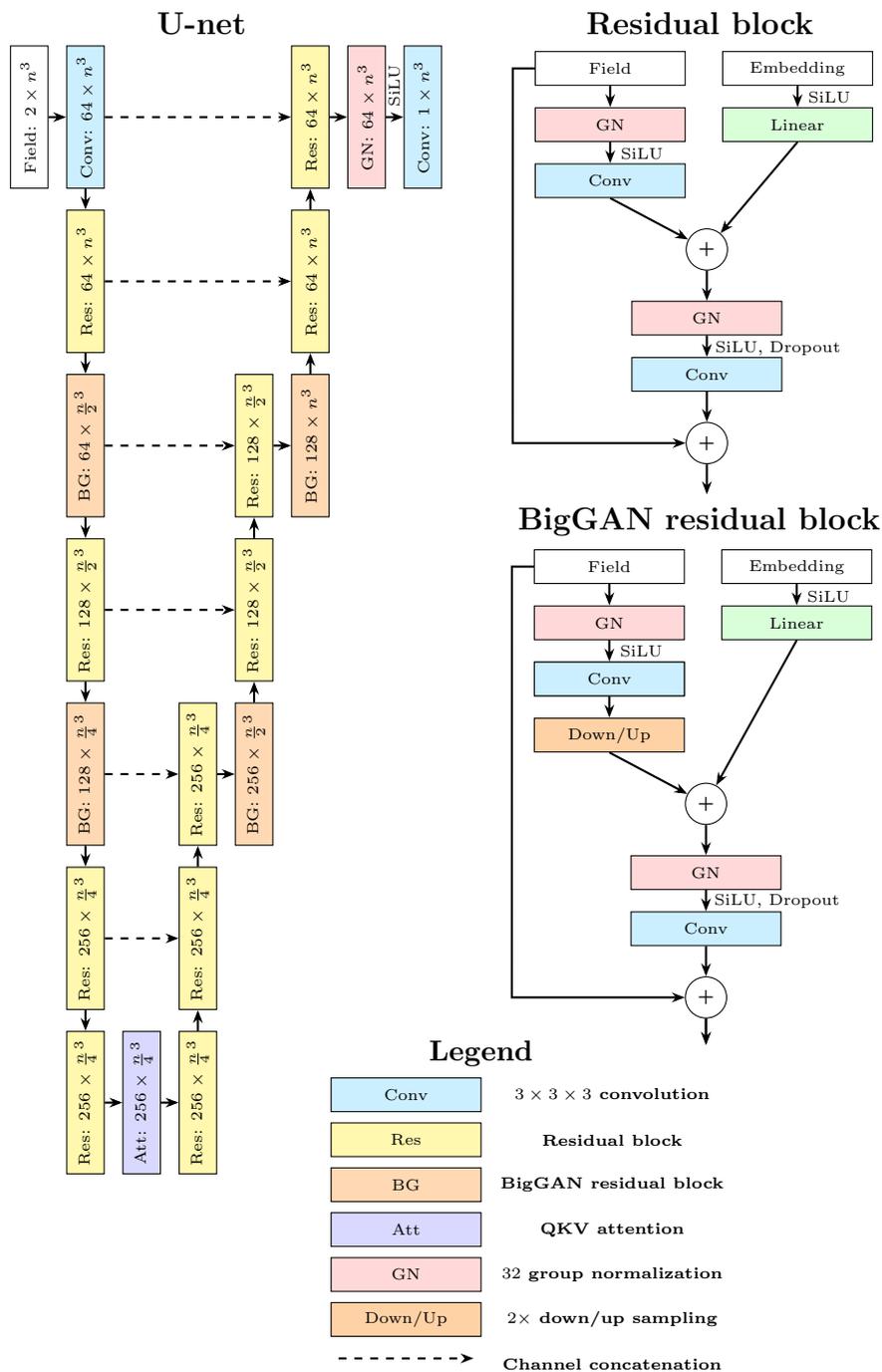

Our model is a modified version of Palette based on code from~\cite{palette_github}. We use a U-net~\cite{Ronneberger2015} to learn the function $f_{\bm{\theta}}{(\bm{x},\bm{y}_t,\gamma_t)}$ that handles the denoising routine from $\bm{y}_t$ to $\bm{y}_{t-1}$. The details of the U-net used here are based on the ``guided diffusion" U-net described in~\cite{dhariwal2021diffusion}, which we minimally modify to operate on 3-dimensional data. A diagram of the U-net is shown in Fig.~\ref{fig:unet}. The U-net takes as input $2$ channels, $\bm{y}_t$ and the conditional $\bm{x}$, and outputs $1$ channel, a prediction for $\bm{\epsilon}$. It has two downsampling steps, with the number of channels being $64$, $128$, and $256$, and similarly two upsampling steps. BigGAN residual blocks~\cite{brock2018large} are used for the downsampling and upsampling, alternating between standard residual blocks at each resolution. The middle section of our U-net has 2 residual blocks and a query-key-value attention layer~\cite{vaswani2017}. We also experimented with the SR3 U-net introduced in~\cite{sr3}, being the network used in the original Palette implementation, and we found similar model performance between these two U-nets.

Our residual blocks contain two $3\times3\times3$ convolutions. To allow the U-net to be dependent on the diffusion step number $t$, each residual block has an embedding as an additional input. Each embedding goes through a linear layer and then gets added in the middle of the block, as shown in the figure. Prior to these convolutions is a group normalization\cite{Wu2020} with sigmoid linear unit (SiLU)~\cite{Elfwing2018} activation function. We also experimented with ReLU activation functions, with which we found similar model performance. We use $0.2$ dropout before the second convolution in each residual block. There is additionally a $1\times1\times1$ convolution in residual skip connections when the number of channels is changed in the residual block. This model has 31.5 million learnable parameters.



\subsection{Model training}

We train the diffusion model on the LR-HR pairs of $48$~px length fields described above. Due to the nature of matter clustering in cosmology, the distribution of values in both the LR and HR data is heavily skewed to the right, and so we preprocess the data by applying a natural logarithm. We want the diffusion model to take as input data values in the range $(-1, 1)$, so we additionally apply a sigmoid to the data. We should not allow the diffusion model to train on data with values well outside of $(-1,1)$, for then the $\bm{y}_T$ fields would not accurately be noisy but rather be spacially correlated with high density values in the truth $\bm{y}_0$.

After each batch of HR-LR pairs is loaded as input to the model, each HR field is randomly assigned a mask, chosen from the 8 mask types shown in Fig.~\ref{fig:masking_demo}. We train with $T=2000$ diffusion time steps with noise variances linearly increasing from $\beta_0=10^{-6}$ to $\beta_T=10^{-2}$. We use the Adam optimizer with a learning rate of $10^{-4}$, and a $0.9999$ exponential moving average learning rate decay. We train on two A100 GPUs with a batch size of $16$ per GPU, and the training loss~(Eq.~\ref{eq:loss}) converges in 70 hours. The validation loss shows no sign of over-training.

\section{Super-resolution results for a large volume}
\label{sec:results}

\begin{figure}
    \begin{center}
    \Large\textbf{Training data}\normalsize
    
    \rotatebox{90}{\small\hspace{-1.6cm}$205\ \text{Mpc}/h$, $264$~px}\begin{tabular}{ccc}
        LR & \hspace{4.2cm} & HR\\
        \includegraphics[width=0.245\textwidth]{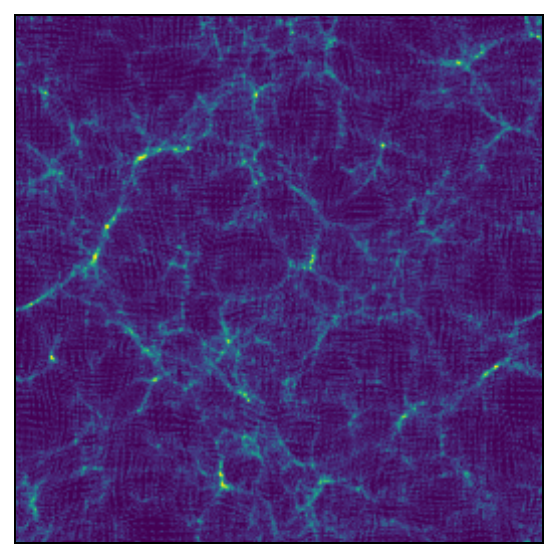} & \hspace{4.2cm} & \includegraphics[width=0.245\textwidth]{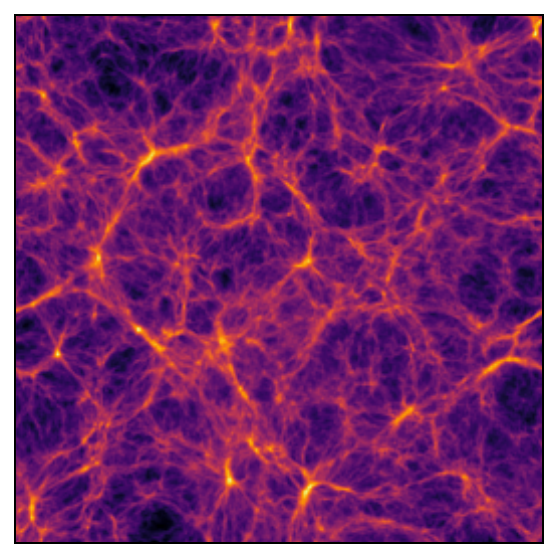}
    \end{tabular}
    
    \Large\textbf{Results}\normalsize
    
    \rotatebox{90}{\small\hspace{-1.6cm}$410\ \text{Mpc}/h$, $528$~px}\begin{tabular}{cc}
        LR test data & SR model output\\
        \includegraphics[width=0.495\textwidth]{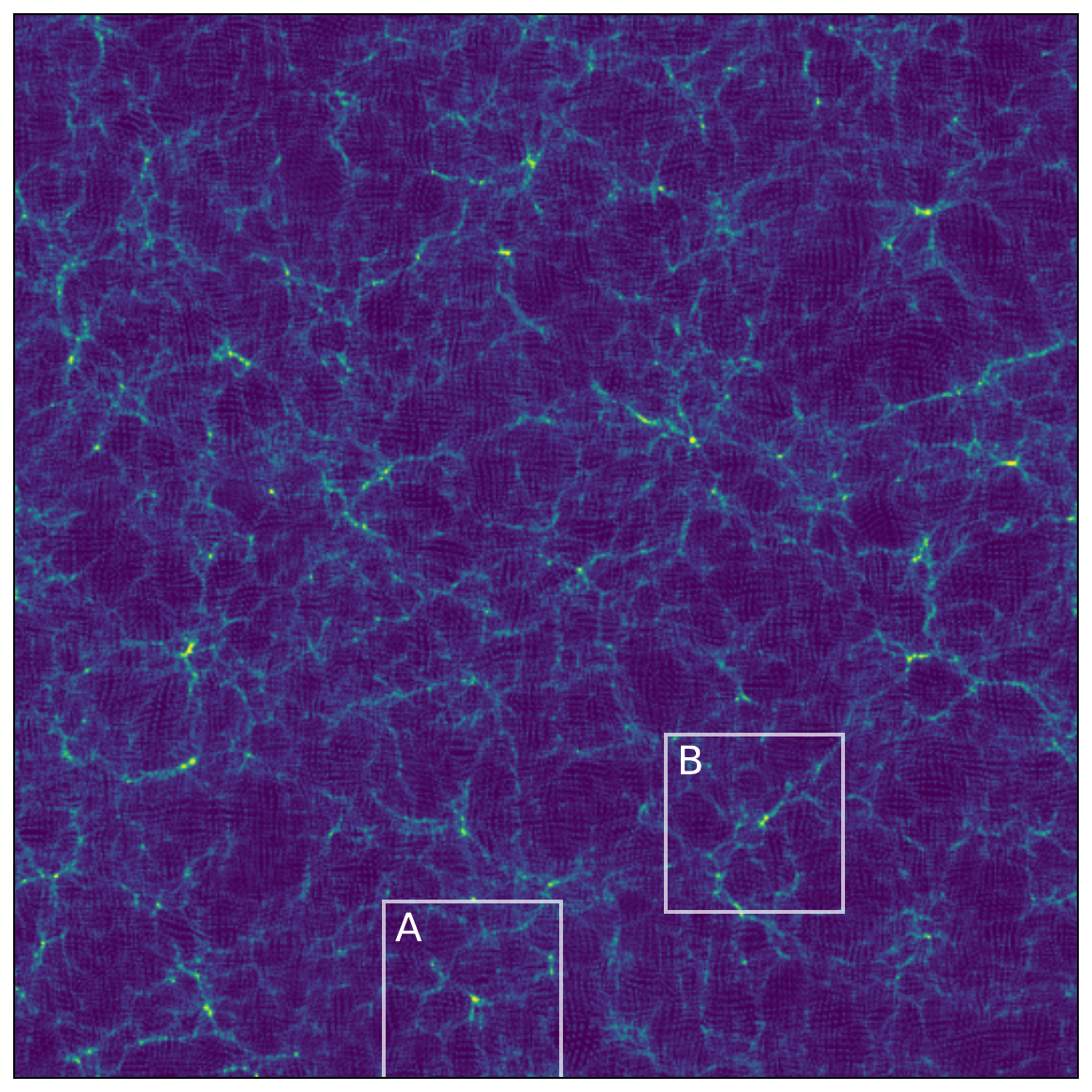} & \includegraphics[width=0.495\textwidth]{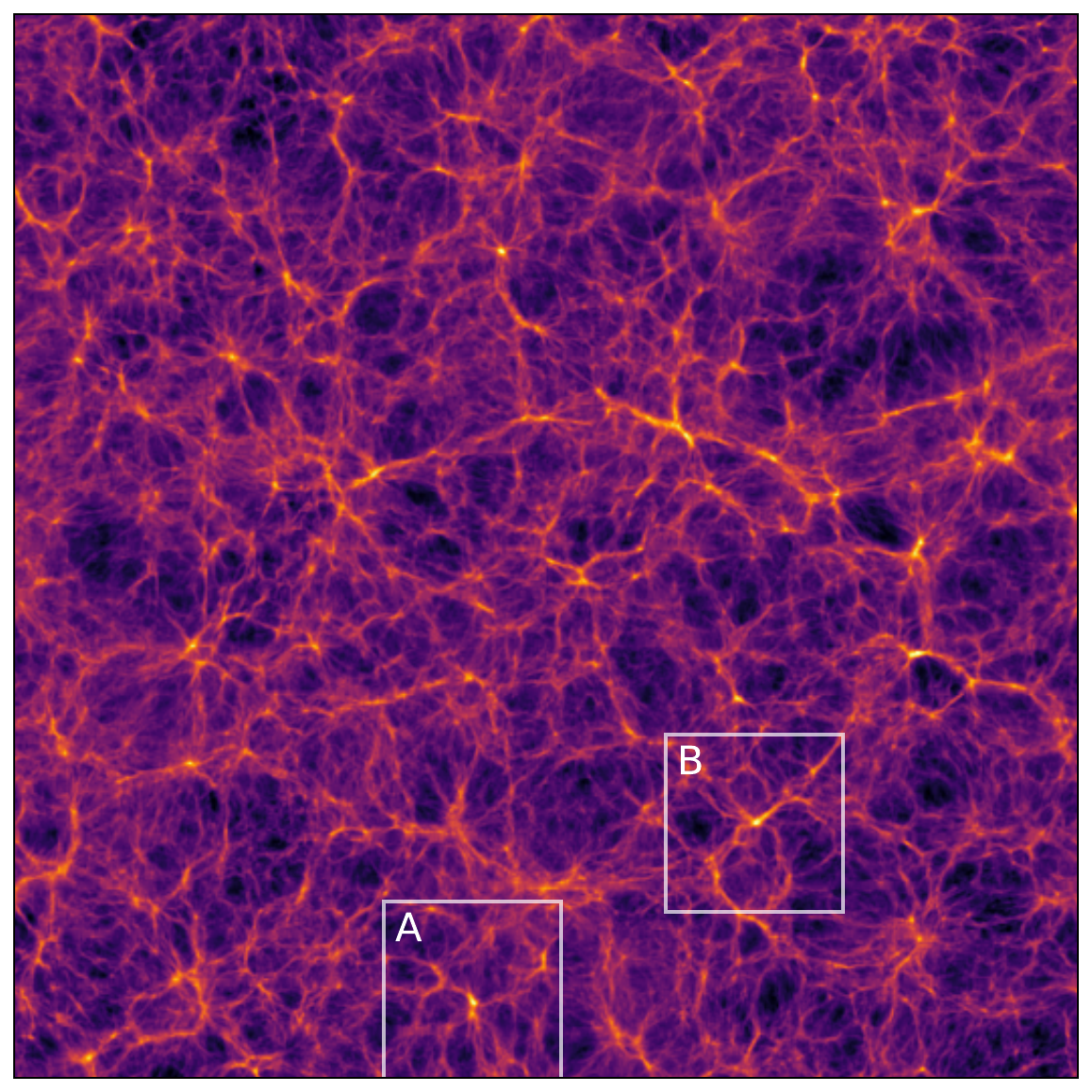}
    \end{tabular}
    
    \rotatebox{90}{\small\hspace{-1.4cm}$68\ \text{Mpc}/h$, $88$~px}\begin{tabular}{cccc}
        LR zoom A & LR zoom B & SR zoom A & SR zoom B\\
        \includegraphics[width=0.245\textwidth]{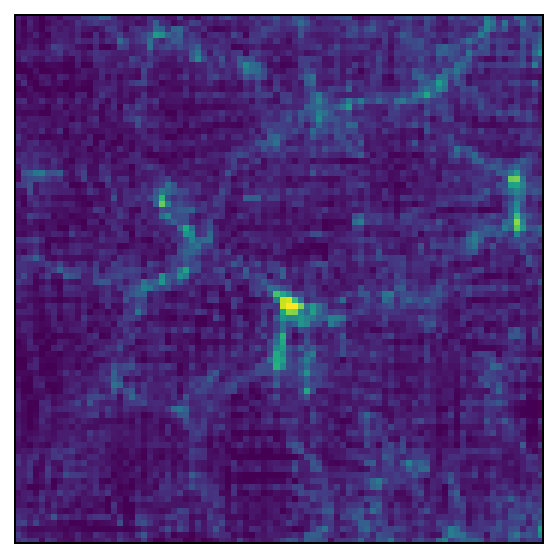} & \includegraphics[width=0.245\textwidth]{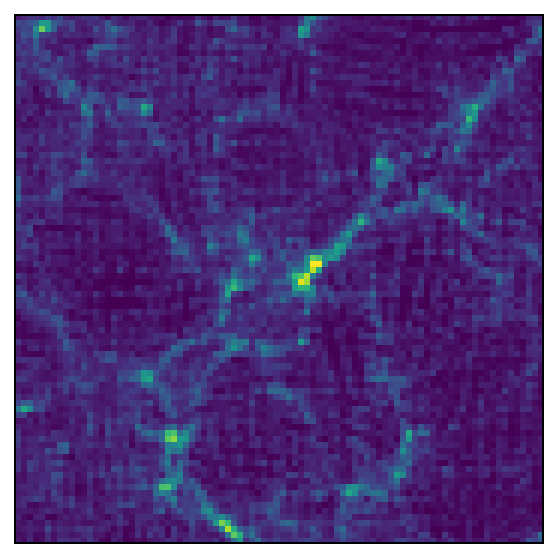} & \includegraphics[width=0.245\textwidth]{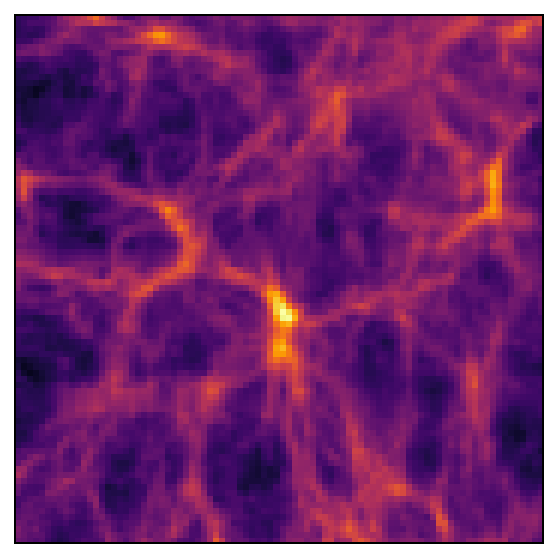} & \includegraphics[width=0.245\textwidth]{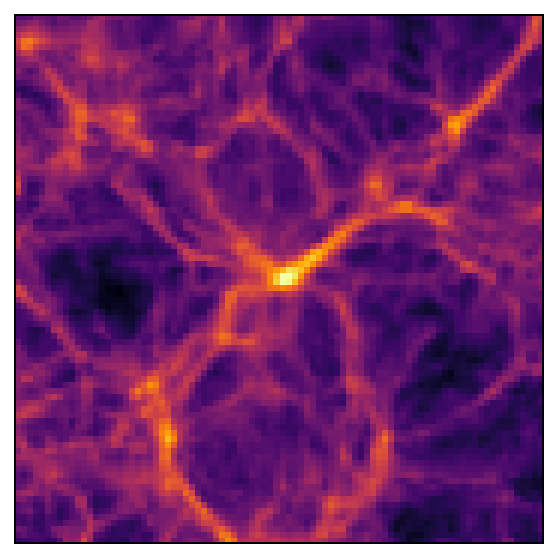}
    \end{tabular}
    \end{center}
    \caption[Diffusion model results for a large volume]{Matter density results for our super-resolution diffusion model in generating a volume larger than the entire training data volume. Images are 2-dimensional projections of depth $19\ \text{Mpc}/h$. (Top~row)~The training data comes from the single pair of boxes shown. The model trains on $48$~px length LR-HR pairs cut out of these boxes. (Center~left)~LR conditional test data. (Center~right)~SR model output generated with $21^3$ outpainting iterations, having 8 times the volume of the training data. (Bottom~row)~Two zoom-ins of the LR and SR fields.}
    \label{fig:results_visual}
\end{figure}

The main result of this work is the SR emulation of a $410\ \text{Mpc}/h\approx600\ \text{Mpc}$ length cube, with $528^3$~px, constructed with $21^3$ outpainting iterations. As the training data came from the $205\ \text{Mpc}$ length TNG300 cube meshed onto $264^3$~px, our SR result thus has an increased volume of the entire training data by a factor of 8. We show our results in Fig.~\ref{fig:results_visual}.

To generate our large SR cube, we used a linear noise schedule of $T=1250$ steps, with $\beta_0=10^{-6}$ and $\beta_T=1.5\times10^{-2}$. We found $1250$ steps to be the lowest possible before losing any measurable amount of quality in the power spectrum; a discussion of potential further speed-ups is in Section~\ref{sec:conclusion}. The time to generate the full cube of $21^3$ outpainting volumes was $120$ hours on a single A100 GPU.

\setlength{\tabcolsep}{0pt}

We measure the accuracy of the diffusion model with several summary statistics familiar to cosmology. Results are computed on mean zero overdensities $\delta(\bm{r})$, its Fourier transform denoted by $\delta(\bm{k})$.

\begin{figure}
    \centering
    
    \begin{tabular}{cc}
        \hspace{1cm}Probability density & \hspace{1cm}Power spectrum\\
        \includegraphics[width=0.495\textwidth]{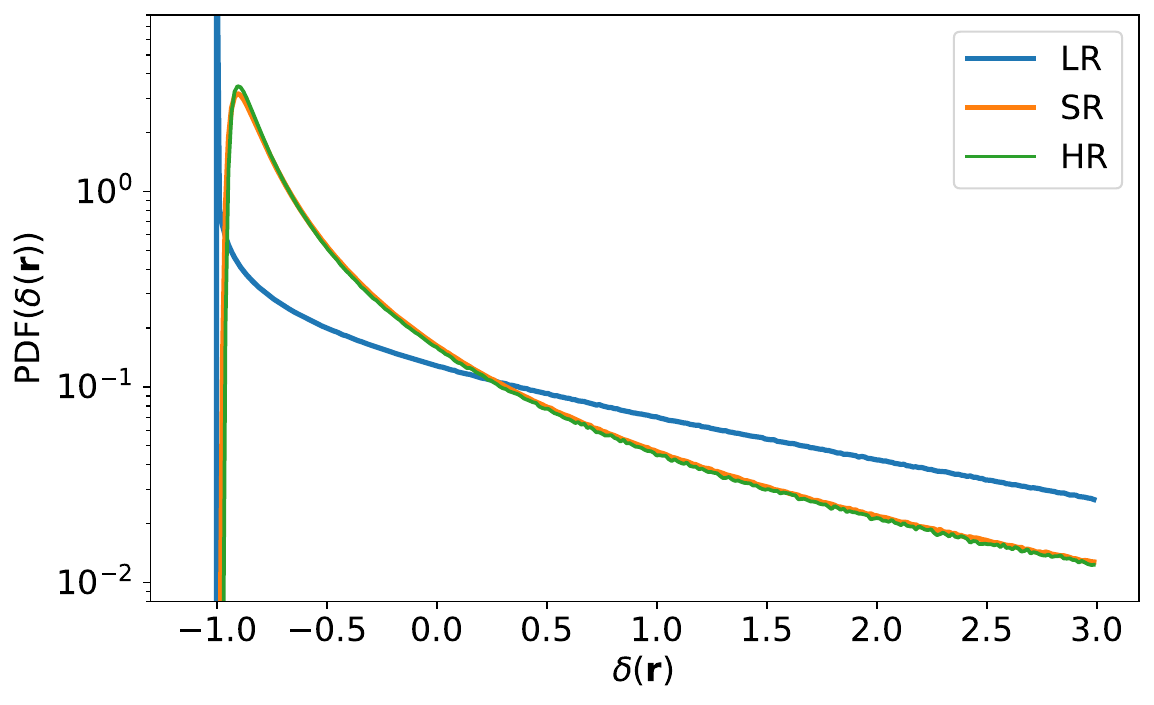} & \includegraphics[width=0.495\textwidth]{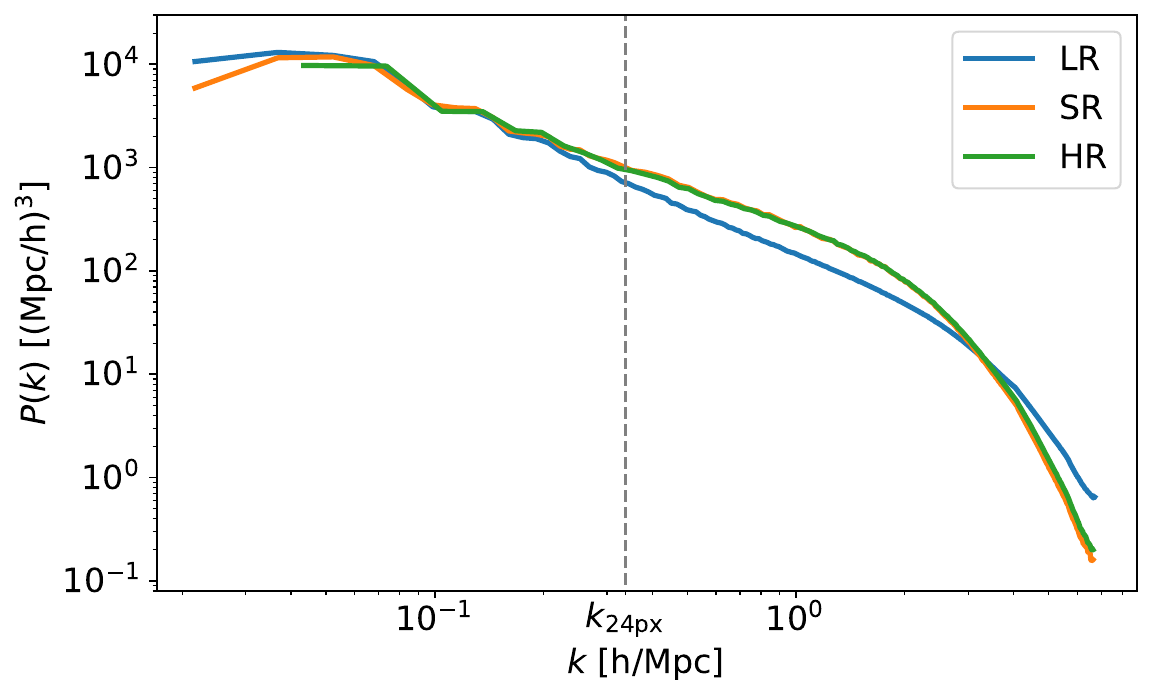}
    \end{tabular}

    \begin{tabular}{cc}
        \hspace{.5cm}Bispectrum $(k_1,k_2)=(0.15\ \frac{h}{\text{Mpc}},0.25\ \frac{h}{\text{Mpc}})$ & \hspace{.85cm}Bispectrum $(k_1,k_2)=(0.6\ \frac{h}{\text{Mpc}},1.0\ \frac{h}{\text{Mpc}})$\\
        \includegraphics[width=0.495\textwidth]{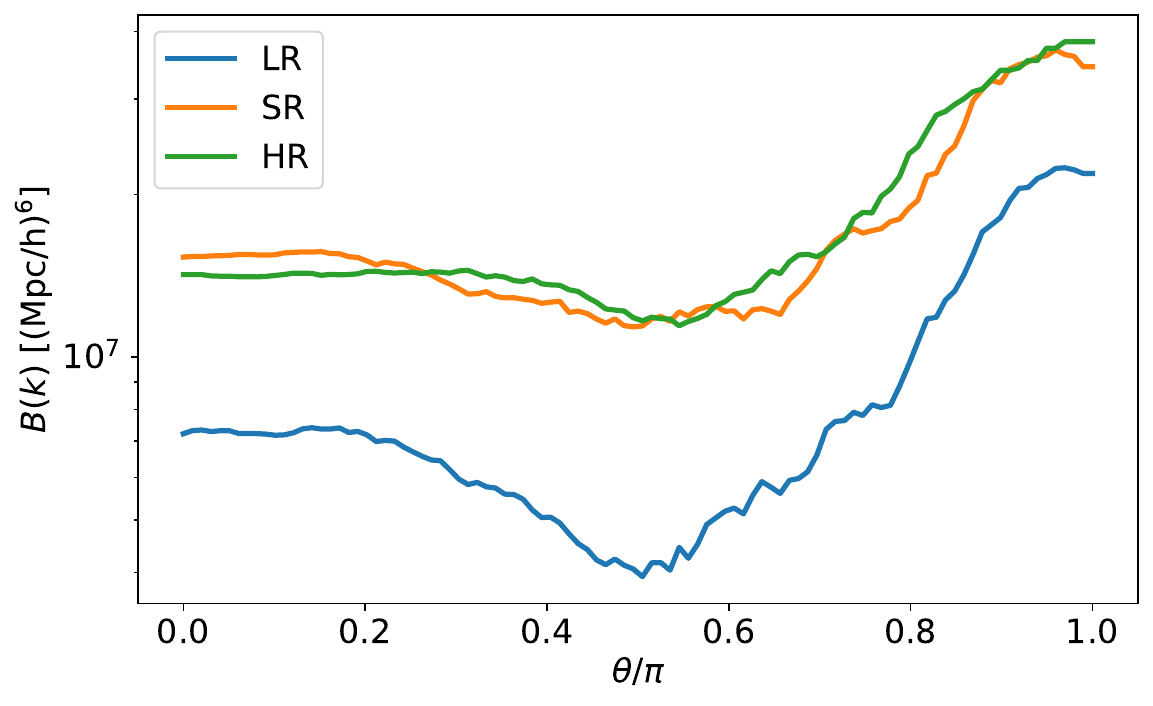} & \includegraphics[width=0.495\textwidth]{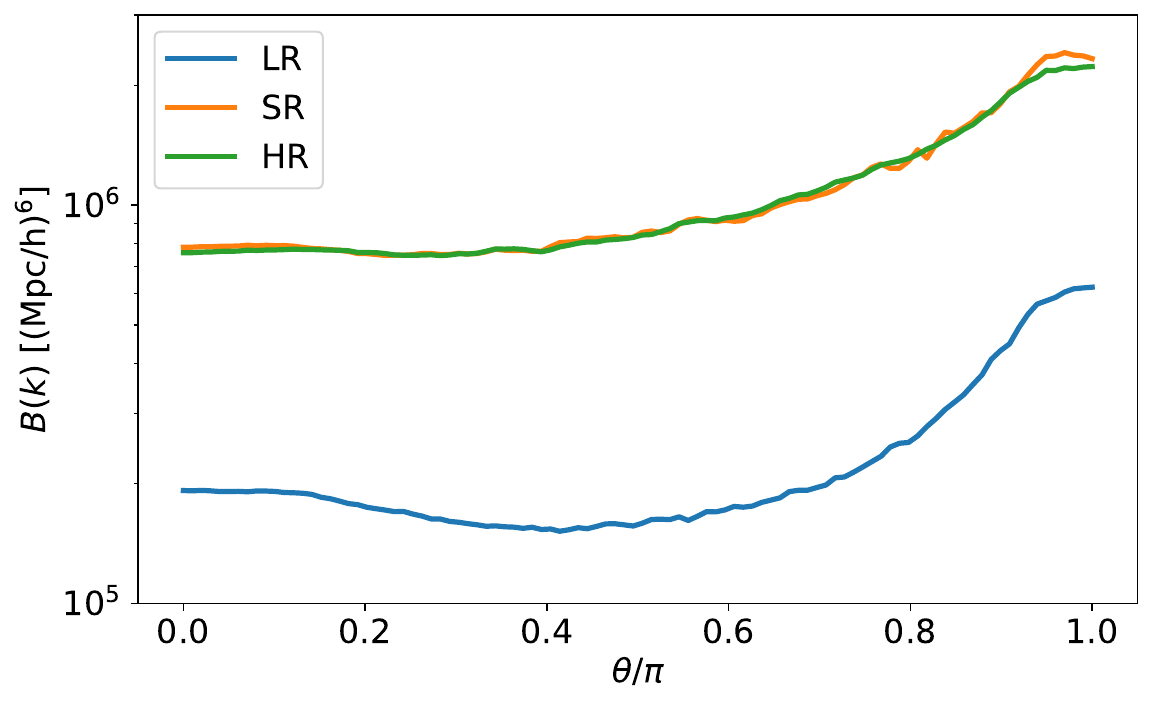}
    \end{tabular}

    \begin{tabular}{c}
        \hspace{1cm}Void size function\\
        \includegraphics[width=0.495\textwidth]{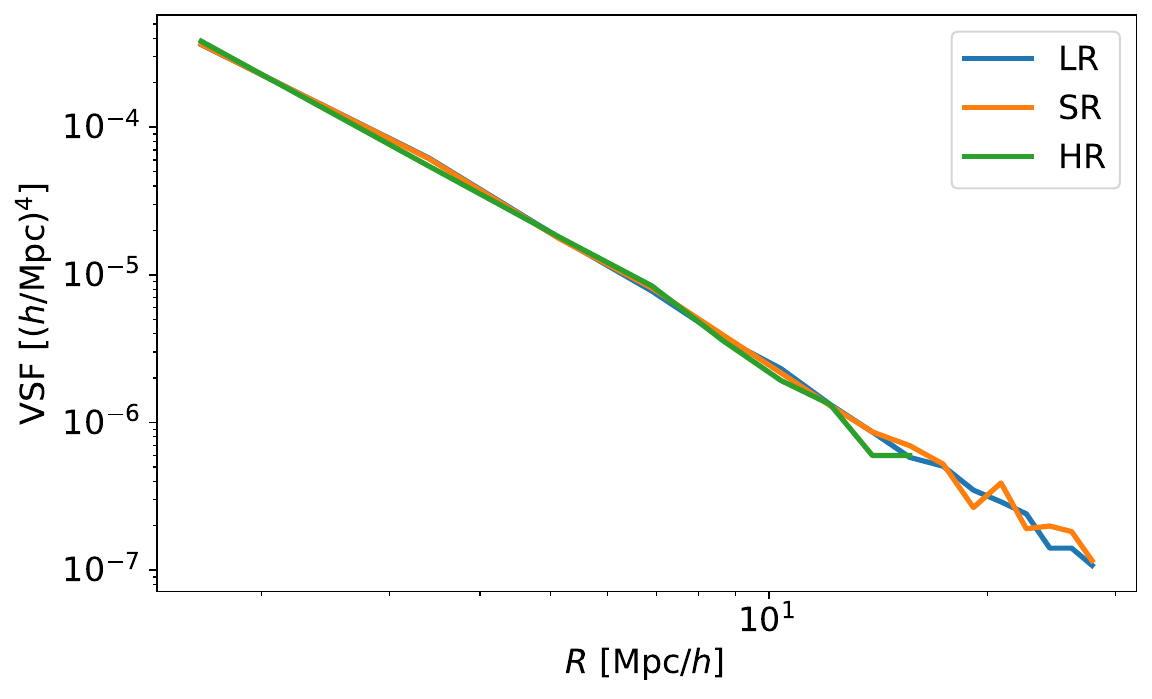}
    \end{tabular}
    
    \caption[Summary statistics comparing the $410\ \text{Mpc}/h$ length super-resolution volume to the $205\ \text{Mpc}/h$ length high-resolution training data]{Summary statistics comparing the $410\ \text{Mpc}/h$ length LR test field and SR model output, and the $205\ \text{Mpc}/h$ length HR training field as a truth comparison.}
    \label{fig:results_statistics}
\end{figure}

\subsubsection*{One-point probability distribution}

We plot the one-point probability density function (PDF) in Fig.~\ref{fig:results_statistics} (top left). The LR test data has many empty voxels, as it was created with $256^3$ particles in a $528^3$~px mesh. The diffusion model accurately generates SR data with the correct HR PDF, which is a smooth skew-Gaussian curve.

\subsubsection*{Power spectrum}


The power spectra comparing the LR, SR, and HR fields are plotted in Fig.~\ref{fig:results_statistics} (top right). As the LR and SR fields have twice the box length of the HR field, they have additional small $k$ Fourier modes around the $k\sim0.2h/\text{Mpc}$ turning point. We see that below the outpainting mode of $k_{24\text{px}}=0.34\ h/\text{Mpc}$, the SR power spectrum is guided by the LR power spectrum, properly emulating the baryon acoustic oscillations (BAOs) in the LR conditional. We further investigate BAOs in Appendix~\ref{sec:BAO_response}. Additionally, the SR field correctly begins to turn over for small $k$ along with the LR conditional. At high $k$, the SR simulation correctly follows the HR power spectrum, which is more suppressed for baryons than for dark matter, as physically expected. Our LR simulation dark matter power spectrum is not physically accurate at high $k$ due to its low particle resolution, but the diffusion model nevertheless learns the correct baryonic power spectrum.

A possibly surprising fact is that in the smallest $k$ power spectrum bin the SR power spectrum is somewhat suppressed with respect to the LR. In principle, the LR conditioner should dominate the power spectrum here. However, it appears that the mean density of the sequentially generated $24$~px length SR volumes can slightly drift, introducing a modulation on the very largest scales. We expect that this problem could be resolved with a multi-scale conditioner, which will be discussed in Section \ref{sec:conclusion}. A simpler solution would be to Fourier filter the SR map and take the largest scales directly from the LR simulation.

\subsubsection*{Bispectrum}

The power spectrum measures only the Gaussian structure, as discussed in Section~\ref{sec:nonlinear_lss} and Appendix~\ref{sec:wick}, so we must also calculate the next higher moment the understand the non-Gaussian structure in our fields. The bispectrum $B{(k_1,k_2,k_3)}$ is a three-point correlation function defined by
\begin{equation}
    (2\pi)^3B{(k_1, k_2, k_3)}\delta_\text{D}{(\bm{k}_1+\bm{k}_2+\bm{k}_3)}=\langle\delta{(\bm{k}_1)}\delta{(\bm{k}_2)}\delta{(\bm{k}_3)}\rangle.
\end{equation}
The bispectrum is a function of three variables, so for easier analysis, we show a lower dimensional  projection of the bispectrum. We pick the first two vectors $(k_1,k_2)$ to have a constant magnitude, and plot the bispectrum as a function of the angle $\theta$ between these two vectors.


\subsubsection*{Void size function}

Voids are an important feature of the highly non-Gaussian large-scale structure at late times. The SR field should retain the overall geometry of the LR conditioner, and so we expect the quantity and size of voids in the LR, SR, and HR fields to be similar. We calculate the void size function with \texttt{Pylians}~\cite{pylians}, which uses the spherical overdensity void finder presented in~\cite{Banerjee_2016}. The void finder smooths the field with a top-hat filter of radius $R$, and then finds underdense volumes of radius $R$ below a given threshold.

The void size function is plotted in Fig.~\ref{fig:results_statistics} (bottom) for a void threshold of $0.0$ (compare to the PDF in Fig.~\ref{fig:results_statistics}). The HR void size function loses resolution at a radius of about $R=15\ \text{Mpc}/h$. The SR and LR void size functions match up to $R=28\ \text{Mpc}/h$, indicating that the SR field respects the structure of the LR field for voids larger than can be accurately measured by the full TNG300 training volume.






\section{SR response to BAOs in the LR conditional}
\label{sec:BAO_response}

Our diffusion model's SR output follows the basic structure of the LR power spectrum at large length scales, even though the SR field is constructed with many smaller conditional outpaintings. We can demonstrate the response of our SR field to LR modes by examining BAOs. As we saw in Section~\ref{sec:results}, the SR output was able to accurately emulate the BAOs around $k\sim0.1\ h/\text{Mpc}$, lower than our outpainting scale of $k_{24\text{px}}=0.34\ h/\text{Mpc}$.

\begin{figure}
    \centering
    \textbf{BAOs emulated from LR to SR}

    \ 
    
    \rotatebox{90}{\small\hspace{-1.6cm}$410\ \text{Mpc}/h$, $528$~px}\begin{tabular}{ccc}
        LR test data & \hspace{0.5cm} & SR model output \\
        \includegraphics[width=0.4\textwidth]{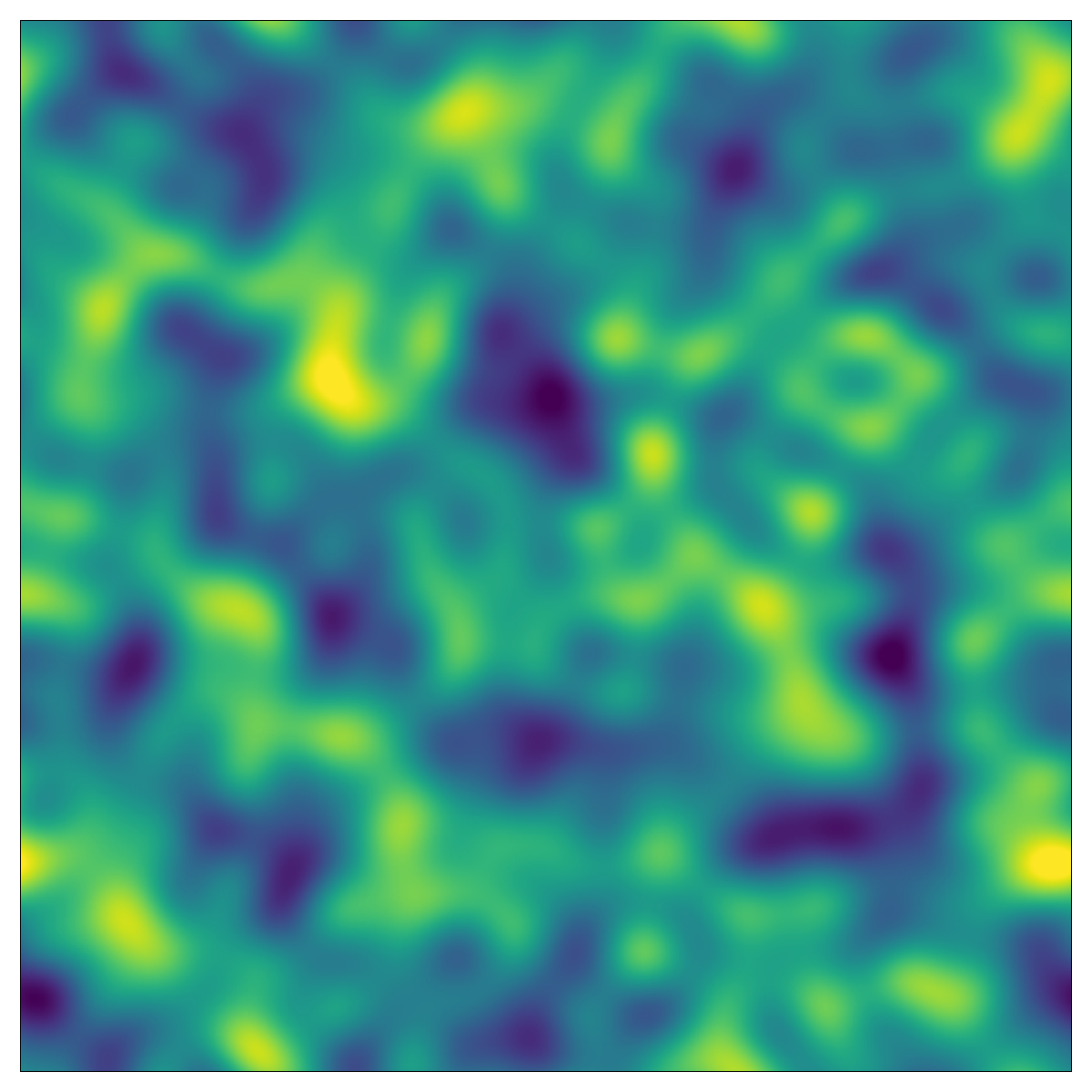} & \hspace{0.5cm} & \includegraphics[width=0.4\textwidth]{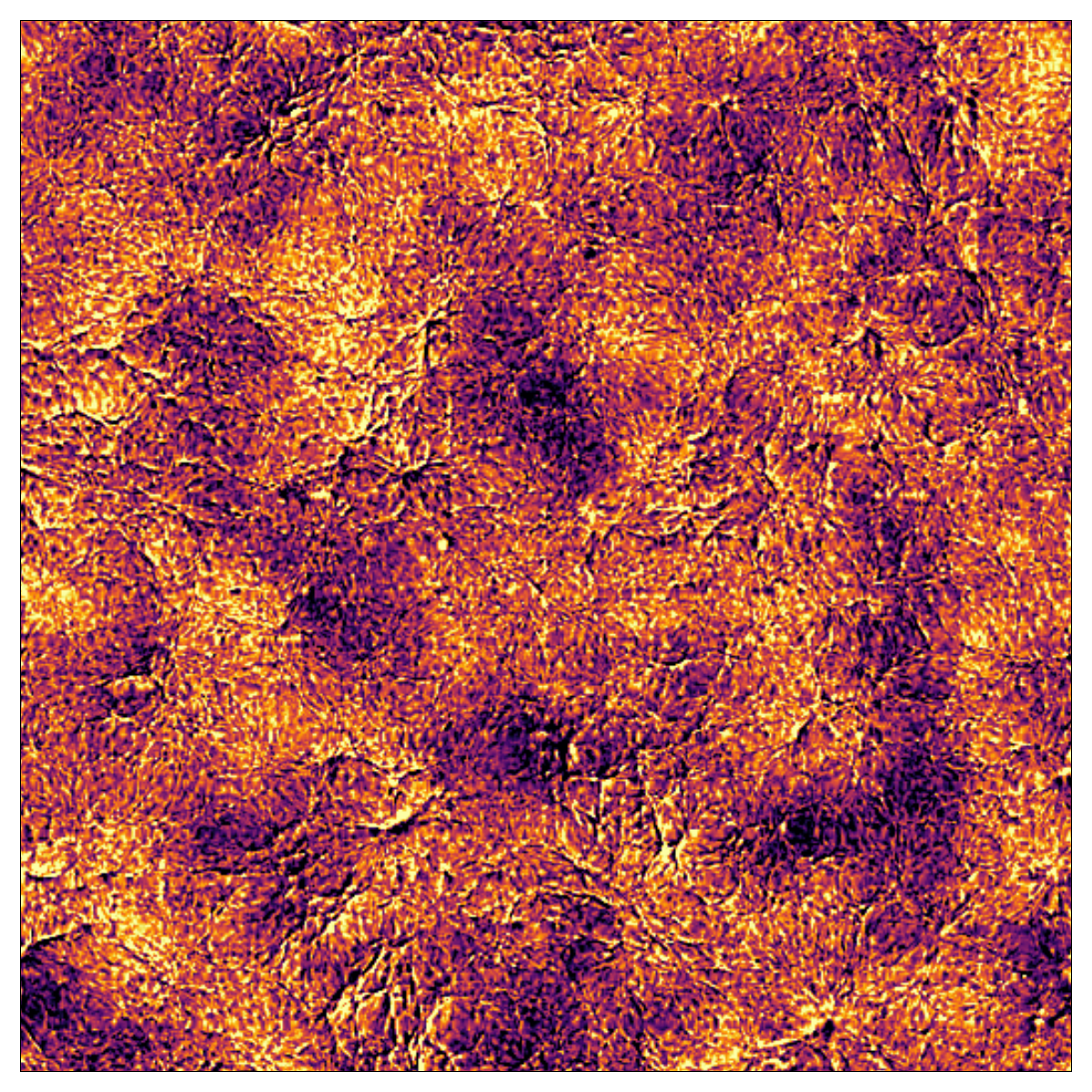}
    \end{tabular}
    \caption[Field-level visualization of BAOs emulated to super-resolution]{Field-level visualization of BAOs in the LR and SR fields, $19\ \text{Mpc}/h$ depth 2-dimensional projections. We constructed two $410\ \text{Mpc}/h$ length LR boxes with the same initial seed, one with BAOs and one without, and we generated their respective SR fields. Shown are the LR and SR BAOs in position space, computed as the difference between fields with and without BAOs. Large-scale fluctuations are emulated from the LR conditional to the SR model output, even though the diffusion model only trained on $37\ \text{Mpc}/h$ length boxes.}
    \label{fig:BAO_dif}
\end{figure}

\begin{figure}
    \centering
    \hspace{0.9cm}Power spectra with and without BAOs
    
    \includegraphics[width=0.495\textwidth]{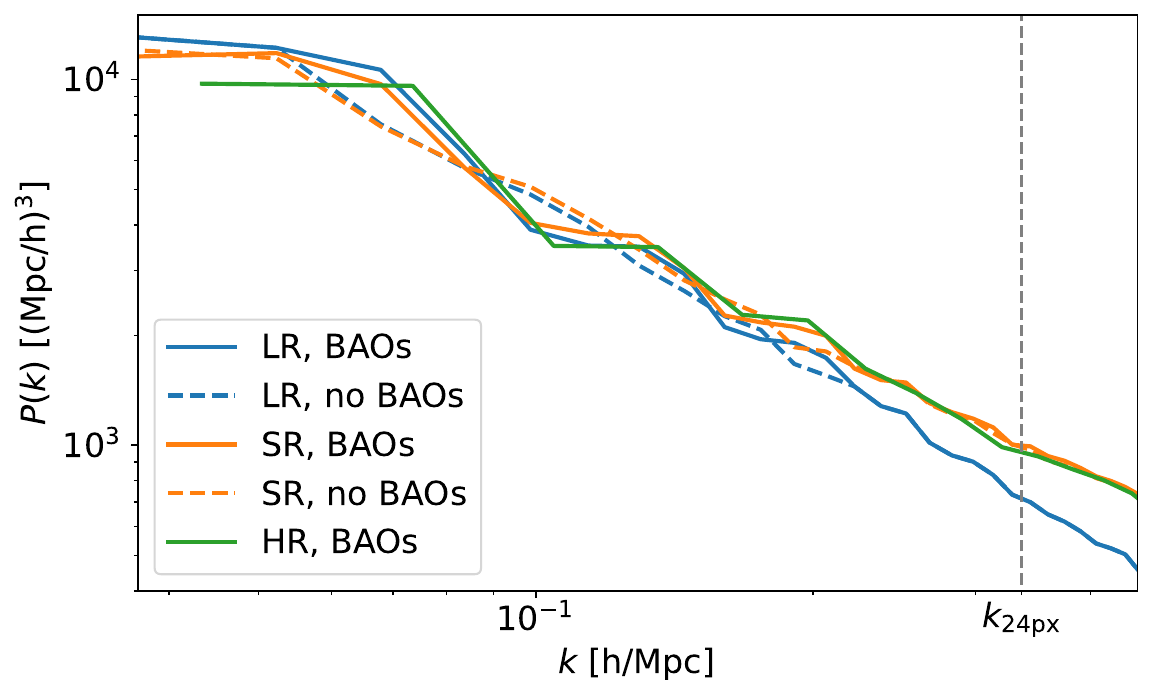}
    \caption[Power spectra for fields with and without BAOs in the low-resolution conditional]{Power spectra for fields with and without BAOs, plotted as solid and dashed lines, respectively. When the LR conditional contains BAOs, they are emulated to SR, matching the HR training data. Conversely, an LR conditional without BAOs generates an SR field without BAOs. The SR model output emulates structure in the LR conditional at Fourier modes smaller than the $k_{24\text{px}}$ outpainting scale.}
    \label{fig:BAO_dif_ps}
\end{figure}

We further illustrate this successful large-scale response of our model by comparing fields with and without BAOs. We constructed two $410\ \text{Mpc}/h$ length LR boxes with the same initial seed, one with BAOs and one without. We have then generated their respective SR fields, and the one containing BAOs was presented in Sec~\ref{sec:results}. We show field-level images of BAOs in Fig.~\ref{fig:BAO_dif}, computed as the difference between the BAO and non-BAO fields. We see that large-scale fluctuations contained in the LR conditional field are emulated to the SR model output. (The SR fields additionally have high-frequency phase differences irrelevant to the BAOs, explained in Section~\ref{sec:results_variety}.) In Fig.~\ref{fig:BAO_dif_ps}, we plot power spectra of our LR and SR fields comparing the existence of BAOs. The SR power spectra follow the same pattern as the LR power spectra below $k_{24\text{px}}$, whether there are BAOs or not. Large length structures are not learned from the HR field with our outpainting method, and must accurately represented in the LR field in order to be generated in the SR field.

\section[Model output variety from a single low-resolution field]{Model output variety from a single\\low-resolution field}
\label{sec:results_variety}

Due to the stochastic nature of denoising diffusion models, our SR emulator is able to sample from the manifold of SR solutions. In this section, we measure the variety of SR samples created from a single LR conditional field. We generate $25$ SR fields of size $144^3$~px (physical length $112\ \text{Mpc}/h$), each requiring $5^3$ outpaintings, conditional on the same LR field. We use the same $1250$ step noise schedule as in the previous section.

\begin{figure}[h]
    \centering
    
    \rotatebox{90}{\small\hspace{-1.7cm}$112\ \text{Mpc}/h$, $144$~px}\begin{tabular}{ccccc}
        LR & SR 1 & SR 2 & SR 3 & SR 4\\
        \includegraphics[width=0.1955\textwidth]{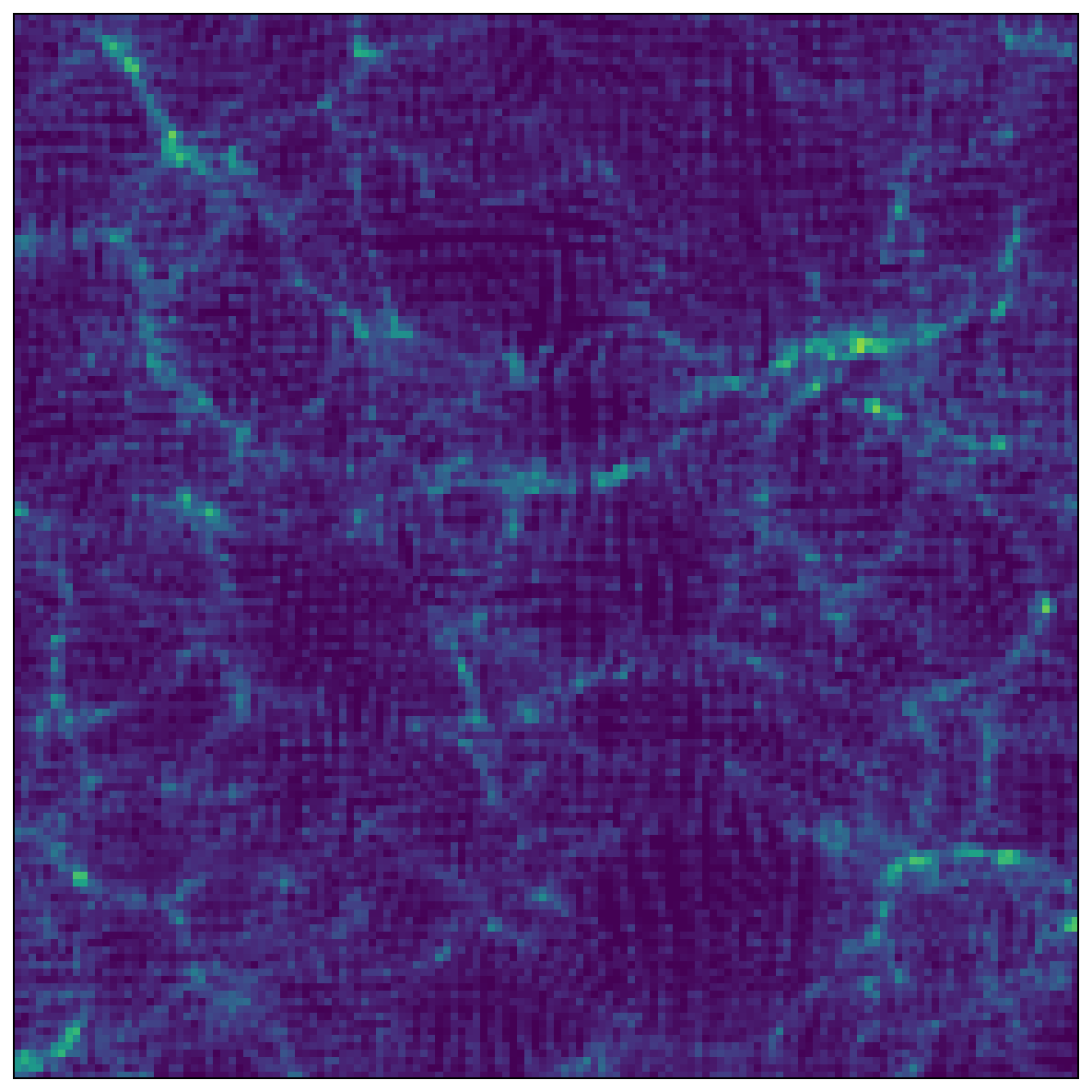} & 
        \includegraphics[width=0.1955\textwidth]{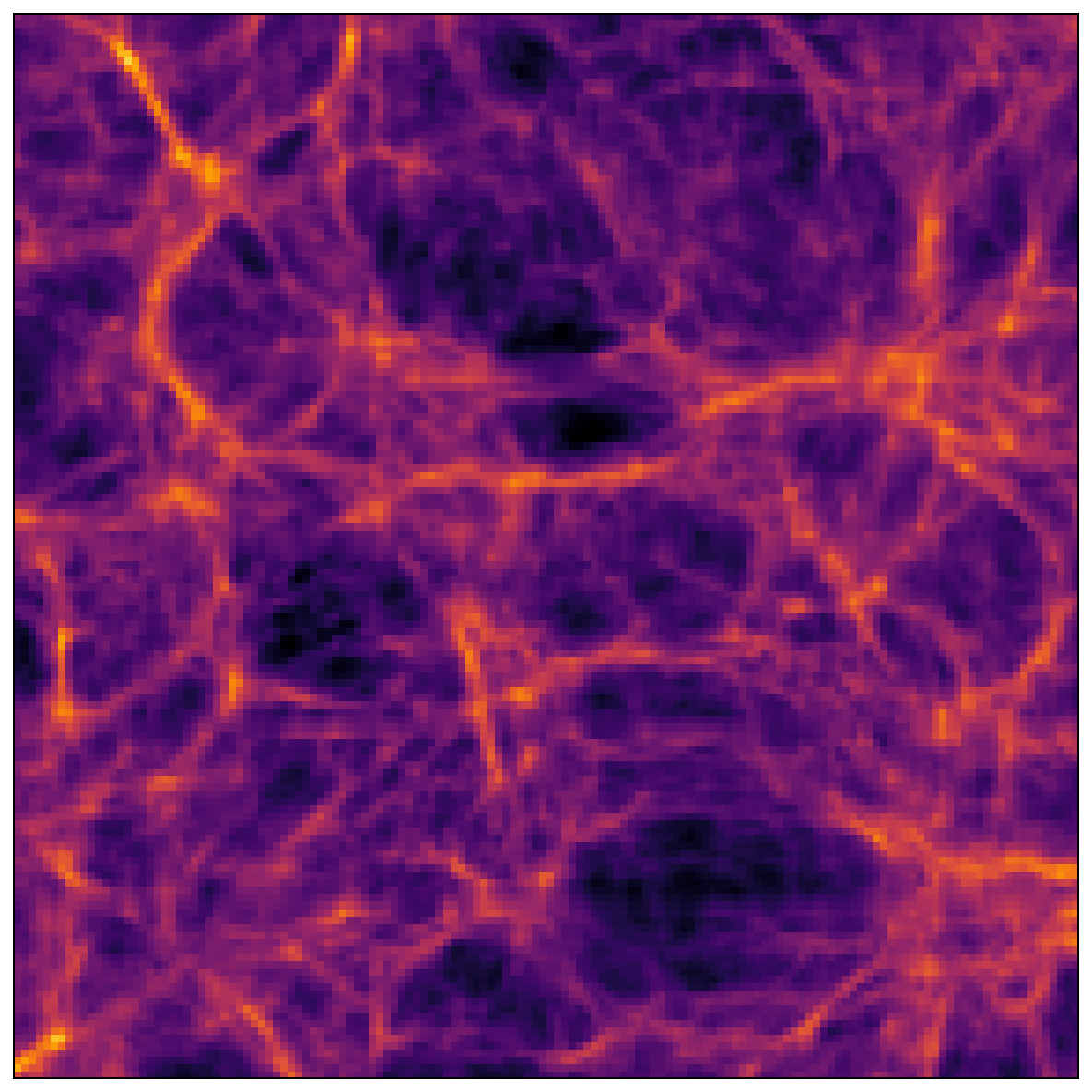} & 
        \includegraphics[width=0.1955\textwidth]{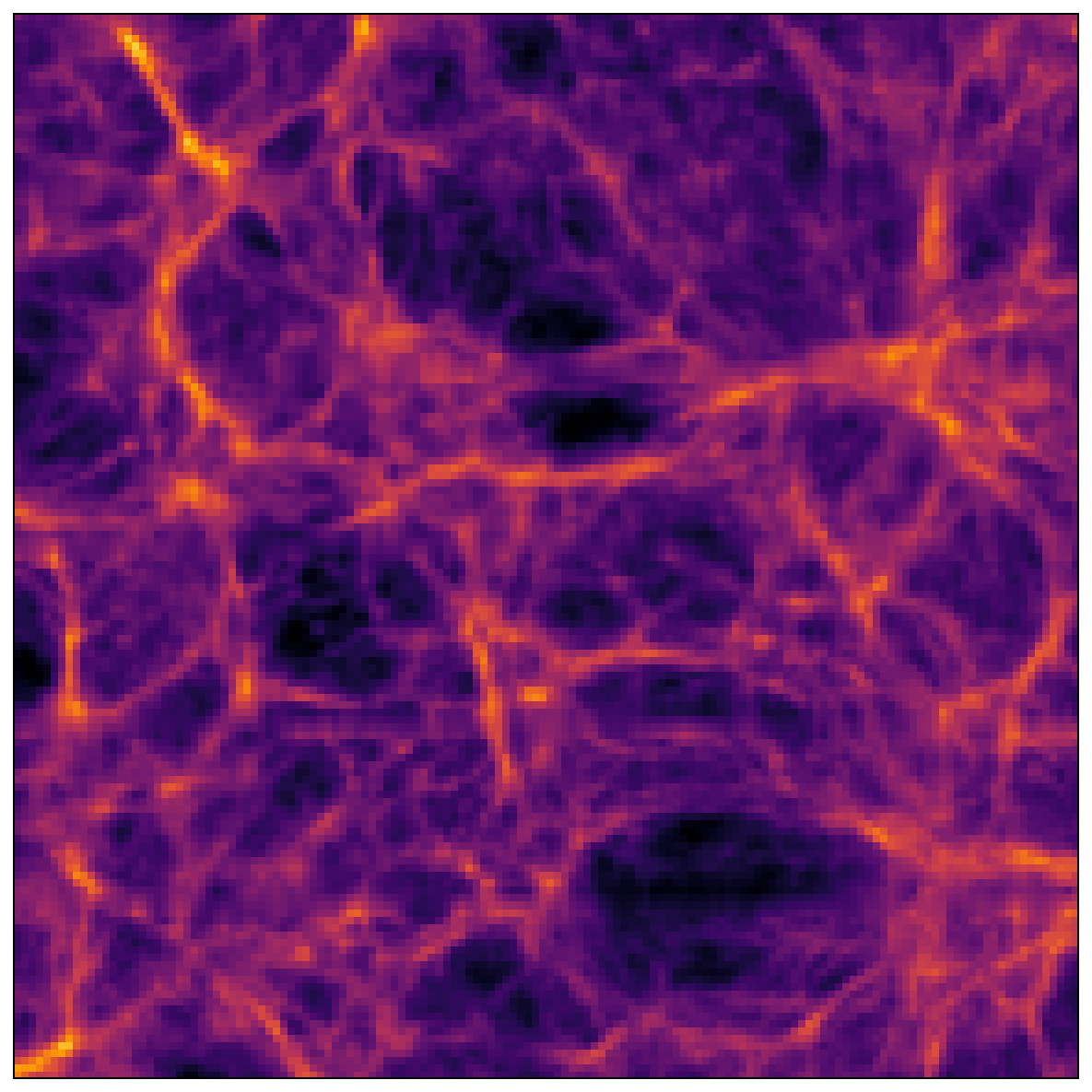} & 
        \includegraphics[width=0.1955\textwidth]{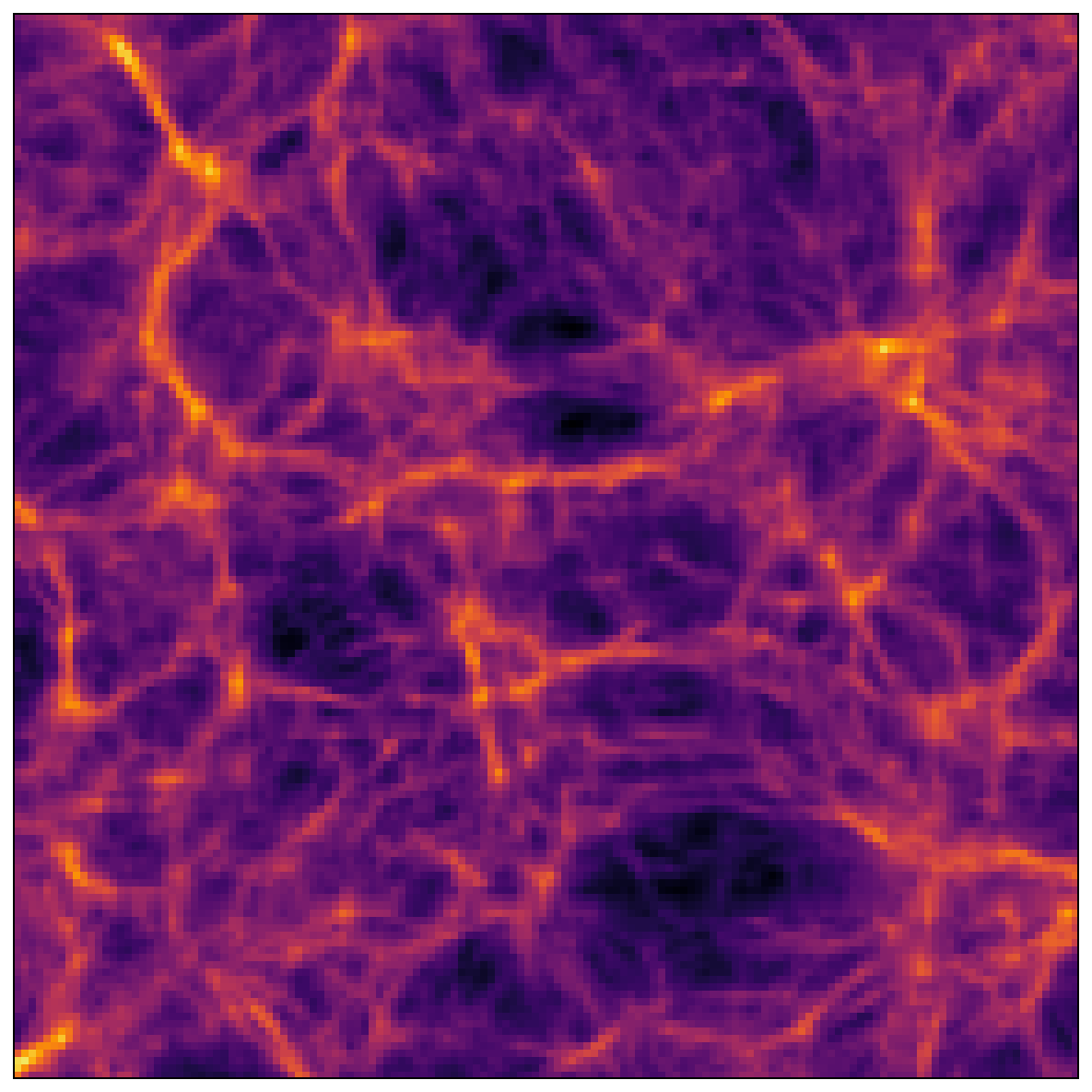} & 
        \includegraphics[width=0.1955\textwidth]{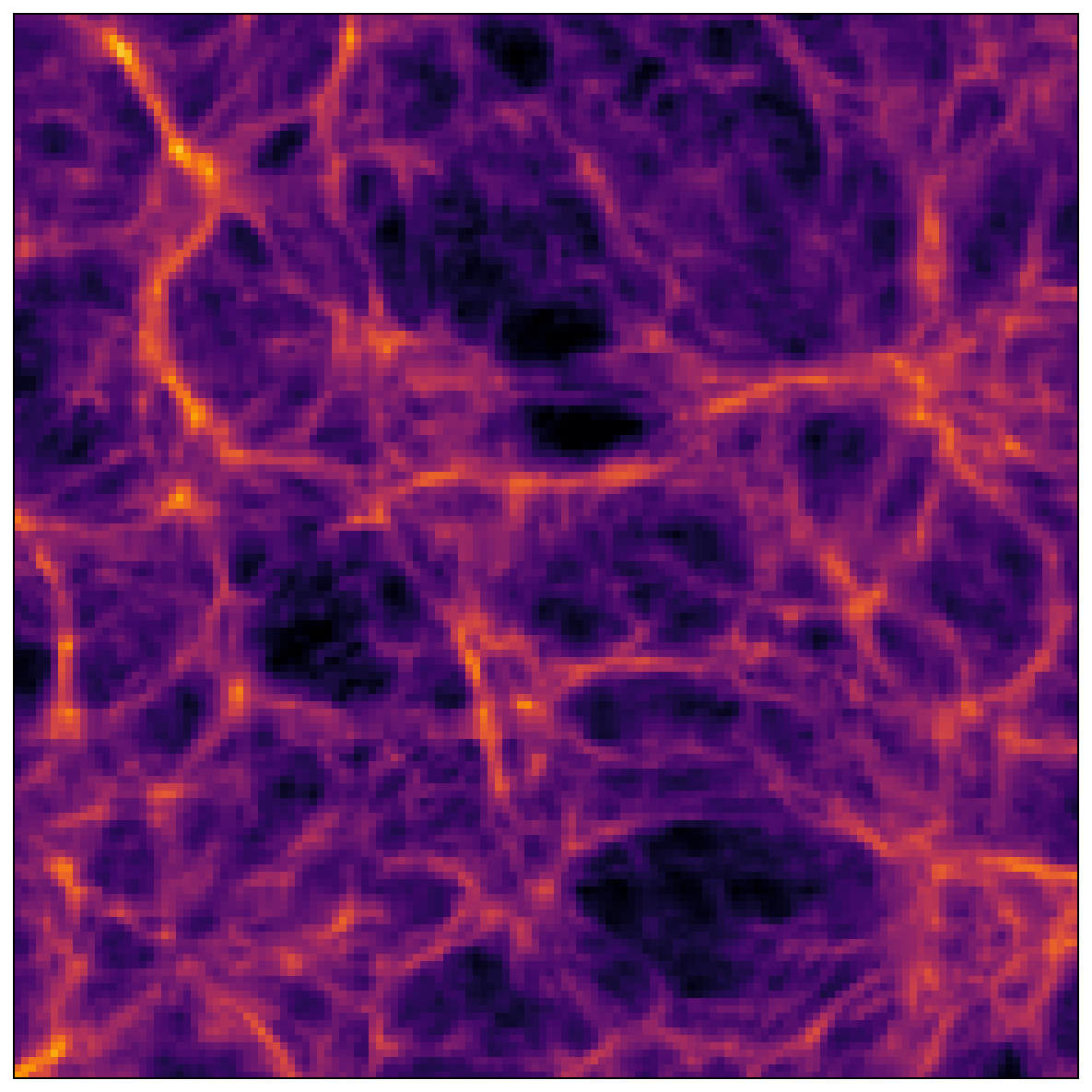}
    \end{tabular}
    
    \ 
    
    \ 
    
    \begin{tabular}{cc}
        SR 1 and SR 2 & SR 3 and SR 4\\
        \includegraphics[width=0.498\textwidth, trim={0 0 0 2.5cm}, clip]{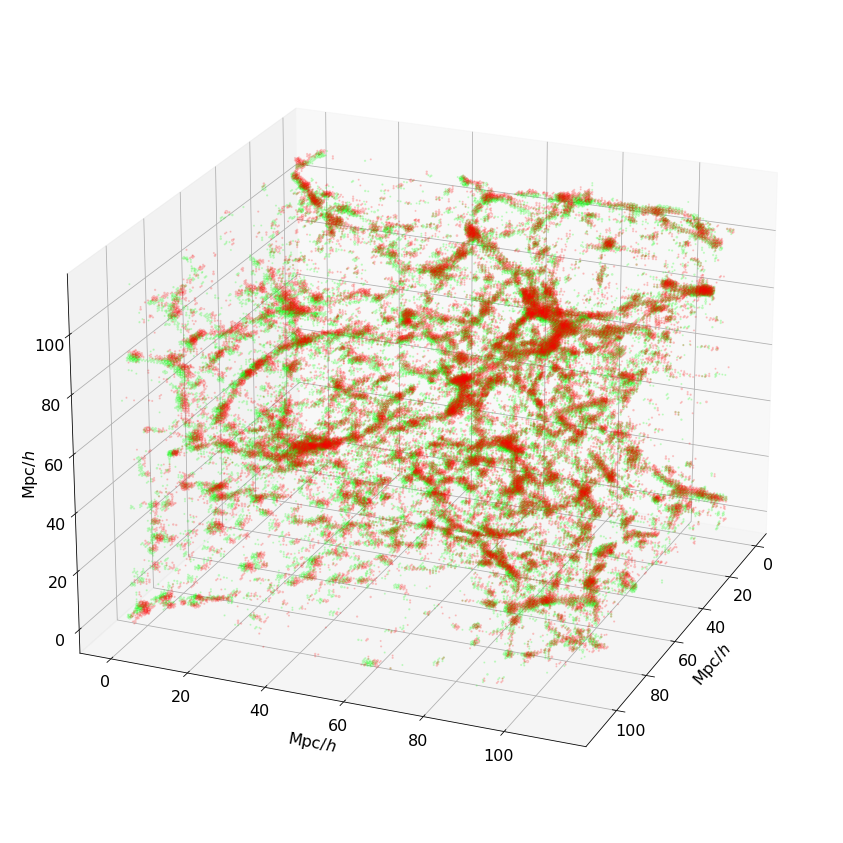} & 
        \includegraphics[width=0.498\textwidth, trim={0 0 0 2.5cm}, clip]{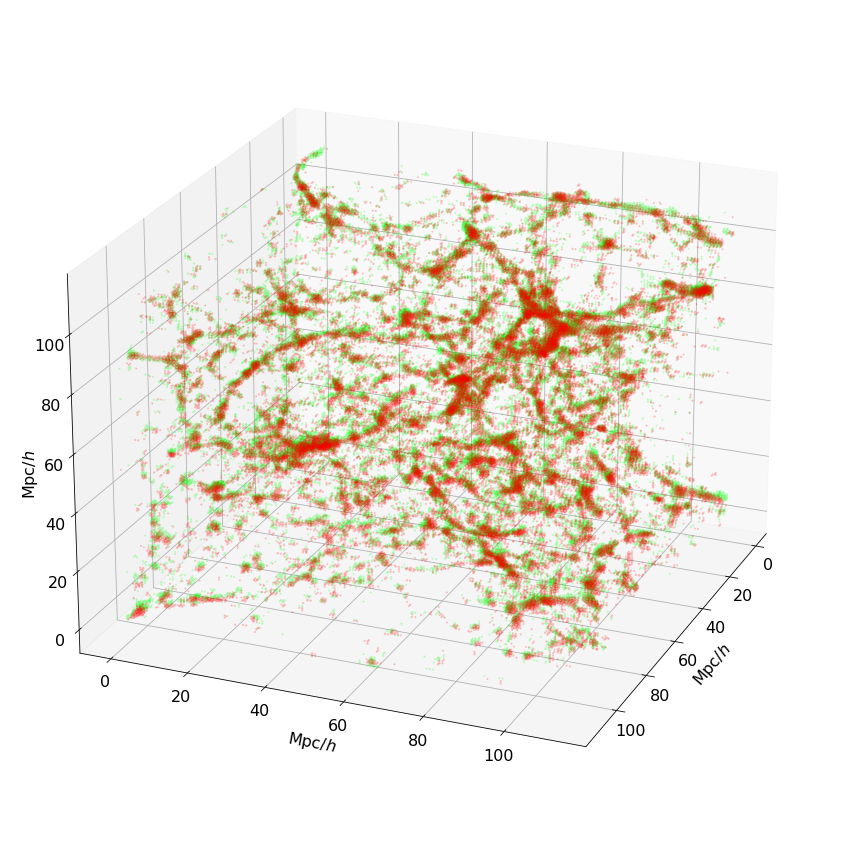}
    \end{tabular}
    
    \caption[Variety of super-resolution fields from a single low-resolution conditional field]{Visual results for a variety of SR fields from a single LR conditional field. (Top row) LR and multiple SR realizations, 2-dimensional projects. (Bottom row) High density features of the above SR fields in 3 dimensions. To better visually discriminate differences between fields, we plot two different SR fields in the same volume.}
    \label{fig:results_visual_variety}
\end{figure}

We show visual results in Fig.~\ref{fig:results_visual_variety} for the LR volume, and four samples of the SR results. It is difficult to discern differences between the SR samples side-by-side, and so we show also two SR fields in the same 3-dimensional plot. These point plots are constructed with the large density features of two SR fields, one field in red and the other in green. When plotted on top of each other, it becomes more apparent that each SR field has a different phase in their placement and shape of halos and filaments.

\begin{figure}
    \centering

    \begin{tabular}{cc}
        \hspace{1cm}Probability density & \hspace{1cm}Power spectrum\\
        \includegraphics[width=0.495\textwidth]{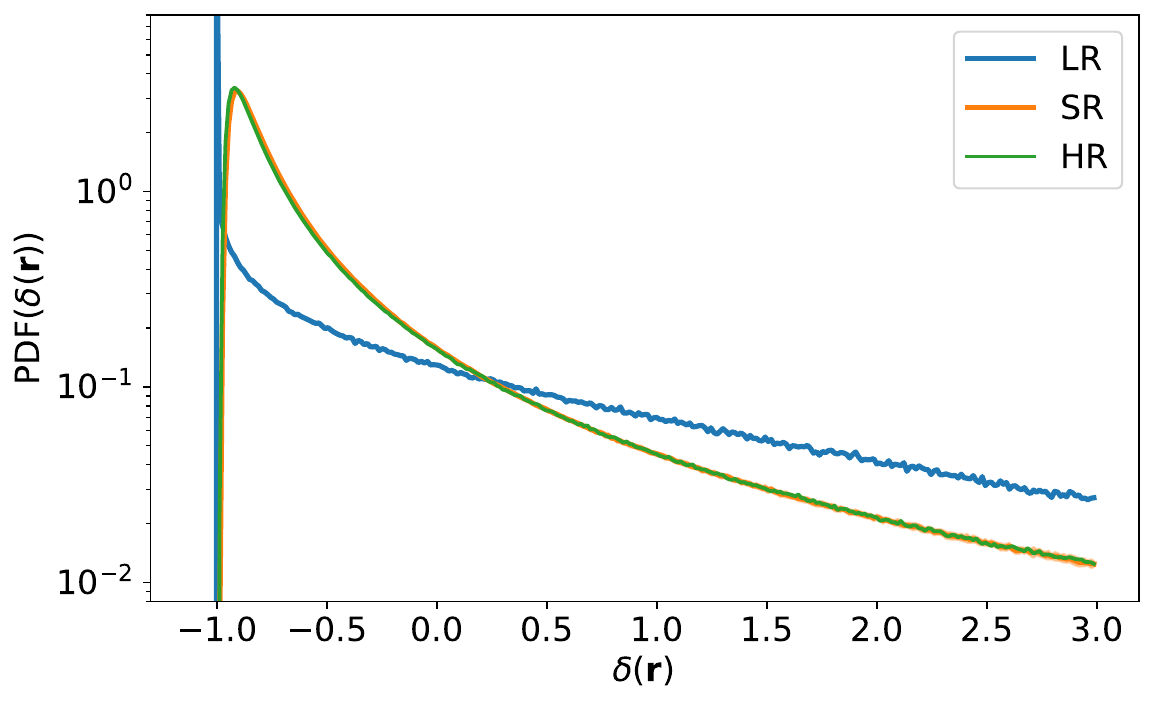} & \includegraphics[width=0.495\textwidth]{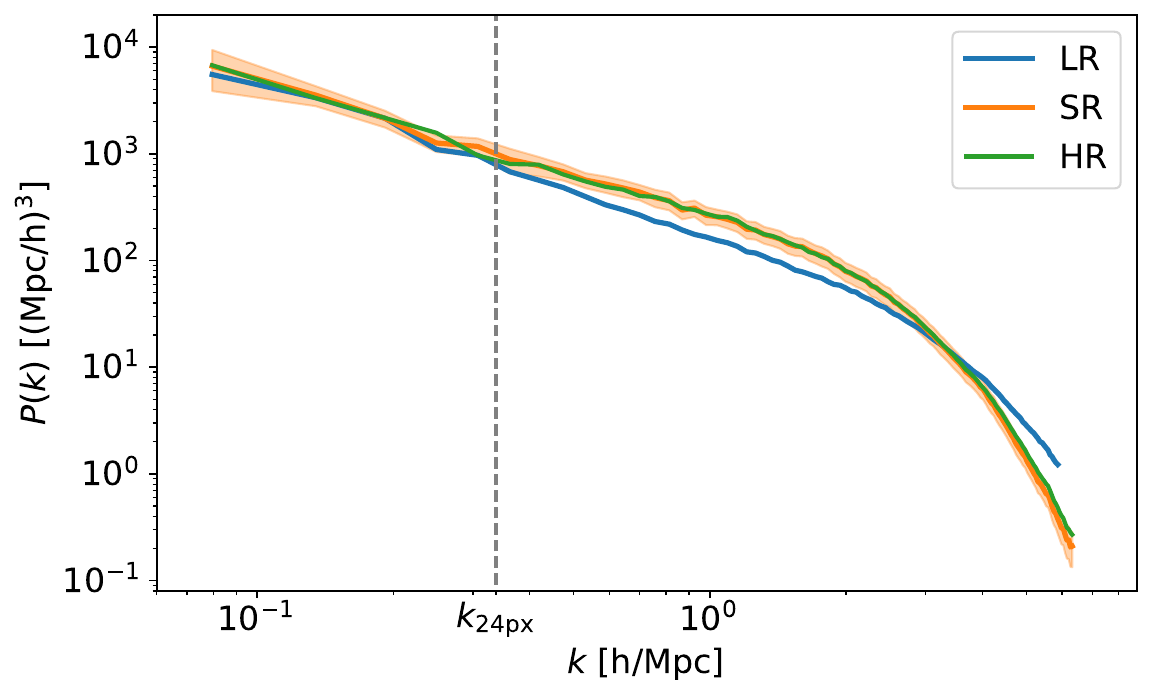}
    \end{tabular}

    \begin{tabular}{cc}
        \hspace{.5cm}Bispectrum $(k_1,k_2)=(0.6\ \frac{h}{\text{Mpc}},1.0\ \frac{h}{\text{Mpc}})$ & \hspace{.68cm}Cross-correlation coefficient between SR samples\\
        \includegraphics[width=0.495\textwidth]{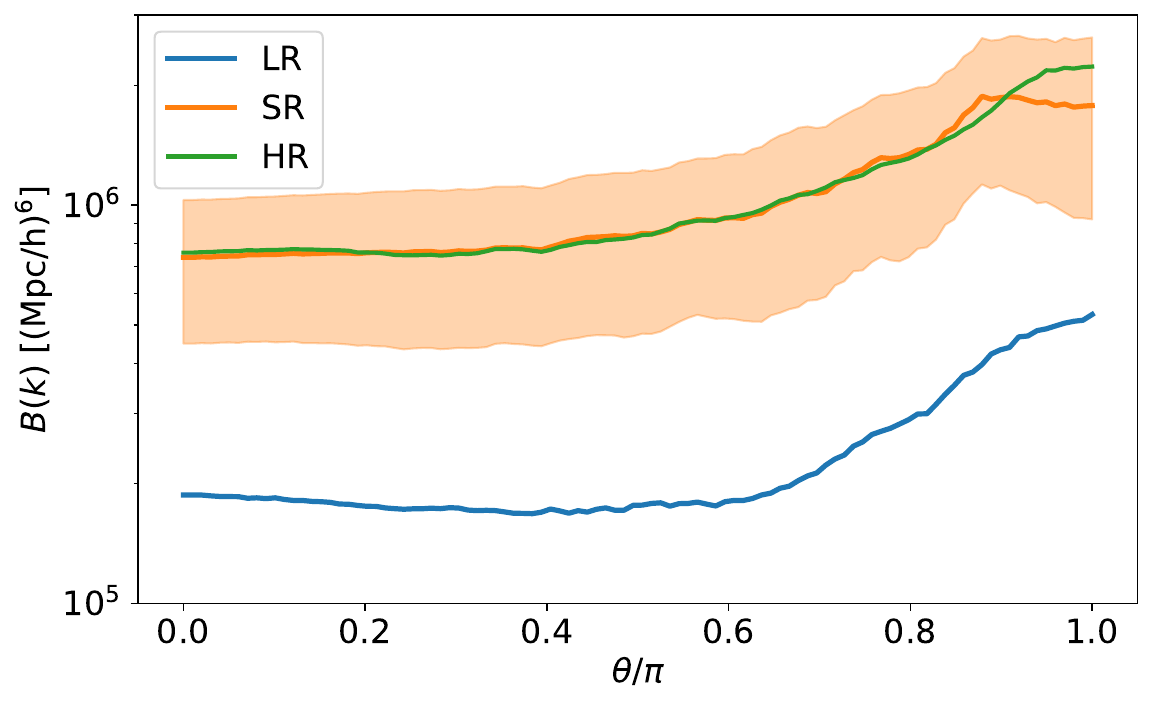} & 
        \includegraphics[width=0.495\textwidth]{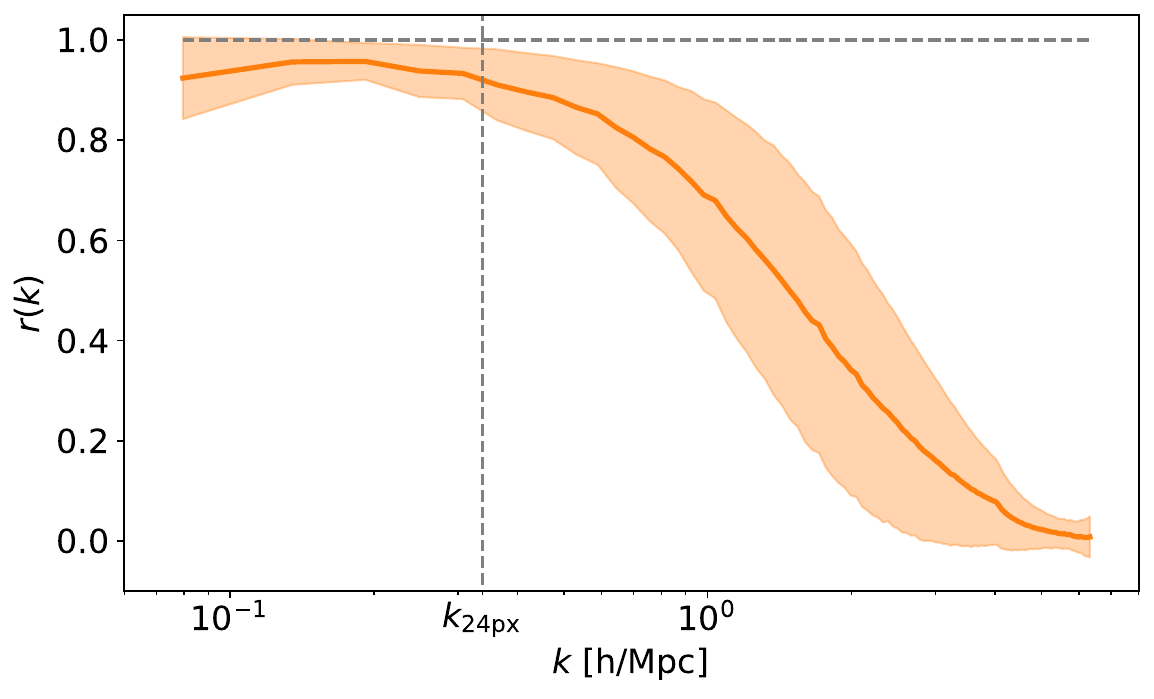}
    \end{tabular}
    
    \caption[Summary statistics for a variety of super-resolution fields]{PDF (top left), power spectrum (top right), and bispectrum (bottom left) for a variety of $25$ SR fields generated from a single LR conditional field, with $1\sigma$ bands in the SR fields. The cross-correlation coefficient of Fourier modes (bottom right) is calculated between all pairs of the 25 SR fields.}
    \label{fig:results_statistics_variety}
\end{figure}

Summary statistics for the LR and $25$ SR samples with $1\sigma$ variation between SR samples are shown in Fig.~\ref{fig:results_statistics_variety}, with the HR training data as a truth comparison. The results are highly accurate between SR and HR for the PDF (top left), power spectrum (top right), and bispectrum with $(k_1,k_2)=(0.6\ h/\text{Mpc},1.0\ h/\text{Mpc})$ (bottom left). Both the power spectrum and the bispectrum in the small SR volumes show a sample variance due to the finite mode number and the freedom of the diffusion model to modify nonlinear modes.

We further quantify the differences in the SR samples with their cross-correlations in Fourier space . For two fields $\delta_i(\bm{k})$ and $\delta_j(\bm{k})$, their cross-correlation coefficient is
\begin{equation}
    r_{ij}(k)=\frac{P_{ij}(k)}{\sqrt{P_i(k)P_j(k)}}.
\end{equation}
Here $P_{ij}(k)$ is the cross-power spectrum
\begin{equation}
    (2\pi)^3P_{ij}(k)\delta_\text{D}(\bm{k}+\bm{k}')=\langle\delta_i(\bm{k})\delta_j(\bm{k}')\rangle.
\end{equation}
We compute $r_{ij}(k)$ for all $300$ different $(i,j)$ pairs for our $25$ SR fields, and plot the result in Fig.~\ref{fig:results_statistics_variety} (bottom right). The cross-correlation coefficient is close to $1$ below the $k_{24\text{px}}$ outpainting scale as expected, as small $k$ modes are guided by the single LR field. At nonlinear scales, the cross-correlation coefficient drops to nearly $0$, signifying little correlation between SR samples. Additionally, there is almost no variance in the largest $k$-bins where the $r_{ij}(k)$ is nearly $0$, and therefore very few SR samples are correlated at the smallest length scales.

\section{Conclusion}
\label{sec:conclusion}

Super-resolution emulators are a promising tool to open the computational bottleneck of HR baryonic simulations in cosmology. In this work we evaluated the performance of a diffusion model on volumetric data, and developed a conditional outpainting scheme that can upgrade large LR volumes. We trained on data from the TNG300 simulation, and generated an SR field with 8 times the volume of the entire training data volume. Our diffusion model is capable of making accurate 3-dimensional SR emulation results, with the resulting SR volume matching the summary statistics of the training HR simulation closely. Our SR field is guided by the LR field at length scales larger than the outpainting scale. We also demonstrated the stochasticity of the diffusion model by generating a variety of SR fields conditional on a single LR field.

Leading up to this work, we found that training probabilistic generative models in 3 dimensions is not always successful, as we intended to train normalizing flows such as Real NVP~\cite{2016arXiv160508803D} and Glow~\cite{2018arXiv180703039K} on the same task. Normalizing flows have been used to learn the PDF of 2-dimensional projections of cosmological fields~\cite{rouhiainen2021,rouhiainen2022}, as was presented in Chapter~\ref{ch:nf_random}, and our 2-dimensional super-resolution results were promising. However, the flows were not accurate in 3 spacial dimensions and we thus moved to the more expressive diffusion models. On the other hand, diffusion models are far slower at inference than normalizing flows, and this makes it challenging to generate very large volumes or many volumes. Our $410\ \text{Mpc}/h$ SR simulation took about 120 hours to generate with a single A100 GPU, and generation time scales linearly with the volume.

Further research with 3-dimensional diffusion models would be greatly aided with a faster denoising algorithm. While this work uses a conditional denoising diffusion probabilistic model (DDPM), recent developments in diffusion research include the significantly faster denoising diffusion implicit model (DDIM)~\cite{song2022denoising} and denoising probabilistic models solver (DPM-Solver)~\cite{lu2022dpmsolver}. The DDIM algorithm and DPM-Solver advertise a factor of $10$ to $100$ reduction in the number of denoising steps with a trade-off of a small loss in accuracy. We tested an implementation of DDIM in preparing this work, but our results were subpar compared to DDPM. Nevertheless, it is likely that with more work the sample generation time can be reduced by a significant factor.

It would be interesting to study different LR conditioners. Our LR conditioner is only $48$~px in field of view, which could be increased, perhaps using a multi-scale conditioner that includes up to the entire simulation. It is plausible that in this way the bispectrum at intermediate scales (see Fig. \ref{sec:results}) could be modelled even more precisely. This would also improve the modelling of SR squeezed limit N-point functions (beyond the response of the LR simulation). One may also wonder about the influence of the order of sample generation in the 3-dimensional volume. Our sample generation process breaks spatial homogeneity in principle, since we need to start generating fields somewhere in space. On the other hand, small-scale structure at some point in space is independent of small-scale structure far away from it, due to the locality of structure formation.

A physical application for our simulations is making simulations for CMB~$\times$~LSS cross-correlation analyses~\cite{kvasiuk2023autodifferentiable}. An interesting case here is kSZ tomography~\cite{Deutsch:2017ybc,smith2018ksz}, which is sensitive to the cross-correlation of the galaxy density and the electron density $P_{ge}(k)$. As shown in~\cite{smith2018ksz}, with SO and DESI, $P_{ge}(k)$ can be probed up to about $k\sim8\ \text{Mpc}^{-1}$. This required resolution is near the Nyquist frequency of the pixelization used in this work. It would thus be important to scale up the resolution of our data for future studies, which TNG300 allows owing to its large $2500^3$ particle count.

Cross-correlation analyses in general require several correlated fields, and the kSZ tomography use case requires a two field output of electron density and galaxy density. Such fields are available in IllustrisTNG, and multiple fields could be generated by our diffusion model. The most apparent way to include multiple fields would be to add additional input and output channels to the diffusion model's U-net. While technically straightforward, this approach becomes computationally challenging, especially in 3 dimensions. Another possibility is to include a second machine learning model on top of the diffusion model output. For example, invertible mappings were recently developed~\cite{andrianomena2023invertible} between different cosmological fields in the CAMELS simulations~\cite{camels2021}. 

Our method could be applied to particle level super-resolution emulation with point cloud probabilistic diffusion models~\cite{luo2021diffusion}. This would be closer to the super-resolution emulator presented in \cite{li2021_1, li2021_2, li2023}. Point cloud diffusion models have been recently used to generate QCD jets with high precision~\cite{Mikuni:2023dvk,Leigh:2023toe}. For some analyses, a hybrid generator that can both make a continuous matter field and a point-cloud galaxy field would be useful.

Finally, it would be useful to be able to vary cosmological and astrophysical parameters. In Appendix~\ref{sec:variable_cosmos} we explored transferring new cosmologies with different cosmological parameters from the LR conditional field to the SR field. Unfortunately, our diffusion model did not behave well with out-of-distribution LR fields. A more straightforward implementation of varied cosmological parameters requires multiple training simulations with different values of such parameters. For the IllustrisTNG simulations, this is not currently available, but the CAMELS project and its planned extensions can be useful. For our approach, we require simulations that both include large linear and small nonlinear scales. In principle, such parameter dependence can be included in the diffusion model with techniques such as style transfer and Low-Rank Adaptation~\cite{hu2022lora}, the latter of which has been successful at concept tuning and concept fusion with text-to-image diffusion models~\cite{gu2023mixofshow}. With larger but still realistic GPU resources, it would be possible to generate a few large volume simulations with different astrophysical parameters for the same cosmology.


\section[Varying cosmologies in the LR conditional]{Appendix:\\Varying cosmologies in the LR conditional}
\label{sec:variable_cosmos}

In this Appendix, we generate new SR simulations with varied cosmological parameters in the LR conditional field as to examine the model's robustness to out-of-distribution conditional data. We vary each of $\sigma_8$ and $\Omega_\text{M}$ by $\pm5\%$ in the LR conditional from their fiducial values of $\sigma_8=0.8159$, $\Omega_\text{M}=0.3089$, giving four experiments to test the sensitivity to out-of-distribution data. For our four varied cosmologies along with a fiducial cosmology, we simulate $205\ \text{Mpc}/h$ length boxes of $128^3$ dark matter particles to build the LR conditional, each with the same initial seed. 

For each varied cosmology, we independently generate 8 SR fields, each of length $112\ \text{Mpc}/h$ ($144$~px) consisting of $5^3$ outpainting iterations. The 8 fields are generated from different volumes of the $205\ \text{Mpc}/h$ conditional. By generating several SR fields, rather than a single large field, we get an estimate of the variance of our diffusion model output.

\begin{figure}
    \centering
    \begin{tabular}{cc}
        \hspace{.6cm}$\sigma_8^-=0.7751$ & \hspace{.6cm}$\sigma_8^+=0.8567$\\
        \includegraphics[width=0.498\textwidth]{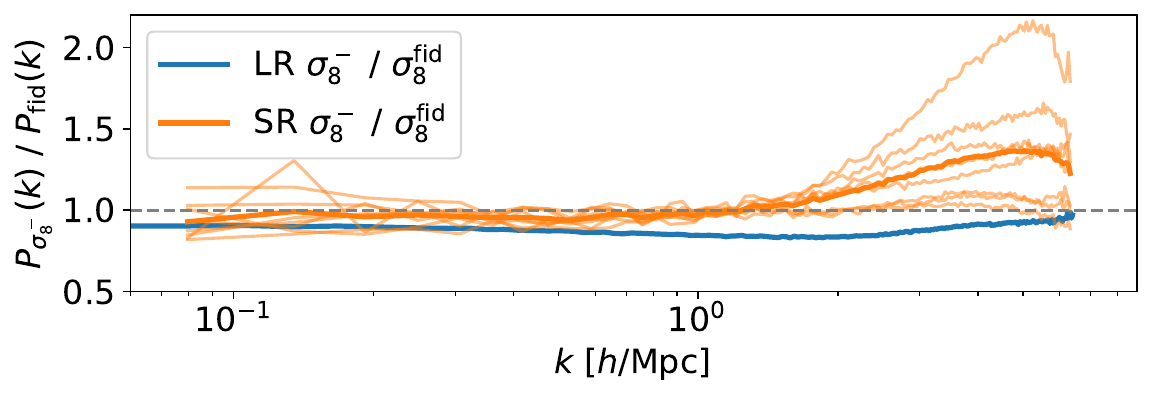} & \includegraphics[width=0.498\textwidth]{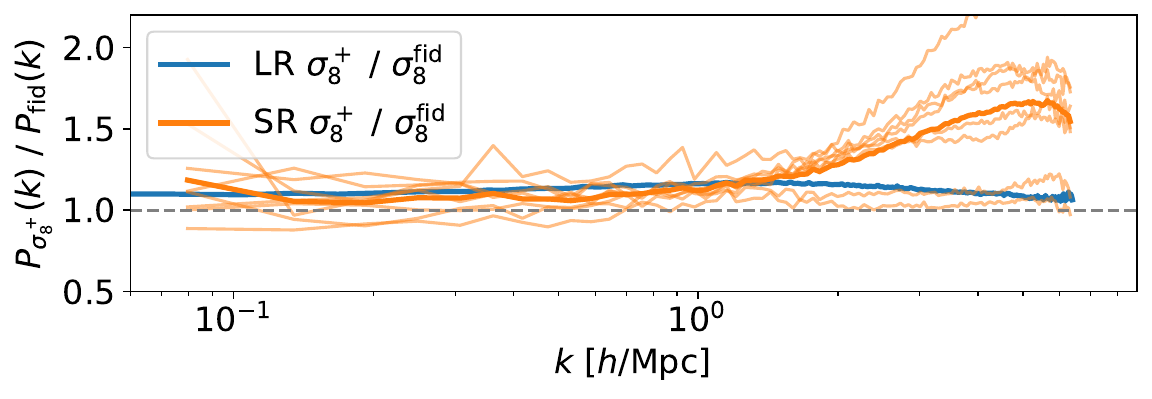}\\
        \hspace{.6cm}$\Omega_\text{M}^-=0.2935$ & \hspace{.6cm}$\Omega_\text{M}^+=0.3243$\\
        \includegraphics[width=0.498\textwidth]{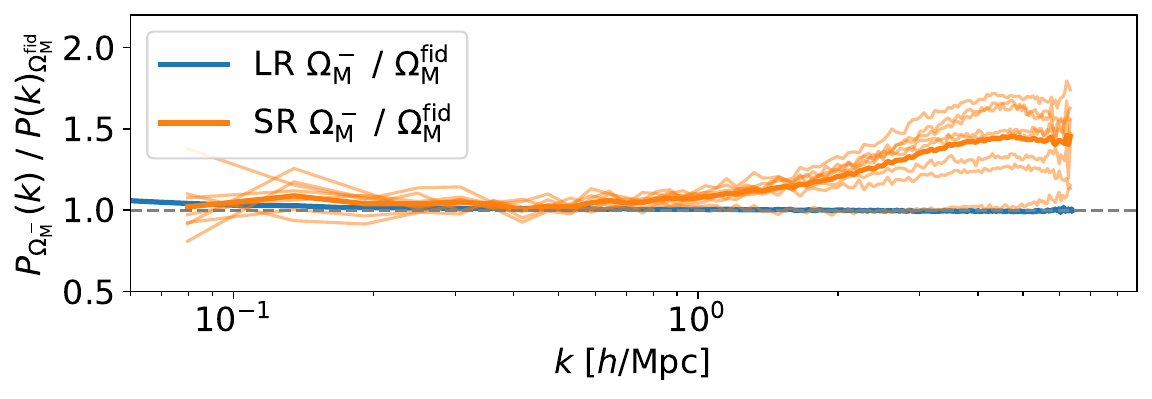} & \includegraphics[width=0.498\textwidth]{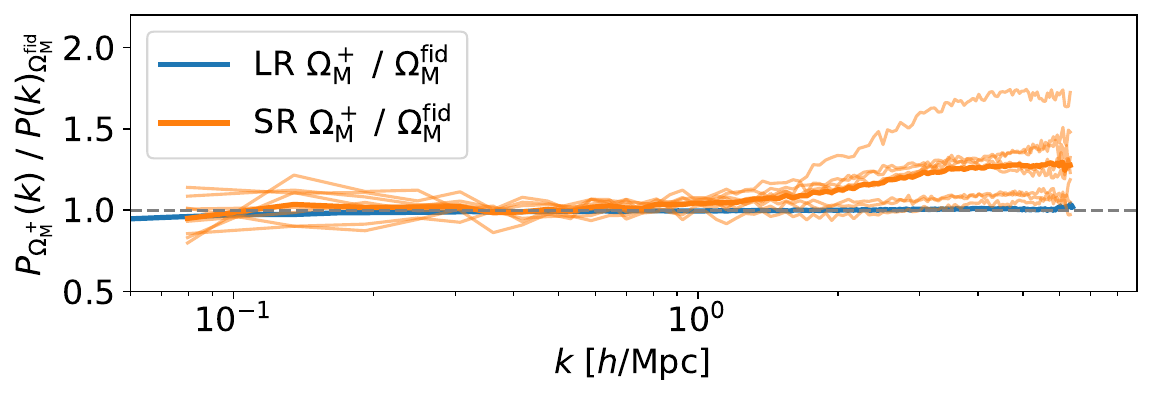}\\
    \end{tabular}
    \caption[Power spectra varying $\sigma_8$ and $\Omega_\text{M}$ in the low-resolution conditional]{Power spectra for varying $\sigma_8$ and $\Omega_\text{M}$ by $\pm5\%$ in the LR conditional field, shown as ratios over the fiducial power spectra with $\sigma_8=0.8159$, $\Omega_\text{M}=0.3089$. In all four varied cosmologies, the SR field tends to have increased power at nonlinear scales. The model seems to be highly sensitive to out-of-distribution conditional fields.}
    \label{fig:ps_ratios}
\end{figure}

Here we use $T=2000$ denoising steps (increased from $1250$ previously) and use a noise schedule with $\beta_0=10^{-6}$, $\beta_T=10^{-2}$. Increasing the number of steps while reducing the $\beta_t$ between steps may possibly reduce errors in this out-of-distribution regime, but we did experiment with either noise schedule and obtained qualitatively similar results. We show power spectra ratios between the varied and fiducial SR fields for the four cosmologies in Fig.~\ref{fig:ps_ratios}, along with the LR power spectra.

For $\sigma_8^-$, we expect the power spectrum to be decreased at nonlinear scales by a factor of about $0.95^2$, but our SR power spectrum only decreases slightly at these scales. There is a bit of success for $\sigma_8^+$, as we can see that the model output has nearly the proper $1.05^2$ increase in power for $k<1\ h/\text{Mpc}$. However, a major issue is that both $\sigma_8^-$ and $\sigma_8^+$ significantly increase in power above $k=1\ h/\text{Mpc}$, but we expect them to move in opposite directions. 
Considering $\Omega_\text{M}^-$ and $\Omega_\text{M}^+$, our SR model output has remained the same at nonlinear scales in the power spectrum. However, the model output again has a significant power increase above $k=1\ h/\text{Mpc}$ for both $\Omega_\text{M}^-$ and $\Omega_\text{M}^+$.

In each of the four cosmologies, the SR model output did not accurately follow the expected power spectra curve compared to the fiducial cosmology, but rather the power at larger $k$ tends to incorrectly increase significantly. We suspect this is a signature of pixel-wise noise not being properly denoised by the diffusion model. The diffusion model seems to be highly sensitive to out-of-distribution conditional data and is not robust to emulating new cosmologies from the LR conditioner. To include cosmological parameter dependence, we will thus need training data that varies these parameters.

\chapter{Conclusion}
\label{ch:conclusion}





Cosmological inference has been greatly aided in the last several years by significant advances in machine learning technology. With a likelihood-free approach, machine learning models may be used to extract information from simulations, or generate simulated data, without being required to carefully model many of the complicated aspects of the cosmological data. Additionally, using field level data to solve these problems allows the maximum amount of information to be passed to the model. 

A number of machine learning models have been developed for probabilistic modelling and inference. This work has considered two such models, normalizing flows and denoising diffusion. Normalizing flows learn high-dimensional PDFs by learning an invertible transformation from a simple base distribution $\bm{u}$ to a complex target distribution $\bm{x}$. In Chapter~\ref{ch:nf_random}, field level data of the non-Gaussian large-scale structure data is learned with the real NVP and Glow normalizing flows, and the accuracy of the PDF is measured by the power spectrum and non-Gaussian estimators. Two base distributions $\bm{u}$ are considered: uncorrelated Gaussian noise and correlated Gaussian fields. We find that for a correlated Gaussian base distribution having the same power spectrum as the training data, the model trains significantly faster than a model flowing from noise.

Cosmological parameter inference with field level data can be performed with a likelihood-free approach. Normalizing flows are an excellent tool for parameter inference, capable of directly learning the matter PDF conditional on the cosmological parameters, $p(\bm{x}|\bm{\phi})$. A conditional version of Glow was used in Chapter~\ref{ch:nf_parameters} to learn $p(\bm{x}|\Omega_\text{M},\sigma_8)$ with dark matter density fields $\bm{x}$. In the setup presented, the parameters are embedded into the CNN that learns the flow's affine transformations, and therefore every step of the flow is conditional on such parameters. In this way, samples generated across the space of $\Omega_\text{M}\in(0.1,0.5)$ and $\sigma_8\in(0.6,1.0)$ accurately learn the physics of the dark matter density. Reversing the PDF with Bayes' theorem, the model accurately gives the full PDF $p(\Omega_\text{M},\sigma_8|\bm{x})$ for random samples $\bm{x}$. The estimates of $\Omega_\text{M}$ and $\sigma_8$ have $0.032$ and $0.012$ root-mean-squared error, and about $0.04$ and $0.01$ standard deviations, respectively. Additionally, by marginalizing the conditional PDF to obtain both $p(\Omega_\text{M}|\bm{x})$ and $p(\sigma_8|\bm{x})$, we have quantified the model's accuracy for 200 random test samples $\bm{x}$ with the $R^2$ coefficient and reduced $\chi^2$. The model's mean values are highly accurate, with $R_{\Omega_\text{M}}^2=0.90$ and $R_{\sigma_8}^2=0.99$. The model is slightly under-confident at the $1\sigma$ level, where we find $\chi_{\Omega_\text{M}}^2=0.76$ and $\chi_{\sigma_8}^2=0.74$, although we see no obvious trends in the parameter inference that would suggest a systematic difficulty on certain subsets of data.

Data reconstruction of cosmological fields is approached with normalizing flows in Chapter~\ref{ch:denoising}. We use real NVP to learn the prior distribution $p(\bm{s})$ for a true signal $\bm{s}$. The reconstructed signal given noisy data $\bm{d}$ may be either found as the maximum a posteriori (MAP) solution by extremizing the posterior $p(\bm{s}|\bm{d})$, or sampled from $p(\bm{s}|\bm{d})$ with Monte Carlo techniques. For the setup considered in Chapter~\ref{ch:denoising} with $512\ \text{Mpc}/h$ length fields projected to 2 dimensions with $128\ \text{Mpc}/h$ depth, the flow MAP gives up to a $30\%$ reduction in mean-squared-error over Wiener filtering. At small length scales, the Fourier space cross-correlation coefficient between the flow MAP and truth is about twice the coefficient between the Wiener filtered solution and truth. Considering the speed of computing probabilities with normalizing flows, these models are well suited for sampling from $p(\bm{d}|\bm{s})$ with Hamiltonian Monte Carlo. Multiple reconstructed samples $\bm{d}$ consistent with the signal $\bm{s}$ may be generated in this way, giving quantitatively similar results to the MAP solutions. Considering the architecture of the real NVP flow, the denoising with a non-Gaussian prior may be applied to local patches of the field, and these local patches may then be combined into a larger field. Edges between patches are made smooth thanks to the large length scale physics being analytically solvable with Wiener filtering, and therefore modes that cross patch boundaries are identical between individual patch solutions. With this patching scheme, a model trained on $128\times128$~px images is capable of accurately denoising $1024\times1024$~px images.

Moving to the highly expressive denoising diffusion models, a super-resolution emulation of cosmological fields in 3 spacial dimensions is presented in Chapter~\ref{ch:sr_diffusion}. The novel development here is an iterative outpainting scheme capable of generating super-resolution volumes much larger than the training data. The model was trained on $48$~px length cubes, and as an application, we generated a $528$~px length cube with $21^3$ outpainting iterations. Of particular note is that the entire training data set was a single $264$~px length cube from the TNG300 simulation, and therefore the model is capable of increasing the volume of a single given simulation run. In this way, the model learns the local physics of a computationally expensive simulation run, and transfers it onto a larger volume. The super-resolution model output is accurate as measured by the one-point PDF, power spectrum, bispectra, and void size function, and closely matches the summary statistics of the training data. BAOs are also accurately emulated, even though the length scales of BAOs is larger than the length of the training data cubes; therefore, the super-resolution volume is successfully guided at the large length scales by the low-resolution conditional field. Additionally, the diffusion model is a stochastic generative model, and generates samples throughout the space of possible solutions to the super-resolution problem. A set of multiple super-resolution samples generated from a single low-resolution has almost no correlations above $k\gtrsim0.3\ h/\text{Mpc}$, indicating that the model is successfully generating a wide variety of samples conditional on the same low-resolution background.

This work has used simulated data throughout. Simulations are a fundamental tool in cosmology for testing models and parameter inference on real data. However, one of the great problems in machine learning is generalizing models to do inference on out-of-distribution data, and there is no guarantee that a model trained on simulated data will perform well on real data. In cosmology, we do not have the luxury of easily training on a diverse distribution of distributions, such as ImageNet~\cite{imagenet}, where any new data sample introduced to the model is surely graphically related to some samples in the training data. Normalizing flows are particularly susceptible at failing to detect out-of-distribution data, which is basically due to the flow learning graphical properties (local pixel correlations) and not semantic properties (the overall structure) of the data~\cite{kirichenko2020normalizing}. The flow architecture is designed to send the target data to a (usually Gaussian) base distribution, and the KL-divergence that measures the accuracy of the flow has been found to have little influence on the detection of out-of-distribution data. In Chapter~\ref{ch:denoising}, denoising very noisy, and therefore highly out-of-distribution data, was less accurate than denoising moderately noisy data; we believe that the model has difficulty in matching a very noisy signal with the prior distribution. Out-of-distribution cosmological parameter inference with normalizing flows is explored with a multi-scale normalizing flow in \cite{Dai:2023lcb}, where the model is divided into several independent flows learning different length scales independently; in this way, specific out-of-distribution length scales can be turned off during inference.

The problem of robustly learning cosmological parameters may be alleviated in an experimental fashion with both careful model selection and training on a variety of different simulations. Such an approach was taken in~\cite{deSanti:2023zzn} at measuring $\Omega_\text{M}$, training graph neural networks on several different large-scale structure simulation suites. This may be seen as a sort of work-around to the out-of-distribution problem, the motivation being that a new volume introduced to the model is more likely to be recognized if the model has seen a wide distribution of samples. The downside with this approach is that that none of the simulations will actually match detector data without considering the various complications and quirks of the detector. A model may be trained on as many independently simulated volumes as we would like, but will still fail at estimating parameters if real data introduced to the model has systematic errors or survey masks not accounted for in the model.

Another approach towards inference on out-of-distribution data would be to rethink the actual data that the model is trained on. A topological representation of the large-scale structure may be more robust to slight variations and errors than the density field or halo positions. Persistence images were introduced as an alternative data structure to dark matter halos in Chapter~\ref{ch:introduction} Sec.~\ref{sec:persistence_images}, and were shown to be viable statistics for accurate estimation of $\Omega_\text{M}$, $\sigma_8$, and $n_s$. Adjusting the features of the halos by slightly adjusting the quantity of halos to account for changes in data resolution, or by introducing local errors in the halo positions (such as foreground removal), would not significantly change the persistence images. In the eyes of topology, a filament of halos with a local undercount of halos in half of the filament still looks like a filament of halos.

Training a model on simulated detector data, as opposed to volumetric or projected N-body simulations, would be a more direct approach at estimating parameters from real survey data. The Vera C. Rubin Observatory~\cite{LSSTSciBook,LSST:2008ijt} is expected to start its first run in 2025, and a realistic simulator of detector data, \texttt{PhoSim}~\cite{Peterson_2015}, has been developed. There are a multitude of errors that \texttt{PhoSim} simulates, including tracking errors, diffraction, lens perturbations, mirror perturbations, dome seeing, atmospheric effects, pixelization, and read noise. Likelihoods can be written down analytically, or estimated, for most of these effects; however, a likelihood-free approach of parameter inference and data reconstruction could be more elegant and more accurate. Clearly, the methods presented in Chapter~\ref{ch:denoising} must be updated if used to reconstruct such complex simulated data. Perhaps the approach of learning a prior $p(\bm{d})$ and writing down a likelihood $p(\bm{d}|\bm{s})$ is not ideal for realistic cosmological data. Rather, it might be more optimal to learn $p(\bm{s}|\bm{d})$ directly, with a conditional probabilistic machine learning model.

While the diffusion model used in Chapter~\ref{ch:sr_diffusion} is conditional on a cosmological field, there have been great advances recently with diffusion models generating images conditional on text~\cite{dhariwal2021diffusion,nichol2022glide,saharia2022photorealistic}. In a similar fashion, diffusion models may be conditional on cosmological parameters, allowing parameter inference and highly expressible data generation conditional on the parameters. Additionally, considering the successes of diffusion models for inpainting and outpainting tasks, a parameter inference scheme that accounts for survey masks could be implemented. Fields with survey masks are out-of-distribution when introduced to a model trained on data without masks, and machine learning models must somehow account for the geometry of real survey data.

The simulated data used throughout this work has been placed onto a grid to obtain density fields of either the dark matter or baryonic matter content. However, assigning particles to a grid necessarily loses information to the location data as 3-dimensional coordinates. In machine learning literature, such particle based data is often referred to as point clouds. A 3-dimensional point cloud normalizing flow is introduced in PointFlow~\cite{PointFlow}, and point cloud diffusion models have been developed~\cite{luo2021diffusion}. These point cloud models replace CNNs with a graph neural network~\cite{battaglia2018relational,gadelha2018multiresolution}. A point cloud diffusion emulator of dark matter halos was recently presented modelling about $17^3$ halos, conditional on $\Omega_\text{M}$ and $\sigma_8$~\cite{Cuesta-Lazaro:2023zuk}. Computational limitations must be carefully considered when choosing between density field and point cloud models, balancing the density of particles with the physical size of the data volume.

Besides normalizing flows and denoising diffusion, there are a number of other probabilistic machine learning models, the most popular being variational autoencoders~\cite{kingma2022autoencoding} and score-based generative models~\cite{song2020improved}. Score-based models are most similar to normalizing flows as these models learn the gradients of a PDF, rather than the entire PDF directly. The methods of normalizing flows, where a likelihood term is computed between steps of the flow, may be combined with a the iterative gradual denoising aspect of diffusion models~\cite{zhang2021diffusion,zhang2023diffflow}. In this way, Bayesian quantities may be directly obtained from the model while accessing the greater expressibility of diffusion models. A bijective transformation between noise and the target distribution was recently discovered as a solution to the Poisson equation~\cite{xu2022poisson}, and such a model may be competitive to diffusion models.

There are a number of potential issues with the $\Lambda\text{CDM}$ model of cosmology. Looking at cosmological parameters, the $\sigma_8$ tension and Hubble tension are the most pressing. There is about a $3\sigma$ tension between early and late time measurements of $\sigma_8$~\cite{Kazantzidis2021,Vivian:2023}. As found in Chapter~\ref{ch:nf_parameters} with simulated data, tight constraints may be placed on field level estimates of $\sigma_8$. This is not surprising, as $\sigma_8$ is by definition related to the amplitude of the linear power spectrum (Eq.~\ref{eq:sigma_8_def}). However, a likelihood-free field level measurement of the late time $\sigma_8$ may sidestep issues related to obtaining the linear power spectrum from late time survey data.

Presently, the most significant problem with the $\Lambda\text{CDM}$ model is the Hubble tension~\cite{DiValentino:2021izs}. Indirect early time measurements from the Planck mission find $h=67.4\pm0.5$~\cite{Planck:2018parameters}. Direct measurements of the Hubble constant from lensed quasars and standard candles such as Cepheids have been combined in~\cite{DiValentino:2020vnx} to arrive at $h=72.94\pm0.75$. There is therefore a nearly $6\sigma$ tension between indirect CMB measurements and direct measurements of $h$. The Hubble constant can also be inferred from gravitational wave events, having recently been measured at $h=68.50_{-15.9}^{+11.8}$~\cite{Gayathri:2020mra} from black hole-black hole and neutron star-neutron star mergers. Unfortunately, in contrast to $\sigma_8$, N-body simulations are not very sensitive to $h$, with information on $h$ only showing up in the initial conditions from the transfer function; this makes it difficult to measure $h$ from dark matter simulations~\cite{Lazanu_2021}.

The $\sigma_8$ and Hubble tensions might be solved simultaneously with modified gravity, allowing for a late time adjustment in the dynamics of expansion. Considering the analytic complexities of modelling modified gravity at late times, a likelihood-free approach would be useful to test how well modified gravity may be detected with late time surveys. Towards this aim, the Quijote dark matter simulations~\cite{Villaescusa-Navarro:2019bje} have been run with the Hu-Sawicki $f(R)$ model~\cite{Hu:2007nk}, varying the amplitude of the $f(R)$ correction.

This work has largely taken a machine learning first approach, using various machine learning models to represent and perform inference on simulations of the late time large-scale structure. As the field of machine learning matures, the field of cosmology should turn towards applying these models to real survey data. The next generation of surveys will record an unprecedented amount of information, we should make sure we are optimally using as much this information as possible to understand our cosmology.

\appendix
\chapter{Mathematics}
\label{ch:mathematics}
\section{Fourier transforms}
\label{sec:fourier_transforms}

The Fourier transform of a continuous function $f(\bm{r})$ that maps from $\mathbb{C}^d$ to $\mathbb{C}$ is
\begin{equation}
    \mathcal{F}[f(\bm{r})]=\int f(\bm{r})e^{2\pi i\bm{k}\cdot\bm{r}}\text{d}^d\bm{r}.
\end{equation}

Discrete tensors of data are frequently used in this work. The discrete Fourier transform on a $d$-dimensional tensor $x$ of dimension lengths $N_i$ is
\begin{equation}
    \mathcal{F}[\bm{x}]=\sum_{n_i=0}^{N_i-1}e^{2\pi ik_1n_1/N_1}\left(...\left(\sum_{n_d=0}^{N_d-1}\bm{x}_{n_1,...,n_d}e^{2\pi ik_dn_d/N_d}\right)\right).
\end{equation}

The asymptotic complexity of the discrete Fourier transform as shown above for a tensor with volume $N^d\equiv\prod_{i=1}^dN_i$ is $O\left(N^{2d}\right)$. There are faster ways of calculating the discrete Fourier transforms, with any of a number of algorithms known as a fast Fourier transform. The fast Fourier transform factorizes the data into a new basis with many zeros, and calculates $N$ frequencies. The complexity of a $1$-dimensional fast Fourier transform is $N\log_2{N}$. The complexity of a $d$-dimensional fast Fourier transform of a tensor with volume $N^d\equiv\prod_{i=1}^dN_i$ is
\begin{equation}
    \sum_{i=1}^d\frac{N^d}{N_i}\mathcal{O}{\left(N_i\log_2{N_i}\right)}=\mathcal{O}{\left(N^d\log_2{N^d}\right)}.
\end{equation}

There is a potentially faster version of the fast Fourier transform that maps $\mathbb{R}^d$ to $\mathbb{R}$. This real fast Fourier transform calculates $(\lfloor N/2\rfloor+1)N^{d-1}$ frequencies. This algorithm has the same big O complexity as the fast Fourier transform, as $\mathcal{O}{\left((\lfloor N/2\rfloor+1)N^{d-1}\log_2{N^d}\right)}=\mathcal{O}{\left(N^d\log_2{N^d}\right)}$. However, computing about half of the values is often faster for larger ($\gtrsim128$~px length) tensors.



\section{Sampling Gaussian random fields}
\label{sec:gaussian_random_fields}

A correlated Guassian field $u_\text{C}(\bm{x})$ can be constructed from an uncorrelated Gaussian (noise) field $u_\text{N}$. A noise field has a constant power spectrum equal to its variance $\sigma_\text{N}^2$,
\begin{equation}
     \langle u_\text{N}(\bm{k})u_\text{N}(\bm{k}')\rangle=\sigma_\text{N}^2\delta_\text{D}(\bm{k}+\bm{k}').
\end{equation}
A correlated Gaussian field in Fourier space can be constructed from the noise field and the desired correlated Gaussian power spectrum $P(k)$ with
\begin{equation}
\label{eq:noise_to_corr}
    u_\text{C}(\bm{k})=\frac{P^\frac{1}{2}(k)}{\sigma_\text{N}}u_\text{N}(\bm{k}).
\end{equation}
That $u_\text{C}$ has the correct power spectrum can be verified by
\begin{equation}
    \langle u_\text{C}(\bm{k})u_\text{C}(\bm{k}')\rangle=\frac{P(k)}{\sigma_\text{N}^2}\langle u_\text{N}(\bm{k})u_\text{N}(\bm{k}')\rangle=P(k)\delta_\text{D}(\bm{k}+\bm{k}').
\end{equation}
A correlated Gaussian field is the inverse Fourier transform of Eq.~\ref{eq:noise_to_corr},
\begin{equation}
\label{eq:fourier_correlated}
    u_\text{C}(\bm{x})=\mathcal{F}^{-1}[\frac{P^\frac{1}{2}(k)}{\sigma_\text{N}}u_\text{N}(\bm{k})].
\end{equation}
To interpret this equation, $u_\text{C}$ has random Fourier modes generated by $u_\text{N}$ scaled to specific amplitudes defined by $P^\frac{1}{2}(k)/\sigma_\text{N}$. This method of constructing correlated Gaussian fields only works because a the characteristic function of a Gaussian distribution with mean $0$ is also a Gaussian distribution; no other function is the Fourier transform of itself. We cannot construct a random field with different statistics in such a straightforward way; if $u_\text{N}$ were an uncorrelated Poisson distribution, the Fourier transform in Eq.~\ref{eq:fourier_correlated} would not return a Poisson distribution.




We may also view sampling correlated Gaussian random fields in the language of stochastic partial differential equations. The Whittle~\cite{Whittle:1954} equation
\begin{equation}
    \tau\left(\kappa^2-\nabla^2\right)^{\frac{\alpha}{2}}u_\text{C}(\bm{x})=u_\text{N}(\bm{x})
\end{equation}
writes a correlated Gaussian field $u_\text{C}(\bm{x})$ from a source Gaussian noise $u_\text{N}$, with some parameters $\tau$, $\kappa$, and $\alpha$. The Fourier transform of the Whittle equation is
\begin{equation}
    \tau\left(\kappa^2+|\bm{k}|^2\right)^{\frac{\alpha}{2}}\mathcal{F}{[u_\text{C}(\bm{x})]}=\mathcal{F}{[u_\text{N}(\bm{x})]}.
\end{equation}
From this, a correlated Gaussian random field can be constructed as
\begin{equation}
    u_\text{C}(\bm{x})=\mathcal{F}^{-1}\bigg[\frac{1}{\tau\left(\kappa^2+|\bm{k}|^2\right)^{\frac{\alpha}{2}}}\mathcal{F}{[u_\text{N}(\bm{x})]}\bigg].
\end{equation}
This expression for $u(\bm{x})$ is similar to the result of Eq.~\ref{eq:fourier_correlated}, where $\alpha$ would control the slope of the power spectrum.

\section{Wick's probability theorem}
\label{sec:wick}

A Gaussian field is fully described by its two-point correlation function. This is known as Wick's theorem in physics, or Isserlis' theorem in mathematics.

A random field $u(\bm{r})$ has a Gaussian probability distribution $p(u)$ if
\begin{equation}
    \langle u(\bm{r}_1)u(\bm{r}_2)...\rangle=0
\end{equation}
for an odd number of products of $u(\bm{r}_i)$, and
\begin{equation}
    \langle u(\bm{r}_1)u(\bm{r}_2)...\rangle=\sum_\text{pairings}\prod_\text{pairs}\langle u(\bm{r}_i)u(\bm{r}_j)\rangle
\end{equation}
for an even number of products of $u(\bm{r}_i)$. Here the product is with all $\langle u(\bm{r}_i)u(\bm{r}_j)\rangle$ using every $u(\bm{r}_i)$ once, and the sum is over all combinations of such terms. To show that a $u(\bm{r})$ with these properties is described by a Gaussian distribution, consider that the sum over all products of $\langle u^{2n}\rangle$ is
\begin{equation}
    \langle u^{2n}\rangle=\frac{(2n)!}{2^nn!}\langle u^2\rangle^n.
\end{equation}
The probability density that solves this equation is the Gaussian distribution
\begin{equation}
p(u)=\frac{1}{\sqrt{2\pi\langle u^2\rangle}} \exp{\left(-\frac{u^2}{2\langle u^2\rangle}\right)},
\end{equation}
as shown by evaluating the integral
\begin{equation}
    \langle u^{2n}\rangle=\int_{-\infty}^\infty u^{2n}p(u)\ \text{d}u=\int_{-\infty}^\infty\frac{u^{2n}}{\sqrt{2\pi\langle u^2\rangle}}e^{-\frac{u^2}{2\langle u^2\rangle}}\ \text{d}u=\frac{(2n)!}{2^nn!}\langle u^2\rangle^n.
\end{equation}

A consequence of Wick's theorem is that a Gaussian distribution $u$ is fully described by $\langle u^2\rangle$.

\chapter{Non-Gaussianity estimation}
\label{sec:nongaussestim}







In Chapter~\ref{ch:nf_random}, we quantify the performance of the flow samples at the level of the power spectrum and the bispectrum. In this Appendix we recall the standard estimator technology for the three-point function used in cosmology (see~\cite{Fergusson:2010ia}), adapted to the simple case of a 2-dimensional periodic field without sky curvature or mask. 

For a statistically isotropic random field $\delta(\bm{r})$, its power spectrum is defined by
\begin{align}
    \langle\delta{(\bm{k})}\delta{(\bm{k}')}\rangle&=(2\pi)^2\delta_\text{D}(\bm{k}+\bm{k}')P(k)\,,
\end{align}
Here $k$ are the 2-dimensional wave vectors associated with the random field in Fourier space. Similarily, the three-point correlator in momentum space, the bispectrum $ B(k_1,k_2,k_3)$, is defined by  
\begin{align}
    \langle\delta{(\bm{k}_1)}\delta{(\bm{k}_2})\delta{(\bm{k}_3)}\rangle&=(2\pi)^2\delta_\text{D}(\bm{k}_1+\bm{k}_2+\bm{k}_3) B(k_1,k_2,k_3),
\end{align}
where $\delta_\text{D}({k})$ is the 2-dimensional Dirac $\delta$-function enforcing a triangle condition on the wavevectors $\bm{k}_i$. Due to statistical isotropy, the 3-point function can be written as a function of the magnitudes $k_1$, $k_2$, $k_3$, rather than their vector value, the notation taken in the equations that follow. However, it is also common to write the bispectrum in terms of magnitudes of $\bm{k}_1$ and $\bm{k}_2$, along with the angle between these two vectors. Further, because it takes 3 numbers to uniquely define a triangle, we do not need any additional scalar terms such as $\bm{k}_1\cdot\bm{k}_2$ or $\bm{k}_1\cdot(\bm{k}_2\times\bm{k}_3)$ in the bispectrum. Note that our real NVP normalizing flow architecture using periodic padding, CNNs, and a periodic prior guarantees statistical isotropy of the model distribution. 

\subsubsection*{Template estimator}

For the power spectrum, which is a function of a single variable, it is reasonable to simply bin it and measure it in the data as we did above. The bispectrum however is a 3-dimensional quantity and is difficult to visualize and computationally expensive to estimate. Further, in our examples, the bispectrum is smaller than the power spectrum so a plot would mostly show noise. A better method to examine the three point function is thus to fit a set of bispectrum templates to the data and measure their amplitude. Assuming a bispectrum template $B(k_1,k_2,k_3)$, the optimal estimator for its amplitude is (see e.g.~\cite{Babich:2005en}) 

\begin{align}\label{eq:bestimator}
\mathcal{E}=\frac{1}{6\mathcal{N}}\int\frac{\text{d}^2\bm{k}_1}{(2\pi)^2}\frac{\text{d}^2\bm{k}_2}{(2\pi)^2}\frac{\text{d}^2\bm{k}_3}{(2\pi)^2}\frac{(2\pi)^2\delta_\text{D}(\bm{k}_1+\bm{k}_2+\bm{k}_3)B(k_1,k_2,k_3)}{P(k_1)P(k_2)P(k_3)}\delta{(\bm{k}_1)}\delta{(\bm{k}_2)}\delta{(\bm{k}_3)}\,.
\end{align}
This equation can be interpreted as follows. Sum over all combinations of three modes $k_i$ that form a triangle, and weight them by their signal-to-noise $\frac{B(k_1,k_2,k_3)}{P(k_1)P(k_2)P(k_3)}$. Here we have assumed that the noise is dominated by the Gaussian contribution, so this estimator is only optimal in the weak non-Gaussian limit. The estimator normalization $\mathcal{N}$ can be calculated analytically in the weak non-Gaussian case, however here we will determine it from simulations to include non-Gaussian variance.

\subsubsection{Local and equilateral templates}

We are interested in two basic templates for non-Gaussianity, so-called local and equilateral non-Gaussianity. Local non-Gaussianity \cite{Komatsu:2001rj} is generated by transforming a correlated Gaussian field $\delta_{G}(\bm{r})$ as
\begin{align}
    \delta_\text{NG}(\bm{r})=\delta_\text{G}(\bm{r})+\tilde{f}_{\text{NL}}^{\text{local}}\left(\delta_\text{G}^2(\bm{r})-\langle\delta_\text{G}^2(\bm{r})\rangle\right).
\end{align}
The resulting bispectrum to first order in the non-Gaussianity is 
\begin{align}
    B^{\text{local}}(k_1,k_2,k_3)=2\tilde{f}_\text{NL}^{\text{local}}\bigg(P(k_1)P(k_2)+P(k_2)P(k_3)+P(k_1)P(k_3)\bigg).
\end{align}
For a scale-invariant power spectrum in 2 dimensions $P(k) = A_s/k^2$ this gives
\begin{align}
    B^{\text{local}}(k_1,k_2,k_3)=2\tilde{f}_\text{NL}^{\text{local}}A_s^2\left(\frac{1}{k_1^2 k_2^2}+\frac{1}{k_2^2 k_3^2}+\frac{1}{k_1^2 k_3^2}\right),
\end{align}
but we will use the former expression in practice. This bispectrum form has the property that it strongly correlates large-scale and small scale modes, meaning $B(k_1,k_2,k_3)$ peaks for squeezed triangles where one $k_i$ is much shorter than the other two. Another case is a bispectrum which peaks when all $k_i$ are about the same length. This template is called equilateral non-Gaussianity~\cite{Fergusson:2010ia} and in 2 dimensions we define it as
\begin{align}
    B^{\text{equi}}(k_1,k_2,k_3)=6\tilde{f}_\text{NL}^{\text{equi}}A_s^2\left(\frac{(k_1+k_2-k_3)(k_2+k_3-k_1)(k_3+k_1-k_2)}{(k_1 k_2 k_3)^{7/3}}\right).
\end{align}
This is an empirical template that is constructed to peak for equilateral configurations, where it probes interactions of modes of about the same spatial scales. Unlike for the local bispectrum, map making is not trivial, so we are not generating maps with this specific template here.

\subsubsection*{Efficient position space estimator}

Both of these templates have the property that they can be written in factorizable form as
\begin{align}
    B(k_1,k_2,k_3)=A(\bm{k}_1)B(\bm{k}_2)C(\bm{k}_3)+5\,\text{permutations}.
\end{align}
The estimator in Eq.~\ref{eq:bestimator} can then be rewritten in a computationally much more efficient but mathematically equivalent position space form. To do this we insert the identity 
\begin{align}
    (2\pi)^2\delta_\text{D}(\bm{k}_1+\bm{k}_2+\bm{k}_3)=\int\text{d}^2\bm{r}\,e^{i(\bm{k}_1+\bm{k}_2+\bm{k}_3)\cdot\bm{r}}
\end{align}
and re-organize the terms into a real space integral over the product of Fourier filtered fields. For local non-Gaussianity we obtain 
\begin{align}
    \mathcal{E}=\frac{1}{\mathcal{N}}\int\text{d}^2\bm{r}\ A^2(\bm{r})B(\bm{r})
\end{align}
with
\begin{align}
    A(\bm{r})&=\int\frac{\text{d}^2\bm{k}}{(2\pi)^2}\delta(\bm{k})e^{i\bm{k}\cdot\bm{x}},\\
    B(\bm{r})&=\int\frac{\text{d}^2\bm{k}}{(2\pi)^2}\frac{\delta(\bm{k})}{P(k)}e^{i\bm{k}\cdot\bm{x}}. 
\end{align}
The expression for equilateral non-Gaussianity is very similar:
\begin{align}
    \mathcal{E}=\frac{1}{\mathcal{N}}\int\text{d}^2\bm{r}\ A_s^{-1}\left[-3\tilde{A}(\bm{r})\tilde{B}^2(\bm{r})+6\tilde{B}(\bm{r})\tilde{C}(\bm{r})\tilde{D}(\bm{r})-2\tilde{D}^3(\bm{r})\right]
\end{align}
with 
\begin{align}
    \tilde{A}(\bm{r})&=\int\frac{\text{d}^2\bm{k}}{(2\pi)^2}\bm{k}^{8/3}\delta(\bm{k})e^{i\bm{k}\cdot\bm{r}},\\
    \tilde{B}(\bm{r})&=\int\frac{\text{d}^2\bm{k}}{(2\pi)^2}\bm{k}^{-1/3}\delta(\bm{k})e^{i\bm{k}\cdot\bm{r}},\\
    \tilde{C}(\bm{r})&=\int\frac{\text{d}^2\bm{k}}{(2\pi)^2}\bm{k}^{5/3}\delta(\bm{k})e^{i\bm{k}\cdot\bm{r}},\\
    \tilde{D}(\bm{r})&=\int\frac{\text{d}^2\bm{k}}{(2\pi)^2}\bm{k}^{2/3}\delta(\bm{k})e^{i\bm{k}\cdot\bm{r}}. 
\end{align}
Note that these expressions are for scale-invariant non-Gaussianity in 2 dimensions, and thus slightly different from their 3-dimensional counterparts more commonly used in cosmology.




\chapter{Machine Learning}
\label{ch:ml}
In this section, a tensor is defined to be a list of numbers arranged in a $d$-dimensional cuboid array. The components $i,...,k$ of a tensor $\bm{x}$ are listed in square brackets as $\bm{x}[i,...k]$.

\section{Fully connected layers}
\label{sec:fc}

A fully connected, or dense, layer with weight matrix $\bm{w}$ and bias $\bm{b}$ maps $\bm{x}$ to $\bm{y}$ with
\begin{equation}
    \bm{y}=\bm{w}^\text{T}\bm{x}+\bm{b}
\end{equation}
There is usually an activation function after every dense layer.

The number of learnable parameters in a fully connected layer is simply the sum number of values in $\bm{w}$ and $\bm{b}$, or equivalently, the length of $\bm{x}$ times the length of $\bm{y}$ plus the length of $\bm{y}$. A detriment of fully connected layers is the massive number of parameters it requires to learn mappings between large collections of data, and this problem is especially apparent in image or volumetric tensors of data. For example, a typical size large-scale structure simulation might be represented by $128^3$ voxels. A fully connected layer between two such maps would have $128^6+128^3\sim4\times10^{12}$ parameters.

\section{Convolutions}
\label{sec:convolutions}

Convolutions are used to correlate nearby pixels in a tensor by generating feature maps with a set of filters, or ``kernels." What is called a ``convolution" in this work, and in the machine learning community, would be called a discrete cross-correlation by the mathematics community. Convolutions can be done on data with any number of spacial dimensions $d$, and any length, and do not have the large parameter issue of fully connected layers. A convolution has weights $\bm{w}$ arranged in a $d$-dimensional cube, called a kernel. The convolution of a 1-, 2-, or 3-dimensional tensor $\bm{x}$ with a kernel $\bm{w}$ of odd kernel size $K$ having elements $[i]$, $[i,j]$, or $[i,j,k]$, has components
\begin{equation}
    (\bm{x}*\bm{w})[i]=\sum_{u=0}^{K-1}\bm{x}[i+u]\bm{w}[u],
\end{equation}
\begin{equation}
    (\bm{x}*\bm{w})[i,j]=\sum_{u=0}^{K-1}\sum_{v=0}^{K-1}\bm{x}[i+u,j+v]\bm{w}[u,v],
\end{equation}
\begin{equation}
    (\bm{x}*\bm{w})[i,j,k]=\sum_{u=0}^{K-1}\sum_{v=0}^{K-1}\sum_{w=0}^{K-1}\bm{x}[i+u,j+v,k+w]\bm{w}[u,v,w],
\end{equation}
respectively. (The mathematics community would call a convolution the same operation except with $i+u$ replaced with $i-u$, etc.) An illustration of a convolution is shown in Fig.~\ref{fig:conv}.

\begin{figure}
\centering
\begin{tikzpicture}
\hspace{-5.5cm}
\matrix [
    matrix of nodes,
    row sep=-\pgflinewidth,
    column sep=-\pgflinewidth,
    nodes={draw, anchor=center, minimum size=0.8cm, fill=azul!75}]
    (w){
    $w_{02}$ & $w_{12}$ & $w_{22}$\\
    $w_{01}$ & $w_{11}$ & $w_{21}$\\
    $w_{00}$ & $w_{10}$ & $w_{20}$\\
    };
    \node[font=\Large,anchor=south,yshift=1.2cm] at (w.north) {$\bm{w}$};
\hspace{5.5cm}
\matrix [
    matrix of nodes,
    row sep=-\pgflinewidth,
    column sep=-\pgflinewidth,
    nodes={draw, anchor=center, minimum size=0.8cm}]
    (x){
    $x_{05}$ & $x_{15}$ & $x_{25}$ & $x_{35}$ & $x_{45}$ & $x_{55}$\\
    $x_{04}$ & |[fill=black!15]|$x_{14}$ & |[fill=black!15]|$x_{24}$ & |[fill=black!15]|$x_{34}$ & $x_{44}$ & $x_{54}$\\
    $x_{03}$ & |[fill=black!15]|$x_{13}$ & |[fill=black!15]|$x_{23}$ & |[fill=black!15]|$x_{33}$ & $x_{43}$ & $x_{53}$\\
    $x_{02}$ & |[fill=black!15]|$x_{12}$ & |[fill=black!15]|$x_{22}$ & |[fill=black!15]|$x_{32}$ & $x_{42}$ & $x_{52}$\\
    $x_{01}$ & $x_{11}$ & $x_{21}$ & $x_{31}$ & $x_{41}$ & $x_{51}$\\
    $x_{00}$ & $x_{10}$ & $x_{20}$ & $x_{30}$ & $x_{40}$ & $x_{50}$\\
    };
    \node[font=\Large,anchor=south] at (x.north) {$\bm{x}$};
\hspace{5.5cm}
\matrix [
    matrix of nodes,
    row sep=-\pgflinewidth,
    column sep=-\pgflinewidth,
    nodes={draw, anchor=center, minimum size=0.8cm}]
    (xconvw){
    \\
    };
    \node[font=\Large,anchor=south,yshift=2.2cm] at (xconvw.north) {$(\bm{x}*\bm{w})[2,3]$};
    \node[anchor=south,yshift=0.5cm] at (xconvw.south) {\hspace{0.75em}$x_{12}w_{00}+x_{22}w_{10}+x_{32}w_{20}$};
    \node[anchor=south] at (xconvw.south) {$+x_{13}w_{01}+x_{23}w_{11}+x_{33}w_{21}$};
    \node[anchor=south,yshift=-0.5cm] at (xconvw.south) {$+x_{14}w_{02}+x_{24}w_{12}+x_{34}w_{22}$};
\end{tikzpicture}
\caption[2-dimensional convolution with a kernel of size 3]{Illustration of a 2-dimensional convolution with a kernel of size $3$. A single calculation, $(\bm{x}*\bm{w})[2,3]$, is shown here.}
\label{fig:conv}
\end{figure}

\begin{figure}
    \centering
\begin{tikzpicture}
\hspace{-5.5cm}
\matrix [
    matrix of nodes,
    row sep=-\pgflinewidth,
    column sep=-\pgflinewidth,
    nodes={draw, anchor=center, minimum size=0.8cm, fill=azul!75}]
    (w){
    $-1$ & $0$ & $1$\\
    $-2$ & $0$ & $2$\\
    $-1$ & $0$ & $1$\\
    };
    \node[font=\Large,anchor=south,yshift=1.2cm] at (w.north) {$\bm{w}$};
\hspace{5.5cm}
\node[inner sep=0pt] (chamberlin) at (0,0)
    {\includegraphics[width=0.38\textwidth]{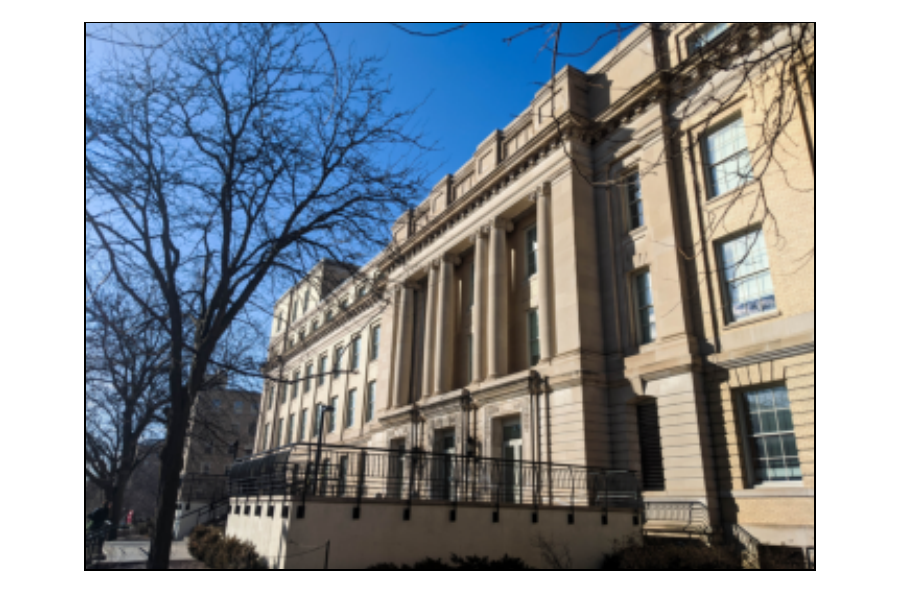}};
    \node[font=\Large,anchor=south] at (x.north) {$\bm{x}$};
\hspace{5.5cm}
\node[inner sep=0pt] (chamberlin) at (0,0)
    {\includegraphics[width=0.38\textwidth]{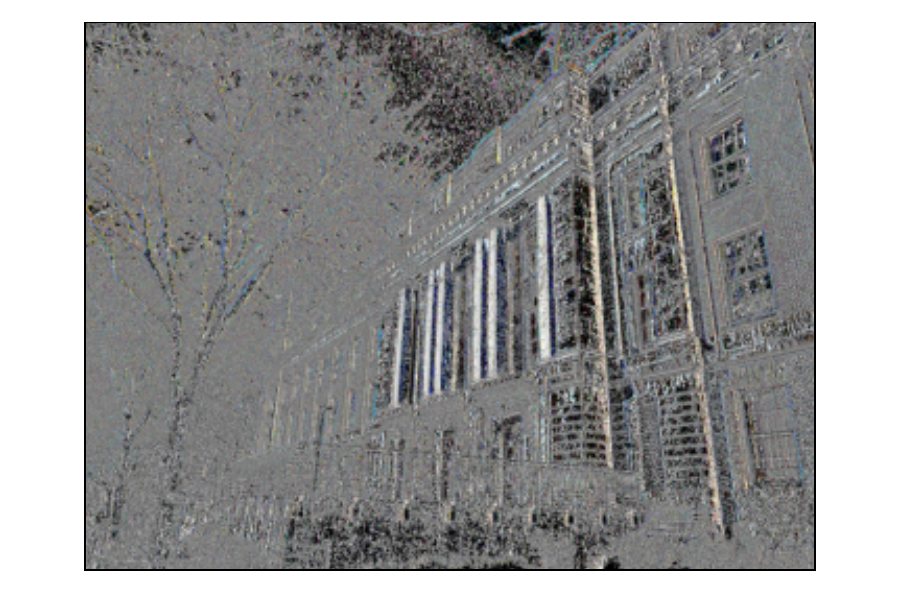}};
    \node[font=\Large,anchor=south,yshift=2.2cm] at (xconvw.north) {$\bm{x}*\bm{w}$};
\end{tikzpicture}
    \caption[Convolution example]{Example of a convolution with a kernel that detects vertical edges.}
    \label{fig:CNN_example}
\end{figure}
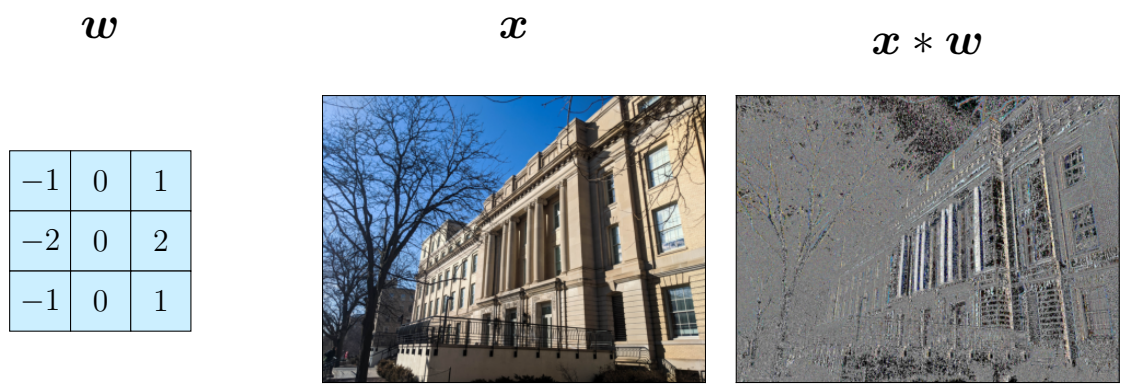

The kernel size is almost always chosen to be an odd integer, as even kernel sizes have no center, and have been found to be suboptimal in machine learning. As the main purpose of a convolution is to correlate or recognize features, we want a kernel size greater than $1$. The most common kernel size is $3$. It often turns out that having more CNN layers is better than having less layers with larger kernel sizes. A kernel size of $5$ has seen some use, an example being in a minority of convolutions in EfficientNet~\cite{tan2020efficientnet} (with the rest of the convolutions having kernel size $3$). A kernel size of $1$ would be used to mix channels. A $2$-dimensional kernel used for vertical edge detection is shown in Fig.~\ref{fig:CNN_example}.


A single valued bias is added after every convolution. As with fully connected layers, there is usually an activation function after every convolutional layer.

The number of learnable parameters in a convolutional layer is the number of weights in each convolution, $K^d$, times the number of convolutions in the layer, $c_ic_{i+1}$, plus the number of bias parameters, $c_{i+1}$:
\begin{equation}
    K^dc_ic_{i+1}+c_{i+1}
\end{equation}
A typical CNN might have $K=3$ and $c_i=c_{i+1}=128$ in a layer, and the number of learnable parameters in this case is about $1.5\times10^5$ in one such layer.

Convolutions are a local operation, and therefore a CNN has a window size, or ``receptive field," defining the maximum distance two elements of the tensor can be while still being connected in the network. For each convolution of kernel size $K_i$, the receptive field has length
\begin{equation}
    1+\sum_{i}(K_i-1).
\end{equation}

\section{Activation functions}
\label{sec:activation_functions}

\begin{figure}[h]
\centering
\includegraphics[width=0.495\textwidth]{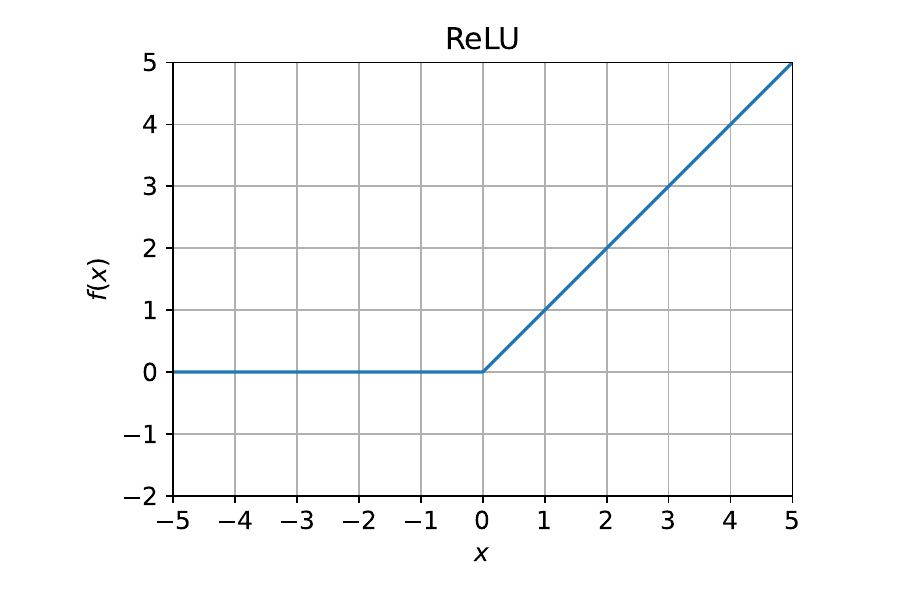}
\includegraphics[width=0.495\textwidth]{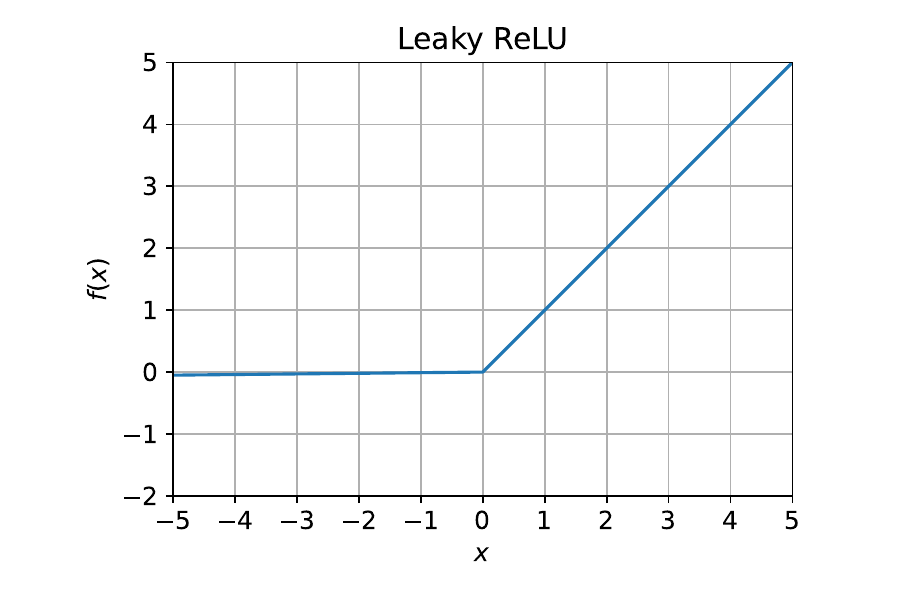}

\includegraphics[width=0.495\textwidth]{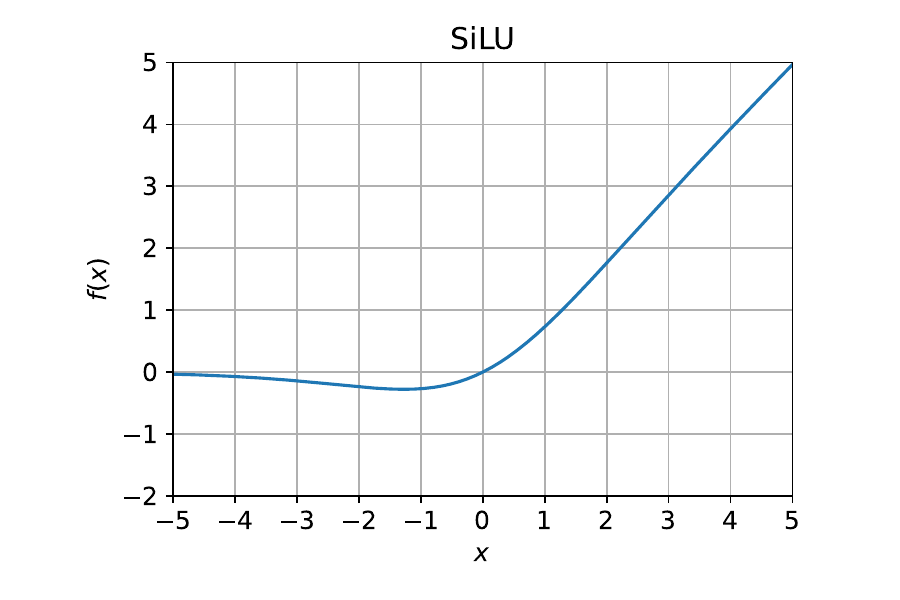}
\includegraphics[width=0.495\textwidth]{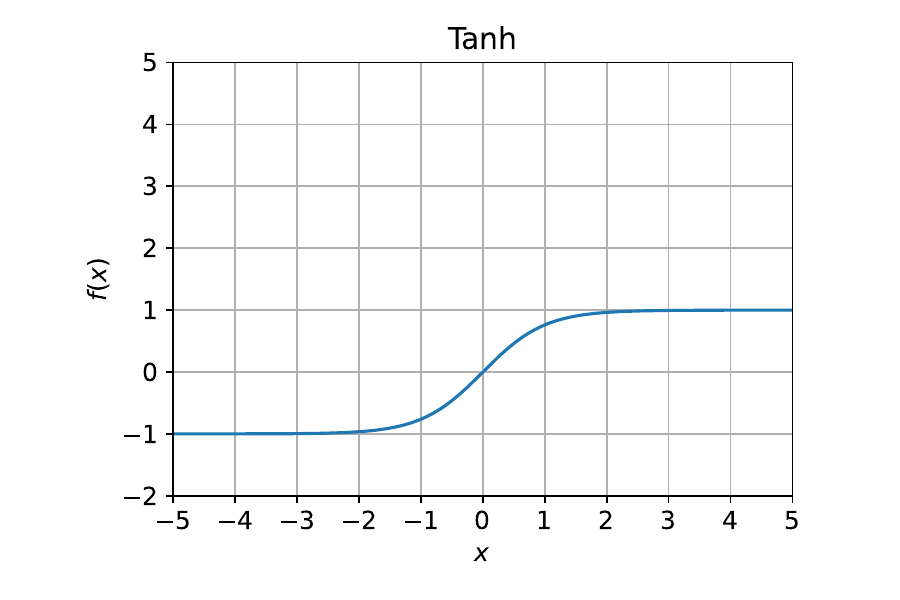}
\caption[Some activation functions]{Some activation functions}
\label{fig:activation_functions}
\end{figure}

Non-linearity is given to the neural network with activation functions. Nearly every fully connected layer and every convolution is sent to an activation function, with an exception often in the final layer of a model. The activation functions mentioned in this work are plotted in Figure~\ref{fig:activation_functions} and described below.

The rectified linear unit (ReLU), introduced in~\cite{Kunihiko:1969}, is a non-smooth activation function, defined by
\begin{equation}
    f(x)= 
    \begin{cases}
        0&\text{for}\ x<1,\\
        x&\text{for}\ x\geq0.
    \end{cases}
\end{equation}
The leaky ReLU, researched in~\cite{Maas2013RectifierNI}, is similar to the ReLU, with a shallow tail on the left half:
\begin{equation}
    f(x)= 
    \begin{cases}
        0.01x&\text{for}\ x<1,\\
        x    &\text{for}\ x\geq0.
    \end{cases}
\end{equation}
The sigmoid linear unit (SiLU), researched as an alternative to ReLU in~\cite{Elfwing2018}, is a smooth activation function that approaches $0$ on the left and $x$ on the right, defined by
\begin{equation}
    f(x)=\frac{x}{1-e^{-x}}.
\end{equation}
The tanh activation function,
\begin{equation}
    f(x)=\frac{e^x-e^{-x}}{e^{x}+e^{-x}},
\end{equation}
is often used at the end of a network when the output needs to be contained in some range. In this work, the real NVP affine transformation parameters $\bm{s}$ and $\bm{t}$ are found with a CNN having a tanh after its final layer.

\section{Pooling}
\label{sec:pooling}

Pooling layers are used to lower the size of data by mapping multiple data values to a single value. A pooling layer with an $n\times...\times n$ window size on $d$-dimensional tensor reduces the amount of data by a factor of $n^{-d}$.

Average pooling is defined as
\begin{equation}
    \bm{y}[i,...,k]=\frac{1}{n^d}\sum_{w_1=0}^{n-1}...\sum_{w_d=0}^{n-1}\bm{x}[ni+w,...,nk+w],
\end{equation}
and max pooling is defined as
\begin{equation}
    \bm{y}[i,...,k]=\text{max}\bigg(\bm{x}[ni,...,nk],\ \bm{x}[ni+1,...,nk],\ ...,\ \bm{x}[ni+n-1,...,nk+n-1]\bigg)
\end{equation}

As an example, an average pooling layer with a $2\times2$ window on a $2$-dimensional tensor has elements
\begin{equation}
    \bm{y}[i,j]=\frac{1}{4}\bigg(\bm{x}[2i,2j]+\bm{x}[2i+1,2j]+\bm{x}[2i,2j+1]+\bm{x}[2i+1,2j+1]\bigg),
\end{equation}
and a max pooling layer has elements
\begin{equation}
    \bm{y}[i,j]=\text{max}{\bigg(\bm{x}[2i,2j],\ \bm{x}[2i+1,2j],\ \bm{x}[2i,2j+1],\ \bm{x}[2i+1,2j+1]\bigg)}.
\end{equation}
An illustration of max pooling is shown in Fig.~\ref{fig:pool_example}.


\begin{figure}
\centering
\begin{tikzpicture}
\hspace{-3cm}
\matrix [
    matrix of nodes,
    row sep=-\pgflinewidth,
    column sep=-\pgflinewidth,
    nodes={draw, anchor=center, minimum size=0.8cm}]
    (x){
    |[fill=red!15]|$3$ & |[fill=red!15]|$7$ & |[fill=orange!15]|$2$ & |[fill=orange!15]|$1$ & |[fill=yellow!15]|$14$ & |[fill=yellow!15]|$5$\\
    |[fill=red!15]|$2$ & |[fill=red!15]|$5$ & |[fill=orange!15]|$5$ & |[fill=orange!15]|$5$ & |[fill=yellow!15]|$8$ & |[fill=yellow!15]|$6$\\
    |[fill=green!15]|$0$ & |[fill=green!15]|$9$ & |[fill=blue!15]|$5$ & |[fill=blue!15]|$2$ & |[fill=purple!15]|$3$ & |[fill=purple!15]|$5$\\
    |[fill=green!15]|$4$ & |[fill=green!15]|$5$ & |[fill=blue!15]|$8$ & |[fill=blue!15]|$11$ & |[fill=purple!15]|$5$ & |[fill=purple!15]|$6$\\
    |[fill=violet!15]|$5$ & |[fill=violet!15]|$1$ & |[fill=brown!15]|$10$ & |[fill=brown!15]|$9$ & |[fill=black!15]|$1$ & |[fill=black!15]|$4$\\
    |[fill=violet!15]|$4$ & |[fill=violet!15]|$8$ & |[fill=brown!15]|$5$ & |[fill=brown!15]|$0$ & |[fill=black!15]|$4$ & |[fill=black!15]|$2$\\
    };
    \node[font=\Large,anchor=south] at (x.north) {$\bm{x}$};
\hspace{6cm}
\matrix [
    matrix of nodes,
    row sep=-\pgflinewidth,
    column sep=-\pgflinewidth,
    nodes={draw, anchor=center, minimum size=0.8cm, fill=azul!75}]
    (w){
    |[fill=red!15]|$7$ & |[fill=orange!15]|$5$ & |[fill=yellow!15]|$14$\\
    |[fill=green!15]|$9$ & |[fill=blue!15]|$11$ & |[fill=purple!15]|$6$\\
    |[fill=violet!15]|$8$ & |[fill=brown!15]|$10$ & |[fill=black!15]|$4$\\
    };
    \node[font=\Large,anchor=south,yshift=1.2cm] at (w.north) {$\bm{y}$};
\end{tikzpicture}
\caption[Max pooling example]{Max pooling example with a $2\times2$ window size. The output of the max pooling $\bm{y}$ has half the length and height of the input data $\bm{x}$.}
\label{fig:pool_example}
\end{figure}

\section{Group normalization}
\label{sec:group_normalization}

Group normalization was introduced in~\ref{sec:group_normalization} as a more effective generalization of batch normalization~\cite{ioffe2015batch}. These normalization layers speed up optimization and improve the accuracy of the network. For learned matrices $\bm{\gamma}$ and $\bm{\beta}$, group normalization re-scales the data $\bm{x}$ as
\begin{equation}
    \bm{y}=\frac{\bm{x}-\mu_{\bm{x}}}{\sqrt{\sigma_{\bm{x}}^2+\epsilon}}\bm{\gamma}+\bm{\beta}
\end{equation}
for mean $\mu_{\bm{x}}$, variance $\sigma_{\bm{x}}^2$, and bias $\bm{\beta}$. The $\epsilon=10^{-5}$ is for numerical stability.

\section{Dropout}
\label{sec:dropout}

Dropout~\cite{hinton2012improving,JMLR:v15:srivastava14a} is a simple way to prevent neural networks from overfitting by preventing co-adaptation of feature detectors. At some specified point in the network with a specified dropout rate, a random subset of the data is discarded from the network. Dropout is used in the U-net used in Chapter~\ref{ch:sr_diffusion} in each residual block, after a group normalization.

\section{Parameter inference with a CNN}
\label{sec:parameter_inference}

\begin{figure}
\centering

\begin{tikzpicture}[node distance=0.65cm]
\footnotesize

\pgfdeclarelayer{bg}
\pgfsetlayers{bg, main}

\node[align=center, font=\bfseries] (title){};
\node (start) [cnn, below of=start] {Image: $1\times n^2$};
\node (cnn) [cnn, below of=start, fill=azul!75] {$3\times3$ conv: $c_1\times(n-2)^2$};
\node (pool) [cnn, below of=cnn, fill=red!15] {$2\times2$ pool: $c_2\times(\lfloor\frac{n-2}{2}\rfloor)^2$};
\node (ghost0) [cnn, below of=pool, draw=red!0] {...};
\node (fc0) [cnn, below of=ghost0, fill=green!15] {Dense: $c_{d_1}$};
\node (ghost1) [cnn, below of=fc0, draw=red!0] {...};
\node (fc1) [cnn, below of=ghost1, fill=green!15] {Dense: $n_\text{parameters}$};
\node (end) [cnn, below of=fc1] {$n_\text{parameters}$};

\draw [arrow] (start.east) to [out=-30,in=30](cnn.east);
\draw [line width=0.1mm] (-7.3,-1.615) -- (-0.45,-1.615);
\draw [arrow] (cnn.east) to [out=-30,in=30](pool.east) node[yshift=0.35cm] {\hspace{1.3cm}ReLU};
\draw [arrow, draw=red!0] (cnn.west) to [out=-120,in=120](ghost0.west) node[yshift=0.65cm, align=center] {\hspace{-2.3cm}Alternating\\\hspace{-2.3cm}convolutions\\\hspace{-2.9cm}and pooling};
\draw [arrow] (pool.east) to [out=-30,in=30](ghost0.east);
\draw [line width=0.1mm] (-7.3,-3.575) -- (-0.45,-3.575);
\draw [arrow] (ghost0.east) to [out=-30,in=30](fc0.east) node[yshift=0.35cm] {\hspace{1.5cm}Flatten};
\draw [arrow] (fc0.east) to [out=-30,in=30](ghost1.east) node[yshift=0.35cm] {\hspace{1.3cm}ReLU};
\draw [arrow, draw=red!0] (fc0.west) to [out=-120,in=120](fc1.west) node[yshift=0.65cm, align=center] {\hspace{-2.3cm}Stack of\\\hspace{-2.3cm}dense layers};
\draw [arrow] (ghost1.east) to [out=-30,in=30](fc1.east) node[yshift=0.35cm] {\hspace{1.3cm}ReLU};
\draw [line width=0.1mm] (-7.3,-5.521) -- (-0.45,-5.521);
\draw [arrow] (fc1.east) to [out=-30,in=30](end.east) node[yshift=0.35cm] {\hspace{1.22cm}Tanh};

\end{tikzpicture}

\caption[CNN for parameter inference]{CNN for parameter inference. Shown are layer names along with output channel $\times$ spacial dimensions. An activation function is used after every convolution and dense layer.}

\label{fig:CNN}
\end{figure}
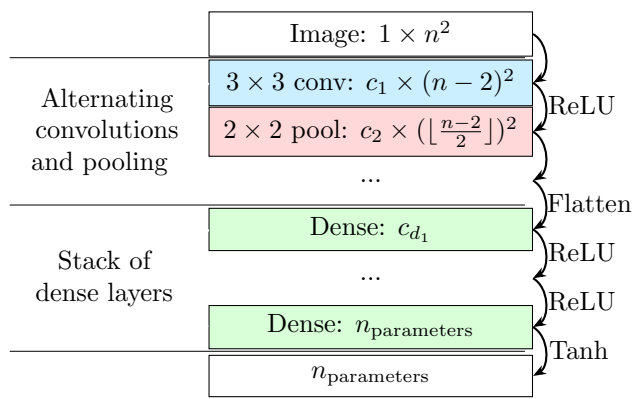

A common neural network task is to map a tensor of high-dimensional data to a single number, or a list of a few numbers, to classify data or perform parameter inference. A CNN can accomplish this with alternating convolutions and pooling layers. The convolutions learn feature maps, and the pooling layers essentially choose the important data from the feature maps. A convolution with zero padding maps a length $n$ tensor to length $n-2$. Combined with a pooling layer, the length of a zero padding convolution and stride $2$ pooling is $\lfloor\frac{n-2}{2}\rfloor$. Once the data is compressed to a manageable number of network parameters, the final layers of the CNN are fully connected. An illustration of a CNN mapping images to parameters is shown in Fig.~\ref{fig:CNN}.

\section{Stochastic gradient descent}
\label{sec:sgd}

A model's parameters $\bm{\theta}$ are updated to extremize, or ``optimize," a loss function $\mathcal{L}$ with stochastic gradient descent. Basic stochastic gradient descent would update the model parameters as
\begin{equation}
\label{eq:basic_sgd}
    \bm{\theta}(t)=\bm{\theta}(t-1)-\eta\nabla_{\bm{\theta}}\mathcal{L}(t),
\end{equation}
for a step size, or learning rate, $\eta$. A momentum term $\alpha\ \delta\bm{\theta}$ can be added to the right side of Eq.~\ref{eq:basic_sgd} as a way for the optimizer to ``remember" previous gradients $\delta\bm{\theta}$~\cite{SGD1999}, with hyperparameter $0<\alpha<1$.




The adaptive moment estimation (Adam) optimizer~\cite{Kingma2014AdamAM} computes biased estimates of first and second order moments of the first-order gradients $\nabla_{\bm{\theta}}\mathcal{L}$. The Adam optimizer has become ubiquitous in machine learning, and this work solely uses of the Adam optimizer.

The first step of the Adam optimization algorithm is to compute the gradients $\nabla_{\bm{\theta}}\mathcal{L}$. With the gradients, the biased first moment estimate is
\begin{equation}
    \bm{m}(t)=\beta_1\bm{m}(t-1)+(1-\beta_1)\nabla_{\bm{\theta}}\mathcal{L}(t)
\end{equation}
for a ``forgetting factor" $\beta_1=0.9$. The biased second moment estimate is
\begin{equation}
    \bm{v}(t)=\beta_2\bm{v}(t-1)+(1-\beta_2)\left(\nabla_{\bm{\theta}}\mathcal{L}\left(t\right)\right)^2
\end{equation}
for $\beta_2=0.999$. The hyperparameters $\beta_1$ and $\beta_2$ define the decay rates of the first and second moments of the gradients $\nabla_{\bm{\theta}}L$. A consequence of introducing the $\beta_1$ and $\beta_2$ is that the moment estimates are biased towards the zero vector, which is due to the $\beta_1$ and $\beta_2$ being set near unity in combination with $\bm{m}$ and $\bm{v}$ being initialized at zeros. The algorithm therefore introduces initialization bias corrections. The bias-corrected first moment estimate is
\begin{equation}
    \hat{\bm{m}}(t)=\frac{\bm{m}(t)}{1-\beta_1},
\end{equation}
and the bias-corrected second moment estimate is
\begin{equation}
    \hat{\bm{v}}(t)=\frac{\bm{v}(t)}{1-\beta_2}.
\end{equation}
Finally, the model parameters are updated as
\begin{equation}
    \bm{\theta}(t)=\bm{\theta}(t-1)+\eta\frac{\hat{\bm{m}}(t)}{\sqrt{\hat{\bm{v}}(t)}+\epsilon}
\end{equation}
where $\epsilon=10^{-8}$ is included for numerical stability. For a standard neural network model with more than $\sim10^5$ parameters, the learning rate $\eta$ is often chosen in the range $10^{-6}\lesssim\eta\lesssim10^{-3}$.


There can be a benefit to progressively lowering the learning rate during training, and two methods of doing so are used in this work. The first method is to lower the learning rate if the value of the loss function does not significantly change over a number of training cycles. The PyTorch class \texttt{ReduceLROnPlateau} accomplishes this with a ``patience" value, after a number of patience steps having taken place, if the loss has not lowered, the learning rate is decreased by a chosen factor. The second method is to lower the learning rate as a function of the number of training cycles, regardless of the value of the loss; one common example would be an exponential moving average learning rate scheduler. 

\singlespacing
\bibliographystyle{unsrturl}
\bibliography{refs}{}

\begin{thebibliography}{100}

\bibitem{rouhiainen2021}
Adam {Rouhiainen}, Utkarsh {Giri}, and Moritz {M\"{u}nchmeyer}.
\newblock {Normalizing flows for random fields in cosmology}.
\newblock In {\em Machine Learning and the Physical Sciences at NeurIPS}, 2021.
\newblock \href {https://arxiv.org/abs/2105.12024} {\path{arXiv:2105.12024}}.

\bibitem{rouhiainen2022}
Adam {Rouhiainen} and Moritz {M\"{u}nchmeyer}.
\newblock {De-noising non-Gaussian fields in cosmology with normalizing flows}.
\newblock In {\em {Machine Learning and the Physical Sciences at NeurIPS}}, 2022.
\newblock \href {https://arxiv.org/abs/2211.15161} {\path{arXiv:2211.15161}}.

\bibitem{Rouhiainen:2023ewv}
Adam Rouhiainen, Michael Gira, Moritz M\"unchmeyer, Kangwook Lee, and Gary Shiu.
\newblock {Super-Resolution Emulation of Large Cosmological Fields with a 3D Conditional Diffusion Model}.
\newblock In {\em {Machine Learning and the Physical Sciences at NeurIPS}}, 2023.
\newblock \href {https://arxiv.org/abs/2311.05217} {\path{arXiv:2311.05217}}.

\bibitem{yip2023learning}
Jacky H.~T. Yip, Adam Rouhiainen, and Gary Shiu.
\newblock Learning from topology: Cosmological parameter estimation from the large-scale structure.
\newblock In {\em The Synergy of Scientific and Machine Learning Modeling at ICML}, 2023.
\newblock \href {https://arxiv.org/abs/2308.02636} {\path{arXiv:2308.02636}}.

\bibitem{Bishop1968}
Richard~L. Bishop and Samuel~I. Goldberg.
\newblock {\em {Tensor Analysis on Manifolds}}.
\newblock Dover Publications, 1980.

\bibitem{Robertson:1935zz}
Howard~P. Robertson.
\newblock {Kinematics and World-Structure}.
\newblock {\em Astrophys. J.}, 82:284--301, 1935.
\newblock \href {https://doi.org/10.1086/143681} {\path{doi:10.1086/143681}}.

\bibitem{Robertson:1935jpx}
Howard~P. Robertson.
\newblock {Kinematics and World-Structure. II}.
\newblock {\em Astrophys. J.}, 83:187--201, 1935.
\newblock \href {https://doi.org/10.1086/143716} {\path{doi:10.1086/143716}}.

\bibitem{Robertson:1936zz}
Howard~P. Robertson.
\newblock {Kinematics and World-Structure. III}.
\newblock {\em Astrophys. J.}, 83:257--271, 1936.
\newblock \href {https://doi.org/10.1086/143726} {\path{doi:10.1086/143726}}.

\bibitem{Walker:1937}
Arthur~G. Walker.
\newblock {On Milne's Theory of World-Structure}.
\newblock {\em Proceedings of the London Mathematical Society}, s2-42(1):90--127, 1937.
\newblock \href {https://doi.org/10.1112/plms/s2-42.1.90} {\path{doi:10.1112/plms/s2-42.1.90}}.

\bibitem{Einstein:1916vd}
Albert Einstein.
\newblock {The foundation of the general theory of relativity}.
\newblock {\em Annalen Phys.}, 49(7):769--822, 1916.
\newblock \href {https://doi.org/10.1002/andp.19163540702} {\path{doi:10.1002/andp.19163540702}}.

\bibitem{Friedman:1922kd}
Alexander Friedman.
\newblock {On the Curvature of space}.
\newblock {\em Z. Phys.}, 10:377--386, 1922.
\newblock \href {https://doi.org/10.1007/BF01332580} {\path{doi:10.1007/BF01332580}}.

\bibitem{Peebles:1982ff}
Phillip J.~E. Peebles.
\newblock {Large scale background temperature and mass fluctuations due to scale invariant primeval perturbations}.
\newblock {\em Astrophys. J. Lett.}, 263:L1--L5, 1982.
\newblock \href {https://doi.org/10.1086/183911} {\path{doi:10.1086/183911}}.

\bibitem{Planck:2018parameters}
N.~Aghanim et~al.
\newblock {Planck 2018 results. VI. Cosmological parameters}.
\newblock {\em Astron. Astrophys.}, 641:A6, 2020.
\newblock [Erratum: Astron.Astrophys. 652, C4 (2021)].
\newblock \href {https://doi.org/10.1051/0004-6361/201833910} {\path{doi:10.1051/0004-6361/201833910}}.

\bibitem{Brout:2022vxf}
Dillon Brout et~al.
\newblock {The Pantheon+ Analysis: Cosmological Constraints}.
\newblock {\em Astrophys. J.}, 938(2):110, 2022.
\newblock \href {https://doi.org/10.3847/1538-4357/ac8e04} {\path{doi:10.3847/1538-4357/ac8e04}}.

\bibitem{DES:2021wwk}
T.~M.~C. Abbott et~al.
\newblock {Dark Energy Survey Year 3 results: Cosmological constraints from galaxy clustering and weak lensing}.
\newblock {\em Phys. Rev. D}, 105(2):023520, 2022.
\newblock \href {https://doi.org/10.1103/PhysRevD.105.023520} {\path{doi:10.1103/PhysRevD.105.023520}}.

\bibitem{SupernovaSearchTeam:1998fmf}
Adam~G. Riess et~al.
\newblock {Observational evidence from supernovae for an accelerating universe and a cosmological constant}.
\newblock {\em Astron. J.}, 116:1009--1038, 1998.
\newblock \href {https://doi.org/10.1086/300499} {\path{doi:10.1086/300499}}.

\bibitem{SupernovaCosmologyProject:1998vns}
S.~Perlmutter et~al.
\newblock {Measurements of $\Omega$ and $\Lambda$ from 42 high redshift supernovae}.
\newblock {\em Astrophys. J.}, 517:565--586, 1999.
\newblock \href {https://doi.org/10.1086/307221} {\path{doi:10.1086/307221}}.

\bibitem{Weinberg2008}
Steven Weinberg.
\newblock {\em {Cosmology}}.
\newblock Oxford University Press, 2008.
\newblock \href {https://doi.org/10.1007/s10714-008-0728-z} {\path{doi:10.1007/s10714-008-0728-z}}.

\bibitem{BBKS_1983}
J.~R. {Bond} and A.~S. {Szalay}.
\newblock {The collisionless damping of density fluctuations in an expanding universe}.
\newblock {\em Astrophys. J.}, 274:443--468, 1983.
\newblock \href {https://doi.org/10.1086/161460} {\path{doi:10.1086/161460}}.

\bibitem{BBKS_1984}
J.~R. {Bond} and G.~{Efstathiou}.
\newblock {Cosmic background radiation anisotropies in universes dominated by nonbaryonic dark matter}.
\newblock {\em Astrophys. J. Lett.}, 285:L45--L48, 1984.
\newblock \href {https://doi.org/10.1086/184362} {\path{doi:10.1086/184362}}.

\bibitem{Eisenstein_1998}
Daniel~J. Eisenstein and Wayne Hu.
\newblock {Baryonic Features in the Matter Transfer Function}.
\newblock {\em Astrophys. J.}, 496(2):605–614, 1998.
\newblock \href {https://doi.org/10.1086/305424} {\path{doi:10.1086/305424}}.

\bibitem{Eisenstein_2005}
{Detection of the Baryon Acoustic Peak in the Large‐Scale Correlation Function of SDSS Luminous Red Galaxies}.
\newblock {\em Astrophys. J.}, 633(2):560–574, 2005.
\newblock \href {https://doi.org/10.1086/466512} {\path{doi:10.1086/466512}}.

\bibitem{Cole_2005}
{The 2dF Galaxy Redshift Survey: power-spectrum analysis of the final data set and cosmological implications}.
\newblock {\em Mon. Not. R. Astron. Soc.}, 362(2):505–534, 2005.
\newblock \href {https://doi.org/10.1111/j.1365-2966.2005.09318.x} {\path{doi:10.1111/j.1365-2966.2005.09318.x}}.

\bibitem{Silk_1968}
Joseph {Silk}.
\newblock {Cosmic Black-Body Radiation and Galaxy Formation}.
\newblock {\em Astrophys. J.}, 151:459, 1968.
\newblock \href {https://doi.org/10.1086/149449} {\path{doi:10.1086/149449}}.

\bibitem{Orjuela-Quintana:2022nnq}
J.~Bayron Orjuela-Quintana, Savvas Nesseris, and Wilmar Cardona.
\newblock {Using machine learning to compress the matter transfer function T(k)}.
\newblock {\em Phys. Rev. D}, 107(8):083520, 2023.
\newblock \href {https://doi.org/10.1103/PhysRevD.107.083520} {\path{doi:10.1103/PhysRevD.107.083520}}.

\bibitem{Dodelson2003}
Scott Dodelson.
\newblock {\em {Modern Cosmology}}.
\newblock Academic Press, Elsevier Science, 2003.
\newblock \href {https://doi.org/10.1016/C2017-0-01943-2} {\path{doi:10.1016/C2017-0-01943-2}}.

\bibitem{Hand:2017pqn}
Nick Hand, Yu~Feng, Florian Beutler, Yin Li, Chirag Modi, Uros Seljak, and Zachary Slepian.
\newblock {nbodykit: an open-source, massively parallel toolkit for large-scale structure}.
\newblock {\em Astron. J.}, 156(4):160, 2018.
\newblock \href {https://doi.org/10.3847/1538-3881/aadae0} {\path{doi:10.3847/1538-3881/aadae0}}.

\bibitem{Bonjean_2020}
V.~Bonjean, N.~Aghanim, M.~Douspis, N.~Malavasi, and H.~Tanimura.
\newblock Filament profiles from wisexscos galaxies as probes of the impact of environmental effects.
\newblock {\em Astron. Astrophys.}, 638:A75, 2020.
\newblock \href {https://doi.org/10.1051/0004-6361/201937313} {\path{doi:10.1051/0004-6361/201937313}}.

\bibitem{Kuchner_2020}
Ulrike Kuchner, Alfonso Aragón-Salamanca, Frazer~R Pearce, Meghan~E Gray, Agustín Rost, Chunliang Mu, Charlotte Welker, Weiguang Cui, Roan Haggar, Clotilde Laigle, Alexander Knebe, Katarina Kraljic, Florian Sarron, and Gustavo Yepes.
\newblock {Mapping and characterization of cosmic filaments in galaxy cluster outskirts: strategies and forecasts for observations from simulations}.
\newblock {\em Mon. Not. R. Astron. Soc.}, 494(4):5473–5491, 2020.
\newblock \href {https://doi.org/10.1093/mnras/staa1083} {\path{doi:10.1093/mnras/staa1083}}.

\bibitem{Hansen_2005}
Sarah~M. Hansen, Timothy~A. McKay, Risa~H. Wechsler, James Annis, Erin~Scott Sheldon, and Amy Kimball.
\newblock Measurement of galaxy cluster sizes, radial profiles, and luminosity functions from sdss photometric data.
\newblock {\em Astrophys. J.}, 633(1):122–137, 2005.
\newblock \href {https://doi.org/10.1086/444554} {\path{doi:10.1086/444554}}.

\bibitem{Philcox_2022}
Oliver H.~E. Philcox and Mikhail~M. Ivanov.
\newblock {The BOSS DR12 Full-Shape Cosmology: $\Lambda$CDM Constraints from the Large-Scale Galaxy Power Spectrum and Bispectrum Monopole}.
\newblock {\em Phys. Rev. D}, 105(4), 2022.
\newblock \href {https://doi.org/10.1103/physrevd.105.043517} {\path{doi:10.1103/physrevd.105.043517}}.

\bibitem{Chen_2022}
Shi-Fan Chen, Martin White, Joseph DeRose, and Nickolas Kokron.
\newblock {Cosmological analysis of three-dimensional BOSS galaxy clustering and Planck CMB lensing cross correlations via Lagrangian perturbation theory}.
\newblock {\em J. Cosmol. Astropart. Phys.}, 2022(07):041, 2022.
\newblock \href {https://doi.org/10.1088/1475-7516/2022/07/041} {\path{doi:10.1088/1475-7516/2022/07/041}}.

\bibitem{Sefusatti_2006}
Emiliano Sefusatti, Martín Crocce, Sebastián Pueblas, and Román Scoccimarro.
\newblock {Cosmology and the bispectrum}.
\newblock {\em Phys. Rev. D}, 74(2), 2006.
\newblock \href {https://doi.org/10.1103/physrevd.74.023522} {\path{doi:10.1103/physrevd.74.023522}}.

\bibitem{Kilbinger_2005}
M.~Kilbinger and P.~Schneider.
\newblock {Cosmological parameters from combined second- and third-order aperture mass statistics of cosmic shear}.
\newblock {\em Astron. Astrophys.}, 442(1):69–83, 2005.
\newblock \href {https://doi.org/10.1051/0004-6361:20053531} {\path{doi:10.1051/0004-6361:20053531}}.

\bibitem{Scoccimarro_1999}
Roman Scoccimarro, H.~M.~P. Couchman, and Joshua~A. Frieman.
\newblock {The Bispectrum as a Signature of Gravitational Instability in Redshift Space}.
\newblock {\em Astrophys. J.}, 517(2):531–540, 1999.
\newblock URL: \url{http://dx.doi.org/10.1086/307220}, \href {https://doi.org/10.1086/307220} {\path{doi:10.1086/307220}}.

\bibitem{Cabass_2023}
Giovanni Cabass, Sadra Jazayeri, Enrico Pajer, and David Stefanyszyn.
\newblock Parity violation in the scalar trispectrum: no-go theorems and yes-go examples.
\newblock {\em Journal of High Energy Physics}, 2023(2), 2023.
\newblock \href {https://doi.org/10.1007/jhep02(2023)021} {\path{doi:10.1007/jhep02(2023)021}}.

\bibitem{Coulton:2023oug}
William~R. Coulton, Oliver H.~E. Philcox, and Francisco Villaescusa-Navarro.
\newblock {Signatures of a Parity-Violating Universe}.
\newblock 2023.
\newblock \href {https://arxiv.org/abs/2306.11782} {\path{arXiv:2306.11782}}.

\bibitem{Carrasco:2012cv}
John Joseph~M. Carrasco, Mark~P. Hertzberg, and Leonardo Senatore.
\newblock {The Effective Field Theory of Cosmological Large Scale Structures}.
\newblock {\em J. High Energy Phys.}, 09:082, 2012.
\newblock \href {https://doi.org/10.1007/JHEP09(2012)082} {\path{doi:10.1007/JHEP09(2012)082}}.

\bibitem{Buchert:1997dr}
Thomas Buchert and Alvaro Dominguez.
\newblock {Modeling multistream flow in collisionless matter: approximations for large scale structure beyond shell crossing}.
\newblock {\em Astron. Astrophys.}, 335:395--402, 1998.
\newblock \href {https://arxiv.org/abs/astro-ph/9702139} {\path{arXiv:astro-ph/9702139}}.

\bibitem{Jelic_Cizmek_2018}
Goran Jelic-Cizmek, Francesca Lepori, Julian Adamek, and Ruth Durrer.
\newblock {The generation of vorticity in cosmological N-body simulations}.
\newblock {\em J. Cosmol. Astropart. Phys.}, 2018(09):006–006, 2018.
\newblock \href {https://doi.org/10.1088/1475-7516/2018/09/006} {\path{doi:10.1088/1475-7516/2018/09/006}}.

\bibitem{Wang_2021}
Peng Wang, Noam~I. Libeskind, Elmo Tempel, Xi~Kang, and Quan Guo.
\newblock Possible observational evidence for cosmic filament spin.
\newblock {\em Nature Astronomy}, 5(8):839–845, 2021.
\newblock \href {https://doi.org/10.1038/s41550-021-01380-6} {\path{doi:10.1038/s41550-021-01380-6}}.

\bibitem{Ma_1995}
Chung-Pei Ma and Edmund Bertschinger.
\newblock {Cosmological Perturbation Theory in the Synchronous and Conformal Newtonian Gauges}.
\newblock {\em Astrophys. J.}, 455:7, 1995.
\newblock \href {https://doi.org/10.1086/176550} {\path{doi:10.1086/176550}}.

\bibitem{Munshi_1994}
Dipak Munshi and Alexei~A. Starobinsky.
\newblock {Nonlinear approximations to gravitational instability: A comparison in second-order perturbation theory}.
\newblock {\em Astrophys. J.}, 428:433, 1994.
\newblock \href {https://doi.org/10.1086/174255} {\path{doi:10.1086/174255}}.

\bibitem{Crocce:2006ve}
M.~Crocce, S.~Pueblas, and R.~Scoccimarro.
\newblock {Transients from Initial Conditions in Cosmological Simulations}.
\newblock {\em Mon. Not. R. Astron. Soc.}, 373:369--381, 2006.
\newblock \href {https://doi.org/10.1111/j.1365-2966.2006.11040.x} {\path{doi:10.1111/j.1365-2966.2006.11040.x}}.

\bibitem{Peebles:1980yev}
James Peebles.
\newblock {\em {The Large-Scale Structure of the Universe}}.
\newblock Princeton University Press, 1980.
\newblock \href {https://doi.org/10.2307/j.ctvxrpz4n} {\path{doi:10.2307/j.ctvxrpz4n}}.

\bibitem{Fry:1983cj}
James~N. Fry.
\newblock {The Galaxy correlation hierarchy in perturbation theory}.
\newblock {\em Astrophys. J.}, 279:499--510, 1984.
\newblock \href {https://doi.org/10.1086/161913} {\path{doi:10.1086/161913}}.

\bibitem{Goroff1986}
M.~H. {Goroff}, B.~{Grinstein}, S.~J. {Rey}, and M.~B. {Wise}.
\newblock {Coupling of modes of cosmological mass density fluctuations}.
\newblock {\em Astrophys. J.}, 311:6--14, 1986.
\newblock \href {https://doi.org/10.1086/164749} {\path{doi:10.1086/164749}}.

\bibitem{Chudaykin_2020}
Anton Chudaykin, Mikhail~M. Ivanov, Oliver H.~E. Philcox, and Marko Simonović.
\newblock {Nonlinear perturbation theory extension of the Boltzmann code CLASS}.
\newblock {\em Phys. Rev. D}, 102(6), 2020.
\newblock \href {https://doi.org/10.1103/physrevd.102.063533} {\path{doi:10.1103/physrevd.102.063533}}.

\bibitem{Takahashi_2012}
Ryuichi Takahashi, Masanori Sato, Takahiro Nishimichi, Atsushi Taruya, and Masamune Oguri.
\newblock {Revising the Halofit Model for the Nonlinear Matter Power Spectrum}.
\newblock {\em Astrophys. J.}, 761(2):152, 2012.
\newblock \href {https://doi.org/10.1088/0004-637x/761/2/152} {\path{doi:10.1088/0004-637x/761/2/152}}.

\bibitem{Diego_Blas_2011}
Diego Blas, Julien Lesgourgues, and Thomas Tram.
\newblock {The Cosmic Linear Anisotropy Solving System (CLASS) Part II: Approximation schemes}.
\newblock {\em J. Cosmol. Astropart. Phys.}, 2011(07):034–034, 2011.
\newblock \href {https://doi.org/10.1088/1475-7516/2011/07/034} {\path{doi:10.1088/1475-7516/2011/07/034}}.

\bibitem{Hand_2018}
Nick Hand, Yu~Feng, Florian Beutler, Yin Li, Chirag Modi, Uroš Seljak, and Zachary Slepian.
\newblock {nbodykit: An Open-source, Massively Parallel Toolkit for Large-scale Structure}.
\newblock {\em The Astronomical Journal}, 156(4):160, 2018.
\newblock \href {https://doi.org/10.3847/1538-3881/aadae0} {\path{doi:10.3847/1538-3881/aadae0}}.

\bibitem{tng2017_1}
Volker Springel, Rüdiger Pakmor, Annalisa Pillepich, Rainer Weinberger, Dylan Nelson, Lars Hernquist, Mark Vogelsberger, Shy Genel, Paul Torrey, Federico Marinacci, and Jill Naiman.
\newblock {First results from the IllustrisTNG simulations: matter and galaxy clustering}.
\newblock {\em Mon. Not. R. Astron. Soc.}, 475(1):676--698, 2017.
\newblock \href {https://doi.org/10.1093/mnras/stx3304} {\path{doi:10.1093/mnras/stx3304}}.

\bibitem{tng2017_2}
Annalisa Pillepich, Dylan Nelson, Lars Hernquist, Volker Springel, Rüdiger Pakmor, Paul Torrey, Rainer Weinberger, Shy Genel, Jill~P Naiman, Federico Marinacci, and Mark Vogelsberger.
\newblock {First results from the IllustrisTNG simulations: the stellar mass content of groups and clusters of galaxies}.
\newblock {\em Mon. Not. R. Astron. Soc.}, 475(1):648--675, 2017.
\newblock \href {https://doi.org/10.1093/mnras/stx3112} {\path{doi:10.1093/mnras/stx3112}}.

\bibitem{tng2017_3}
Jill~P Naiman, Annalisa Pillepich, Volker Springel, Enrico Ramirez-Ruiz, Paul Torrey, Mark Vogelsberger, Rüdiger Pakmor, Dylan Nelson, Federico Marinacci, Lars Hernquist, Rainer Weinberger, and Shy Genel.
\newblock {First results from the IllustrisTNG simulations: a tale of two elements – chemical evolution of magnesium and europium}.
\newblock {\em Mon. Not. R. Astron. Soc.}, 477(1):1206--1224, 2018.
\newblock \href {https://doi.org/10.1093/mnras/sty618} {\path{doi:10.1093/mnras/sty618}}.

\bibitem{tng2017_4}
Federico Marinacci, Mark Vogelsberger, Rüdiger Pakmor, Paul Torrey, Volker Springel, Lars Hernquist, Dylan Nelson, Rainer Weinberger, Annalisa Pillepich, Jill Naiman, and Shy Genel.
\newblock {First results from the IllustrisTNG simulations: radio haloes and magnetic fields}.
\newblock {\em Mon. Not. R. Astron. Soc.}, 480(4):5113--5139, 2018.
\newblock \href {https://doi.org/10.1093/mnras/sty2206} {\path{doi:10.1093/mnras/sty2206}}.

\bibitem{tng2017_5}
Dylan Nelson, Annalisa Pillepich, Volker Springel, Rainer Weinberger, Lars Hernquist, Rüdiger Pakmor, Shy Genel, Paul Torrey, Mark Vogelsberger, Guinevere Kauffmann, Federico Marinacci, and Jill Naiman.
\newblock {First results from the IllustrisTNG simulations: the galaxy colour bimodality}.
\newblock {\em Mon. Not. R. Astron. Soc.}, 475(1):624--647, 2017.
\newblock \href {https://doi.org/10.1093/mnras/stx3040} {\path{doi:10.1093/mnras/stx3040}}.

\bibitem{pylians}
Francisco Villaescusa-Navarro.
\newblock {Pylians}, 2020.
\newblock URL: \url{https://github.com/franciscovillaescusa/Pylians3}.

\bibitem{Scoccimarro:1997st}
Roman Scoccimarro, Stephane Colombi, James~N. Fry, Joshua~A. Frieman, Eric Hivon, and Adrian Melott.
\newblock {Nonlinear evolution of the bispectrum of cosmological perturbations}.
\newblock {\em Astrophys. J.}, 496:586, 1998.
\newblock \href {https://doi.org/10.1086/305399} {\path{doi:10.1086/305399}}.

\bibitem{Lesgourgues:2012uu}
Julien Lesgourgues and Sergio Pastor.
\newblock {Neutrino mass from Cosmology}.
\newblock {\em Adv. High Energy Phys.}, 2012:608515, 2012.
\newblock \href {https://doi.org/10.1155/2012/608515} {\path{doi:10.1155/2012/608515}}.

\bibitem{Nojiri_2011}
Shin’ichi Nojiri and Sergei~D. Odintsov.
\newblock {Unified cosmic history in modified gravity: from $F(R)$ theory to Lorentz non-invariant models}.
\newblock {\em Physics Reports}, 505(2–4):59–144, 2011.
\newblock \href {https://doi.org/10.1016/j.physrep.2011.04.001} {\path{doi:10.1016/j.physrep.2011.04.001}}.

\bibitem{Renyi:1961}
Alfr{\'e}d R{\'e}nyi.
\newblock On measures of entropy and information.
\newblock In {\em Proceedings of the Fourth Berkeley Symposium on Mathematical Statistics and Probability, Volume 1: Contributions to the Theory of Statistics}, volume~4, pages 547--562. University of California Press, 1961.

\bibitem{Kullback:1951}
S.~Kullback and R.~A. Leibler.
\newblock {On Information and Sufficiency}.
\newblock {\em The Annals of Mathematical Statistics}, 22(1):79 -- 86, 1951.
\newblock \href {https://doi.org/10.1214/aoms/1177729694} {\path{doi:10.1214/aoms/1177729694}}.

\bibitem{Springel_2010}
Volker Springel.
\newblock {E pur si muove: Galilean-invariant cosmological hydrodynamical simulations on a moving mesh}.
\newblock {\em Mon. Not. R. Astron. Soc.}, 401(2):791--851, 2010.
\newblock \href {https://doi.org/10.1111/j.1365-2966.2009.15715.x} {\path{doi:10.1111/j.1365-2966.2009.15715.x}}.

\bibitem{Feng:2016yqz}
Yu~Feng, Man-Yat Chu, Uros Seljak, and Patrick McDonald.
\newblock {FastPM: a new scheme for fast simulations of dark matter and haloes}.
\newblock {\em Mon. Not. R. Astron. Soc.}, 463(3):2273--2286, 2016.
\newblock \href {https://doi.org/10.1093/mnras/stw2123} {\path{doi:10.1093/mnras/stw2123}}.

\bibitem{gadget2}
Volker Springel.
\newblock {The cosmological simulation code gadget-2}.
\newblock {\em Mon. Not. R. Astron. Soc.}, 364(4):1105--1134, 2005.
\newblock \href {https://doi.org/10.1111/j.1365-2966.2005.09655.x} {\path{doi:10.1111/j.1365-2966.2005.09655.x}}.

\bibitem{Ronneberger2015}
Olaf Ronneberger, Philipp Fischer, and Thomas Brox.
\newblock {U-Net: Convolutional Networks for Biomedical Image Segmentation}.
\newblock In {\em Medical Image Computing and Computer-Assisted Intervention -- MICCAI 2015}, pages 234--241. Springer International Publishing, 2015.
\newblock \href {https://doi.org/10.1007/978-3-319-24574-4_28} {\path{doi:10.1007/978-3-319-24574-4_28}}.

\bibitem{he2015deep}
Kaiming He, Xiangyu Zhang, Shaoqing Ren, and Jian Sun.
\newblock Deep residual learning for image recognition.
\newblock In {\em CVPR}, 2016.
\newblock \href {https://arxiv.org/abs/1512.03385} {\path{arXiv:1512.03385}}.

\bibitem{Gupta:2018eev}
Arushi Gupta, Jos\'e Manuel~Zorrilla Matilla, Daniel Hsu, and Zolt\'an Haiman.
\newblock {Non-Gaussian information from weak lensing data via deep learning}.
\newblock {\em Phys. Rev. D}, 97(10):103515, 2018.
\newblock \href {https://doi.org/10.1103/PhysRevD.97.103515} {\path{doi:10.1103/PhysRevD.97.103515}}.

\bibitem{Ribli:2018kwb}
Dezs\H{o} Ribli, B\'alint~\'Armin Pataki, and Istv\'an Csabai.
\newblock {An improved cosmological parameter inference scheme motivated by deep learning}.
\newblock {\em Nature Astron.}, 3(1):93--98, 2019.
\newblock \href {https://doi.org/10.1038/s41550-018-0596-8} {\path{doi:10.1038/s41550-018-0596-8}}.

\bibitem{Fluri:2018hoy}
Janis Fluri, Tomasz Kacprzak, Alexandre Refregier, Adam Amara, Aurelien Lucchi, and Thomas Hofmann.
\newblock {Cosmological constraints from noisy convergence maps through deep learning}.
\newblock {\em Phys. Rev. D}, 98(12):123518, 2018.
\newblock \href {https://doi.org/10.1103/PhysRevD.98.123518} {\path{doi:10.1103/PhysRevD.98.123518}}.

\bibitem{ravanbakhsh2017estimating}
Siamak Ravanbakhsh, Junier Oliva, Sebastien Fromenteau, Layne~C. Price, Shirley Ho, Jeff Schneider, and Barnabas Poczos.
\newblock {Estimating Cosmological Parameters from the Dark Matter Distribution}, 2017.
\newblock \href {https://arxiv.org/abs/1711.02033} {\path{arXiv:1711.02033}}.

\bibitem{Mathuriya:2018luj}
Amrita Mathuriya et~al.
\newblock {CosmoFlow: Using Deep Learning to Learn the Universe at Scale}.
\newblock August 2018.
\newblock \href {https://arxiv.org/abs/1808.04728} {\path{arXiv:1808.04728}}.

\bibitem{Lazanu_2021}
Andrei Lazanu.
\newblock {Extracting cosmological parameters from N-body simulations using machine learning techniques}.
\newblock {\em J. Cosmol. Astropart. Phys.}, 2021(09):039, 2021.
\newblock \href {https://doi.org/10.1088/1475-7516/2021/09/039} {\path{doi:10.1088/1475-7516/2021/09/039}}.

\bibitem{Abdalla:2022yfr}
Elcio Abdalla et~al.
\newblock {Cosmology intertwined: A review of the particle physics, astrophysics, and cosmology associated with the cosmological tensions and anomalies}.
\newblock {\em {J. High Energy Astrophys.}}, 34:49--211, 2022.
\newblock \href {https://doi.org/10.1016/j.jheap.2022.04.002} {\path{doi:10.1016/j.jheap.2022.04.002}}.

\bibitem{Jeffrey:2020itg}
Niall Jeffrey and Benjamin~D. Wandelt.
\newblock {Solving high-dimensional parameter inference: marginal posterior densities \& Moment Networks}.
\newblock In {\em {NeurIPS}}, 2020.
\newblock \href {https://arxiv.org/abs/2011.05991} {\path{arXiv:2011.05991}}.

\bibitem{Villaescusa_Navarro_2022}
Villaescusa-Navarro et~al.
\newblock {The CAMELS Multifield Data Set: Learning the Universe’s Fundamental Parameters with Artificial Intelligence}.
\newblock {\em Astrophys. J. Suppl. Ser.}, 259(2):61, 2022.
\newblock \href {https://doi.org/10.3847/1538-4365/ac5ab0} {\path{doi:10.3847/1538-4365/ac5ab0}}.

\bibitem{yipinprogress}
Jacky H.~T. Yip, Adam Rouhiainen, and Gary Shiu.
\newblock {In progress}.
\newblock 2024.

\bibitem{Edelsbrunner2002}
Herbert Edelsbrunner, David Letscher, and Afra Zomorodian.
\newblock {Topological Persistence and Simplification}.
\newblock {\em Discrete \& Comput. Geom.}, 28:511–533, 2002.
\newblock \href {https://doi.org/10.1007/s00454-002-2885-2} {\path{doi:10.1007/s00454-002-2885-2}}.

\bibitem{Zomorodian2004}
Afra Zomorodian and Gunnar Carlsson.
\newblock {Computing Persistent Homology}.
\newblock {\em Discrete \& Comput. Geom.}, 33:249–274, 2005.
\newblock \href {https://doi.org/10.1007/s00454-004-1146-y} {\path{doi:10.1007/s00454-004-1146-y}}.

\bibitem{carlsson_2014}
Gunnar Carlsson.
\newblock {Topological pattern recognition for point cloud data}.
\newblock {\em Acta Numerica}, 23:289–368, 2014.
\newblock \href {https://doi.org/10.1017/S0962492914000051} {\path{doi:10.1017/S0962492914000051}}.

\bibitem{Cole_2018}
Alex Cole and Gary Shiu.
\newblock {Persistent homology and non-Gaussianity}.
\newblock {\em J. Cosmol. Astropart. Phys.}, 2018(03), 2018.
\newblock \href {https://doi.org/10.1088/1475-7516/2018/03/025} {\path{doi:10.1088/1475-7516/2018/03/025}}.

\bibitem{Biagetti_2021}
Matteo Biagetti, Alex Cole, and Gary Shiu.
\newblock {The persistence of large scale structures. Part I. Primordial non-Gaussianity}.
\newblock {\em J. Cosmol. Astropart. Phys.}, 2021(04), 2021.
\newblock \href {https://doi.org/10.1088/1475-7516/2021/04/061} {\path{doi:10.1088/1475-7516/2021/04/061}}.

\bibitem{Biagetti_2022}
Matteo Biagetti, Juan Calles, Lina Castiblanco, Alex Cole, and Jorge Noreña.
\newblock Fisher forecasts for primordial non-gaussianity from persistent homology.
\newblock {\em J. Cosmol. Astropart. Phys.}, 2022(10), 2022.
\newblock \href {https://doi.org/10.1088/1475-7516/2022/10/002} {\path{doi:10.1088/1475-7516/2022/10/002}}.

\bibitem{Villaescusa-Navarro:2019bje}
Villaescusa-Navarro et~al.
\newblock {The Quijote Simulations}.
\newblock {\em Astrophys. J. Suppl. Ser.}, 250(1):2, 2020.
\newblock \href {https://doi.org/10.3847/1538-4365/ab9d82} {\path{doi:10.3847/1538-4365/ab9d82}}.

\bibitem{yip2024}
Jacky H.~T. Yip, Matteo Biagetti, Alex Cole, et~al.
\newblock {Cosmology with persistent homology: A Fisher forecast}.
\newblock {\em To appear}, 2024.

\bibitem{Li_2023}
Shun-Sheng Li, Henk Hoekstra, Konrad Kuijken, Marika Asgari, Maciej Bilicki, Benjamin Giblin, Catherine Heymans, Hendrik Hildebrandt, Benjamin Joachimi, Lance Miller, Jan~Luca van~den Busch, Angus~H. Wright, Arun Kannawadi, Robert Reischke, and HuanYuan Shan.
\newblock {KiDS-1000: Cosmology with improved cosmic shear measurements}.
\newblock {\em Astron. Astrophys.}, 679:A133, 2023.
\newblock URL: \url{http://dx.doi.org/10.1051/0004-6361/202347236}, \href {https://doi.org/10.1051/0004-6361/202347236} {\path{doi:10.1051/0004-6361/202347236}}.

\bibitem{kingma2022autoencoding}
Diederik~P Kingma and Max Welling.
\newblock Auto-encoding variational bayes.
\newblock In {\em ICLR}, 2013.
\newblock \href {https://arxiv.org/abs/1312.6114} {\path{arXiv:1312.6114}}.

\bibitem{NEURIPS2019_3001ef25}
Yang Song and Stefano Ermon.
\newblock {Generative Modeling by Estimating Gradients of the Data Distribution}.
\newblock In {\em NeurIPS}, 2019.
\newblock \href {https://arxiv.org/abs/1907.05600} {\path{arXiv:1907.05600}}.

\bibitem{song2020improved}
Yang Song and Stefano Ermon.
\newblock {Improved Techniques for Training Score-Based Generative Models}.
\newblock In {\em NeurIPS}, 2020.
\newblock \href {https://arxiv.org/abs/2006.09011} {\path{arXiv:2006.09011}}.

\bibitem{score_matching_2020}
Benjamin Remy et~al.
\newblock {Probabilistic Mapping of Dark Matter by Neural Score Matching}.
\newblock In {\em Machine Learning and the Physical Sciences at NeurIPS}, 2019.
\newblock \href {https://arxiv.org/abs/2011.08271} {\path{arXiv:2011.08271}}.

\bibitem{2019arXiv191202762P}
George Papamakarios, Eric Nalisnick, Danilo~Jimenez Rezende, Shakir Mohamed, and Balaji Lakshminarayanan.
\newblock {Normalizing Flows for Probabilistic Modeling and Inference}.
\newblock {\em J. Mach. Learn. Res.}, 22(57):1--64, 2021.
\newblock URL: \url{http://jmlr.org/papers/v22/19-1028.html}.

\bibitem{2018arXiv180703039K}
Diederik~P. Kingma and Prafulla Dhariwal.
\newblock {Glow: Generative Flow with Invertible $1\times1$ Convolutions}.
\newblock In {\em NeurIPS}, 2018.
\newblock \href {https://arxiv.org/abs/1807.03039} {\path{arXiv:1807.03039}}.

\bibitem{2016arXiv160508803D}
Laurent Dinh, Jascha Sohl-Dickstein, and Samy Bengio.
\newblock {Density estimation using Real NVP}.
\newblock In {\em ICLR}, 2017.
\newblock \href {https://arxiv.org/abs/1605.08803} {\path{arXiv:1605.08803}}.

\bibitem{rezende2016variational}
Danilo~Jimenez Rezende and Shakir Mohamed.
\newblock Variational inference with normalizing flows.
\newblock In {\em ICML}, 2015.
\newblock \href {https://arxiv.org/abs/1505.05770} {\path{arXiv:1505.05770}}.

\bibitem{2017arXiv170507057P}
George Papamakarios, Theo Pavlakou, and Iain Murray.
\newblock {Masked Autoregressive Flow for Density Estimation}.
\newblock In {\em NeurIPS}, 2017.
\newblock \href {https://arxiv.org/abs/1705.07057} {\path{arXiv:1705.07057}}.

\bibitem{2019arXiv191101429C}
Kyle Cranmer, Johann Brehmer, and Gilles Louppe.
\newblock {The frontier of simulation-based inference}.
\newblock {\em Proc. Natl. Acad. Sci. U.S.A.}, 117(48):30055--30062, 2020.
\newblock \href {https://doi.org/10.1073/pnas.1912789117} {\path{doi:10.1073/pnas.1912789117}}.

\bibitem{Alsing:2019xrx}
Justin Alsing, Tom Charnock, Stephen Feeney, and Benjamin Wandelt.
\newblock {Fast likelihood-free cosmology with neural density estimators and active learning}.
\newblock {\em Mon. Not. R. Astron. Soc.}, 488(3):4440--4458, 2019.
\newblock \href {https://doi.org/10.1093/mnras/stz1960} {\path{doi:10.1093/mnras/stz1960}}.

\bibitem{DiazRivero:2020oai}
Ana Diaz~Rivero and Cora Dvorkin.
\newblock {Flow-Based Likelihoods for Non-Gaussian Inference}.
\newblock {\em Phys. Rev. D}, 102(10):103507, 2020.
\newblock \href {https://doi.org/10.1103/PhysRevD.102.103507} {\path{doi:10.1103/PhysRevD.102.103507}}.

\bibitem{Hyvarinen:1999}
Aapo Hyvärinen and Petteri Pajunen.
\newblock {Nonlinear independent component analysis: Existence and uniqueness results}.
\newblock {\em Neural Networks}, 12(3):429--439, 1999.
\newblock \href {https://doi.org/10.1016/S0893-6080(98)00140-3} {\path{doi:10.1016/S0893-6080(98)00140-3}}.

\bibitem{Lewis:2006fu}
Antony Lewis and Anthony Challinor.
\newblock {Weak gravitational lensing of the CMB}.
\newblock {\em Phys. Rept.}, 429:1--65, 2006.
\newblock \href {https://doi.org/10.1016/j.physrep.2006.03.002} {\path{doi:10.1016/j.physrep.2006.03.002}}.

\bibitem{2016arXiv160100670B}
Alp~Kucukelbir David M.~Blei and Jon~D. McAuliffe.
\newblock {Variational Inference: A Review for Statisticians}.
\newblock {\em J. Am. Stat. Assoc.}, 112(518):859--877, 2017.
\newblock \href {https://doi.org/10.1080/01621459.2017.1285773} {\path{doi:10.1080/01621459.2017.1285773}}.

\bibitem{Carron:2017mqf}
Julien Carron and Antony Lewis.
\newblock {Maximum a posteriori CMB lensing reconstruction}.
\newblock {\em Phys. Rev. D}, 96(6):063510, 2017.
\newblock \href {https://doi.org/10.1103/PhysRevD.96.063510} {\path{doi:10.1103/PhysRevD.96.063510}}.

\bibitem{Millea:2020iuw}
M.~Millea et~al.
\newblock {Optimal Cosmic Microwave Background Lensing Reconstruction and Parameter Estimation with SPTpol Data}.
\newblock {\em Astrophys. J.}, 922(2):259, 2021.
\newblock \href {https://doi.org/10.3847/1538-4357/ac02bb} {\path{doi:10.3847/1538-4357/ac02bb}}.

\bibitem{2014arXiv1410.6460S}
Tim {Salimans}, Diederik~P. {Kingma}, and Max {Welling}.
\newblock {Markov Chain Monte Carlo and Variational Inference: Bridging the Gap}.
\newblock {\em arXiv e-prints}, page arXiv:1410.6460, 2014.
\newblock \href {https://arxiv.org/abs/1410.6460} {\path{arXiv:1410.6460}}.

\bibitem{Thorne:2021nux}
Ben Thorne, Lloyd Knox, and Karthik Prabhu.
\newblock {A generative model of galactic dust emission using variational autoencoders}.
\newblock {\em Mon. Not. Roy. Astron. Soc.}, 504(2):2603--2613, 2021.
\newblock \href {https://doi.org/10.1093/mnras/stab1011} {\path{doi:10.1093/mnras/stab1011}}.

\bibitem{2014arXiv14108516D}
Laurent Dinh, David Krueger, and Yoshua Bengio.
\newblock Nice: Non-linear independent components estimation.
\newblock In {\em ICLR}, 2014.
\newblock \href {https://arxiv.org/abs/1410.8516} {\path{arXiv:1410.8516}}.

\bibitem{Albergo:2021vyo}
Michael~S. Albergo, Denis Boyda, Daniel~C. Hackett, Gurtej Kanwar, Kyle Cranmer, S\'ebastien Racani\`ere, Danilo~Jimenez Rezende, and Phiala~E. Shanahan.
\newblock {Introduction to Normalizing Flows for Lattice Field Theory}.
\newblock 2021.
\newblock \href {https://arxiv.org/abs/2101.08176} {\path{arXiv:2101.08176}}.

\bibitem{NEURIPS2019_9015}
Adam Paszke et~al.
\newblock {PyTorch: An Imperative Style, High-Performance Deep Learning Library}.
\newblock In {\em NeurIPS}. 2019.
\newblock \href {https://arxiv.org/abs/1912.01703} {\path{arXiv:1912.01703}}.

\bibitem{trenf}
Biwei Dai and Uroš Seljak.
\newblock {Translation and rotation equivariant normalizing flow (TRENF) for optimal cosmological analysis}.
\newblock {\em Mon. Not. R. Astron. Soc.}, 516(2):2363--2373, 2022.
\newblock \href {https://doi.org/10.1093/mnras/stac2010} {\path{doi:10.1093/mnras/stac2010}}.

\bibitem{Komatsu:2001rj}
Eiichiro Komatsu and David~N. Spergel.
\newblock {Acoustic signatures in the primary microwave background bispectrum}.
\newblock {\em Phys. Rev. D}, 63:063002, 2001.
\newblock \href {https://doi.org/10.1103/PhysRevD.63.063002} {\path{doi:10.1103/PhysRevD.63.063002}}.

\bibitem{Byrnes:2010em}
Christian~T. Byrnes and Ki-Young Choi.
\newblock {Review of local non-Gaussianity from multi-field inflation}.
\newblock {\em Adv. Astron.}, 2010:724525, 2010.
\newblock \href {https://doi.org/10.1155/2010/724525} {\path{doi:10.1155/2010/724525}}.

\bibitem{1981csup.book.....H}
R.~W. {Hockney} and J.~W. {Eastwood}.
\newblock {\em {Computer Simulation Using Particles}}.
\newblock 1981.

\bibitem{pmlr-v97-hoogeboom19a}
Emiel Hoogeboom, Rianne Van Den~Berg, and Max Welling.
\newblock {Emerging Convolutions for Generative Normalizing Flows}.
\newblock In {\em Theoretical Foundations and Applications of Deep Generative Models at ICML}, 2019.
\newblock \href {https://arxiv.org/abs/1901.11137} {\path{arXiv:1901.11137}}.

\bibitem{GlowGithub}
Joost van Amersfoort.
\newblock Glow-pytorch.
\newblock \url{https://github.com/y0ast/Glow-PyTorch}, 2019.

\bibitem{Deutsch:2017ybc}
Anne-Sylvie Deutsch, Emanuela Dimastrogiovanni, Matthew~C. Johnson, Moritz M\"unchmeyer, and Alexandra Terrana.
\newblock {Reconstruction of the remote dipole and quadrupole fields from the kinetic Sunyaev Zel\textquoteright{}dovich and polarized Sunyaev Zel\textquoteright{}dovich effects}.
\newblock {\em Phys. Rev. D}, 98(12):123501, 2018.
\newblock \href {https://doi.org/10.1103/PhysRevD.98.123501} {\path{doi:10.1103/PhysRevD.98.123501}}.

\bibitem{Smith:2018bpn}
Kendrick~M. Smith, Mathew~S. Madhavacheril, Moritz M\"unchmeyer, Simone Ferraro, Utkarsh Giri, and Matthew~C. Johnson.
\newblock {KSZ tomography and the bispectrum}.
\newblock 2018.
\newblock \href {https://arxiv.org/abs/1810.13423} {\path{arXiv:1810.13423}}.

\bibitem{hypernetworks}
David Ha, Andrew Dai, and Quoc~V. Le.
\newblock Hypernetworks.
\newblock 2017.
\newblock URL: \url{https://openreview.net/pdf?id=rkpACe1lx}.

\bibitem{SDSS:2008tqn}
Kevork~N. Abazajian et~al.
\newblock {The Seventh Data Release of the Sloan Digital Sky Survey}.
\newblock {\em Astrophys. J. Suppl.}, 182:543--558, 2009.
\newblock \href {https://doi.org/10.1088/0067-0049/182/2/543} {\path{doi:10.1088/0067-0049/182/2/543}}.

\bibitem{germain2015made}
Mathieu Germain, Karol Gregor, Iain Murray, and Hugo Larochelle.
\newblock {MADE: Masked Autoencoder for Distribution Estimation}.
\newblock In {\em ICML}, 2015.
\newblock \href {https://arxiv.org/abs/1502.03509} {\path{arXiv:1502.03509}}.

\bibitem{Hassan:2021ymv}
Sultan Hassan et~al.
\newblock {HIFlow: Generating Diverse HI Maps Conditioned on Cosmology using Normalizing Flow}.
\newblock 2021.
\newblock \href {https://arxiv.org/abs/2110.02983} {\path{arXiv:2110.02983}}.

\bibitem{Dai:2023lcb}
Biwei Dai and Uros Seljak.
\newblock {Multiscale Flow for Robust and Optimal Cosmological Analysis}.
\newblock In {\em Machine Learning for Astrophysics at ICML}, 2023.
\newblock \href {https://arxiv.org/abs/2306.04689} {\path{arXiv:2306.04689}}.

\bibitem{Huang_2019}
Hung-Jin Huang, Tim Eifler, Rachel Mandelbaum, and Scott Dodelson.
\newblock {Modelling baryonic physics in future weak lensing surveys}.
\newblock {\em Mon. Not. R. Astron. Soc.}, 488(2):1652–1678, 2019.
\newblock URL: \url{http://dx.doi.org/10.1093/mnras/stz1714}, \href {https://doi.org/10.1093/mnras/stz1714} {\path{doi:10.1093/mnras/stz1714}}.

\bibitem{camels2021}
Francisco Villaescusa-Navarro et~al.
\newblock {The CAMELS Project: Cosmology and Astrophysics with Machine-learning Simulations}.
\newblock {\em Astrophys. J.}, 915(1):71, 2021.
\newblock \href {https://doi.org/10.3847/1538-4357/abf7ba} {\path{doi:10.3847/1538-4357/abf7ba}}.

\bibitem{score_matching_2022}
B.~Remy, F.~Lanusse, N.~Jeffrey, J.~Liu, J.-L. Starck, K.~Osato, and T.~Schrabback.
\newblock Probabilistic mass-mapping with neural score estimation.
\newblock {\em Astron. Astrophys.}, 672:A51, 2023.
\newblock \href {https://doi.org/10.1051/0004-6361/202243054} {\path{doi:10.1051/0004-6361/202243054}}.

\bibitem{Carlstrom_2002}
John~E. Carlstrom, Gilbert~P. Holder, and Erik~D. Reese.
\newblock {Cosmology with the Sunyaev-Zel'dovich Effect}.
\newblock {\em Annual Review of Astron. Astrophys.}, 40(1):643--680, September 2002.
\newblock \href {https://doi.org/10.1146/annurev.astro.40.060401.093803} {\path{doi:10.1146/annurev.astro.40.060401.093803}}.

\bibitem{Akaike:1974}
H.~Akaike.
\newblock A new look at the statistical model identification.
\newblock {\em IEEE Transactions on Automatic Control}, 19(6):716--723, 1974.
\newblock \href {https://doi.org/10.1109/TAC.1974.1100705} {\path{doi:10.1109/TAC.1974.1100705}}.

\bibitem{Munchmeyer:2019kng}
Moritz M\"unchmeyer and Kendrick~M. Smith.
\newblock {Fast Wiener filtering of CMB maps with Neural Networks}.
\newblock In {\em Machine Learning and the Physical Sciences at NeurIPS}, 2019.
\newblock \href {https://arxiv.org/abs/1905.05846} {\path{arXiv:1905.05846}}.

\bibitem{LSSTSciBook}
Paul~A. Abell et~al.
\newblock {LSST Science Book, Version 2.0}.
\newblock 2009.
\newblock \href {https://arxiv.org/abs/0912.0201} {\path{arXiv:0912.0201}}.

\bibitem{simons_observatory}
Peter Ade et~al.
\newblock {The Simons Observatory: science goals and forecasts}.
\newblock {\em J. Cosmol. Astropart. Phys.}, 2019(02):056--056, 2019.
\newblock \href {https://doi.org/10.1088/1475-7516/2019/02/056} {\path{doi:10.1088/1475-7516/2019/02/056}}.

\bibitem{nf_ood}
Polina Kirichenko, Pavel Izmailov, and Andrew~Gordon Wilson.
\newblock Why normalizing flows fail to detect out-of-distribution data.
\newblock In {\em NeurIPS}, 2020.
\newblock \href {https://arxiv.org/abs/2006.08545} {\path{arXiv:2006.08545}}.

\bibitem{hamiltorch}
Adam~D. Cobb, Atılım~G. Baydin, and Brian Jalaian.
\newblock hamiltorch, 2019.
\newblock URL: \url{https://github.com/AdamCobb/hamiltorch}.

\bibitem{NUTSHMC}
Matthew~D. Hoffman and Andrew Gelman.
\newblock {The No-U-Turn Sampler: Adaptively Setting Path Lengths in Hamiltonian Monte Carlo}.
\newblock {\em J. Mach. Learn. Res.}, 15(47):1593--1623, 2014.
\newblock URL: \url{http://jmlr.org/papers/v15/hoffman14a.html}.

\bibitem{reconstructVBS}
Chirag Modi, Yin Li, and David Blei.
\newblock Reconstructing the universe with variational self-boosted sampling.
\newblock {\em J. Cosmol. Astropart. Phys.}, 2023(03):059, 2023.
\newblock \href {https://doi.org/10.1088/1475-7516/2023/03/059} {\path{doi:10.1088/1475-7516/2023/03/059}}.

\bibitem{neal_mcmc}
Radford~M. Neal.
\newblock {MCMC using Hamiltonian dynamics}.
\newblock In Steve Brooks, Andrew Gelman, Galin Jones, and Xiao-Li Meng, editors, {\em Handbook of Markov Chain Monte Carlo}, chapter~5. Chapman and Hall/{CRC}, 2011.
\newblock \href {https://doi.org/10.1201/b10905} {\path{doi:10.1201/b10905}}.

\bibitem{2009PhRvL.103i1303S}
Uro{\v{s}} {Seljak}, Nico {Hamaus}, and Vincent {Desjacques}.
\newblock {How to Suppress the Shot Noise in Galaxy Surveys}.
\newblock {\em Phys. Rev. Lett.}, 103(9):091303, 2009.
\newblock \href {https://arxiv.org/abs/0904.2963} {\path{arXiv:0904.2963}}, \href {https://doi.org/10.1103/PhysRevLett.103.091303} {\path{doi:10.1103/PhysRevLett.103.091303}}.

\bibitem{Schmittfull:2017ffw}
Marcel Schmittfull and Uros Seljak.
\newblock {Parameter constraints from cross-correlation of CMB lensing with galaxy clustering}.
\newblock {\em Phys. Rev. D}, 97(12):123540, 2018.
\newblock \href {https://doi.org/10.1103/PhysRevD.97.123540} {\path{doi:10.1103/PhysRevD.97.123540}}.

\bibitem{Munchmeyer:2018eey}
Moritz M\"unchmeyer, Mathew~S. Madhavacheril, Simone Ferraro, Matthew~C. Johnson, and Kendrick~M. Smith.
\newblock {Constraining local non-Gaussianities with kinetic Sunyaev-Zel\textquoteright{}dovich tomography}.
\newblock {\em Phys. Rev. D}, 100(8):083508, 2019.
\newblock \href {https://doi.org/10.1103/PhysRevD.100.083508} {\path{doi:10.1103/PhysRevD.100.083508}}.

\bibitem{sohl2015}
Jascha Sohl-Dickstein, Eric Weiss, Niru Maheswaranathan, and Surya Ganguli.
\newblock {Deep Unsupervised Learning using Nonequilibrium Thermodynamics}.
\newblock In {\em {ICML}}, 2015.
\newblock \href {https://arxiv.org/abs/1503.03585} {\path{arXiv:1503.03585}}.

\bibitem{ho2020denoising}
Jonathan Ho, Ajay Jain, and Pieter Abbeel.
\newblock {Denoising Diffusion Probabilistic Models}.
\newblock In {\em NeurIPS}, 2020.
\newblock \href {https://arxiv.org/abs/2006.11239} {\path{arXiv:2006.11239}}.

\bibitem{palette}
Chitwan Saharia, William Chan, Huiwen Chang, Chris Lee, Jonathan Ho, Tim Salimans, David Fleet, and Mohammad Norouzi.
\newblock {Palette: Image-to-Image Diffusion Models}.
\newblock In {\em ACM SIGGRAPH 2022 Conference Proceedings}, 2022.
\newblock \href {https://doi.org/10.1145/3528233.3530757} {\path{doi:10.1145/3528233.3530757}}.

\bibitem{dhariwal2021diffusion}
Prafulla Dhariwal and Alexander~Quinn Nichol.
\newblock {Diffusion Models Beat GANs on Image Synthesis}.
\newblock In {\em NeurIPS}, 2021.
\newblock \href {https://arxiv.org/abs/2105.05233} {\path{arXiv:2105.05233}}.

\bibitem{kong2020expressive}
Zhifeng Kong and Kamalika Chaudhuri.
\newblock {The Expressive Power of a Class of Normalizing Flow Models}, 2020.
\newblock \href {https://arxiv.org/abs/2006.00392} {\path{arXiv:2006.00392}}.

\bibitem{kodiramanah2020}
Doogesh Kodi~Ramanah, Tom Charnock, Francisco Villaescusa-Navarro, and Benjamin~D Wandelt.
\newblock {Super-resolution emulator of cosmological simulations using deep physical models}.
\newblock {\em Mon. Not. R. Astron. Soc.}, 495(4):4227--4236, 2020.
\newblock \href {https://doi.org/10.1093/mnras/staa1428} {\path{doi:10.1093/mnras/staa1428}}.

\bibitem{li2021_1}
Yin Li, Yueying Ni, Rupert A.~C. Croft, Tiziana~Di Matteo, Simeon Bird, and Yu~Feng.
\newblock Ai-assisted superresolution cosmological simulations.
\newblock {\em Proc. Natl. Acad. Sci. U.S.A.}, 118(19):e2022038118, 2021.
\newblock \href {https://doi.org/10.1073/pnas.2022038118} {\path{doi:10.1073/pnas.2022038118}}.

\bibitem{li2021_2}
Yueying Ni, Yin Li, Patrick Lachance, Rupert A.~C. Croft, Tiziana Di~Matteo, Simeon Bird, and Yu~Feng.
\newblock {AI-assisted superresolution cosmological simulations – II. Halo substructures, velocities, and higher order statistics}.
\newblock {\em Mon. Not. R. Astron. Soc.}, 507(1):1021--1033, 2021.
\newblock \href {https://doi.org/10.1093/mnras/stab2113} {\path{doi:10.1093/mnras/stab2113}}.

\bibitem{li2023}
Xiaowen Zhang, Patrick Lachance, Yueying Ni, Yin Li, Rupert A.~C. Croft, Tiziana~Di Matteo, Simeon Bird, and Yu~Feng.
\newblock {AI-assisted super-resolution cosmological simulations III: Time evolution}, 2023.
\newblock \href {https://arxiv.org/abs/2305.12222} {\path{arXiv:2305.12222}}.

\bibitem{schanz2023stochastic}
Andreas Schanz, Florian List, and Oliver Hahn.
\newblock {Stochastic Super-resolution of Cosmological Simulations with Denoising Diffusion Models}, 2023.
\newblock \href {https://arxiv.org/abs/2310.06929} {\path{arXiv:2310.06929}}.

\bibitem{Mudur:2022gfq}
Nayantara Mudur and Douglas~P. Finkbeiner.
\newblock {Can denoising diffusion probabilistic models generate realistic astrophysical fields?}
\newblock In {\em {NeurIPS}}, 2022.
\newblock \href {https://arxiv.org/abs/2211.12444} {\path{arXiv:2211.12444}}.

\bibitem{Zhao:2023giv}
Xiaosheng Zhao, Yuan-Sen Ting, Kangning Diao, and Yi~Mao.
\newblock {Can diffusion model conditionally generate astrophysical images?}
\newblock {\em Mon. Not. R. Astron. Soc.}, 526(2):1699--1712, 2023.
\newblock \href {https://doi.org/10.1093/mnras/stad2778} {\path{doi:10.1093/mnras/stad2778}}.

\bibitem{Karchev:2022ycy}
Konstantin Karchev, Noemi Anau~Montel, Adam Coogan, and Christoph Weniger.
\newblock {Strong-Lensing Source Reconstruction with Denoising Diffusion Restoration Models}.
\newblock In {\em {NeurIPS}}, 2022.
\newblock \href {https://arxiv.org/abs/2211.04365} {\path{arXiv:2211.04365}}.

\bibitem{10.1093/mnrasl/slad152}
Ronan Legin, Matthew Ho, Pablo Lemos, Laurence Perreault-Levasseur, Shirley Ho, Yashar Hezaveh, and Benjamin Wandelt.
\newblock {Posterior Sampling of the Initial Conditions of the Universe from Non-linear Large Scale Structures using Score-Based Generative Models}.
\newblock {\em Mon. Not. R. Astron. Soc. Lett.}, 527(1):L173--L178, 2023.
\newblock \href {https://doi.org/10.1093/mnrasl/slad152} {\path{doi:10.1093/mnrasl/slad152}}.

\bibitem{Leigh:2023toe}
Matthew Leigh, Debajyoti Sengupta, Guillaume Qu\'etant, John~Andrew Raine, Knut Zoch, and Tobias Golling.
\newblock {PC-JeDi: Diffusion for Particle Cloud Generation in High Energy Physics}.
\newblock 2023.
\newblock \href {https://arxiv.org/abs/2303.05376} {\path{arXiv:2303.05376}}.

\bibitem{Mikuni:2023dvk}
Vinicius Mikuni, Benjamin Nachman, and Mariel Pettee.
\newblock {Fast Point Cloud Generation with Diffusion Models in High Energy Physics}.
\newblock 2023.
\newblock \href {https://arxiv.org/abs/2304.01266} {\path{arXiv:2304.01266}}.

\bibitem{Shmakov:2023kjj}
Alexander Shmakov, Kevin Greif, Michael Fenton, Aishik Ghosh, Pierre Baldi, and Daniel Whiteson.
\newblock {End-To-End Latent Variational Diffusion Models for Inverse Problems in High Energy Physics}.
\newblock 2023.
\newblock \href {https://arxiv.org/abs/2305.10399} {\path{arXiv:2305.10399}}.

\bibitem{Mikuni:2023tqg}
Vinicius Mikuni and Benjamin Nachman.
\newblock {CaloScore v2: Single-shot Calorimeter Shower Simulation with Diffusion Models}.
\newblock 2023.
\newblock \href {https://arxiv.org/abs/2308.03847} {\path{arXiv:2308.03847}}.

\bibitem{Imani:2023blb}
Zeviel Imani, Shuchin Aeron, and Taritree Wongjirad.
\newblock {Score-based Diffusion Models for Generating Liquid Argon Time Projection Chamber Images}.
\newblock 2023.
\newblock \href {https://arxiv.org/abs/2307.13687} {\path{arXiv:2307.13687}}.

\bibitem{song2022denoising}
Jiaming Song, Chenlin Meng, and Stefano Ermon.
\newblock {Denoising Diffusion Implicit Models}.
\newblock In {\em ICLR}, 2021.
\newblock \href {https://arxiv.org/abs/2010.02502} {\path{arXiv:2010.02502}}.

\bibitem{lu2022dpmsolver}
Cheng Lu, Yuhao Zhou, Fan Bao, Jianfei Chen, Chongxuan Li, and Jun Zhu.
\newblock {DPM-Solver: A Fast ODE Solver for Diffusion Probabilistic Model Sampling in Around 10 Steps}.
\newblock In {\em NeurIPS}, 2022.
\newblock \href {https://arxiv.org/abs/2206.00927} {\path{arXiv:2206.00927}}.

\bibitem{improved_diffusion}
Alexander~Quinn Nichol and Prafulla Dhariwal.
\newblock {Improved Denoising Diffusion Probabilistic Models}.
\newblock In {\em ICML}, 2021.
\newblock \href {https://arxiv.org/abs/2102.09672} {\path{arXiv:2102.09672}}.

\bibitem{palette_github}
Liangwei Jiang.
\newblock {Palette: Image-to-Image Diffusion Models by Pytorch}.
\newblock \url{https://github.com/Janspiry/Palette-Image-to-Image-Diffusion-Models}, 2022.

\bibitem{brock2018large}
Andrew Brock, Jeff Donahue, and Karen Simonyan.
\newblock {Large Scale GAN Training for High Fidelity Natural Image Synthesis}.
\newblock In {\em ICLR}, 2019.
\newblock \href {https://arxiv.org/abs/1809.11096} {\path{arXiv:1809.11096}}.

\bibitem{vaswani2017}
Ashish Vaswani, Noam Shazeer, Niki Parmar, Jakob Uszkoreit, Llion Jones, Aidan~N. Gomez, Lukasz Kaiser, and Illia Polosukhin.
\newblock {Attention Is All You Need}.
\newblock In {\em NeurIPS}, 2017.
\newblock \href {https://arxiv.org/abs/1706.03762} {\path{arXiv:1706.03762}}.

\bibitem{sr3}
Chitwan Saharia, Jonathan Ho, William Chan, Tim Salimans, David~J. Fleet, and Mohammad Norouzi.
\newblock {Image Super-Resolution via Iterative Refinement}.
\newblock {\em IEEE Transactions on Pattern Analysis and Machine Intelligence}, 45(4):4713--4726, 2023.
\newblock \href {https://doi.org/10.1109/TPAMI.2022.3204461} {\path{doi:10.1109/TPAMI.2022.3204461}}.

\bibitem{Wu2020}
Yuxin Wu and Kaiming He.
\newblock {Group Normalization}.
\newblock {\em Int. J. Comput. Vis.}, 128(3):742--755, 2020.
\newblock \href {https://doi.org/10.1007/s11263-019-01198-w} {\path{doi:10.1007/s11263-019-01198-w}}.

\bibitem{Elfwing2018}
Stefan Elfwing, Eiji Uchibe, and Kenji Doya.
\newblock {Sigmoid-weighted linear units for neural network function approximation in reinforcement learning}.
\newblock {\em Neural Networks}, 107:3--11, 2018.
\newblock Special issue on deep reinforcement learning.
\newblock \href {https://doi.org/10.1016/j.neunet.2017.12.012} {\path{doi:10.1016/j.neunet.2017.12.012}}.

\bibitem{Banerjee_2016}
Arka Banerjee and Neal Dalal.
\newblock {Simulating nonlinear cosmological structure formation with massive neutrinos}.
\newblock {\em J. Cosmol. Astropart. Phys.}, 2016(11):015, 2016.
\newblock \href {https://doi.org/10.1088/1475-7516/2016/11/015} {\path{doi:10.1088/1475-7516/2016/11/015}}.

\bibitem{kvasiuk2023autodifferentiable}
Yurii Kvasiuk and Moritz Münchmeyer.
\newblock {An Auto-Differentiable Likelihood Pipeline for the Cross-Correlation of CMB and Large-Scale Structure due to the Kinetic Sunyaev-Zeldovich Effect}, 2023.
\newblock \href {https://arxiv.org/abs/2305.08903} {\path{arXiv:2305.08903}}.

\bibitem{smith2018ksz}
Kendrick~M. Smith, Mathew~S. Madhavacheril, Moritz Münchmeyer, Simone Ferraro, Utkarsh Giri, and Matthew~C. Johnson.
\newblock {KSZ tomography and the bispectrum}, 2018.
\newblock \href {https://arxiv.org/abs/1810.13423} {\path{arXiv:1810.13423}}.

\bibitem{andrianomena2023invertible}
Sambatra Andrianomena, Sultan Hassan, and Francisco Villaescusa-Navarro.
\newblock {Invertible mapping between fields in CAMELS}, 2023.
\newblock \href {https://arxiv.org/abs/2303.07473} {\path{arXiv:2303.07473}}.

\bibitem{luo2021diffusion}
Shitong Luo and Wei Hu.
\newblock {Diffusion Probabilistic Models for 3D Point Cloud Generation}.
\newblock In {\em CVPR}, 2021.
\newblock \href {https://arxiv.org/abs/2103.01458} {\path{arXiv:2103.01458}}.

\bibitem{hu2022lora}
Edward~J Hu, Yelong Shen, Phillip Wallis, Zeyuan Allen-Zhu, Yuanzhi Li, Shean Wang, Lu~Wang, and Weizhu Chen.
\newblock {LoRA: Low-Rank Adaptation of Large Language Models}.
\newblock In {\em ICLR}, 2022.
\newblock \href {https://arxiv.org/abs/2106.09685} {\path{arXiv:2106.09685}}.

\bibitem{gu2023mixofshow}
Yuchao Gu, Xintao Wang, Jay~Zhangjie Wu, Yujun Shi, Yunpeng Chen, Zihan Fan, Wuyou Xiao, Rui Zhao, Shuning Chang, Weijia Wu, Yixiao Ge, Ying Shan, and Mike~Zheng Shou.
\newblock {Mix-of-Show: Decentralized Low-Rank Adaptation for Multi-Concept Customization of Diffusion Models}.
\newblock In {\em NeurIPS}, 2023.
\newblock \href {https://arxiv.org/abs/2305.18292} {\path{arXiv:2305.18292}}.

\bibitem{imagenet}
Jia Deng, Wei Dong, Richard Socher, Li-Jia Li, Kai Li, and Li~Fei-Fei.
\newblock {ImageNet: A large-scale hierarchical image database}.
\newblock In {\em CVPR}, pages 248--255, 2009.
\newblock \href {https://doi.org/10.1109/CVPR.2009.5206848} {\path{doi:10.1109/CVPR.2009.5206848}}.

\bibitem{kirichenko2020normalizing}
Polina Kirichenko, Pavel Izmailov, and Andrew~Gordon Wilson.
\newblock {Why Normalizing Flows Fail to Detect Out-of-Distribution Data}.
\newblock In {\em NeurIPS}, 2020.
\newblock \href {https://arxiv.org/abs/2006.08545} {\path{arXiv:2006.08545}}.

\bibitem{deSanti:2023zzn}
Natal\'\i{} S.~M. de~Santi et~al.
\newblock {Robust Field-level Likelihood-free Inference with Galaxies}.
\newblock {\em Astrophys. J.}, 952(1):69, 2023.
\newblock \href {https://doi.org/10.3847/1538-4357/acd1e2} {\path{doi:10.3847/1538-4357/acd1e2}}.

\bibitem{LSST:2008ijt}
\v{Z}eljko Ivezi\'c et~al.
\newblock {LSST: from Science Drivers to Reference Design and Anticipated Data Products}.
\newblock {\em Astrophys. J.}, 873(2):111, 2019.
\newblock \href {https://doi.org/10.3847/1538-4357/ab042c} {\path{doi:10.3847/1538-4357/ab042c}}.

\bibitem{Peterson_2015}
J.~R. Peterson, J.~G. Jernigan, S.~M. Kahn, A.~P. Rasmussen, E.~Peng, Z.~Ahmad, J.~Bankert, C.~Chang, C.~Claver, D.~K. Gilmore, E.~Grace, M.~Hannel, M.~Hodge, S.~Lorenz, A.~Lupu, A.~Meert, S.~Nagarajan, N.~Todd, A.~Winans, and M.~Young.
\newblock Simulation of astronomical images from optical survey telescopes using a comprehensive photon monte carlo approach.
\newblock {\em The Astrophysical Journal Supplement Series}, 218(1):14, 2015.
\newblock \href {https://doi.org/10.1088/0067-0049/218/1/14} {\path{doi:10.1088/0067-0049/218/1/14}}.

\bibitem{nichol2022glide}
Alex Nichol, Prafulla Dhariwal, Aditya Ramesh, Pranav Shyam, Pamela Mishkin, Bob McGrew, Ilya Sutskever, and Mark Chen.
\newblock {GLIDE: Towards Photorealistic Image Generation and Editing with Text-Guided Diffusion Models}.
\newblock In {\em ICML}, 2022.
\newblock \href {https://arxiv.org/abs/2112.10741} {\path{arXiv:2112.10741}}.

\bibitem{saharia2022photorealistic}
Chitwan Saharia, William Chan, Saurabh Saxena, Lala Li, Jay Whang, Emily Denton, Seyed Kamyar~Seyed Ghasemipour, Burcu~Karagol Ayan, S.~Sara Mahdavi, Rapha~Gontijo Lopes, Tim Salimans, Jonathan Ho, David~J Fleet, and Mohammad Norouzi.
\newblock {Photorealistic Text-to-Image Diffusion Models with Deep Language Understanding}.
\newblock In {\em NeurIPS}, 2022.
\newblock \href {https://arxiv.org/abs/2205.11487} {\path{arXiv:2205.11487}}.

\bibitem{PointFlow}
G.~Yang, X.~Huang, Z.~Hao, M.~Liu, S.~Belongie, and B.~Hariharan.
\newblock {PointFlow: 3D Point Cloud Generation With Continuous Normalizing Flows}.
\newblock In {\em ICCV}, pages 4540--4549, 2019.
\newblock \href {https://doi.org/10.1109/ICCV.2019.00464} {\path{doi:10.1109/ICCV.2019.00464}}.

\bibitem{battaglia2018relational}
Peter~W. Battaglia et~al.
\newblock {Relational inductive biases, deep learning, and graph networks}, 2018.
\newblock \href {https://arxiv.org/abs/1806.01261} {\path{arXiv:1806.01261}}.

\bibitem{gadelha2018multiresolution}
Matheus Gadelha, Rui Wang, and Subhransu Maji.
\newblock {Multiresolution Tree Networks for 3D Point Cloud Processing}.
\newblock In {\em Computer Vision -- ECCV 2018}, pages 105--122. Springer International Publishing, 2018.
\newblock \href {https://doi.org/10.1007/978-3-030-01234-2_7} {\path{doi:10.1007/978-3-030-01234-2_7}}.

\bibitem{Cuesta-Lazaro:2023zuk}
Carolina Cuesta-Lazaro and Siddharth Mishra-Sharma.
\newblock {A point cloud approach to generative modeling for galaxy surveys at the field level}.
\newblock 2023.
\newblock \href {https://arxiv.org/abs/2311.17141} {\path{arXiv:2311.17141}}.

\bibitem{zhang2021diffusion}
Qinsheng Zhang and Yongxin Chen.
\newblock {Diffusion Normalizing Flow}.
\newblock In {\em NeurIPS}, 2021.
\newblock \href {https://arxiv.org/abs/2110.07579} {\path{arXiv:2110.07579}}.

\bibitem{zhang2023diffflow}
Jingwei Zhang, Han Shi, Jincheng Yu, Enze Xie, and Zhenguo Li.
\newblock {DiffFlow: A Unified SDE Framework for Score-Based Diffusion Models and Generative Adversarial Networks}, 2023.
\newblock \href {https://arxiv.org/abs/2307.02159} {\path{arXiv:2307.02159}}.

\bibitem{xu2022poisson}
Yilun Xu, Ziming Liu, Max Tegmark, and Tommi Jaakkola.
\newblock {Poisson Flow Generative Models}.
\newblock In {\em NeurIPS}, 2022.
\newblock \href {https://arxiv.org/abs/2209.11178} {\path{arXiv:2209.11178}}.

\bibitem{Kazantzidis2021}
Lavrentios Kazantzidis and Leandros Perivolaropoulos.
\newblock {${\sigma}_{8}$ Tension. Is Gravity Getting Weaker at Low z? Observational Evidence and Theoretical Implications}.
\newblock In {\em Modified Gravity and Cosmology}, page 507–537. Springer International Publishing, 2021.
\newblock \href {https://doi.org/10.1007/978-3-030-83715-0_33} {\path{doi:10.1007/978-3-030-83715-0_33}}.

\bibitem{Vivian:2023}
Vivian Poulin, Jos\'e~Luis Bernal, Ely~D. Kovetz, and Marc Kamionkowski.
\newblock Sigma-8 tension is a drag.
\newblock {\em Phys. Rev. D}, 107:123538, 2023.
\newblock \href {https://doi.org/10.1103/PhysRevD.107.123538} {\path{doi:10.1103/PhysRevD.107.123538}}.

\bibitem{DiValentino:2021izs}
Eleonora Di~Valentino, Olga Mena, Supriya Pan, Luca Visinelli, Weiqiang Yang, Alessandro Melchiorri, David~F. Mota, Adam~G. Riess, and Joseph Silk.
\newblock {In the realm of the Hubble tension\textemdash{}a review of solutions}.
\newblock {\em Class. Quant. Grav.}, 38(15):153001, 2021.
\newblock \href {https://doi.org/10.1088/1361-6382/ac086d} {\path{doi:10.1088/1361-6382/ac086d}}.

\bibitem{DiValentino:2020vnx}
Eleonora Di~Valentino.
\newblock {A combined analysis of the $H_0$ late time direct measurements and the impact on the Dark Energy sector}.
\newblock {\em Mon. Not. R. Astron. Soc.}, 502(2):2065--2073, 2021.
\newblock \href {https://arxiv.org/abs/2011.00246} {\path{arXiv:2011.00246}}, \href {https://doi.org/10.1093/mnras/stab187} {\path{doi:10.1093/mnras/stab187}}.

\bibitem{Gayathri:2020mra}
V.~Gayathri, J.~Healy, J.~Lange, B.~O'Brien, M.~Szczepanczyk, I.~Bartos, M.~Campanelli, S.~Klimenko, C.~O. Lousto, and R.~O'Shaughnessy.
\newblock {Measuring the Hubble Constant with GW190521 as an Eccentric black hole Merger and Its Potential Electromagnetic Counterpart}.
\newblock {\em Astrophys. J. Lett.}, 908(2):L34, 2021.
\newblock \href {https://arxiv.org/abs/2009.14247} {\path{arXiv:2009.14247}}, \href {https://doi.org/10.3847/2041-8213/abe388} {\path{doi:10.3847/2041-8213/abe388}}.

\bibitem{Hu:2007nk}
Wayne Hu and Ignacy Sawicki.
\newblock {Models of $f(R)$ Cosmic Acceleration that Evade Solar-System Tests}.
\newblock {\em Phys. Rev. D}, 76:064004, 2007.
\newblock \href {https://arxiv.org/abs/0705.1158} {\path{arXiv:0705.1158}}, \href {https://doi.org/10.1103/PhysRevD.76.064004} {\path{doi:10.1103/PhysRevD.76.064004}}.

\bibitem{Whittle:1954}
Peter Whittle.
\newblock {On Stationary Processes in the Plane}.
\newblock {\em Biometrika}, 41(3/4):434--449, 1954.
\newblock \href {https://doi.org/10.2307/2332724} {\path{doi:10.2307/2332724}}.

\bibitem{Fergusson:2010ia}
J.~R. Fergusson, D.~M. Regan, and E.~P.~S. Shellard.
\newblock {Rapid Separable Analysis of Higher Order Correlators in Large Scale Structure}.
\newblock {\em Phys. Rev. D}, 86:063511, 2012.
\newblock \href {https://doi.org/10.1103/PhysRevD.86.063511} {\path{doi:10.1103/PhysRevD.86.063511}}.

\bibitem{Babich:2005en}
Daniel Babich.
\newblock {Optimal estimation of non-Gaussianity}.
\newblock {\em Phys. Rev. D}, 72:043003, 2005.
\newblock \href {https://doi.org/10.1103/PhysRevD.72.043003} {\path{doi:10.1103/PhysRevD.72.043003}}.

\bibitem{tan2020efficientnet}
Mingxing Tan and Quoc~V. Le.
\newblock {EfficientNet: Rethinking Model Scaling for Convolutional Neural Networks}.
\newblock In {\em ICML}, 2019.
\newblock \href {https://arxiv.org/abs/1905.11946} {\path{arXiv:1905.11946}}.

\bibitem{Kunihiko:1969}
Kunihiko Fukushima.
\newblock {Visual Feature Extraction by a Multilayered Network of Analog Threshold Elements}.
\newblock {\em IEEE Transactions on Systems Science and Cybernetics}, 5(4):322--333, 1969.
\newblock \href {https://doi.org/10.1109/TSSC.1969.300225} {\path{doi:10.1109/TSSC.1969.300225}}.

\bibitem{Maas2013RectifierNI}
Andrew L., Awni~Y. Hannun, and Andrew~Y. Ng.
\newblock Rectifier nonlinearities improve neural network acousic models.
\newblock 2013.
\newblock URL: \url{https://ai.stanford.edu/~amaas/papers/relu_hybrid_icml2013_final.pdf}.

\bibitem{ioffe2015batch}
Sergey Ioffe and Christian Szegedy.
\newblock {Batch Normalization: Accelerating Deep Network Training by Reducing Internal Covariate Shift}.
\newblock In {\em ICML}, 2015.
\newblock \href {https://arxiv.org/abs/1502.03167} {\path{arXiv:1502.03167}}.

\bibitem{hinton2012improving}
Geoffrey~E. Hinton, Nitish Srivastava, Alex Krizhevsky, Ilya Sutskever, and Ruslan~R. Salakhutdinov.
\newblock {Improving neural networks by preventing co-adaptation of feature detectors}, 2012.
\newblock \href {https://arxiv.org/abs/1207.0580} {\path{arXiv:1207.0580}}.

\bibitem{JMLR:v15:srivastava14a}
Nitish Srivastava, Geoffrey Hinton, Alex Krizhevsky, Ilya Sutskever, and Ruslan Salakhutdinov.
\newblock {Dropout: A Simple Way to Prevent Neural Networks from Overfitting}.
\newblock {\em J. Mach. Learn. Res.}, 15(56):1929--1958, 2014.
\newblock URL: \url{http://jmlr.org/papers/v15/srivastava14a.html}.

\bibitem{SGD1999}
L\'{e}on Bottou.
\newblock {On-Line Learning and Stochastic Approximations}.
\newblock In David Saad, editor, {\em On-Line Learning in Neural Networks}, Publications of the Newton Institute, page 9–42. Cambridge University Press, 1999.
\newblock \href {https://doi.org/10.1017/CBO9780511569920.003} {\path{doi:10.1017/CBO9780511569920.003}}.

\bibitem{Kingma2014AdamAM}
Diederik~P. Kingma and Jimmy Ba.
\newblock {Adam: A Method for Stochastic Optimization}.
\newblock In {\em ICLR}, 2014.
\newblock \href {https://arxiv.org/abs/1412.6980} {\path{arXiv:1412.6980}}.

\end{thebibliography}

\end{document}